\PassOptionsToPackage{dvipsnames}{xcolor}
\documentclass[12pt,titlepage,oneside]{book}
\usepackage[T1]{fontenc}
\usepackage[english]{babel}
\usepackage[top=1.8cm, bottom=1.8cm,left=4.0cm,right=1.5cm]{geometry}

\usepackage{libertine}
\usepackage{amsmath,amssymb,amsfonts}
\usepackage{FiraMono}

\usepackage[babel]{csquotes}
\usepackage[
backend=biber,
sorting=none,
block=ragged
]{biblatex}
\usepackage{graphicx}
\graphicspath{{./media/}}
\PassOptionsToPackage{hyphens}{url}
\usepackage[breaklinks]{hyperref}
\usepackage{caption}
\usepackage{subcaption}
\usepackage{enumitem}
\usepackage{listings}
\usepackage{xcolor}
\usepackage{colortbl}
\usepackage[noabbrev, capitalise]{cleveref}
\usepackage{bm}
\usepackage{comment}
\usepackage{wrapfig}
\usepackage{multirow}
\usepackage{pifont} 
\usepackage[onehalfspacing]{setspace}  
\usepackage{quiver}
\usepackage{stmaryrd}
\usepackage{mathtools}

\usepackage{tikz}
\newcommand\tikzmark[1]{\tikz[remember picture,overlay]\node[below] (#1) {};}
\usetikzlibrary{tikzmark, positioning, arrows, calc}

\crefname{lstlisting}{listing}{listings}
\Crefname{lstlisting}{Listing}{Listings}

\definecolor{halide}{HTML}{882255}
\definecolor{elevate}{HTML}{332288}
\definecolor{scala}{HTML}{13729c}
\definecolor{rise}{HTML}{117733}
\definecolor{DPIA}{HTML}{c23210}
\definecolor{sketch}{HTML}{1c4ca6}
\definecolor{highlight}{HTML}{c7143e}
\definecolor{pseudo-text}{HTML}{666666}

\lstdefinestyle{halide}{
  language=C,
  morekeywords=[2]{Func, Var, Expr, split, parallel, vectorize, store_at, compute_at, compute_with},
  keywordstyle=[2]{\bfseries\color{halide}}
}
\lstnewenvironment{halide}[1][]{\lstset{style=halide, #1}}{}
\newcommand{\inlineHalide}{\lstinline[style=halide]}

\lstdefinestyle{elevate-rise}{
  morekeywords = [0]{def, type, enum, record, case},
  keywordstyle= [0]{\bfseries\color{black}},
  morekeywords = [1]{Strategy, RewriteResult, Success, Failure, List},
  keywordstyle= [1]{\bfseries\color{scala}},
  otherkeywords = {;, <+, @},
  morekeywords= [2]{ 
    rule, strategy, where,
    ;, <+, @,
    id, fail,
    try, repeat,
    topDown, bottomUp, normalize, repeatNTimes
  },
  keywordstyle= [2]{\bfseries\color{elevate}},
  morekeywords= [5]{
    nat, data, addr
  },
  keywordstyle= [5]{\bfseries\color{black}},
  morekeywords= [6]{ 
    let, app, fun,
    map, mapSeq, mapSeqUnroll,
    mapGlobal, mapWorkGroup, mapLocal, mapVec, mapStream,
    mapPar,
    toMem, oclRun,
    zip, fst, snd,
    reduce, reduceSeq, reduceSeqUnroll,
    asVector, asScalar, vectorFromScalar,
    padEmpty, take,
    slide, split, join, transpose,
    circularBuffer, rotateValues, iterateStream,
    global, local, private 
  },
  keywordstyle= [6]{\bfseries\color{rise}},
	basicstyle=\ttfamily\footnotesize,
	commentstyle=\itshape, 
	numbers=left, 
	breaklines=true, 
  mathescape=true,
  moredelim=[is][\bfseries\color{highlight}]{*}{*}
}
\lstnewenvironment{elevate-rise}[1][]{\lstset{style=elevate-rise, #1}}{}
\newcommand{\inlineElevateRise}{\lstinline[style=elevate-rise]}

\lstdefinestyle{DPIA}{
  morecomment = [l][\itshape\color{gray}]{|},
  morekeywords=[1]{
    exp, acc, comm, access, read, write, nat, addr, local, global, private, data,
    assign, seq, new, for, parFor, newRegRot, f32,
    idx, idxAcc, split, join, splitAcc, joinAcc,
    zip, fst, snd, zipAcc1, zipAcc2,
    mapSeq, reduceSeq, toMem,
    barrier,
    kernelCallCmd, hostExecution, newBuffer,
    circularBuffer, rotateValues, iterateStream, mapStream
  },
  keywordstyle=[1]{\bfseries\color{DPIA}},
  morekeywords=[2]{record, def, if, then, else, match},
  keywordstyle=[2]{\bfseries\color{black}},
  mathescape=true,
  moredelim=[is][\itshape\color{pseudo-text}]{~}{~},
}
\lstnewenvironment{DPIA}[1][]{\lstset{style=DPIA, #1}}{}
\newcommand{\inlineDPIA}{\lstinline[style=DPIA]}

\lstdefinestyle{rise}{
  morekeywords = {def, for},
  otherkeywords={|},
  commentstyle=\itshape\color{gray},
  morekeywords= [5]{
    nat, data, addr
  },
  keywordstyle= [5]{\bfseries\color{black}},
  morekeywords= [6]{ 
    let, app, 
    map, mapSeq, mapSeqUnroll,
    mapGlobal, mapWorkGroup, mapLocal, mapVec, mapStream,
    mapPar,
    toMem, oclRun, store, kernelCall,
    zip, fst, snd, generate, unzip,
    reduce, reduceSeq, reduceSeqUnroll,
    add, sub,
    asVector, asScalar, vectorFromScalar,
    padEmpty, padClamp, take,
    slide, split, join, transpose,
    circularBuffer, rotateValues, iterateStream,
    global, local, private 
  },
  keywordstyle= [6]{\bfseries\color{rise}},
	basicstyle=\ttfamily\footnotesize,
  commentstyle=\itshape, 
  numbers=left, 
  breaklines=true, 
  mathescape=true,
  moredelim=[is][\bfseries\color{highlight}]{*}{*},
  moredelim=[is][\color{gray}]{[}{]}
}
\lstnewenvironment{rise}[1][]{\lstset{style=rise, #1}}{}
\newcommand{\inlineRise}{\lstinline[style=rise]}

\lstdefinestyle{rise-sketch}{
  morekeywords = {def, for},
  otherkeywords={?, ::, |},
  morecomment = [l][\itshape\color{gray}]{|},
  commentstyle=\itshape\color{gray},
  morekeywords= [5]{ 
    ?, contains, ::,
    containsMap, containsMapPar,
    containsReduceSeq, containsReduceSeqUnroll,
    containsAddMul, containsAddMulVec,
    slide, circularBuffer,
    isSlide, isCircularBuffer,
    containsGrayLine, containsSobelLine, containsCoarsityLine,
    containsToMem,
  },
  keywordstyle= [5]{\bfseries\color{sketch}},
  morekeywords= [6]{ 
    app, 
    map, mapSeq, mapSeqUnroll,
    mapGlobal, mapVec, mapStream,
    toMem,
    zip, fst, snd, generate, unzip,
    reduce, reduceSeq, reduceSeqUnroll,
    asVector, asScalar, vectorFromScalar,
    padEmpty, padClamp, take,
    split, join, transpose,
    rotateValues, iterateStream,
    global, local, private 
  },
  keywordstyle= [6]{\bfseries\color{rise}},
	basicstyle=\ttfamily\footnotesize,
  commentstyle=\itshape, 
  numbers=left, 
  breaklines=true, 
  mathescape=true,
  moredelim=[is][\bfseries\color{highlight}]{*}{*},
}
\lstnewenvironment{rise-sketch}[1][]{\lstset{style=rise-sketch, #1}}{}
\newcommand{\inlineRiseSketch}{\lstinline[style=rise-sketch]}

\lstnewenvironment{c-code}[1][]{\lstset{
  language=C,
	basicstyle=\ttfamily\footnotesize,
	numbers=left, 
#1}}{}

\lstnewenvironment{opencl}[1][]{\lstset{
  language=C,
  morekeywords={kernel, global, float4},
	basicstyle=\ttfamily\footnotesize,
	numbers=left, 
#1}}{}
\newcommand{\inlineOpencl}{\lstinline[style=rise-sketch]}
\renewcommand{\mkbegdispquote}[2]{\textooquote}

\newcommand{\chapquote}[3]{\clearpage
~\\
\vfill
\begin{quotation}\itshape #1 \end{quotation}
\begin{flushright} --- #2\end{flushright}
\vfill
~\\
\clearpage}

\begin{document}

\setcounter{secnumdepth}{3}
\newcommand{\Lift}{\textsc{Lift}}
\newcommand{\Rise}{\textsc{Rise}}
\newcommand{\Shine}{\textsc{Shine}}
\newcommand{\Elevate}{\textsc{Elevate}}
\newcommand{\HIPAcc}{HIPA\textsuperscript{CC}}
\newcommand{\kles}{\textsc{Risegg}}

\newcommand{\then}{\vartriangleright}
\newcommand{\rewritesTo}{{\color{elevate}\bm{\mapsto}}}
\newcommand{\rewritesToM}{\longmapsto}

\newcommand{\fun}[1]{\bm{\mathsf{fun}}~#1.}
\newcommand{\arr}[2]{#1.#2}
\newcommand{\stream}[2]{\left[#1\right]#2}
\newcommand{\vecT}[2]{\left<#1\right>#2}
\newcommand{\tup}[2]{(#1 \times #2)}

\newcommand{\float}{\mathsf{f32}}

\newcommand{\nat}{\mathsf{nat}}
\newcommand{\data}{\mathsf{data}}
\newcommand{\addr}{\mathsf{addr}}

\newcommand{\prim}[1]{\bm{\mathsf{\color{rise}#1}}}

\newcommand{\code}[1]{{\small\texttt{#1}}}

\newcommand{\yes}{\ding{51}}
\newcommand{\no}{\ding{55}}

\newcommand{\goalStyle}[1]{\textit{\textsf{#1}}}


\begin{titlepage}
\centering
\vspace*{3cm}  
\bfseries\Large
A Domain-Extensible Compiler with Controllable Automation of Optimisations\\
\vspace{3cm}
\normalfont\large
Thomas K{\oe}hler\\
\vspace{2cm}
Submitted in fulfilment of the requirements for the\\
Degree of Doctor of Philosophy\\
\vspace{2cm}
School of Computing Science\\
College of Science and Engineering\\
University of Glasgow\\
\vspace{1cm}
\includegraphics[scale=0.125]{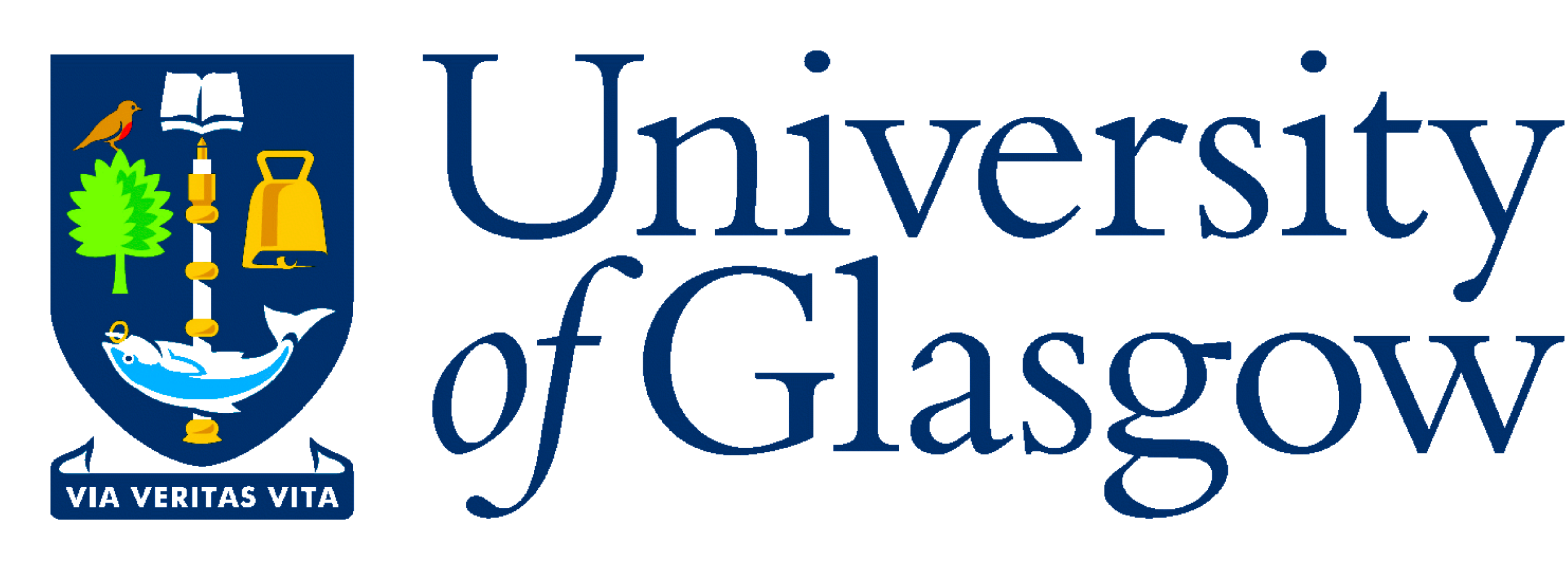}
\\
\vspace{1cm}
2022
\end{titlepage}
\frontmatter  
\chapter{Abstract}

In high performance domains like image processing, physics simulation or machine learning, program performance is critical.
Programmers called \emph{performance engineers} are responsible for the challenging task of optimising programs.
Two major challenges prevent modern compilers targeting heterogeneous architectures from reliably automating optimisation.
First, domain-specific compilers such as Halide for image processing and TVM for machine learning are difficult to
extend with the new optimisations required by new algorithms and hardware.
Second, automatic optimisation is often unable to achieve the required performance, and performance engineers often fall back to painstaking manual optimisation.

This thesis shows the potential of the \Shine{} compiler to achieve domain-extensibility, controllable automation, and generate high performance code.
\emph{Domain-extensibility} facilitates adapting compilers to new algorithms and hardware.
\emph{Controllable automation} enables performance engineers to gradually take control of the optimisation process.

The first research contribution is to add 3 code generation features to \Shine{}, namely: synchronisation barrier insertion, kernel execution, and storage folding.
Adding these features requires making novel design choices in terms of compiler extensibility and controllability.
The rest of this thesis builds on these features to generate code with competitive runtime compared to established domain-specific compilers.

The second research contribution is to demonstrate how extensibility and controllability are exploited to optimise a standard image processing pipeline for corner detection.
\Shine{} achieves 6 well-known image processing optimisations, 2 of them not being supported by Halide.
Our results on 4 ARM multi-core CPUs show that the code generated by \Shine{} for corner detection runs up to 1.4$\times$ faster than the Halide code.
However, we observe that controlling rewriting is tedious, motivating the need for more automation.

The final research contribution is to introduce \emph{sketch-guided equality saturation}, a semi-automated technique that allows performance engineers to guide program rewriting by specifying rewrite goals as sketches: program patterns that leave details unspecified.
We evaluate this approach by applying 7 realistic optimisations of matrix multiplication.
Without guidance, the compiler fails to apply the 5 most complex optimisations even given an hour and 60GB of RAM.
With the guidance of at most 3 sketch guides, each 10 times smaller than the complete program, the compiler applies the optimisations in seconds using less than 1GB.

\setcounter{tocdepth}{1}
\tableofcontents
\chapter{Acknowledgements}


This thesis is the result of 3 short years of research, and 6 long months of write-up.
I am grateful to those who helped me throughout this journey.

Thanks to my supervisors, Michel Steuwer and Phil Trinder, for their encouragements, guidance, and constructive feedback.
Michel, you always cared about my personal growth, I owe you for attending many interesting events and learning to confidently promote my work.
Phil, your helicopter view helped me to focus on the bigger picture, strengthened my research and writing.
I am still using the mug from your last supply drop.

Thanks to everybody in the School of Computing Science in Glasgow for creating such a welcoming environment.
The pandemic made me realise how much I missed the social lunches, coffee breaks and pub nights.
I had the pleasure of many refreshing interactions with Adrian Ramsingh, Dejice Jacob, Cristian Urlea and Kyle Simpson in our shared F101 office.

Thanks to the Halide developers and egg developers for maintaining open-source code, to Max Willsey for sharing his insights on equality saturation.

Thanks to the entire \Lift{}, \Rise{} and \Elevate{} teams for their work and our shared moments.
Bastian Köpcke and Federico Pizzuti, you developed the \Shine{} compiler at my side, and shared many pre-pandemic whiteboard sessions.
Christof Schlaak, you unexpectedly shared the post-pandemic write-up experience with me (or was it a table tennis competition?).

Above all, thanks to my loved ones, who gave me the strength to withstand this journey.
Gözel Shakeri, my partner, you gave meaning to work-life balance, celebrated my successes, and relativized my failures.
Madeleine N{\oe}uveglise and François K{\oe}hler, my parents, you supported me unconditionally as I moved far away from home.
My friends, stay who you are.
My levain, may you live a long and sour life.

\chapter{Declaration}


The work reported in this thesis is primarily my own unless otherwise explicitly stated, and has not been submitted for any other degree or qualification.
This thesis unifies and expands work reported in the following publications:
\preto\fullcite{\AtNextCite{\defcounter{maxnames}{99}}}
\begin{itemize}
    \item[\cite{koehler2021-elevate-imgproc}] \fullcite{koehler2021-elevate-imgproc}
    \item[\cite{koehler2021-sketch-guided}] \fullcite{koehler2021-sketch-guided}
\end{itemize}
During the work on this thesis, I also co-authored the following publications:
\begin{itemize}
    \item[\cite{hagedorn2020-elevate}] \fullcite{hagedorn2020-elevate}
    \item[\cite{steuwer2022-rise-shine}] \fullcite{steuwer2022-rise-shine}
    \item[\cite{woodruff2022-cgra}] \fullcite{woodruff2022-cgra}
\end{itemize}


\mainmatter 

\chapter{Introduction}
\label{ch:introduction}
\begin{figure}
    \centering
    \includegraphics[width=0.9\linewidth]{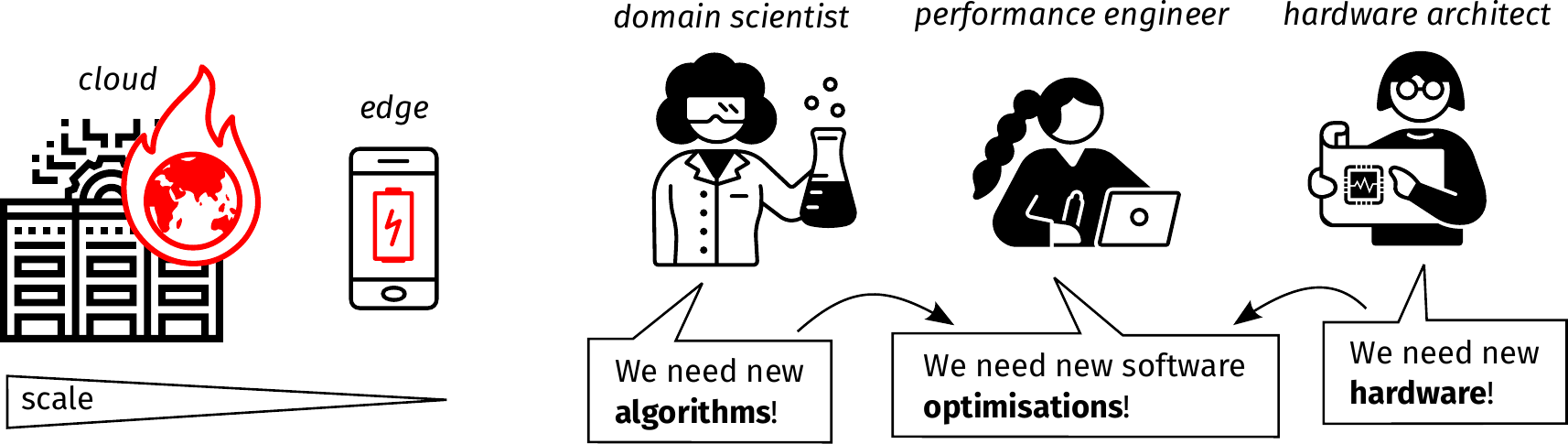}
    \caption{(Left-to-Right) Physical resources are limited at all scales: from the impact of the cloud on our planet, to battery limitations at the edge. Different actors optimise resource usage: while domain scientists design new algorithms, hardware architects design new hardware, and performance engineers design new program optimisations.}
    \label{fig:alg-opt-hard}
\end{figure}

Breakthroughs in artificial intelligence keep making our machines smarter: they become better at diagnosing diseases, driving vehicles, predicting the weather, understanding natural languages, and more.
Part of this success is due to machine learning models like artificial neural networks that have increasingly more neural connections (aka parameters).
Networks went from millions of parameters (e.g. ResNet) to billions of parameters (e.g. Microsoft Turing), and are anticipated to reach trillions of parameters.
Such models require massive computing power, which is a major issue \cite{2020-green-ai} since physical resources are limited (left of \cref{fig:alg-opt-hard}).
In the cloud, huge data centres are taking an increasing toll on our planet: a recent study estimates that training a single natural language processing neural network has a carbon footprint equivalent to driving 4,311 km in a European car and would take 823 tree-months to offset \cite{2020-green-algorithms}.
At the edge, running advanced neural networks on small devices with low energy supply is not yet possible.
To sustain progress, we need to use resources more efficiently.

In high performance domains like image processing, physics simulation or machine learning, resources are used more efficiently through continuous algorithm, hardware and program optimisation (right of \cref{fig:alg-opt-hard}).
Here domain scientists are responsible for optimising algorithms, hardware architects optimise hardware, and \emph{performance engineers} are programmers responsible for optimising programs.
The three optimisation aspects are inter-dependent: in particular, for each new algorithm and hardware architecture, new specialised program optimisations are required.

Optimising programs is challenging.
On one side, skilled performance engineers perform optimisations manually on low-level code (e.g. C, OpenCL), which takes months and risks introducing bugs, slowing down algorithm and hardware innovation.
On the other side, modern compilers targeting heterogeneous and parallel hardware do not reliably automate program optimisation, facing the two following major challenges.



\begin{center}
\emph{Challenge 1) Extending compilers with new program optimisations is hard.}
\end{center}


Traditional \emph{domain-agnostic} compilers such as LLVM~\cite{llvm-2004} or GCC only support a fixed, generic set of program abstractions and optimisations.
As a result, they do not automate optimisations for specific domains and performance engineers must perform them manually (\cref{sec:hpc-programming}).

Established \emph{domain-specific} compilers such as Halide~\cite{halide-2012} for image processing or TVM~\cite{tvm-2018} for machine learning only support a fixed, specialised set of program abstractions and optimisations.
Although they successfully automate many optimisations, extending them with new optimisations as the domain evolves is difficult \cite{barham2019-ml-rut}.
Domain evolution includes changes in algorithms, changes in program optimisations, but also changes in specialised hardware (e.g. machine learning accelerators \cite{2017-ml-hardware}).
Performance engineers are not necessarily compiler writers, and falling back to manual optimisation is often easier for them than modifying a compiler to achieve their goals.
In consequence, many state-of-the-art high-performance kernel libraries (e.g., BLAS, cuDNN, MKL) are still predominantly written by hand \cite{2022-exo}.

Recently, \emph{domain-extensible} compilers such as Delite~\cite{delite-2011}, \Lift{}~\cite{lift-rewrite-2015, lift-stencil-2018}, or AnyDSL~\cite{anydsl-2018} provide an extensible set of program abstractions and optimisations, showing potential to mitigate \emph{Challenge 1} (\cref{sec:extending-compilers}).
However, domain-extensible compilers remain relatively immature and lack adoption compared to established domain-specific compilers.
For example, previous work does not investigate whether \Lift{} is able to generate code with competitive runtime performance compared to established domain-specific compilers.


\begin{center}
\emph{Challenge 2) Manual optimisation is tedious, automatic optimisation is unsatisfactory.}
\end{center}

Compilers such as \Lift{}~\cite{lift-rewrite-2015} or PolyMage~\cite{polymage-2015} aim for \emph{full automation} of optimisations, removing performance engineers from the optimisation process.
This is highly desirable in scenarios where performance engineers are not available.
However, full automation of optimisations is not always feasible or even desirable as it may result in poor performance or may be too time-consuming \cite{maleki2011-vectorizing-compilers, parello2004-pragmatic-opt, chambers2002-staged-compilation}.
When compiler optimisations are unsatisfactory, performance engineers often fall back to manual optimisation in order to achieve their performance goals \cite{niittylahti2002-high-perf, hlt-lacassagne-2014, lemaitre2017-cholesky}.


Compilers such as Halide offer \emph{control} of optimisations through schedules \cite{halide-2012, tvm-2018, fireiron-2020}.
However, schedules are challenging to write \cite{ikarashi2021-guided-scheduling}.
Successful research goes into automatically generating schedules \cite{halide-autosched-2016, halide-autosched-2019, anderson2021-autosched}, but sacrifices control for full automation of optimisations.

This thesis explores \emph{controllable automation} (\cref{controllable-automation}) of optimisations that embraces trade-offs between full automation and precise control of optimisations.
Controllable automation enables performance engineers to gradually take control, instead of abruptly falling back to manual optimisation when compiler optimisations are unsatisfactory, and has the potential to mitigate \emph{Challenge 2} (\cref{sec:controlling-compilers}).



\paragraph{Thesis Statement}
This thesis shows the potential of a novel compiler design to achieve domain-extensibility, controllable automation, and generate high performance code.
Domain-extensibility is combined with controlled rewriting in an image processing case study to generate faster code than the established domain-specific compiler Halide.
Controllable automation is exploited in a linear algebra case study to automatically explore an optimisation space while providing the performance engineer with control over the optimisation outcome.

~\\
This thesis makes the following research contributions:
\begin{enumerate}
    \item \textbf{Enhancing Code Generation in a Domain-Extensible Compiler} (\cref{ch:imperative-code}).\\
    Three important code generation features are added to a domain-extensible compiler with controllable automation of optimisations called \Shine{}, making it possible to:
    \begin{itemize}
        \item generate correct and efficient synchronisation barriers
        \item generate multiple computation kernels and the host code launching them
        \item generate folded storage for temporary arrays
    \end{itemize}
    Adding these features requires making novel design choices in terms of compiler extensibility and controllability.
    Crucially the following chapters require these features to generate imperative code with competitive runtime performance compared to established domain-specific compilers such as Halide.


    \item \textbf{Going Beyond Halide Scheduling with Controlled Rewriting} \cite{koehler2021-elevate-imgproc} (\cref{ch:imgproc}).\\
    Domain-extensibility is combined with controlled rewriting to optimise a standard image processing pipeline: the Harris corner detection \cite{harris-1988}.
    We show how rewriting is controlled in \Shine{} to achieve 6 well-known image processing pipeline optimisations, including 2 optimisations that are not supported by Halide schedules.
    Our results on four mobile ARM multi-core CPUs, often used for image processing tasks, show that the code generated using \Shine{} for the Harris corner detection 
    is up to 1.4$\times$ (geomean of 1.27$\times$) faster than Halide.
    However, we also observe that controlling rewriting is tedious, motivating the following chapter that aims to lower performance engineer effort with semi-automation.

    \item \textbf{Proposing a Novel Semi-Automatic Optimisation Technique} \cite{koehler2021-sketch-guided} (\cref{ch:guided-rewriting}).\\
    A new semi-automatic optimisation technique called \emph{sketch-guided equality saturation} is developed, based on a fully automated technique called equality saturation.
    Sketch-guiding allows performance engineers to guide program rewriting by specifying rewrite goals as \emph{sketches}: program patterns that leave details unspecified.
    We evaluate sketch-guided equality saturation by applying 7 realistic optimisations of matrix multiplication.
    Unguided equality saturation alone does not scale to the 5 most complex optimisations, even given an hour and 60GB of RAM.
    With the guidance of at most 3 sketch guides, each 10 times smaller than the complete program, the compiler applies the optimisations in seconds using less than 1GB of RAM.
    We also explore how to efficiently encode a polymorphically typed lambda calculus for equality saturation.
    The runtime and memory consumption of unguided equality saturation over lambda terms is reduced by orders of magnitude, which is also beneficial for sketch-guided equality saturation.

\end{enumerate}


\chapter{Motivating Background}
\label{ch:background}
This chapter motivates the contributions of this thesis by presenting technical background and related work.
The remaining background will be introduced separately in the technical chapters.
Similarly, more specific related work will be presented throughout the thesis and discussed in \cref{ch:discussion}.

\Cref{sec:hpc-programming} provides background on how performance engineers traditionally program and optimise performance-demanding applications in low-level languages like C and OpenCL.\linebreak
Performance engineers typically achieve orders of magnitude performance improvements by manually exploring potential optimisations using their expert knowledge.
Unfortunately this optimisation process is time consuming, risks introducing bugs, and needs to be repeated for each new algorithm and hardware target.

To improve over this traditional optimisation process, we would like to combine higher-level programming models with compilers that automate program optimisations, but this is a challenging task.

\Cref{sec:extending-compilers} motivates research on \emph{domain-extensible} compilers.
The compiler \emph{extensibility challenge} is introduced: extending compilers with new optimisations is hard.
Existing compilers are categorised into traditional domain-agnostic compilers; established domain-specific compilers; and emerging domain-extensible compilers.
The advantages and disadvantages of each category of compilers are highlighted.

\Cref{sec:controlling-compilers} motivates research on \emph{controllable automation} of optimisations.
The compiler \emph{controllability challenge} is introduced: even though manual optimisation is tedious for performance engineers, automatic optimisation is not always satisfactory.
The advantages and disadvantages of automated optimisation and controlled optimisation are discussed before introducing the concept of controllable automation of optimisations.


\section{High Performance Programming}
\label{sec:hpc-programming}



\subsection{Hardware Architectures are Evolving}
\label{hardware}


In the quest to maximise runtime performance and energy efficiency, hardware architects continuously come up with significant architectural changes.
First, hardware architectures became highly parallel \cite{processor-scaling-2004, hill2008-amdahl}.
Nowadays, hardware architectures are increasingly heterogeneous \cite{dark-silicon-2011, 2016-heterogeneous-computing}.
Such hardware evolution seriously impacts high performance programming practices.
Performance engineers must consider hardware parallelism and heterogeneity to achieve high performance when optimising software \cite{2017-perf-prog}.

In particular, various forms of parallelism are exploited:
\begin{itemize}
     \item \emph{instruction-level parallelism}, where multiple instructions are executed simultaneously.
     The hardware commonly exploits this parallelism implicitly at runtime (e.g. instruction pipelining, superscalar execution), but some hardware explicitly exposes this parallelism to software instead (e.g. Very Long Instruction Word).
     \item \emph{vector-level parallelism}, where multiple vector elements are processed simultaneously.
     This is also referred to as Single Instruction, Multiple Data (SIMD).
     Many processors explicitly expose this parallelism.
     As a result, software must use explicit instructions to exploit this parallelism.
     \item \emph{thread-level parallelism}, where multiple software threads are executed simultaneously.
     A thread is a programmed sequence of instructions that can be scheduled independently to other threads on the available hardware cores.
     Care must be taken when developing explicitly multi-threaded software, as it is a difficult and error-prone task \cite{2010-concurrency-bugs}.
\end{itemize}

Two common hardware architectures are often combined in heterogeneous systems: multi-core Central Processing Units (CPUs) and many-core Graphics Processing Units (GPUs).


\paragraph{Multi-Core CPUs}
Optimised for individual thread performance and low latency, multi-core CPUs have a few sophisticated cores able to run relatively heavyweight threads.
CPUs dynamically adapt to various usage patterns through complex execution, control flow and memory management.
To exploit locality of memory accesses, CPUs have a hierarchy of caches that automatically stores recently accessed data closer to the cores.
Three levels of caches named L1, L2 and L3 are commonly used; where 
L1 is the closest to the cores.
The closer to a core a cache is, the smaller and faster it is.
Even though data is moved implicitly between caches, program optimisations can still improve cache usage by improving the locality of accesses.

While CPUs focus on individual thread performance, they still expose significant parallelism.
It is standard for CPUs to execute 8 to 16 threads simultaneously and to provide 128 to 256 bits of vector parallelism (i.e. 4 to 8 32-bit values) on top of instruction-level parallelism.

\paragraph{Many-Core GPUs}
Optimised for high throughput, many-core GPUs hide high latency operations by overlapping the execution of many relatively lightweight threads.
Originally specifically designed to accelerate 3D graphics rendering, GPUs have been generalised to support more use cases and are becoming increasingly flexible \cite{gpgpu-2004}.
GPUs are now popular for massively parallel and regular computations in many performance demanding domains.

It is standard for GPUs to contain hundreds or thousands of cores, grouped hierarchically to share hardware resources during execution. 
For example, while GPUs have a cache hierarchy, they also typically provide scratchpad memory which is shared by a group of threads.
A scratchpad is similar to a cache, but is less sophisticated because data transfers are explicitly programmed instead of implicit.
In general, GPUs tend to avoid sophisticated hardware techniques with high hardware cost and instead favor simpler techniques with low hardware cost.
This shift from implicit hardware logic to explicit software programming makes program optimisation both more important and difficult.


\paragraph{Other Processors}
While this thesis focuses on CPUs and GPUs, heterogeneous systems also include reconfigurable hardware such as Field-Programmable Gate Arrays (FPGAs) or Coarse-Grained Reconfigurable Architectures (CGRAs); as well as highly specialised hardware such as Digital Signal Processors (DSPs) or Tensor Processing Units (TPUs).
The contributions of this thesis may be combined with related work that tackles such hardware \cite{2015-TyTra, siefke2022-tpus, 2022-schlaak-shir}.
For example, the D2A methodology \cite{2022-D2A} addresses compilation challenges that are specific to accelerators, such as the mismatch between fine-grained compiler IR (Intermediate Representation) operations and coarse-grained accelerator operations, and the need for a formal hardware specification.
In the future, processors are likely to become even more diverse, complicating program optimisation tasks: performance engineers may need to exploit quantum \cite{2013-ion-trap-quantum}, neuromorphic \cite{2018-loihi-neuromorphic} or optical processors \cite{2018-optical}.

\subsection{The OpenCL Programming Model is too Low-Level}
\label{low-level-programming}

To achieve high performance on parallel and heterogeneous hardware, the industry standard is to use relatively low-level programming models.
In this thesis, \emph{low-level code} is code that is specialised to a given hardware for performance reasons.
By extension, a \emph{low-level programming model} is a programming model where non-specialised code performs poorly, i.e. the programming model fails to provide what is also called performance portability \cite{steuwer2015-thesis}.
We specifically provide more background on the OpenCL programming model \cite{munshi2009-opencl, stone2010-opencl}, an open standard targeting diverse processors that we will use in this thesis.

The OpenCL standard consists of a C API to orchestrate computation from the \emph{host} and a programming language to express computation on \emph{OpenCL devices}.

\paragraph{Compute Model}
Programs executed on a device are called \emph{kernels}.
A kernel is a program that each thread executes in a \emph{Single Program, Multiple Data} (SPMD) fashion.
The host program may submit a kernel for execution on a device by specifying an N-dimensional index space (where N goes from 1 to 3).
For each point in the index space, an instance of the kernel program called \emph{work-item} is executed.
Each point in the index space is a global identifier for the corresponding work-item and can be queried by the kernel program to inform its execution.

The index space is decomposed into \emph{work-groups} of multiple work-items that may share hardware resources for cooperation.
Each work-group has a corresponding identifier, and each work-item has a local identifier within its work-group.
This thread hierarchy closely corresponds to the compute hierarchy found in most GPU architectures, allowing efficient runtime scheduling, and is compatible with execution on other architectures such as CPUs.



\paragraph{Memory Model}
The OpenCL memory model also closely corresponds to the memory hierarchy of common GPU architectures, exposing different address spaces:
\begin{itemize}
  \item \emph{Global memory} is accessible to all work-items in all work-groups, and also to the host program.
  On CPUs and GPUs, it typically corresponds to device RAM.
  \item \emph{Local memory} is only accessible to work-items of the same work-group.
  It usually corresponds to RAM on CPUs, and faster scratchpad memory on GPUs.
  \item \emph{Private memory} is only accessible to a single work-item.
  On CPUs and GPUs, it typically corresponds to registers, the fastest possible memory.
\end{itemize}

OpenCL also provides \emph{constant memory}, read-only global memory that remains constant during kernel execution, however we will not exploit this type of memory in this thesis.

\paragraph{Kernel Programming}
The OpenCL C programming language, used to define kernels, is derived from the C99 specification.
On one hand, OpenCL C disallows pointers to functions, recursion, and dynamic memory allocation.
On the other hand, OpenCL C introduces address space qualifiers, custom data types and built-in functions.

\Cref{fig:opencl-add} shows a simple implementation of point-wise addition of two arrays using an OpenCL kernel.
In line \ref{l:add-sig}, the kernel signature consists of pointers referring to arrays in global memory (\inlineOpencl{a}, \inlineOpencl{b}, \inlineOpencl{output}) and an integer representing the size of these arrays (\inlineOpencl{n}).
The kernel program is executed by multiple threads, and starts by querying the global identifier of the current thread in line \ref{l:add-loop} (\inlineOpencl{get_global_id(0)}).
Depending on the number of available threads (\inlineOpencl{get_global_size(0)}), the program may loop depending on how many elements the current thread needs to compute.
An element is computed inside the loop using indexing in line \ref{l:add-compute}.

\clearpage
\begin{opencl}[label={fig:opencl-add}, caption={Point-wise addition of two arrays in OpenCL}, basicstyle=\ttfamily\scriptsize, mathescape]
kernel void add_v1(global float* a, global float* b, global float* output, int n) {$\label{l:add-sig}$
  for (int gid = get_global_id(0); gid < n; gid += get_global_size(0)) {$\label{l:add-loop}$
    output[gid] = a[gid] + b[gid];$\label{l:add-compute}$
  }
}
\end{opencl}

In practice, optimised kernel implementations are significantly more complex.
\Cref{fig:opencl-add-opt} shows another implementation of the same computation.
The program leverages vector-level parallelism using \inlineOpencl{vload4} and \inlineOpencl{vstore4} to transfer vectors of 4 single-precision floating-point values, as well as a vectorised addition.
The program loops are tiled to increase single-thread workload and potentially improve data locality: chunks of 4 vectors are computed together. 

\begin{opencl}[label={fig:opencl-add-opt}, caption={Point-wise addition of two arrays in OpenCL: tiled, vectorized.}, basicstyle=\ttfamily\scriptsize]
kernel void add_v2(global float* a, global float* b, global float* output, int n) {
  for (int gid = get_global_id(0); gid < n/16; gid += get_global_size(0)) {
    for (int i = 0; i < 4; i += 1) {
      float4 av = vload4(gid * 4 + i, a);
      float4 bv = vload4(gid * 4 + i, b);
      vstore4(av + bv, gid * 4 + i, output);
    }
  }
}
\end{opencl}

Optimised code is typically orders of magnitude faster than naive code, but is also more complex and specialised to a given hardware.
For performance engineers, manually optimising low-level code using techniques such as tiling or vectorisation is time-consuming and risks introducing bugs.
This is aggravated by the fact that many optimisation decisions require global thinking: naive compositions of locally optimised code often perform poorly.

\paragraph{Other Industrial Programming Standards}
The \emph{C} and \emph{C++} languages are commonly used to program CPUs, in combination with primitives such as \emph{POSIX threads} for multi-threading and \emph{SIMD instructions} (e.g. SSE, AVX, NEON) for vectorisation.
\emph{CUDA} is a popular closed framework developed by NVIDIA for NVIDIA GPUs.
It is also common to build convenient abstractions on top of these standards.
\emph{SYCL} allows single-source programming in C++ for OpenCL, avoiding the need to write separate OpenCL C code, and providing implicit memory transfers.
\emph{OpenMP} and \emph{OpenACC} are used for quick code acceleration with compiler directives, such as parallelising a for loop by preceding it with \lstinline{#pragma omp parallel for}.
All of these programming models suffer from similar performance engineering productivity and portability issues as OpenCL: non-specialised code performs poorly compared to specialised code.

\subsection{Higher-Level Programming Models are Challenging to Compile}

Higher-level programming models and languages are a promising way to increase productivity, performance and portability.
We give concrete examples with domain-specific languages and array programming languages before mentioning common compilation challenges.

\paragraph{Domain-Specific Languages}
Domain-Specific Languages (DSLs) enable combining convenient, hardware-agnostic programming with high-performance for specific domains.
This is a proven success for domains such as image processing \cite{halide-2012, darkroom-2014, polymage-2015}, signal processing \cite{spiral-2018}, linear algebra \cite{linnea-2017}, machine learning \cite{tvm-2018} or partial differential equations \cite{2015-stella-pde, 2016-firedrake-pde}.

For example, the point-wise addition of two arrays can be defined in the Halide image processing DSL by writing a high-level \emph{algorithm} as in \cref{halide-add-algorithm}.
Crucially, DSLs separate concerns: domain scientists can focus on writing high-level programs (what to compute) while compilers and performance engineers optimise programs for performance (how to compute).
For example, \cref{halide-add-algorithm} could be compiled by Halide into an OpenCL implementation (e.g. \cref{fig:opencl-add}, \cref{fig:opencl-add-opt}) or into some other optimised implementation, depending on the target hardware.

\begin{halide}[label={halide-add-algorithm}, caption={Point-wise addition of two arrays in Halide}]
Var i; Func output;

output(i) = a(i) + b(i);
\end{halide}


\paragraph{Array Programming Languages}
Array programming languages such as Accelerate \cite{accelerate-2011}, SaC \cite{2006-SaC}, NOVA \cite{2014-nova}, \Lift{} \cite{lift-rewrite-2015}, Futhark \cite{futhark-2017}, or Dex \cite{2021-Dex} provide a middle-ground between highly specialised languages and general-purpose languages.
Multi-dimensional arrays are a key abstraction for many performance-demanding application domains (e.g. physics, chemistry, mechanics, image processing, data processing, linear algebra, machine learning), as they can be used to represent widely used tensors \cite{2020-dagstuhl-tensor, 2017-taco}.

Some array programming languages focus on index-based notations \cite{2006-SaC, 2021-Dex}, similar to how the \code{i} index is used in the Halide algorithm of \cref{halide-add-algorithm}.
Other array programming languages focus on collective array operations such as map or reduce \cite{accelerate-2011, 2014-nova, lift-rewrite-2015}, that can also be seen as algorithmic skeletons \cite{algorithmic-skeletons-cole2004}.
For example, the point-wise addition of two arrays can be defined in the \Lift{} language as in \cref{fig:lift-add}.
\inlineRise{zip a b} combines two arrays \inlineRise{a} and \inlineRise{b} whose elements are added pairwise using \inlineRise{map}, and no explicit indexing is required.

\begin{rise}[label={fig:lift-add}, caption={Point-wise addition of two arrays in \Lift{}}, numbers=none]
map ($\lambda$x. (fst x) + (snd x)) (zip a b)
\end{rise}

Array programming languages offer the potential to share common infrastructure across multiple application domains while still being convenient and hardware-agnostic.

\paragraph{Developing Compilers for High-Level Programs}
Whether high-level programs are written in domain-specific or array languages, building and maintaining compilers that generate high-performance code for a variety of algorithmic domains and hardware architectures is challenging.

Industrial-strength compilers such as Halide for image processing and TVM for machine learning are the result of years of engineering effort from multiple contributors.
The Halide repository contains 278k lines of code written by 167 contributors.\footnote{\scriptsize\url{https://github.com/halide/Halide/commit/820ec1f963f06f53a1808eb9b4631f2031be7468}}
The TVM repository contains 794k lines of code written by 752 contributors.\footnote{\scriptsize\url{https://github.com/apache/tvm/commit/178f82dc481bf31961206412c22dd5519a245b49}}
Further, both use other compiler technologies such as LLVM as backends.
LLVM 8.0.1 comprises 6,887k lines of code written by 1,210 contributors, with an estimated development cost of 529M\$ \cite{cummins2020-thesis}.

The following sections motivate the need to address two specific compiler challenges: the \emph{extensibility challenge} and the \emph{controllability challenge}.


\section{The Compiler Extensibility Challenge}
\label{sec:extending-compilers}

\subsection{Domain-Agnostic Compilers}
\label{domain-agnostic-compilers}

Traditional, general-purpose compilers such as LLVM \cite{llvm-2004} or GCC are \emph{domain-agnostic}: they only support a fixed, generic set of program abstractions and optimisations.

LLVM is a widely adopted compiler framework consisting of production-quality libraries for modular compiler construction.
The LLVM IR (Intermediate Representation) provides many generic abstractions that are useful across multiple programming languages and hardware targets, such as types, functions and exceptions.
LLVM also provides many powerful optimisation passes, such as dead code elimination, common subexpression elimination, function inlining, strength reduction or target-specific instruction selection.
LLVM is successfully used as part of compilers for the C/C++ (clang compiler), Rust and Swift languages, to name a few.

While the LLVM framework is modular and extensible with custom compiler analyses and compiler passes, the LLVM IR is generic and even lower-level than languages like C or OpenCL.
Higher-level program constructs must be lowered by compilers targeting LLVM, erasing domain-specific information which can be hard or impossible to recover.
As a result, domain-agnostic compilers typically do not automate optimisations for specific domains.
Instead, performance engineers must perform them manually as seen in \cref{low-level-programming}.

\paragraph{Optimised Libraries}
Numerous \emph{libraries of optimised functions} (e.g. BLAS, cuDNN, MKL, NVIDIA Performance Primitives, ARM Compute Library, Intel IPP, OpenCV) provide optimised implementations of common computations.
However, they are developed by performance engineers at high cost.
Limited engineering budget leads to limited functionality and limited support for certain use cases or hardware.
Moreover, the composition of optimised functions through library calls is often far from optimal because many optimisations cannot be applied across library calls.
For example, Weld \cite{weld-2017} achieves order of magnitude performance improvements by optimising across libraries and functions using a common intermediate representation.



\subsection{Domain-Specific Compilers}
\label{domain-specific-compilers}

Established \emph{domain-specific} compilers only support a fixed, specialised set of program abstractions and optimisations.
Examples include Halide \cite{halide-2012} and PolyMage \cite{polymage-2015} for image processing; TVM \cite{tvm-2018} for machine learning; SPIRAL for signal processing \cite{spiral-2018}; Firedrake for partial differential equations \cite{2016-firedrake-pde}.

Domain-specific compilers are successfully established in industry where they automate many domain-specific optimisations and achieve impressive performance results.
Halide is used for some processing tasks in the Google Pixel camera, Adobe Photoshop, and YouTube.
TVM is used and developed by companies like AMD, ARM, Microsoft, NVIDIA, and Samsung.

However, extending domain-specific compilers with new optimisations as the domain evolves is difficult \cite{barham2019-ml-rut}.
Domain evolution includes changes in algorithms, changes in program optimisations, but also changes in specialised hardware.

Compiler extensions are particularly difficult when they have an impact on the entire compilation stack, requiring simultaneous expertise of high-level algorithms, multiple compilation aspects, and low-level hardware targets.
Separating concerns facilitates extension, for example by separating the definition of correct transformations from the search for best transformations as shown later in this thesis.

\paragraph{Extending Halide is Hard}
The Halide Development Roadmap from 2020\footnote{\url{https://github.com/halide/Halide/issues/5055}} highlights that extending Halide is hard, listing unsolved questions such as:

\begin{itemize}[nosep]
\item "How do we make Halide easier to use for researchers wanting to [...] extend it [...]?"
\item "How do we make Halide more useful on current and upcoming hardware?"
\item "How do we make Halide more useful for new types of application?"
\end{itemize}

In fact, solutions to the extensibility challenge have been independently researched in the course of this thesis by Halide authors \cite{2022-POPL-verified-tensor-rewrite, 2022-exo}, showing that our research direction is valuable and our motivations shared by the broader community.

\paragraph{Extending TVM is Hard}
The TVM Unity vision for 2022\footnote{\url{https://tvm.apache.org/2021/12/15/tvm-unity}} demonstrates that, currently, extending TVM is hard.
Two problematic boundaries are identified in the compilation stack.
First, vertical boundaries between successive abstraction layers.
Each local decision made within a layer has a global impact on the following layers.
As a result, introducing support for new hardware in the last layer requires re-thinking how decisions are made in all prior layers, requiring global instead of local engineering efforts.
Second, horizontal boundaries between lowering strategies.
For example, TVM may decide to target optimised libraries or to generate code from scratch, but typically cannot combine both approaches.

\begin{displayquote}[TVM Unity vision]
{[\dots]} boundaries are slowing down the pace of innovation in machine learning. New hardware accelerators are emerging with new levels of capability and performance, but harnessing them will require fluid collaboration between ML scientists, ML engineers, hardware vendors that these boundaries prevent. To cope with the rapid pace of change in ML systems, frameworks need to support incremental evolution: Incorporating new capabilities should require effort proportional to the change, not wholesale re-engineering at each level.
\end{displayquote}

\subsection{Domain-Extensible Compilers}

Building DSLs and their compilers has long been recognized as a highly complex task, which is why many projects aim to simplify the development of DSLs.
For example, Spoofax~\cite{2010-spoofax} is a language workbench simplyfing the development of DSLs.
The idea is to produce parsers, type checkers, compilers, interpreters and even IDE support from declarative language specifications.
Gammars are used to declare syntax, rewrite rules to declare semantics.
With Spoofax, multiple domain-specific compilers are still constructed even if the task is simplified.
The constructed compilers may still be hard to adapt as their domain evolves, as seen for established domain-specific compilers in \cref{domain-specific-compilers}.
Another approach is to build \emph{domain-extensible} compilers providing an extensible set of program abstractions and optimisations.

Delite \cite{delite-2011} is a DSL framework providing a fixed set of generic parallel patterns that DSLs can target.
Domain-specific optimisations can be defined using staging, a technique used to eliminate the cost of DSL abstractions while lowering them to generic parallel patterns.
Delite demonstrated that multiple DSLs can be composed together while sharing common infrastructure \cite{delite-2013-composition}.
The Delite projet also identified that its staging mechanism is not sufficient, and showed that using term rewriting techniques is beneficial for certain use cases \cite{delite-extensible-staging-2013}.

AnyDSL~\cite{anydsl-2018} is a more recent approach leveraging partial evaluation to replace DSL compilers with self-specialising library code written in a language called Impala.
The idea of implementing languages as libraries has a long strand of research predating AnyDSL, such as the Racket language that leverages macros instead of partial evaluation \cite{2011-langs-as-libs}.

\Lift{}~\cite{lift-rewrite-2015} combines a high-level functional language with an extensible rewrite system to define an optimisation space and generate high-performance code.
\Lift{} has demonstrated its ability to generate high-performance code by deriving multiple low-level implementations from a single portable high-level program, obtaining performance on par with highly tuned platform-specific libraries on various GPUs~\cite{lift-performance-portability-2016}.
\Lift{} is extensible with rewrite rules and language primitives by design, and has been successfully extended for stencil computations \cite{lift-stencil-2018} which are common in multiple domains including image processing.

Overall, domain-extensible compilers show a lot of potential, but they remain relatively immature and lack adoption compared to established domain-specific compilers.
This thesis builds on the foundations laid by \Lift{} and addresses some of its shortcomings.
For example, previous work does not investigate whether \Lift{} is able to generate code with competitive runtime performance compared to Halide for image processing as we do in \cref{ch:imgproc}.
\Lift{} also does not address the compiler controllability challenge that we discuss in the next section.

While we follow a rewrite-based approach, multiple extensible techniques can complement each other, as demonstrated by Delite for staging and rewriting.
We believe that research on each technique is useful on its own.


\paragraph{Extending Compilers with Rewrite Rules}
Many compilers before \Lift{} already allowed programmers to express domain-specific optimisations as rewrite rules.
The Glasgow Haskell Compiler (GHC) allows this, but only applies rewrite rules according to a simple strategy, assuming that the right-hand side of a rule is always preferable to the left \cite{jones2001-ghc-rules}.
The following pragma would instruct GHC to always rewrite \lstinline{map f (map g xs)} into \lstinline{map (f . g) xs}:
\begin{lstlisting}
{-# RULES
  "map/map" forall f g xs.
    map f (map g xs) = map (f . g) xs
#-}
\end{lstlisting}
While such a simple strategy can be effective in optimising programs, it falls short for use cases where deciding which rewrite rule is beneficial when is hard.

Stratego \cite{visser1998-strategies} is a language for defining customised rewriting strategies, which is included in the Spoofax language workbench.
Phobos is a compiler frontend enabling the development of domain-specific languages by combining an open term language with term rewriting \cite{2003-Phobos-extensible-compilers}.

On one hand, tools like GHC, Stratego and Phobos allow domain-extensbility via rewrite rules but do not focus on high-performance code generation.
On the other hand, tools like Spiral \cite{spiral-2005, spiral-2018} successfully generate high-performance code for heterogeneous hardware using rewrite rules, but are domain-specific and not extensible.
This thesis follows \Lift{}'s ambition of combining domain-extensibility with high-performance code generation for heterogeneous hardware using rewrite rules.

\section{The Compiler Controllability Challenge}
\label{sec:controlling-compilers}

\subsection{Automatic Optimisation}

Many compilers aim for fully automated optimisation, some of them lean towards \emph{greedy optimisation}, while others lean towards \emph{explorative optimisation}.

\paragraph{Greedy Optimisation}
Greedy optimisation is often achieved using a fixed pipeline of optimisation \emph{passes}, relying on expert heuristics to make local decisions that are assumed to be beneficial for final performance.

For example, the LLVM framework \cite{llvm-2004} provides many transformation and optimisation passes, and allows defining custom passes.
In GHC, user-defined rewrite rules are applied greedily \cite{jones2001-ghc-rules}.
Many other projects rely on greedy optimisation passes, such as Accelerate \cite{mcdonell2013-accelerate} and the \textsc{Shir} compiler that adopts ideas from \Lift{} to target FPGAs \cite{2022-schlaak-shir}.

Typically, greedy optimisation is fast, but may result in poor performance as it gets stuck in local minima and may poorly predict performance benefits.
The problem of deciding when to apply which optimisation pass is known as the \emph{phase ordering problem}, and has a huge impact on the performance of the rewritten program.

\paragraph{Explorative Optimisation}
Explorative optimisation is a more holistic approach, where a space of possible implementations arising from different combinations of optimisations is explored, in order to make a decision.

Random sampling, or Monte Carlo methods, can be used to navigate the space of possible implementations.
Using this approach, finding an implementation with high enough performance can be very time consuming.
\Lift{}~\cite{lift-rewrite-2015} uses random sampling, and optimising a single convolution takes hours to reach peak performance on a GPU \cite{lift-stencil-2018, mogers2022-constraint-satisfaction}.

Performance models can be used to choose from the space of possible implementations.
Some performance models are analytical, describing performance as an equation, as done in TyTra \cite{2015-TyTra, 2021-urlea-TyTra}, SPIRAL \cite{spiral-2005, spiral-2018} and Telamon \cite{telamon-2017}.
Analytical performance models are typically fast to evaluate, estimating a program's performance without executing it.
Creating an accurate performance model for complex systems is a challenging task, which is why empirical performance models are also used, predicting future performance by learning from previous performance data \cite{gysi2019-absinthe}.
Although performance modelling is making great progress, it remains a standing challenge that is the focus of multiple papers each year \cite{2019-TyTra-cost, wang2020-RAML-cost, 2020-pricing-python-model, 2021-DL-model, 2022-COMPOFF-model}.


Automatic, explorative optimisation is an active research field and novel techniques are constantly developed to explore various optimisation spaces, such as iterative compilation \cite{kisuki2000-iterative-compilation}, adaptive compilers \cite{cooper2002-adaptive-compilation}, or equality saturation \cite{tate2009-equality-saturation, willsey2021-egg}.


\paragraph{Advantages of Automatic Optimisation}
Fully automated optimisation is invaluable if performance engineers are not available, or are allocated to other tasks.
State-of-the-art automation of tedious, low-level implementation decisions such as register allocation or instruction selection delivers satisfying performance for most applications, and frees up development time for higher-level optimisations where a performance engineer can have higher impact.
Valuable human time is saved, trading it for machine time. 
Both inside and outside of computer science, automation has the potential to increase task feasibility, productivity and quality \cite{nof2009automation}.

\paragraph{Disadvantages of Automatic Optimisation}
Fully automated optimisation may result in poor performance or require too much compilation time \cite{maleki2011-vectorizing-compilers, parello2004-pragmatic-opt, chambers2002-staged-compilation}.
Crucially, control is usually sacrificed.
The compilation flags offered by compilers like clang, that uses LLVM for C-like languages, only provide very limited control.
Therefore, when compiler optimisations are unsatisfactory, performance engineers often fall back to manual optimisation in order to achieve their performance goals \cite{niittylahti2002-high-perf, hlt-lacassagne-2014, lemaitre2017-cholesky}.


\subsection{Controlled Optimisation}


Instead of manually optimising low-level code, or entrusting optimisation to a black box automatic compiler, performance engineers may control semi-automatic compilers that effectively act as optimisation assistants.
Techniques for controlled optimisation include transformation scripts, scheduling APIs, and rewriting strategies.


\paragraph{Transformation scripts}
The polyhedral framework \cite{polyhedral-model-2004} is a powerful technique modelling loop nests as polyhedra, enabling sophisticated dependency analysis, and providing tools to explore the space of valid transformations.
URUK \emph{transformation scripts} were proposed to define and apply sequences of polyhedral transformations \cite{polyhedral-search-2005, girbal2006-semi-auto-poly}.
The CHiLL framework extended this idea to more complex loop transformations \cite{chen2008-chill}.
An example of CHiLL script that applies loop transformations such as \lstinline{tile(statement_id, loop_level, tile_size)} is given in \cref{fig:chill-trans-script}.
Loo.py \cite{2014-loopy} is a more recent programming system embedded in Python and inspired by the polyhedral model, that provides a library of loop transformations.

\begin{lstlisting}[label={fig:chill-trans-script}, caption={CHiLL transformation script used to optimise matrix multiplication in \cite{chen2008-chill}.}]
permute([3,1,2])
tile(0,2,TJ)
tile(0,2,TI)
tile(0,5,TK)
datacopy(0,3,2,[1])
datacopy(0,4,3)
unroll(0,4,UI)
unroll(0,5,UJ)
\end{lstlisting}


\paragraph{Scheduling APIs}
Halide \cite{halide-2012} has popularised the development of compilers that offer scheduling APIs.
In this setting, a \emph{schedule} defines how to optimise an \emph{algorithm} that defines what to compute at a high level of abstraction, separating the two concerns.
The Halide scheduling API focuses on exposing important trade-offs between parallelism, redundancy, and locality; by defining when and where functions are computed and stored.
\Cref{halide-add-schedule} gives a simple example of Halide schedule.
Following the success of Halide, the TVM \cite{tvm-2018}, Fireiron \cite{fireiron-2020} and Tiramisu \cite{tiramisu-2019} compilers apply the same principle to different domains.

\begin{halide}[label=halide-add-schedule, caption={Schedule for the point-wise addition Halide algorithm from \cref{halide-add-algorithm}, loosely comparable to the OpenCL implementation from \cref{fig:opencl-add-opt}.}]
output.split(i, gid, i, 16)
  .parallel(gid)
  .vectorize(i, 4);
\end{halide}

\paragraph{Rewriting Strategies}
As discussed in the previous section, languages such as Stratego \cite{visser1998-strategies} enable the definition of custom \emph{rewriting strategies} \cite{kirchner2005-strategic-rewriting}.
As this thesis follows a rewrite-based approach to tackle the extensibility challenge, rewriting strategies are of particular interest to tackle the controllability challenge.
\Cref{ch:imgproc} will demonstrate how rewriting strategies, written in a language called \Elevate{} \cite{hagedorn2020-elevate}, can be used to apply transformations that are beyond what is possible with Halide schedules on a case study.

\paragraph{Advantages of Controlled Optimisation}
Transformation scripts, schedules and rewriting strategies all enable performance engineers to take control over the optimisation process.
They expose a structured optimisation space to the performance engineer, who explores it in order to achieve his performance goals.
Compared to manual optimisation of low-level code, controlled optimisation saves development time, and given a correct compiler implementation, avoids introducing bugs during optimisation.

\paragraph{Disadvantages of Controlled Optimisation}
Controlled optimisation remains challenging, as argued for schedules in \cite{ikarashi2021-guided-scheduling}.
Further, systems built around these techniques typically do not allow smooth trade-offs between automation and control, as this trade-off is built-in.
Therefore, controlled optimisation may be tedious and take too much performance engineering effort compared to automatic optimisation.

\subsection{Controllable Automation of Optimisations}
\label{controllable-automation}

This thesis explores \emph{controllable automation} of optimisations that promotes trade-offs between automation and control of optimisations.
If a compiler supports both extremes, as well as a spectrum in-between, performance engineers should be able to gradually take control of optimisations depending on their performance requirements and time budget.

Properties like domain-extensibility and controllable automation are hard to quantify on a scale, as they represent informal design principles.
However, it is possible to identify whether different approaches provide controllable automation of specific choices or not, as in \cref{fig:controllable-automation-comparison}.

\begin{table}[th]
  \centering
  \begin{tabular}{|c|c|c|} \hline
  \textbf{Approach} & \textbf{Parameter Choice} (tile size) & \textbf{Optimisation Choice} (tiling) \\ \hline
  Halide Schedule & \no & \no \\ \hline
  TVM Schedule Template & \yes & \no \\ \hline
  This Thesis & \yes & \yes \\ \hline
  \end{tabular}
  \caption{This thesis allows controllable automation of optimisation choices like tiling, as opposed to current Halide or TVM approaches.}
  \label{fig:controllable-automation-comparison}
\end{table}

Halide schedules neither provide controllable automation of parameters (e.g. tile size), nor of optimisations (e.g. tiling).
Successful research goes into \emph{autoscheduling}, the challenge of automatically generating schedules, both for Halide \cite{halide-autosched-2016, halide-autosched-2019, anderson2021-autosched} and TVM \cite{tvm-ansor-2020, tvm-bansor-2021}.
However to the best of our knowledge, autoscheduling techniques offer no controllable automation: performance engineers cannot constrain autoscheduling to apply a specific tiling optimisation, or not use specific tile sizes.
This contrasts with the conclusion of the first Halide paper: \emph{"the ultimate solution must allow a
smooth trade off between inference when it is sufficient, and sparse
programmer control when it is necessary"} \cite{halide-2012}.

TVM \emph{schedule templates} provide controllable automation of parameters, but not of optimisations.
For example, performance engineers may delegate the tuning of some numerical parameters, such as tile sizes, while manually specifying other parameters in a schedule template \cite{tvm-2018}.
However, this approach does not allow delegating optimisations that significantly change code structure, such as deciding whether to use tiling, or how to fuse operators \cite{tvm-ansor-2020}.

By contrast, the techniques presented in this thesis enable controllable automation of both parameters and optimisations (\cref{ch:guided-rewriting}).

In the polyhedral community, there exists multiple trade-offs between automation and control.
Transformation scripts can be used to structure automatic empirical optimisation \cite{2007-POET, 2013-chill-auto}.
Powerful heuristics have also been developed to automate optimisation, without using transformation scripts, in PLuTo \cite{bondhugula2008-polyhedral}, PPCG \cite{2013-ppcg} and PolyMage \cite{polymage-2015, jangda2020-effective-fusion}.
Ways to increase interactivity have been explored, \emph{"enabling users to examine, refine, replay and even design complex optimizations semi-automatically in partnership with the compiler"} \cite{2016-poly-box}.
Program optimisation tactics, loosely comparable to rewriting strategies, have been developed independently during this thesis for the polyhedral model \cite{2018-polyhedral-tactics, loop-tactics-2019}.
While the polyhedral framework is amenable to controllable automation, we focus on term rewriting techniques instead in this thesis.
Term rewriting is more flexible than the polyhedral framework \cite{2014-poly-rewrite}, which gets its strength from a restricted but carefully structured transformation space.

In the previous section, we already identified term rewriting as a compelling technique to tackle the extensibility challenge.
Term rewriting is also a compelling technique to tackle the controllability challenge.
Indeed, the definition of rewrite rules can be clearly separated from their application during rewriting.
Given rewrite rules, we can decide to fully automate their application as done in \Lift{} with random sampling, to precisely control their application as done in Stratego with rewriting strategies as we will leverage in \cref{ch:imgproc}, or trade-off between the two as we will explore in \cref{ch:guided-rewriting}.


\section{Conclusion}

This chapter motivated the contributions of this thesis by presenting technical background and related work.

\Cref{sec:hpc-programming} discussed how, in a context of evolving hardware architectures, the programming models commonly used in industry, such as OpenCL, are too low-level.
Higher-level programming models such as domain-specific languages or array programming languages have been successful at increasing productivity and portability.
However, building and maintaining compilers that generate high-performance code for a variety of algorithmic domains and hardware architectures, while minimising engineering efforts, is challenging.

\Cref{sec:extending-compilers} discussed the compiler \emph{extensibility challenge}.
Domain-agnostic compilers such as LLVM are typically not able to automate important domain-specific optimisations.
Extending domain-specific compilers such as Halide or TVM with new optimisations is hard.
Domain-extensible compilers show potential to mitigate the extensibility challenge, but remain relatively immature compared to domain-specific compilers.
Research is needed to further develop domain-extensible compiler technology.

Finally, \cref{sec:controlling-compilers} discussed the compiler \emph{controllability challenge}.
Automatic optimisation is invaluable if performance engineers are not available, but may result in poor performance or be too time consuming for some applications.
Alternatively, performance engineers may take control of the optimisation process.
However, controlling the optimisation process is challenging as there is typically no smooth transition between automation and control of optimisations.
Research is needed to enable performance engineers to gradually take control instead of abruptly falling back to manual optimisation.

\chapquote{Imagination is the beginning of creation. You imagine what you desire, you will what you imagine and at last you create what you will.}{George Bernard Shaw}

\chapter{Code Generation in a Domain-Extensible Compiler}
\label{ch:imperative-code}
The previous chapter identified term rewriting as a compelling technique to tackle the compiler extensibility and controllability challenges.
\Lift{} \cite{lift-rewrite-2015} addresses the \emph{extensibility challenge} by combining a high-level functional language with an extensible term rewriting system \cite{lift-performance-portability-2016}.
However, \Lift{} does not address the \emph{controllability challenge}, and optimising a single convolution takes hours to reach peak performance on a GPU \cite{lift-stencil-2018, mogers2022-constraint-satisfaction}.
\Shine{} is a novel compiler inspired by \Lift{}, with the additional goal to provide controllability by exploring trade-offs between automation and control of rewrite rule applications.
A key concern for \Shine{} is how to generate imperative code from rewritten functional programs, with competitive runtime with established domain-specific compilers such as Halide and TVM.
Code generation in \Shine{} does not aim to be extensible and controllable, but rather to predictably preserve the optimisation choices encoded during rewriting.

This chapter presents the design and implementation of three important code generation features.
These features are crucial for \Shine{} to generate faster code than Halide on an image processing case study via controlled rewriting (\cref{ch:imgproc}), and similarly fast code as TVM on a linear algebra case study via semi-automated rewriting (\cref{ch:guided-rewriting}).

\Cref{sec:rise-shine} introduces the \Shine{} compiler and its \Rise{} language, both the result of collaboration between multiple researchers \cite{steuwer2022-rise-shine, dpia-2017}.
The rest of this chapter presents work that is solely my own.
\Cref{sec:barrier-insertion} contributes a synchronisation \emph{barrier insertion} algorithm that does not need to be modified when extending \Rise{} patterns, contrasting with the barrier elimination algorithm of \Lift{} \cite{lift-ir-2017}.
While barrier insertion is implicit and automated, the next two features add new \Rise{} patterns explicitly exposing implementation choices to rewriting, allowing design space exploration and external control.
\Cref{sec:host-codegen} demonstrates how a \Rise{} pattern is added for explicit \emph{kernel execution}, and how \Shine{} is modified to generate corresponding imperative code.
\Cref{sec:storage-folding} discusses how \Rise{} patterns are added for explicit \emph{storage folding}, and how \Shine{} is modified to generate corresponding imperative code.
\Cref{sec:codegen-conclusion} concludes.

\section{The \Rise{} Language \& \Shine{} Compiler}
\label{sec:rise-shine}

\begin{figure}
    \centering
    \includegraphics[width=\linewidth]{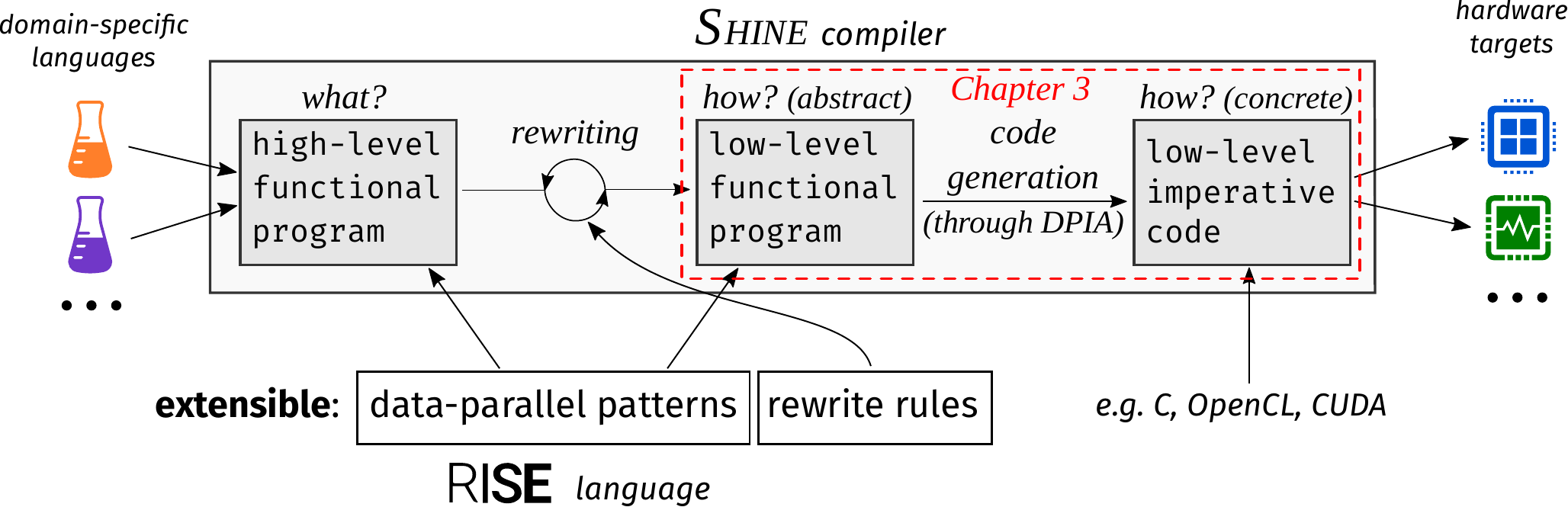}
    \caption{\Rise{} \& \Shine{}: A Domain-Extensible Compiler Design}
    \label{fig:rise-shine}
\end{figure}

\Shine{} is a domain-extensible compiler inspired by the \Lift{} compiler, its implementation is open-source.\footnote{\url{https://github.com/rise-lang/shine}}
\Cref{fig:rise-shine} gives an overview of the \Shine{} compiler, which is meant to be a bridge between domain-specific languages (left of \cref{fig:rise-shine}) and hardware targets (right of \cref{fig:rise-shine}).
A program written in a domain-specific language is translated into a \emph{high-level \Rise{} program} (this step is left for future work and not the topic of this thesis).
Then, the high-level program is rewritten into a \emph{low-level \Rise{} program} where implementation decisions are explicitly encoded (\cref{sec:rewriting}).
Finally, imperative code is generated and can be executed on a given hardware target (\cref{sec:codegen}).
The focus of this chapter (\cref{ch:imperative-code}) is to enhance imperative code generation from low-level \Rise{} programs (red rectangle in \cref{fig:rise-shine}).

\Rise{} is a functional data-parallel language inspired by the \Lift{} language.
More precisely, \Rise{} is a lambda calculus extended with higher-order functions called \emph{patterns} (e.g. \inlineRise{map} or \inlineRise{reduce}) and a restricted form of dependent typing (\cref{sec:rise}).
Functional languages are referentially transparent, meaning that an expression is equivalent to its value.
There are no side effects, and semantics-preserving rewrite rules can be easily defined to encode program optimisations.

The rest of this section gives a complete overview of the compilation stack, to contextualise the following sections.


\paragraph{Relationship to \Lift{}}
\Rise{} \& \Shine{} started as an effort to re-implement \Lift{} in a more principled way by following the typed lambda calculi formalism \cite{barendregt1992lambda}.
In \Shine{}, patterns are implemented as higher-order functions, and functions as expressions.
Adding a pattern only requires providing a name and a type.
This contrasts from the \Lift{} implementation, where adding a pattern also requires defining how to infer its type (\cref{fig:lift-rise}).
Both implementations then evolved separately as the original research group forked to explore two different directions.
There are no plans to merge the implementations, although it would be possible.

\begin{table}
  \begin{tabular}{|c|c|c|c|} \hline
  \textbf{language} & \textbf{functions are expressions} & \textbf{function type} & \textbf{type inference} \\ \hline
  \Lift{} & no & no & pattern-specific code \\ \hline
  \Rise{} & yes & yes & generic for all functions \\ \hline
  \end{tabular}
  \caption{\Rise{} follows the typed lambda calculus formalism more closely than \Lift{}.}
  \label{fig:lift-rise}
\end{table}

\subsection{Optimising Programs via Rewriting}
\label{sec:rewriting}

The \Shine{} compiler takes as input a high-level \Rise{} program describing \emph{what} to compute, rather than \emph{how} to compute.
For example, a vector dot product is represented as the high-level \Rise{} program \inlineRise{dot}:

\begin{rise}[numbers=none, xleftmargin=0pt, label={dot-rise}, caption={\inlineRise{dot} program in \Rise{}}]
def dot a b = zip a b $\then$ map ($\lambda$x. (fst x) $\times$ (snd x)) $\then$ reduce + 0
\end{rise}

The \inlineRise{zip a b} pattern combines two vectors \inlineRise{a} and \inlineRise{b} whose elements are multiplied pairwise using \inlineRise{map} before they are summed using \inlineRise{reduce + 0}.
The triangle symbol (\inlineRise{x $\then$ y}) indicates the chaining of operations via function application (\inlineRise{y x}) or composition (\inlineRise{$\lambda$z. y (x z)}).

The high-level \inlineRise{dot} program does not encode how it is executed. We could parallelise the \inlineRise{map}, store the intermediate result, and perform a sequential \inlineRise{reduce}.
Alternatively, we could avoid intermediate storage by fusing \inlineRise{map} and \inlineRise{reduce} into a single sequential reduction.
Many more options are possible, and such choices are encoded explicitly by applying rewrite rules.

To achieve a fused version avoiding the intermediate results, a compiler writer or performance engineer writes the \inlineElevateRise{reduceMapFusion} rewrite rule.
Currently, the rule is written in an informal syntax, future work may design a formal language for \Rise{} rewrite rules.
This rule states that mapping a function \inlineRise{f} over an array before reducing the array is equivalent to reducing the array while applying \inlineRise{f} on the go.
Note that the reduction must be performed sequentially because the reduction operator is not commutative anymore.

\begin{elevate-rise}[numbers=none, xleftmargin=0pt, label={reduceMapFusion}, caption={\inlineElevateRise{reduceMapFusion} rewrite rule.}]
rule reduceMapFusion = map f $\then$ reduce g init
                    $\rewritesTo$ reduceSeq ($\lambda$acc x. g acc (f x)) init
\end{elevate-rise}

Deciding which rewrite rule to apply where is challenging, and will be discussed in \cref{ch:imgproc,ch:guided-rewriting}.
For now, consider that \inlineElevateRise{reduceMapFusion} is applied to \inlineRise{map $\times$ $\then$ reduce + 0} in the high-level dot product program, replacing it with the right hand side of the rewrite and producing the following low-level program:

\begin{rise}[numbers=none, xleftmargin=0pt, label={dot-seq-rise}, caption={\inlineRise{dotSeq} program in \Rise{}, equivalent to \inlineRise{dot} from \cref{dot-rise}}.]
def dotSeq a b = zip a b $\then$ reduceSeq ($\lambda$acc x. acc + (fst x) $\times$ (snd x)) 0
\end{rise}

Generating the equivalent C or OpenCL code from this representation is conceptually straightforward, but has some technical challenges that have been explored in prior work~\cite{lift-ir-2017,dpia-2017} and are discussed in \cref{sec:codegen}.
Some patterns generate loops (e.g. \inlineRise{reduceSeq}), others generate scalar expressions (e.g. \inlineRise{+}, \inlineRise{$\times$}, \inlineRise{0}), and others affect indexing (e.g. \inlineRise{zip}, \inlineRise{fst}, \inlineRise{snd}).
The C function \lstinline{dotSeqC} is generated from \inlineRise{dotSeq}, implementing the dot product with a sequential reduction loop as expected:

\begin{c-code}[label=dot-seq-c-code, caption={\lstinline{dotSeqC} prorgam in C, generated from \cref{dot-seq-rise}}]
void dotSeqC(float* output, int n, float* a, float* b) {
    float acc;
    acc = 0.0f;
    for (int i = 0; i < n; i++) {
        acc = acc + (a[i] * b[i]);
    }
    output[0] = acc;
}
\end{c-code}

\subsection{High-Level and Low-Level \Rise{} Programs}
\label{sec:rise}

\Rise{} is a functional language with anonymous functions (written as \inlineRise{$\lambda$x. e} where \inlineRise{x} is a variable name and \inlineRise{e} a \Rise{} expression), familiar function application (written as \inlineRise{e e}), identifiers and literals.
The language is embedded in Scala, allowing meta-programming and the definition of macros, i.e. Scala code that will generate \Rise{} programs.
\Rise{} also defines a set of high-level patterns to describe computations as shown in \cref{fig:high-level-patterns} together with their types.
Formally, patterns are higher-order functions that can depend on natural numbers (\inlineRise{nat}) and data types (\inlineRise{data}).
\Rise{} has no support for general recursion or iteration, and instead relies on extensible and higher-level array patterns that explicitly expose data parallelism \cite{dean2008mapreduce, structured-parallel-programming}: \inlineRise{map} applies a function to each element of an array; \inlineRise{reduce} reduces all elements of an array to a single value given a binary reduction function.
Multi-dimensional array reshaping is common in array languages \cite{2006-SaC}, \Rise{} has patterns such as \inlineRise{split}, \inlineRise{join}, or \inlineRise{transpose} for this purpose.

\begin{rise}[caption={Selection of \Rise{} high-level patterns and their type},label={fig:high-level-patterns}]
$+$ $\mid$ $\times$        : t $\to$ t $\to$ t
map       : (s $\to$ t) $\to$ n.s $\to$ n.t
reduce    : (t $\to$ t $\to$ t) $\to$ t $\to$ n.t $\to$ t
split     : (n: nat) $\to$ (n $\times$ m).t $\to$ m.n.t
join      : n.m.t $\to$ (n $\times$ m).t
transpose : n.m.t $\to$ m.n.t
slide     : (sz sp: nat) $\to$ (sp $\times$ n + sz - sp).t $\to$ n.sz.t
zip       : n.s $\to$ n.t $\to$ n.(s $\times$ t)
fst       : (s $\times$ t) $\to$ s
snd       : (s $\times$ t) $\to$ t
\end{rise}

We write \inlineRise{s $\to$ t} for a function type with input of type $s$ and output of type $t$, \inlineRise{n.t} for an array type with $n$ elements of type $t$, \inlineRise{s $\times$ t} for a pair type with component types $s$ and $t$.
To avoid the challenge of storing functions (or closures) in heterogeneous hardware memory, the type system ensures that function types cannot be stored in memory, only data types (e.g. it is challenging for GPUs).

The \Shine{} compiler rewrites a high-level \Rise{} program into a low-level \Rise{} program that describes \emph{how} the result is computed, encoding implementation decisions explicitly.
\Rise{}'s low-level patterns (\cref{fig:low-level-patterns}) indicate specific implementation decisions.

\begin{rise}[caption={Selection of \Rise{} low-level patterns and their type},label={fig:low-level-patterns}]
mapSeq           : (s $\to$ t) $\to$ n.s $\to$ n.t
reduceSeq        : (s $\to$ t $\to$ s) $\to$ s $\to$ n.t $\to$ s
mapGlobal(dim)   : (s $\to$ t) $\to$ n.s $\to$ n.t
toMem            : (a: addr) $\to$ t $\to$ t
asVector         : (m: nat) $\to$ (n $\times$ m).t $\to$ n.<m>t
asScalar         : n.<m>t $\to$ (n $\times$ m).t
vectorFromScalar : t $\to$ <m>t
\end{rise}

For example, \inlineRise{mapSeq} and \inlineRise{reduceSeq} respectively implement \inlineRise{map} and \inlineRise{reduce} with sequential loops.
Some low-level patterns are specific to the target programming model (such as OpenCL) or hardware architecture (such as SIMD vector support).
For OpenCL, \inlineRise{mapGlobal} introduces parallelism by parallelising across the dimension \inlineRise{dim} of global threads.
The \inlineRise{toMem} pattern is used to explicitly encode storing an expression in the given address space in memory \inlineRise{(a: addr)}.
Other patterns enable SIMD vectorisation (e.g. \inlineRise{asVector}).
A vector type with $m$ elements of type $t$ is written as \inlineRise{<m>t}.


\subsection{Code Generation through DPIA}
\label{sec:codegen}

To generate imperative code (e.g. C or OpenCL) from a low-level \Rise{} program, a hybrid functional-imperative language called DPIA is used as an intermediate language.
The DPIA language \cite{dpia-2017} is a variation of idealised ALGOL \cite{reynolds1997-algol} that separates program \emph{phrases} into three categories: functional \emph{expressions}, imperative \emph{commands}, and imperative \emph{acceptors}.
One can think of commands as C statements, and of acceptors as C pointers that can be manipulated and ultimately used for writing to a memory location.
DPIA basically uses the same functional constructs and data types as \Rise{}.
Imperative constructs are added, such as assignment (\inlineDPIA{p = q}), sequential composition (\inlineDPIA{;}), memory allocation (\inlineDPIA{new}), and loops (\inlineDPIA{for}).
The type system reflects the separation into expressions (\inlineDPIA{exp[dt, rw]}), acceptors (\inlineDPIA{acc[dt]}), and commands (\inlineDPIA{comm}).
Functions and pairs can freely combine phrases of all categories.

A selection of DPIA patterns is shown in \cref{fig:dpia-functional,fig:dpia-imperative}.
While \Rise{} introduces patterns as higher-order functions with implicit type-level arguments more suitable for rewriting, DPIA introduces patterns as fully applied values with explicit type-level arguments more suitable for translation and code generation.

\begin{DPIA}[label={fig:dpia-functional}, caption={Selection of functional DPIA patterns}]
zip(n: nat, s: data, t: data, w: access,
    a: exp[n.s, w], b: exp[n.t, w]): exp[n.(s x t), w]
fst(s: data, t: data, x: exp[(s x t), read]): exp[s, read]
snd(s: data, t: data, x: exp[(s x t), read]): exp[t, read]
mapSeq(n: nat, s: data, t: data, f: exp[s, read] $\to$ exp[t, write],
       x: exp[n.s, read]): exp[n.t, write]
reduceSeq(n: nat, a: addr, s: data, t: data,
          f: exp[s, read] $\to$ exp[t, read] $\to$ exp[s, write],
          init: exp[s, write], x: exp[n.t, read]): exp[s, read]
toMem(a: addr, t: data, x: exp[t, write]): exp[t, read]
idx(n: nat, t: data, i: exp[idx[n], read], x: exp[n.t, read]): exp[t, read]
// ^ written as x[i]
\end{DPIA}

\begin{DPIA}[label={fig:dpia-imperative}, caption={Selection of imperative DPIA patterns}]
assign(t: data, p: acc[t], q: exp[t, read]): comm // written as p = q
seq(c1: comm, c2: comm): comm                     // written as c1 ; c2
new(a: addr, t: data, body: (exp[t, read] x acc[t]) $\to$ comm): comm
for(n: nat, body: exp[idx[n], read] $\to$ comm): comm
zipAcc1(n: nat, s: data, t: data, x: acc[n.(s x t)]): acc[n.s]
zipAcc2(n: nat, s: data, t: data, x: acc[n.(s x t)]): acc[n.t]
idxAcc(n: nat, t: data, i: exp[idx[n], read], a: acc[n.t]): acc[t]
\end{DPIA}

The type system of DPIA enforces important invariants and assumptions.
Expression types include an \inlineDPIA{access} annotation, which can either be \inlineDPIA{read} or \inlineDPIA{write} and enforces that DPIA programs must explicit encode how to write in what memory.
A \inlineDPIA{read} annotation signifies that the value can be read from memory while a \inlineDPIA{write} annotation indicates that the value must first be written to memory before being readable.
To convert a \inlineDPIA{write} expression into a \inlineDPIA{read} expression, the \inlineDPIA{toMem} pattern can be used and specifies a choice of address space.

Loop patterns like \inlineDPIA{for} iterate over bounded indices of type \inlineDPIA{idx[n]}, meaning that the index value belongs to the interval $[0; n[$.
This is important to statically enforce safe array accesses and to remove the need for runtime checks.

\paragraph{From \Rise{} to DPIA}
Going from low-level \Rise{} to functional DPIA requires two steps which we will not detail: (1) inferring access annotations to create DPIA types from \Rise{} types, the work of Bastian Köpcke; (2) translating patterns to be fully applied values with explicit type-level arguments, a personal engineering contribution.

To illustrate the second step, \inlineDPIA{$\lambda$f x. mapSeq(n, s, t, f, x)} would result from translating \inlineRise{mapSeq}, extracting \inlineDPIA{n, s, t} from the instantiated \Rise{} type.

\newcommand{\accT}{\mathcal{A}}
\newcommand{\conT}{\mathcal{C}}
\paragraph{Translation to Imperative}
Translating functional DPIA into imperative DPIA is performed by two mutually recursive translation functions (\cref{fig:types-acc-con-trans}).
\begin{DPIA}[label={fig:types-acc-con-trans}, caption={Type of the acceptor and continuation translations}]
$\accT$(e: exp[dt, write], out: acc[dt]): comm
$\conT$(e: exp[dt, read], k: exp[dt, read] $\to$ comm): comm
\end{DPIA}
The \emph{acceptor translation} $\accT$ produces a command which writes a functional \inlineDPIA{write} expression to the memory represented by an acceptor (\cref{fig:acceptor-trans}).
The \emph{continuation translation} $\conT$ produces a command which reads from a functional \inlineDPIA{read} expression and calls a \emph{continuation} function to continue the translation as required (\cref{fig:continuation-trans}).
To start the overall translation process, we need to know where to write the program result.
For this, we generate an output acceptor according to the data type computed by the functional expression before invoking the $\accT$ translation: \inlineDPIA{imperative = $\accT$(functional, output)}.
The \inlineDPIA{output} will later correspond to a runtime parameter in the generated C function (as in \cref{sec:rewriting}), or OpenCL kernel.

\begin{DPIA}[label={fig:acceptor-trans}, caption={Selection of acceptor translations}]
$\accT$(zip(n, s, t, write, a, b), out)  =
  $\accT$(a, zipAcc1(n, s, t, out)) ;
  $\accT$(b, zipAcc2(n, s, t, out))

$\accT$(mapSeq(n, s, t, f, x), out) = $\conT$(x, $\lambda$xT.
  for(n $\lambda$i. $\accT$(f(idx(n, s, i, xT), idxAcc(n, t, i, out)))))

$\accT$(reduceSeq(n, a, s, t, f, init, x), out) =
  $\conT$(reduceSeq(n, a, s, t, f, init, x))($\lambda$r. $\accT$(r)(out))
\end{DPIA}

\begin{DPIA}[label={fig:continuation-trans}, caption={Selection of continuation translations}]
$\conT$(zip(n, s, t, read, a, b), k) =
  $\conT$(a, $\lambda$aT. $\conT$(b, $\lambda$bT. k(zip(n, s, t, read, aT, bT))))

$\conT$(fst(s, t, read, x), k) = $\conT$(x, $\lambda$xT. k(fst(s, t, read, xT)))

$\conT$(snd(s, t, read, x), k) = $\conT$(x, $\lambda$xT. k(snd(s, t, read, xT)))

$\conT$(reduceSeq(n, a, s, t, f, init, x), k) =
 $\conT$(x, $\lambda$xT. new(a, s, $\lambda$(accE, accA).
   $\accT$(init, accA) ;
   for(n, $\lambda$i. $\accT$(f(accE, idx(n, t, i, xT)), accA)) ;
   k(accE)))

$\conT$(toMem(a, t, x), k) = new(a, t, $\lambda$(tmpE, tmpA). $\accT$(e, tmpA) ; k(tmpE))
\end{DPIA}

For example, translating the \inlineRise{dotSeq} program (\cref{dot-seq-rise}) from \Rise{} to functional DPIA and further to imperative DPIA leads to the following translation steps (with syntactic sugar):

\begin{DPIA}[label={dot-seq-imperative}]
$\accT$(reduceSeq ($\lambda$a x. a + (fst x) $\times$ (snd x)) 0 (zip a b), output)
=
$\conT$(reduceSeq ($\lambda$a x. a + (fst x) $\times$ (snd x)) 0 (zip a b))($\lambda$r. $\accT$(r)(output))
=
$\conT$(zip a b, $\lambda$z. new (exp[f32, read] x acc[f32]) in $\lambda$(accE, accA). (
  $\accT$(0, accA);
  for n $\lambda$i.
    $\accT$(accE + (fst z[i]) $\times$ (snd z[i]), accA);
  output = accE))
=
new (exp[f32, read] x acc[f32]) in $\lambda$(accE, accA). (
  $\accT$(0, accA);
  for n $\lambda$i.
    $\accT$(accE + (fst (zip a b)[i]) $\times$ (snd (zip a b)[i]), accA);
  output = accE)
=
new (exp[f32, read] x acc[f32]) in $\lambda$(accE, accA). (
  accA = 0.0f;
  for n $\lambda$i.
    (accA = accE + ((fst (zip a b)[i]) * (snd (zip a b)[i])));
  output = accE)
\end{DPIA}

\paragraph{Imperative Passes}
The use of DPIA in the \Shine{} compiler enables transformation passes to be applied at the imperative DPIA level, as we will show in \cref{sec:barrier-insertion}.
This is valuable because imperative DPIA has more convenient abstractions and richer semantics compared to C or OpenCL: for example, multi-dimensional arrays can be used and precise type information has not been erased yet.

\paragraph{Code Generation}
Finally, code such as C or OpenCL is generated from the imperative DPIA program.
One important transformation which happens at this stage is the translation of indexing patterns (e.g. \inlineDPIA{fst}, \inlineDPIA{snd}, \inlineDPIA{zip}) into array indexing, and the flattening \cite{blelloch1990vector} (or linearisation \cite{maslov1992delinearization}) of multi-dimensional arrays into one-dimensional arrays.
The C code generated for the \lstinline{dotSeq} program (\cref{dot-seq-rise}) was shown in \cref{dot-seq-c-code}.
\section{Implicit Barrier Insertion for \Shine{}}
\label{sec:barrier-insertion}

When multiple threads access the same memory concurrently and at least one thread is writing, synchronisation is required.
Without synchronisation, operation ordering is non-deterministic, and the computation may produce incorrect results: this type of bug is called a \emph{data race} \cite{1992-race-conditions}.

\newcommand{\barinsalg}{DPIA\textsubscript{BI}}
This section contributes \barinsalg{}, a synchronisation barrier insertion algorithm for \Shine{}.
\barinsalg{} transforms imperative DPIA programs and does not need to be modified when extending functional \Rise{} patterns, as opposed to the barrier elimination algorithm of \Lift{} \cite{lift-ir-2017, remmelg2019-thesis}.
Following \Lift{}'s design, barrier insertion is implicit and not controllable by rewriting, allowing \Rise{} programs to ignore low-level synchronisation requirements.

\subsection{Synchronisation in OpenCL}
\label{opencl-sync}

In OpenCL kernels, calling \code{barrier(flag)} synchronises work-items from the same work-group.
The \code{flag} argument specifies which address spaces require a memory fence to ensure correct ordering of memory operations (\code{CLK\_LOCAL\_MEM\_FENCE} for local memory and \code{CLK\_GLOBAL\_MEM\_FENCE} for global memory).
We must also ensure that all work-items execute the barrier, to avoid undefined behaviour and potential \emph{deadlocks} \cite{isloor1980deadlock} where some work-items are indefinitely waiting for others to encounter the barrier:

\begin{displayquote}[OpenCL 1.2 Specification]
All the work-items of a work-group must execute the barrier before any are allowed to continue execution beyond the barrier. Note that the work-group barrier must be encountered by all work-items of a work-group executing the kernel or by none at all.
\end{displayquote}

OpenCL does not support synchronisation across work-groups inside a kernel, instead multiple kernels must be launched.
We will discuss how to add support for launching multiple kernels in \cref{sec:host-codegen} and focus on work-item barriers in this section.

\subsection{Barrier Insertion as an Imperative \barinsalg{} Transformation}

Because functional \Rise{} programs ignore low-level synchronisation requirements, code generation must take care of inserting work-item barriers.
Barriers are relatively expensive \cite{oboyle1995synchronization, 2011-barrier-overhead, 2018-warp-consolidation}, to maximise performance of the generated code, we seek to minimise the number of barriers executed at runtime while ensuring correct synchronisation.

Previous work on \Lift{} follows a pessimistic approach, first inserting barriers at the end of all parallel map patterns to avoid any data-race (e.g. after \inlineRise{mapGlobal}, \inlineRise{mapWorkGroup} and \inlineRise{mapLocal} patterns), then eliminating barriers as an optimisation \cite{lift-ir-2017, remmelg2019-thesis}.
There are two problems with this approach.
First, the pessimistic approach is not sufficient to guarantee correctness because it does not ensure that all work-items execute the barriers.
Following the pessimistic approach without any optimisation generates programs with undefined behaviour, as expanded on in \cref{sec:barrier-insertion-limitations}.
Second, \Lift{}'s barrier elimination is analysing how functional patterns such as \inlineRise{mapLocal} are composed, and their interaction with other patterns that might result in data sharing between work-items (\inlineRise{split}, \inlineRise{join}, etc).
We argue that this requires the definition of too many special cases (for example, special treatment is required when encountering several \inlineRise{mapGlobal} inside the arguments of a \inlineRise{zip}, or nested \inlineRise{mapLocal} patterns).
Additionally, such an approach may require modifying the barrier elimination optimisation when new functional patterns are added, reducing the ease of compiler extension.

Therefore, we follow instead an optimistic approach, inserting barriers as required in imperative DPIA programs.
Before generating OpenCL kernel code from imperative DPIA, the following imperative passes are applied in sequence:
\begin{enumerate}[nosep]
  \item Inject work item sizes. The OpenCL number of groups, local size and global size values are inlined if they are statically known.
  \item Flag private array loops. Loops iterating over private arrays are flagged to be unrolled, which helps OpenCL perform register allocation.
  \item Unroll loops. The unrolled loops have been flagged in the previous pass, or explicitly during rewriting. 
  \item Simplify nats. Tries to simplify certain arithmetic expressions (e.g. $x \times 0 = 0$), such simplifications are not only applied here but also throughout the compilation process.
  \item \textbf{Insert memory barriers}. The topic of this section realised in the \barinsalg{} algorithm.
  \item Hoist memory allocations.
  In OpenCL, global and local memory cannot be allocated while the kernel is running, it must be allocated upfront.
  This pass hoists such memory allocations as required (e.g. outside of loops). 
  \item Adapt kernel parameters.
  Kernel parameters in global or local memory which have a scalar type in DPIA are represented as arrays of size 1 for OpenCL.
\end{enumerate}

\subsection{Definition of \barinsalg{}}

\barinsalg{} is defined by analysing reads and writes to memory, making conservative barrier insertions to ensure synchronisation according to the observed data dependencies.

The \barinsalg{} analysis records memory reads and writes in two mutable dictionaries from memory identifier to address space, one representing reads from outer scope allocations, and one representing work-group parallel writes to outer scope allocations.
This corresponds to the following record structure (a case class is used in Scala):
\begin{DPIA}
record D(reads: MutMap[Ident, Address], wg_writes: MutMap[Ident, Address])
\end{DPIA}

\newcommand{\barinsalgfrec}{$\text{DPIA}_\text{BI}^\text{fresh\_rec}$}
\newcommand{\barinsalgrec}{$\text{DPIA}_\text{BI}^\text{rec}$}
\newcommand{\barinsalglrec}{$\text{DPIA}_\text{BI}^\text{loop\_body}$}

\barinsalg{} is defined recursively in \cref{lst:barrier-insertion}.
The \inlineDPIA|$\text{\barinsalg{}}$| function is the entry point and calls \inlineDPIA|$\text{\barinsalgfrec{}}$|, which creates empty data of type \inlineDPIA{D} before calling \inlineDPIA|$\text{\barinsalgrec{}}$|.
Barriers are represented by the \inlineDPIA{barrier} DPIA command and may be inserted in-between sequences (inserted for \inlineDPIA{seq} in \cref{line:seq-barrier}), or at the end of loop bodies (inserted for \inlineDPIA{for} or \inlineDPIA{parFor} in \cref{line:loop-barrier}).

\clearpage
\begin{DPIA}[escapechar=|,numbers=left,label={lst:barrier-insertion},caption={Pseudo-code for the \barinsalg{} barrier insertion algorithm}]
def |\barinsalg{}|(p: DPIA$_{comm}$): DPIA$_{comm}$ = |\barinsalgfrec{}|(p, Map())$_1$

def |\barinsalgfrec{}|(p: DPIA$_{comm}$,
             allocs: Map[Ident, Address]): (DPIA$_{comm}$, D) =
  def d = D(MutMap(), MutMap())
  (|\barinsalgrec{}|(p, allocs, d), d)

def |\barinsalgrec{}|(p: DPIA$_{comm}$, allocs: Map[Ident, Address], d: D): DPIA$_{comm}$ =
  match p
  for(n, $\lambda$x. body) =>
    for(n, $\lambda$x. |\barinsalglrec{}|(body, allocs, d, MutMap()))
  parFor(level, ..., out, $\lambda$x o. body) =>
    parFor(level, ..., out, $\lambda$x o. |\barinsalglrec{}|(body, allocs, d, 
      if level == local then ~collect writes by inspecting~ out |\label{line:collect-writes}|
      else MutMap()))
  new(adr, dt, $\lambda$x. body) =>
    def allocs2 = if adr != private then allocs + (x -> adr) else allocs |\label{line:collect-allocs}|
    new(adr, dt, $\lambda$x. |\barinsalgrec{}|(body, allocs2, d))
  assign(dt, lhs, rhs) =>
    ~collect reads into~ d.reads ~by inspecting~ rhs |\label{line:collect-reads}|
    assign(dt, lhs, rhs)
  seq(a, b) =>
    def (a2, ad) = |\barinsalgfrec{}|(a, allocs)
    def (b2, bd) = |\barinsalgfrec{}|(b, allocs)
    def dependencies = (ad.reads.keys $\cap$ bd.wg_writes.keys)
      $\cup$ (bd.reads.keys $\cap$ ad.wg_writes.keys)
    ~extend~ d.reads ~with~ ad.reads ~and~ bd.reads
    ~extend~ d.wg_writes ~with~ ad.wg_writes ~and~ bd.wg_writes
    if dependencies ~non empty~ then
      seq(a2, seq(~make~ barrier ~according to~ dependencies, b2)) |\label{line:seq-barrier}|
    else
      seq(a2, b2)
  [...]

def |\barinsalglrec{}|(p: DPIA$_{comm}$, allocs: Map[Ident, Address], d: D,
         outer_wg_writes: MutMap[Ident, Address]): DPIA$_{comm}$ =
  def p2 = |\barinsalgrec{}|(p, allocs, d)
  def dependencies = d.wg_writes.keys $\cap$ d.reads.keys
  if dependencies ~non empty~ then
    ~clear~ d.reads ~and~ d.wg_writes
    seq(p2, ~make~ barrier ~according to~ dependencies) |\label{line:loop-barrier}|
  else
    ~extend~ d.wg_writes ~with~ outer_wg_writes
    p2
\end{DPIA}

To analyse dependencies, reads and work-group parallel writes are collected in lines \ref{line:collect-reads} and \ref{line:collect-writes}.
The \inlineDPIA{parFor} pattern corresponds to parallel loops (e.g. \inlineDPIA{parFor(local, ...)} corresponds to \inlineRise{mapLocal}), and differently from sequential loops, explicitly mentions its global output (\inlineDPIA{out}) and its local output (\inlineDPIA{o}), making analysis easier.
Information about memory allocations and their address space is also collected in line \ref{line:collect-allocs}, since this information is critical to select the right barrier flags (\inlineDPIA{CLK_LOCAL_MEM_FENCE} and \inlineDPIA{CLK_GLOBAL_MEM_FENCE}).

\subsection{Evaluating the Correctness and Efficiency of \barinsalg{}}
\label{barinsalg-eval}

\paragraph{Experimental Setup}
The correctness and efficiency of barrier insertion is evaluated on 48 programs (38 unit tests and 10 benchmarks), by observing and comparing the code generated by \Lift{} and \Shine{}.
The generated code is considered correct if it respects the OpenCL 1.2 specification.
The generated code is considered more efficient if it executes less barriers at runtime, or uses less flags.

The 38 unit tests mainly come from the \Lift{} repository\footnote{\scriptsize\url{https://github.com/lift-project/lift/blob/5e8a18df48ab791ae66016007e18826047a015f7/src/test/opencl/generator/TestBarrier.scala}}, with a few additions.
The unit tests compose various patterns in various ways: parallel and sequential map patterns to introduce threads and loops, \inlineRise{toMem} to introduce temporary memory, and indexing patterns to introduce data sharing between threads.

The 10 benchmarks come from the experimental evaluation of \Lift{} presented in \cite{lift-ir-2017}.
The benchmarks target GPU hardware and represent different domains including physics simulations, statistics, and linear algebra.
Two benchmarks were ported but elided from the results, as they are not fairly comparable: "NBody NVIDIA" and "MM NVIDIA".
\Shine{} allocates additional memory for these programs due to orthogonal issues.

\paragraph{Correctness and Efficiency Results}
\Cref{fig:barrier-insertion-eval} summarises the results, identifying 6 differences between the code generated by \Shine{} and the one generated by \Lift{}:

\newcommand{\valper}[2]{\makebox[3ex][r]{#1} (\makebox[3ex][r]{#2}\%)}
\begin{table}[bh]
  \begin{tabular}{|l|l|c|c|c|} \hline
  \textbf{Impact} & \textbf{Difference} & \textbf{Unit tests} & \textbf{Benchmarks} & \textbf{Total} \\ \hline
  \textit{\color{red}\Lift{} incorrect} & Incorrect flags & \valper{13}{34} & \valper{1}{10} & \valper{14}{29} \\ \hline
  \multirow{2}{*}{\Lift{} more efficient} & Better analysis & \valper{13}{34} & \valper{0}{0} & \valper{13}{27} \\ \cline{2-5}
                                          & Single iteration loops & \valper{6}{16} & \valper{1}{10} & \valper{7}{15} \\ \hline
  \multirow{2}{*}{\Rise{} more efficient} & Better analysis & \valper{0}{0}  & \valper{2}{20} & \valper{2}{4} \\ \cline{2-5}
                                          & Barrier position & \valper{1}{2}  & \valper{0}{0} & \valper{1}{2} \\ \hline
  Trade-off & Sync. vs alloc. & \valper{7}{18} & \valper{0}{0} & \valper{7}{15} \\ \hline
  \multicolumn{2}{|l|}{\textbf{Total}} & \valper{38}{100} & \valper{10}{100} & \valper{48}{100} \\ \hline
  \end{tabular}
  \caption{Number and percentage of unit tests and benchmarks that exhibit each difference identified in the code generated by \Shine{} and \Lift{}.
  \Shine{} generates correct barriers in all cases, which is not the case of \Lift{}.}
  \label{fig:barrier-insertion-eval}
\end{table}

\begin{enumerate}
  \item \emph{Incorrect flags (\Lift{} incorrect).} \label{bi-correctness}
  While \Shine{} generates correct barriers for all programs, \Lift{} generates barriers with incorrect flags for 14 programs (29\%).
  For the test shown in \cref{fig:barrier-insertion-outcome-1}, a barrier on global memory is generated instead of a barrier on local memory.
  In practice, the program still computes the expected result, however the OpenCL 1.2 specification is not respected and this could lead to undefined behaviour with a different OpenCL implementation.
  With \barinsalg{}, \Shine{} inserts barriers with correct flags: one in line \ref{line:test1-barrier1} to protect the read-after-write dependency on local memory, and one in line \ref{line:test1-barrier2} to protect the write-after-read dependency between outer loop iterations.

\begin{figure}
\begin{rise}
$\lambda$in : n.m.f32. (
in $\then$ mapWorkGroup(0) (
  mapLocal(0) ($\lambda$x. x) $\then$
  toMem local $\then$
  slide 3 1 $\then$
  mapLocal(0) sum
))
\end{rise}
\begin{c-code}[escapechar=|,morecomment={[f][\color{Green}]{+\ }},morecomment={[f][\color{OrangeRed}]{-\ }},basicstyle=\ttfamily\scriptsize]
for (int wg0 = get_group_id(0); wg0 < n; wg0 += get_num_groups(0)) {
  for (int l0 = get_local_id(0); l0 < m; l0 += get_local_size(0)) {
    [...] // read from global input; write to local memory
  }
  barrier(CLK_LOCAL_MEM_FENCE); |\label{line:test1-barrier1}|
  for (int l1 = get_local_id(0); l1 < m-2; l1 += get_local_size(0)) {
    [...] // read from local memory; write to global output
  }
- barrier(CLK_GLOBAL_MEM_FENCE);
+ barrier(CLK_LOCAL_MEM_FENCE); |\label{line:test1-barrier2}|
}
\end{c-code}
\caption{Example low-level \Rise{} program and its generated OpenCL code. In red (-), barrier that would be generated by \Lift{}. In green (+), barrier generated by \Shine{}. The second barrier flag is incorrect with \Lift{}, but correct with \Shine{}.}
\label{fig:barrier-insertion-outcome-1}
\end{figure}
  
  \item \emph{Better analysis (\Lift{} more efficient).} \label{bi-analysis-precision}
  \barinsalg{} only tracks memory accesses for entire allocations.
  \Lift{} generates fewer barriers for 13 programs (27\%) by reasoning about data sharing in the functional program.
  This difference is not visible in the benchmarks that exclusively use local memory to share data between work-items.
  However, integrating a more precise dependency analysis into \barinsalg{}, for example by tracking accessed intervals using symbolic natural numbers, would enable \Shine{} to generate as many or fewer barriers than \Lift{} in all observed cases.

  \item \emph{Single iteration loops (\Lift{} more efficient).} \label{bi-single-loops}
  \Lift{} avoids generating barriers for loops whose body is only executed once in 7 programs (15\%), while \Shine{} generates superfluous barriers.
  \Shine{} could avoid generating such barriers by eliminating single iteration loops before barrier insertion, detecting such loops during barrier insertion, or by differentiating between two scenarios which are currently conflated both in \Lift{} and \Rise{}: (1) the loop will be executed again; or (2) the loop will be exited.

  \item \emph{Better analysis (\Rise{} more efficient)}.
  \Rise{} avoids generating barriers that \Lift{} generates, or uses more precise flags, in 2 programs (4\%).
  This is because \Lift{} fails to reason about specific combinations of functional patterns, while \Shine{} handles these combinations naturally at the imperative level, without any special treatment.

  \item \emph{Barrier position (\Rise{} more efficient)}.
  \Lift{} only inserts barriers at the end of parallel map patterns.
  For 1 program (2\%) this strategy is not flexible enough and leads to inefficient barriers compared to \Shine{}.
  \Cref{fig:barrier-insertion-outcome-2} shows an example where \Shine{} is able to generate barriers at the end of sequential loops, which results in a program executing $2 \times (m - 1) \times n$ times fewer barriers at runtime ($m - 1$ less per sequential loop execution).

  \item \emph{Synchronisation vs allocation (Trade-off)}.
  For 7 (15\%) programs, \Lift{} allocates more memory than \Rise{}.
  As a result, \Lift{} generates less barriers because the memory is not re-used while \Rise{} generates more barriers because the memory is re-used.
  It is unclear which trade-off is best without benchmarking a specific program on a specific hardware target.
  It may even be that a fine-grain combination of different options for different allocations is optimal.
  Future work may explore this trade-off.
\end{enumerate}

\begin{figure}[h]
\begin{rise}
$\lambda$in : n.m.o.f32. (
  in $\then$ mapWorkGroup(0) (
    mapSeq (mapLocal(0) ($\lambda$x. x)) $\then$
    toMem local $\then$
    map (slide 3 1) $\then$
    mapSeq (mapLocal(0) sum)
  ))
\end{rise}
\begin{c-code}[escapechar=|,morecomment={[f][\color{Green}]{+\ }},morecomment={[f][\color{OrangeRed}]{-\ }},basicstyle=\ttfamily\scriptsize]
for (int wg0 = get_group_id(0); wg0 < n; wg0 += get_num_groups(0)) {
  for (int i = 0; < m; i += 1) {
    for (int l0 = get_local_id(0); l0 < o; l0 += get_local_size(0)) {
      [...] // read from global input; write to local memory
    }
-   barrier(CLK_LOCAL_MEM_FENCE);
  }
+ barrier(CLK_LOCAL_MEM_FENCE); |\label{line:test2-barrier1}|
  for (int i = 0; < m; i += 1) {
    for (int l1 = get_local_id(0); l1 < o-2; l1 += get_local_size(0)) {
      [...] // read from local memory; write to global output
    }
-   barrier(CLK_GLOBAL_MEM_FENCE);
  }
+ barrier(CLK_LOCAL_MEM_FENCE); |\label{line:test2-barrier2}|
}
\end{c-code}
\caption{Example low-level \Rise{} program and its generated OpenCL code.
In red (-), barrier that would be generated by \Lift{}.
In green (+), barrier generated by \Shine{}.
The second barrier flag is only correct with \Shine{}.
The code generated by \Shine{} is also more efficient: $2 \times (m - 1) \times n$ times less barriers will be executed at runtime.}
\label{fig:barrier-insertion-outcome-2}
\end{figure}



\subsection{Limitations of \barinsalg{}}
\label{sec:barrier-insertion-limitations}

In addition to the "\emph{Better analysis}" and "\emph{Single iteration loops}" issues (\cref{bi-analysis-precision,bi-single-loops} above), there are four other limitations worth mentioning.

\paragraph{Barrier Reachability}
In some cases, the generated barriers might not be reached by all work-items, which would result in undefined behaviour.
By optimising barrier placement on the 48 programs from \cref{barinsalg-eval}, this issue is avoided in 11 unit tests and 3 benchmarks compared to following the naive pessimistic approach of \Lift{} with barrier elimination optimisations turned off.
However, even when optimising barrier placement, the issue remains in 2 unit tests.
An example is given in \cref{fig:barrier-insertion-divergence-limitation}, where part of the work-items might not enter the loop in line \ref{line:partial-wg-loop}.
Previous work on \Lift{} suffered from similar limitations and implemented a mix of compilation time and runtime checks to report the issue. 
For \Shine{}, an additional imperative DPIA pass could be implemented to fix the code as illustrated in \cref{fig:barrier-insertion-divergence-limitation}.

\begin{figure}[th]
\begin{rise}
$\lambda$in : n.m.o.p.f32. (
  in $\then$ mapWorkGroup(1) (mapWorkGroup(0) (
    mapLocal(1) (
      mapLocal(0) ($\lambda$x. x) $\then$
      toMem(local) $\then$
      slide 3 1 $\then$
      mapLocal(0) sum
    ))))
\end{rise}
\begin{c-code}[escapechar=|,morecomment={[f][\color{Green}]{+\ }},morecomment={[f][\color{OrangeRed}]{-\ }},basicstyle=\ttfamily\scriptsize]
for (int wg0 = get_group_id(1); wg0 < n; wg0 += get_num_groups(1)) {
  for (int wg1 = get_group_id(0); wg1 < m; wg1 += get_num_groups(0)) {
-   for (int l0 = get_local_id(1); l0 < o; l0 += get_local_size(1)) { |\label{line:partial-wg-loop}|
+   for (int l0 = get_local_id(1); l0 < ctt(o); l0 += get_local_size(1)) {
+     if (l0 < o) {
      for (int l1 = get_local_id(0); l1 < p; l1 += get_local_size(0)) {
        [...] // read from global input; write to local memory
      }
+     }
      barrier(CLK_LOCAL_MEM_FENCE);
+     if (l0 < o) {
      for (int l2 = get_local_id(0); l2 < p-2; l2 += get_local_size(0)) {
        [...] // read from local memory; write to global output
      }
+     }
      barrier(CLK_LOCAL_MEM_FENCE);
    }
  }
}
\end{c-code}
\caption{Example low-level \Rise{} program and its generated OpenCL code where barriers may not be encountered by all work-items. In red (-), buggy code that would be generated by both \Lift{} and \Shine{}. In green (+), a potential fix where the \code{ctt} function rounds up a number to a multiple of the involved work-items.}
\label{fig:barrier-insertion-divergence-limitation}
\end{figure}

\paragraph{Extensibility of Indexing Patterns}
To simplify \barinsalg{} further and separate concerns, it would be valuable to eliminate indexing patterns (e.g. \inlineDPIA{split}, \inlineDPIA{join}) beforehand.
This would enable extending indexing patterns without having to adjust the "collect reads" operation, which would only need to deal with \inlineDPIA{idx} patterns (line \ref{line:collect-reads} of \cref{lst:barrier-insertion}).

\paragraph{Extensibility of Imperative Patterns}
Adding a new imperative DPIA pattern requires adding a new case in \cref{lst:barrier-insertion}, reducing the ease of extensibility.
However, we argue that this improves on previous \Lift{} work where barrier elimination happened at the functional level.
In \Shine{}, it is possible to add new functional patterns without having to adapt \barinsalg{}, as long as no new imperative patterns are required.

\paragraph{Formal Verification}
No proof of correctness is provided. Future work could provide an algorithm that provably prevents any data race or undefined behaviour for any well-formed low-level \Rise{} program.

\subsection{Related Work}

\paragraph{Barriers for Collective Patterns}
When using data-parallel collective patterns like \inlineRise{map}, a common approach is to insert synchronisation barriers after collective operations.
This is the approach taken by the array-based languages \Lift{} \cite{lift-rewrite-2015}, Futhark \cite{futhark-2017}, and SaC \cite{2005-SaC-sync}.




\paragraph{Barriers for Imperative Programs}
When transforming imperative code, using dependency analyses to reason about synchronisation barriers is common \cite{2013-barrier}.
Both barrier insertion \cite{oboyle1995synchronization, 2005-barrier-placement, 2010-polyhedral-barriers, 2010-polyhedral-C-CUDA-barrier, 2017-polyhedral-parallel-barriers} and barrier elimination \cite{1995-barrier-elimination, 2002-barrier-minimization, 2017-polyehdral-barrier-lifting, 2019-Triton} have been studied, often using advanced dependency analyses, such as polyhedral analyses.
The Tiramisu compiler explicitly exposes synchronisation decisions to the performance engineer, checking validity using the polyhedral model \cite{2018-tiramisu-sync}.

\paragraph{Barriers in Domain-Specific Compilers}
The problem of inserting correct and efficient barriers is barely discussed by the Halide and TVM papers.
For example, the TVM paper \cite{tvm-2018} simply mentions that "memory synchronization barriers must be properly inserted to guarantee that shared loaded data is visible to consumers".
Following papers show that, for some use cases, the TVM barrier insertion pass is unsatisfactory, leading to poor performance and forcing users to modify the compiler \cite{2021-Cortex-barrier}.

~\\
The novelty of \barinsalg{} is to insert barriers using imperative dependency analyses in a domain-extensible compiler for a functional language
based on data-parallel collective patterns.
We showed how, in the \Shine{} domain-extensible compiler, the design of \barinsalg{} facilitates extending functional \Rise{} patterns and can even lead to generating more efficient barriers (\cref{fig:barrier-insertion-outcome-2}).


\subsection{Summary}

This section contributes \barinsalg{}, a synchronisation barrier insertion algorithm for \Shine{} that transforms imperative DPIA programs.
Crucially, \barinsalg{} does not need to be modified when extending \Rise{} patterns, as opposed to the barrier elimination algorithm of \Lift{} \cite{lift-ir-2017, remmelg2019-thesis}.
Following \Lift{}'s design, barrier insertion is implicit and not controllable by rewriting, allowing \Rise{} programs to ignore low-level synchronisation requirements.


The correctness and efficiency of \barinsalg{} is evaluated on 38 unit tests and 10 benchmarks, mostly taken from prior \Lift{} work.
We identify 6 differences in the code generated by \Shine{} and \Lift{} (\cref{fig:barrier-insertion-eval}), and observe that \barinsalg{} fixes \Lift{} bugs in 13 unit tests and 1 benchmark.
There is only 1 benchmark where \Shine{} inserts a barrier that \Lift{} eliminates, and we provide a clear pathway to improve \barinsalg{} to generate more efficient barriers than \Lift{} on all 48 unit tests and benchmarks.
\section{Explicit Kernel Execution for \Shine{}}
\label{sec:host-codegen}

Any high-level language that allows offloading computations using the OpenCL or CUDA programming models must deal with executing kernels from the host.
\Shine{} allows \Rise{} programs to be compiled for CPUs by generating C code, or to various hardware by generating OpenCL kernel code.
However, so far \Rise{} programs cannot be compiled for a host CPU while offloading computations to various hardware devices.
Even if no computations are performed on the host, host-side calls to the OpenCL C API are tightly coupled to the use of OpenCL kernels.
The host is responsible for dynamically allocating global/local memory, launching kernels, and transferring data between host and device.
Launching multiple kernels is also the only way to synchronise across work-groups, if strictly following the OpenCL 1.2 specification (\cref{opencl-sync}).

This section enables explicit kernel execution in \Rise{} by adding the \inlineRise{oclRun} pattern.
Crucially, this design makes kernel execution controllable by rewriting in the domain-extensible \Shine{} compiler, enabling to explore different kernel decompositions of a high-level \Rise{} program via rewriting.
The \inlineRise{oclRun} pattern allows a single low-level \Rise{} program to express both host-side and device-side computations.
\Shine{} is modified to generate imperative code for multiple OpenCL kernels, as well as the necessary C host code to launch them.
To simplify host code generation, we implement a thin runtime called LRA that abstracts over some of the OpenCL runtime details.
The focus is to support executing kernels on a single OpenCL device, support for multiple devices is left for future work.




\subsection{Adding a \Rise{} Pattern for Kernel Execution}

To make device execution explicit, while keeping the interface as simple as possible, we introduce an \inlineRise{oclRun} pattern.
This pattern is an identity function which takes 3D local and global thread counts as additional input, it has the following type:
\begin{rise}[numbers=none]
oclRun : (ls0, ls1, ls2, gs0, gs1, gs2 : nat) $\to$ t $\to$ t
\end{rise}
The input expression of \inlineRise{oclRun} will be computed on the device by generating and calling a corresponding kernel.
We also provide syntactic sugar to enable providing 1D, 2D or 3D local and global thread counts, where the thread counts of unspecified dimensions are set to 1:
\begin{rise}[numbers=none]
oclRun(ls: LocalSize, gs: GlobalSize) = oclRun ls.x ls.y ls.z gs.x gs.y gs.z
\end{rise}

This new pattern can be used to write \Rise{} programs such as the one in \cref{fig:kernels-example}, where \inlineRise{store} is similar to \inlineRise{toMem} without the OpenCL address space parameter, and with a function parameter to provide name binding.
First, the value 3 is added to each element of \inlineRise{in} in parallel on the device (line \ref{line:device-exec1}), storing the result into \inlineRise{x1}.
Second, the value 1 is subtracted to each element of \inlineRise{x1} sequentially on the host (line \ref{line:host-exec}), storing the result into \inlineRise{x2}.
Finally, the value 1 is added to each element of \inlineRise{x2} in parallel on the device (line \ref{line:device-exec2}), producing the final output.

\begin{rise}[label={fig:kernels-example}, caption={Example low-level \Rise{} program mixing host computation (middle \inlineRise{mapSeq} call) and device computation (both \inlineRise{oclRun} calls).}]
store : (s $\to$ t) $\to$ s $\to$ t

$\Lambda$n : nat. $\lambda$in : n.i32. (
  oclRun(LocalSize(2), GlobalSize(n/2)) (mapGlobal (add 3) in) $\then$ store ($\lambda$x1.$\label{line:device-exec1}$
  mapSeq ($\lambda$y. y-1) x1 $\then$ store ($\lambda$x2.$\label{line:host-exec}$
  oclRun(LocalSize(4), GlobalSize(n/4)) (mapGlobal (add 1) x2)))$\label{line:device-exec2}$
)
\end{rise}

\subsection{Creating a Lightweight Runtime Abstraction (LRA)}

Low-level OpenCL host code needs to explicitly manage devices, kernels and memory in great detail.
To simplify code generation, a lightweight runtime abstraction called LRA is designed, delegating some implementation decisions to the runtime implementation.
This runtime is meant to be simplistic and to enable quick prototyping of our ideas, future work may implement or re-use more sophisticated runtime libraries, if more advanced features are required.

\paragraph{Buffer Abstraction}
The most important abstraction provided by LRA is the \lstinline{Buffer} abstraction which offers a unified and consistent view on memory that may reside on the host or on the device, as required during program execution (\cref{fig:buffer-abs}).
This is not a new idea, as libraries such as SPOC \cite{spoc-2014}, SkePU \cite{skepu-2010} and SkelCL \cite{skelcl-2011} provide a similar abstraction.

A \lstinline{Buffer} is created with \lstinline{createBuffer} and deleted with \lstinline{destroyBuffer}.
Creating a buffer requires specifying its size in bytes and the type of accesses that it should support.
\lstinline{AccessFlags} express a combination of reads/writes from the host/device.
To access a buffer, explicit synchronisation functions must be called.
\lstinline{hostBufferSync} allows accessing a buffer from the host, returning a \lstinline{void*} pointer that may be used in regular C code.
\lstinline{deviceBufferSync} allows accessing a buffer from the device, returning a \lstinline{DeviceBuffer} structure that may be used as a kernel parameter.
The runtime user (e.g. the \Shine{} compiler) is responsible for avoiding data races while using this interface.\footnote{reminder: multiple threads should not access the same memory concurrently if at least one thread is writing}

\begin{c-code}[label={fig:buffer-abs},caption={Interface for the \lstinline{Buffer} abstraction.}]
typedef enum {
  HOST_READ = 1 << 0,
  HOST_WRITE = 1 << 1,
  DEVICE_READ = 1 << 2,
  DEVICE_WRITE = 1 << 3,
} AccessFlags;

Buffer createBuffer(Context ctx, size_t byte_size, AccessFlags access);
void destroyBuffer(Context ctx, Buffer b);
void* hostBufferSync(Context ctx, Buffer b,
                     size_t byte_size, AccessFlags access);
DeviceBuffer deviceBufferSync(Context ctx, Buffer b,
                              size_t byte_size, AccessFlags access);
\end{c-code}

We provide two different \lstinline{Buffer} implementations: one with memory copies, and one without.
With memory copies, the buffer is allocated both in host memory and in device memory.
Memory copies are issued between both allocations as required to maintain consistency.
Without memory copies, the buffer is allocated in memory that is accessible by both host and device.
Memory synchronizations are issued as required to maintain consistency.
In both cases, memory copies and synchronisations are performed lazily by keeping track of dependencies between accesses at runtime in a very simple and coarse-grain manner.
Each implementation requires no more than 80 lines of C code, illustrating how lightweight the abstractions are.

\paragraph{Context Abstraction}
As already visible in \cref{fig:buffer-abs}, LRA uses a \lstinline{Context} abstraction (\cref{fig:context-abs}).
The context is used to manage global state required for OpenCL kernel execution.
A \lstinline{Context} may be created using \lstinline{createDefaultContext} or \lstinline{createContext}, and deleted using \lstinline{destroyContext}.
\lstinline{createContext} enables specifying the OpenCL platform and OpenCL device desired for execution, if multiple ones are available.

\begin{c-code}[label={fig:context-abs},caption={Interface for the \lstinline{Context} abstraction.}]
Context createDefaultContext();
Context createContext(const char* platform_subname,
                      const char* device_type_str);
void destroyContext(Context ctx);
\end{c-code}

\paragraph{Kernel Abstraction}
Finally, LRA provides a \lstinline{Kernel} abstraction (\cref{fig:kernel-abs}).
A \lstinline{Kernel} is loaded from OpenCL kernel source code (\lstinline{loadKernelFrom...}) encoded in a string, or in a file.
A \lstinline{Kernel} is launched using \lstinline{launchKernel}, which requires specifying global and local thread counts as well as providing kernel arguments (e.g. scalars, device buffers).
It is deleted using \lstinline{destroyKernel}.

\begin{c-code}[label={fig:kernel-abs},caption={Interface for the \lstinline{Kernel} abstraction.}]
Kernel loadKernelFromSource(Context ctx, const char* name,
                            const char* source, size_t length);
Kernel loadKernelFromFile(Context ctx, const char* name, const char* path);
void launchKernel(Context ctx, Kernel k,
                  const size_t* global_size, const size_t* local_size,
                  size_t arg_count, const KernelArg* args);
void destroyKernel(Context ctx, Kernel k);
\end{c-code}

~\\
In addition to the two possible \lstinline{Buffer} implementations, we provide a single implementation for \lstinline{Context} and \lstinline{Kernel}, which consists of about 300 lines of boilerplate C code.

\paragraph{Code Generation Objective}
Given LRA, the code we want to generate from \cref{fig:kernels-example} is depicted in \cref{fig:kernels-code}.
The kernel source codes are provided as string constants (e.g. \lstinline{k0_source}), each one corresponding to an \inlineRise{oclRun} call.
A \lstinline{add3_t} structure holds the execution environment for the computation (i.e. the two \lstinline{Kernel}s), a \lstinline{add3_init} function initializes this environment and a \lstinline{add3_destroy} function deletes it.
Finally, a \lstinline{add3_run} function runs the actual computation: allocating \lstinline{Buffer}s, performing synchronisations, launching kernels and performing host-side computations.

Crucially, if the \Rise{} program is rewritten differently, the interface of the generated code does not change (i.e. the signature of \lstinline{add3_init}, \lstinline{add3_destroy} and \lstinline{add3_run}).
The \lstinline{Buffer} abstraction is particularly important to enable this: inputs and outputs may freely reside on either host or device without having to generate different code, as the runtime will take care of both cases seamlessly.

\clearpage
\begin{c-code}[label={fig:kernels-code}, caption={Code using LRA that should be generated from \cref{fig:kernels-example}},basicstyle=\ttfamily\scriptsize]
const char k0_source[] =
"__kernel __attribute__ ((reqd_work_group_size(2, 1, 1)))"
"void k0(global int* restrict output, int n, const global int* restrict in){"
"  for (int gl_id = get_global_id(0); gl_id < n; gl_id = gl_id + (n / 2)) {"
"    output[gl_id] = ((int)3) + in[gl_id];"
"  }"
"}";

const char k1_source[] =
"__kernel __attribute__ ((reqd_work_group_size(4, 1, 1)))"
"void k1(global int* restrict output, int n, const global int* restrict in){"
"  for (int gl_id = get_global_id(0); gl_id < n; gl_id = gl_id + (n / 2)) {"
"    output[gl_id] = ((int)1) + in[gl_id];"
"  }"
"}";

typedef struct add3_t {
    Kernel k0;
    Kernel k1;
} add3_t;

void add3_init(Context ctx, add3_t* self){
    (*self).k0 = loadKernelFromSource(ctx, "k0", k0_source, sizeof(k0_source) - 1);
    (*self).k1 = loadKernelFromSource(ctx, "k1", k1_source, sizeof(k1_source) - 1);
}

void add3_destroy(Context ctx, add3_t* self){
    destroyKernel(ctx, (*self).k0);
    destroyKernel(ctx, (*self).k1);
}

void add3_run(Context ctx, add3_t* self, Buffer moutput, int n, Buffer min){
  Buffer mx1 = createBuffer(ctx, ..., HOST_READ | DEVICE_WRITE);
  {
    DeviceBuffer b0 = deviceBufferSync(ctx, mx1, ..., DEVICE_WRITE);
    DeviceBuffer b2 = deviceBufferSync(ctx, min, ..., DEVICE_READ);
    launchKernel(ctx, (*self).k0, {n / 2, 1, 1}, {2, 1, 1}, 3, {b0, n, b2});
  }
  Buffer mx2 = createBuffer(ctx, ..., HOST_WRITE | DEVICE_READ);
  {
    int32_t* x2 = (int32_t*)hostBufferSync(ctx, mx2, ..., HOST_WRITE);
    int32_t* x1 = (int32_t*)hostBufferSync(ctx, mx1, ..., HOST_READ);
    for (int i = 0; i < n; i = 1 + i) {
        x2[i] = x1[i] - ((int32_t)1);
    }
  }
  {
    DeviceBuffer b0 = deviceBufferSync(ctx, moutput, ..., DEVICE_WRITE);
    DeviceBuffer b2 = deviceBufferSync(ctx, mx2, ..., DEVICE_READ);
    launchKernel(ctx, (*self).k1, {n / 2, 1, 1}, {4, 1, 1}, 3, {b0, n, b2});
  }
  destroyBuffer(ctx, mx2);
  destroyBuffer(ctx, mx1);   
}
\end{c-code}

\subsection{Compiling Kernel Executing Programs in \Shine{}}

\begin{figure}[h]
  \centering
  \includegraphics[width=0.8\linewidth]{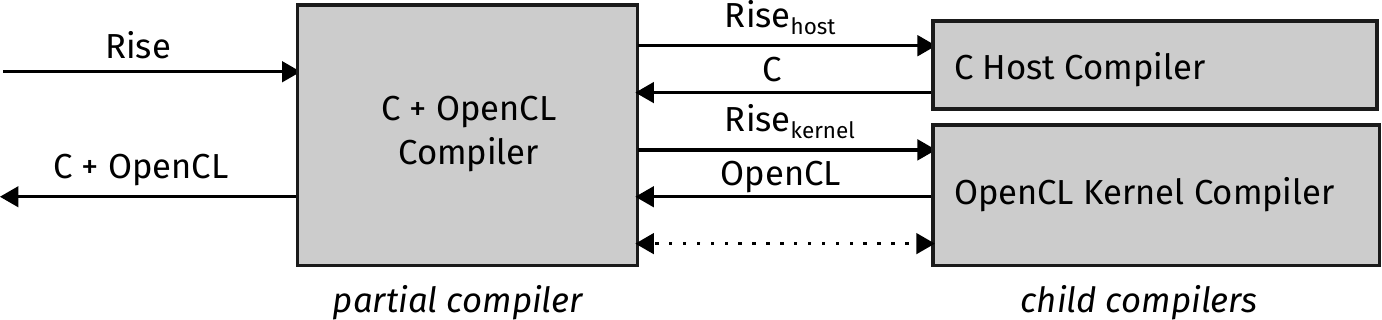}
  \caption{The partial C+OpenCL \Shine{} compiler delegates tasks to child compilers.}
  \label{fig:partial-c-ocl-compiler}
\end{figure}

To generate code targeting LRA from kernel executing \Rise{} programs, we take inspiration from partial compilers \cite{the-compiler-forest-2013} to modify \Shine{}.
A \emph{partial compiler} delegates intermediate compilation tasks to child (partial) compilers before combining their generated intermediate code into the final code.
We implement the partial C+OpenCL \Shine{} compiler that delegates C host code generation and OpenCL kernel code generation to dedicated child compilers (\cref{fig:partial-c-ocl-compiler}).

\paragraph{Partial C+OpenCL Compiler}
Programs with explicit kernel executions such as the one from \cref{fig:kernels-example} can be partially compiled by decomposing them into two simpler compilation tasks: host code generation and kernel code generation.
In the initial program, a kernel definition with name k$i$ is generated for the $i$-th \inlineRise{oclRun} pattern through \emph{lambda lifting} \cite{1985-lambda-lifting}: all free variables in the input expression of \inlineRise{oclRun} are eliminated by introducing explicit parameters (\cref{fig:kernels-k0,fig:kernels-k1}).
Each k$i$ can be compiled separately, while the associated \inlineRise{oclRun} pattern is replaced with a \inlineRise{kernelCall} pattern that treats k$i$ as a regular callable function: producing the host program which can also be compiled separately (\cref{fig:kernels-host}).

Once separately compiled by child compilers, kernel code and host code are put together in a single file as done in \cref{fig:kernels-code}.

\begin{rise}[label={fig:kernels-k0}, caption={\Rise{} k0 kernel program for \cref{fig:kernels-example}}]
$\Lambda$n : nat. $\lambda$in : n.i32. mapGlobal (add 3) in
\end{rise}
\begin{rise}[label={fig:kernels-k1}, caption={\Rise{} k1 kernel program for \cref{fig:kernels-example}}]
$\Lambda$n : nat. $\lambda$x2 : n.i32. mapGlobal (add 1) x2
\end{rise}
\begin{rise}[label={fig:kernels-host}, caption={\Rise{} host program for \cref{fig:kernels-example}}]
$\Lambda$n : nat. $\lambda$in : n.i32. (
  kernelCall("k0", LocalSize(2), GlobalSize(n/2), n, in) $\then$ store (
  mapSeq ($\lambda$y. y-1) x1 $\then$ store ($\lambda$x2.
  kernelCall("k1", LocalSize(4), GlobalSize(n/4), n, x2)))
)
\end{rise}

\paragraph{OpenCL Kernel Compiler}
Generating kernel code does not require anything new: the existing \Shine{} OpenCL kernel compiler is simply re-used.

\paragraph{C Host Compiler}
To generate host code, the existing \Shine{} C compiler requires extensions to target LRA and support compiling the \inlineRise{kernelCall} patterns (which become \inlineDPIA{kernelCallCmd} imperative DPIA commands after a straightforward translation to imperative).

First, we add an imperative DPIA compilation pass that, given a host program which uses plain arrays, introduces LRA \lstinline{Buffer}s as needed, using the \inlineDPIA{newBuffer} pattern:
\begin{DPIA}
newBuffer(af: AccessFlags, dt: data,
          k: (exp[buffer[dt], read] $\times$ acc[buffer[dt]]) $\to$ comm): comm
\end{DPIA}
Sub-programs that only perform host-side computation are kept as-is, they can still use plain arrays and the pre-existing C compilation, we call them \emph{host execution sections} and introduce a corresponding \inlineDPIA{hostExecution} imperative DPIA pattern:
\begin{DPIA}
hostExecution(afs: Map[Ident, AccessFlags], body: comm): comm
\end{DPIA}
To identify host execution sections, the algorithm used is fairly similar to the barrier insertion algorithm from \cref{lst:barrier-insertion}.
Host and device side reads/writes are tracked to insert \inlineDPIA{hostExecution} patterns as required.

Finally, C code generation from imperative DPIA is extended to support the three new patterns.
For each \inlineDPIA{kernelCallCmd}, \lstinline{deviceBufferSync} and \lstinline{launchKernel} LRA calls are generated in a C block for scoping (\lstinline|{}|).
For each \inlineDPIA{newBuffer}, \lstinline{createBuffer} and \lstinline{destroyBuffer} calls are generated in a block.
For each \inlineDPIA{hostExecution}, \lstinline{hostBufferSync} calls are generated in a block also containing the code generated for the host execution body.
As desired, the code generated for the \Rise{} program in \cref{fig:kernels-example} corresponds to the C+OpenCL program in \cref{fig:kernels-code}, modulo small syntactic details.

\subsection{Evaluating the Boilerplate Avoided through Kernel Execution}
\label{sec:host-codegen-eval}

\paragraph{Experimental Setup}
The impact of using the \inlineRise{oclRun} \Rise{} pattern for kernel execution is evaluated by measuring how many lines of host code are automatically generated to launch the kernels, rather than written by hand.
As a case study, we consider optimising the Harris corner (and edge) detector \cite{harris-1988}, a well established image processing pipeline.

\begin{figure}[b]
  \centering
  \includegraphics[width=0.9\linewidth]{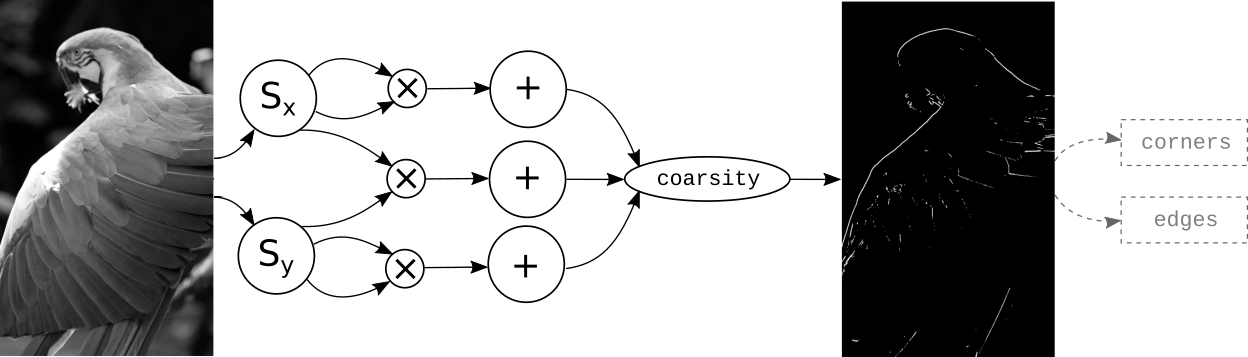}
  \caption{Harris corner detection computation flow, example image taken from Halide.}
  \label{fig:harris-bis}
\end{figure}

\Cref{fig:harris-bis} depicts the Harris algorithm for a grayscale input image.
Given the input on the left, point-to-point operators (multiplications $\times$, coarsity) and $3 \times 3$ convolutions (sobel operators $S_x$ and $S_y$, sums $+$) are combined to detect corners and edges highlighted in the output on the right.
As a composition of point wise and stencil operators, the Harris detector is more complex than its individual parts, and exposes more optimisation opportunities.

We consider well-known optimisations presented in \cite{hlt-lacassagne-2014}.
\Cref{fig:harris-kernels} shows how many different OpenCL kernels need to be implemented to explore a space of important optimisations.
In particular, different decisions for an optimisation called operator fusion lead to different kernel groupings.
4 different ways to fuse operators are considered, each variant offering a different trade-off between memory accesses and arithmetic complexity.
5 different ways to optimise each resulting kernel are also considered (e.g. tiling, vectorisation).

In total, benchmarking this relatively small optimisation space requires implementing 50 different kernels, each associated with slightly different host code to launch them.

\paragraph{Avoided Boilerplate Results}
\Cref{fig:harris-host-code} shows how many lines of code are handwritten and generated with or without the kernel execution feature (i.e. the \inlineRise{oclRun} pattern).
Without \inlineRise{oclRun}, 1K lines of host code are handwritten to launch the generated kernels.
This corresponds to about 20 lines per kernel launch, and does not even count the code required to compose the kernels into the entire Harris processing pipeline.
With \inlineRise{oclRun}, 1.2K lines of host code are automatically generated.
Therefore, the kernel execution feature enables exploring the same optimisation space without having to write 1K lines of boilerplate code.





\begin{table}
  \centering
  \begin{tabular}{c|c|c}
  \textbf{kernel grouping} & \textbf{variants per group} & \textbf{kernel variants} \\ \hline
  $[S_x], [S_y], [\times], [+], [coarsity]$ & 5 & 25 \\
  $[S_x, S_y, \times], [+, coarsity]$ & 5 & 10 \\
  $[S_x, S_y], [\times, +, coarsity]$ & 5 & 10 \\
  $[S_x, S_y, \times, +, coarsity]$ & 5 & 5 \\ \hline
  \emph{total} & \emph{total} & 50 \\
  \end{tabular}
  \caption{Four different kernel groupings (a group is represented as $[]$), and the number of differently optimised kernel variants that are considered (e.g. vectorised, tiled).}
  \label{fig:harris-kernels}
\end{table}

\begin{table}
  \centering
  \begin{tabular}{|c|c|c|c|} \hline
  \textbf{\inlineRise{oclRun}?} & \textbf{handwritten lines} & \textbf{generated lines} \\ \hline
  \color{red}\no & >1000 & 0 \\ \hline
  \color{green}\yes & 0 & >1200 \\ \hline
  \end{tabular}
  \caption{Without \inlineRise{oclRun}, 1K lines of host code are handwritten to launch each kernel for the Harris corner detection design space exploration case study.}
  \label{fig:harris-host-code}
\end{table}

\subsection{Related work}

\paragraph{Kernel Execution in Domain-Specific Compilers}
Halide and PolyMage support kernel executions, including exploring different kernel groupings \cite{halide-2013, jangda2020-effective-fusion}, but are domain-specific compilers for image processing.

\paragraph{Kernel Execution in General-Purpose Languages}
Many general-purpose languages support executing kernels to offload computations on GPUs.
OCaml has the SPOC library \cite{spoc-2014}, Haskell has Accelerate \cite{accelerate-2011}, C/C++ has SYCL, OpenMP and OpenACC.

Typically, the host code is directly written in such languages, leveraging runtime libraries without necessarily generating C host code.
The \Rise{} language is not general-purpose, which is why we generate C host code.
Additionally, \Rise{} and the \inlineRise{oclRun} pattern are designed for rewrite-based design space exploration, which is not a design goal for the acceleration libraries mentioned above.

\paragraph{Implicit Kernel Execution in High-Level Languages}
Many high-level languages support executing kernels to accelerate computations.
The common strategy is to make kernel execution implicit and automated by heuristics, as in MapCG \cite{2010-MapReduce-CPU-GPU}, NOVA \cite{2014-nova} SaC \cite{2011-GPU-SaC, viessmann2020-host-gpu-codegen}, X10 \cite{2011-X10-CUDA} and Lime \cite{2016-Lime}.
MATLAB is an exception, as it allows the manual insertion of compiler directives in imperative code, similar to OpenACC \cite{2015-MATLAB-OpenCL, 2017-MATLAB-OpenCL}.
Both approaches differ from our explicit and functional \inlineRise{oclRun} pattern, designed for rewrite-based design-space exploration.

\paragraph{\Lift{}}
Host code generation support has been independently added to \Lift{} in \cite{stoltzfus2021code}.
However, the approach taken introduces two challenges for rewriting without discussing them, as the paper focuses on manually written low-level programs.
First, the introduced equivalent to our \inlineRise{oclRun} pattern (called \lstinline{OclKernel}) does not take a single input expression, but rather a function as well as $n$ explicit arguments.
It is unclear how a rewrite system would introduce such a function and its arguments.
By contrast, our design abstracts over the kernel function, inferring it via lambda lifting, and allowing a simple rewrite rule definition:
\begin{elevate-rise}[numbers=none]
rule introOclRun(ls: LocalSize, gs: GlobalSize) = 
  (x : dt) $\rewritesTo$ oclRun(ls, gs) x
  where dt : data
\end{elevate-rise}
Second, explicit data transfers (as well as in-place computations) are introduced.
By contrast, our design keeps data transfers implicit at the functional level for simplicity, and rewriting is not burdened with introducing efficient and correct transfers.

\clearpage
\subsection{Summary}

In this section, we demonstrate how the \inlineRise{oclRun} pattern is added to \Rise{} to enable explicit kernel execution.
To the best of our knowledge, \Shine{} is the first domain-extensible compiler that makes kernel execution controllable by rewriting high-level functional programs.
Different kernel decompositions of a high-level \Rise{} program can be explored via rewriting, producing a single low-level \Rise{} program representing both host-side and device-side computations.
\Shine{} is modified to generate imperative code for multiple OpenCL kernels, as well as the necessary C host code to launch them.
Using \inlineRise{oclRun} in a relatively simple Harris corner detection case study enables replacing 1K lines of previously handwritten host code with 1.2K lines of automatically generated code (\cref{fig:harris-host-code}).
The kernel execution feature saves performance engineers from writing boilerplate code and provides a stable interface for the generated code, which is not the case if \Shine{} generates a single OpenCL kernel without host code.
\section{Explicit Storage Folding for \Shine{}}
\label{sec:storage-folding}

Stencil computations are used in a wide range of domains.
A stencil computation processes each element of an N-dimensional input array by accessing a fixed neighborhood pattern (the stencil) to produce an output element.
When optimising stencils, it is important to leverage spatio-temporal locality \cite{hlt-lacassagne-2014}.
For example, to compute the next output of a 1D stencil during iteration $i$ of sequential execution, only the last $m$ stencil inputs are required.
Instead of naively storing all stencil inputs in a temporary storage $T$, storage folding optimisations only store the last $m$ inputs, improving memory usage.
We consider two storage folding optimisations:
\begin{itemize}
  \item \emph{Circular buffering} stores the last $m$ temporary results in memory, using modulo indexing: $T[j]$ is stored in $M[j \mod m]$ with $j \in [i;i+m-1]$.
  \item \emph{Register rotation} stores the last $m$ temporary results in registers, rotating them between computation iterations: $T[i], \ldots, T[i+m-1]$ is stored in registers $r_0, \ldots, r_{m-1}$.
\end{itemize}
Circular buffering is supported by most image processing specific compilers, including Halide.
However, register rotation is not supported by Halide schedules.\footnote{\url{https://github.com/halide/Halide/issues/2905}}

This section enables explicit storage folding in \Rise{} by adding the \inlineRise{circularBuffer} and \inlineRise{rotateValues} patterns, along with 2 additional ones.
As with kernel execution in the last section, introducing these patterns makes storage folding controllable and explorable by rewriting in the domain-extensible \Shine{} compiler.
\Cref{ch:imgproc} will introduce storage folding patterns during rewriting to generate faster code than Halide on a case study.

\subsection{Adding \Rise{} Patterns for Storage Folding}

To explicitly encode storage folding decisions in functional programs, we introduce two new functional patterns, called \inlineRise{circularBuffer} and \inlineRise{rotateValues}, and with following types:
\begin{rise}[numbers=none]
circularBuffer : (a: addr) $\to$ (alloc m: nat) $\to$
                   (s $\to$ t) $\to$ (n + m - 1).s $\to$ n.m.t
rotateValues   : (a: addr) $\to$ (m: nat) $\to$
                   (t $\to$ t) $\to$ (n + m - 1).t $\to$ n.m.t
\end{rise}
Both patterns have similar types, as they derive from \inlineRise{slide m 1 : (n + m - 1).t $\to$ n.m.t}, a sliding window pattern that was introduced to support stencils in \Lift{} \cite{lift-stencil-2018}.
Both patterns take an address space \inlineRise{a} as parameter, controlling where temporary memory will be allocated.

\inlineRise{circularBuffer} takes an additional \inlineRise{alloc} parameter, allowing to allocate memory for more than $m$ values (e.g. using a power of two makes modulo indexing cheaper).
It takes an \inlineRise{s $\to$ t} function specifying how to load values into the circular buffer, potentially performing computations.

\inlineRise{rotateValues} takes a \inlineRise{t $\to$ t} function specifying how to write values into memory.
This function must be an identity function as it will be used to rotate values between iterations.
We talk about value rotation instead of register rotation because the use of registers is not guaranteed.
Regular variables are used in the generated OpenCL code, and the OpenCL compiler may decide to allocate registers for them if the address space is \inlineRise{private}.

Both patterns implement a finite \emph{stream}, where a sequence of values is made available one at a time, sequentially.
We add two other stream based functional patterns, with following types:
\begin{rise}[numbers=none]
iterateStream : (s $\to$ t) $\to$ n.s $\to$ n.t
mapStream     : (s $\to$ t) $\to$ n.s $\to$ n.t
\end{rise}

Like \inlineRise{mapSeq}, both patterns represent concrete implementations of \inlineRise{map} and have exactly the same type.
However, while \inlineRise{mapSeq} operates on arrays, \inlineRise{mapStream} operates on streams, and \inlineRise{iterateStream} transforms streams into arrays.
The lack of stream type in \Rise{} is discussed in \cref{storage-folding-limitations}.

Two simple examples using the new storage folding patterns in \Rise{} are shown in \cref{fig:storage-folding-rise}.
Both examples compute a simple stencil summing windows of 3 input values, corresponding to the high-level program \inlineRise{slide 3 1 $\then$ map (reduce + 0)}.
The code that we want to generate from the two example \Rise{} programs is depicted in \cref{fig:storage-folding-c}.
\clearpage

\begin{figure}[th]
\begin{minipage}{0.44\linewidth}
\begin{rise}
rotateValues private 3 ($\lambda$x. x) $\then$
iterateStream (reduceSeq + 0)
\end{rise}
\end{minipage}\hfill%
\begin{minipage}{0.48\linewidth}
\begin{rise}
circularBuffer private 3 3 ($\lambda$x. x) $\then$
iterateStream (reduceSeq + 0)
\end{rise}
\end{minipage}
\caption{Example \Rise{} programs using storage folding patterns.}
\label{fig:storage-folding-rise}
\end{figure}

\begin{figure}[th]
\begin{minipage}[t]{0.44\linewidth}
\begin{c-code}[basicstyle=\ttfamily\scriptsize]
int tmp[3];
for (int i = 0; i < 2; i += 1) {
  tmp[i] = input[i];
}
  
for (int i = 0; i < n-2; i += 1) {
  tmp[2] = input[2 + i];
  
  int acc = 0;
  for (int j = 0; j < 3; j += 1) {
    acc += tmp[j];
  }
     
  output[i] = acc;
  tmp[0] = tmp[1];
  tmp[1] = tmp[2];
}
\end{c-code}
\end{minipage}\hfill%
\begin{minipage}[t]{0.48\linewidth}
\begin{c-code}[basicstyle=\ttfamily\scriptsize]
int tmp[3];
for (int i = 0; i < 2; i += 1) {
  tmp[i] = input[i];
}

for (int i = 0; i < n-2; i += 1) {
  tmp[(2 + i) 

  int acc = 0;
  for (int j = 0; j < 3; j += 1) {
    acc += tmp[(i + j) 
  }
    
  output[i] = acc;
}
\end{c-code}
\end{minipage}
\caption{Example OpenCL programs corresponding to the \Rise{} programs of \cref{fig:storage-folding-rise}.}
\label{fig:storage-folding-c}
\end{figure}

\subsection{Adding a New Stream Translation to DPIA}

\newcommand{\strT}{\mathcal{S}}

To generate code for storage folding patterns, we extend DPIA with a new translation from functional to imperative.
The \emph{stream translation} $\strT$ produces a command which reads from a functional read expression \inlineDPIA{e} and calls a stream continuation function \inlineDPIA{sk} to continue the translation as required (\cref{fig:stream-trans-ty}).
The stream continuation function is called by providing a function of type \inlineDPIA{(i: nat) $\to$ (exp[dt, read] $\to$ comm) $\to$ comm}, that should stream the $i$-th value by calling the inner continuation function of type \inlineDPIA{exp[dt, read] $\to$ comm}.

\begin{DPIA}[label={fig:stream-trans-ty}, caption={Types of translations, including the new stream translation}]
$\accT$(e: exp[dt, write], out: acc[dt]): comm
$\conT$(e: exp[dt, read], k: exp[dt, read] $\to$ comm): comm
$\strT$(e: exp[n.dt, read],
  sk: ((i: nat) $\to$ (exp[dt, read] $\to$ comm) $\to$ comm) $\to$ comm): comm
\end{DPIA}

The acceptor translation of \inlineDPIA{iterateStream} triggers a stream translation (\cref{fig:iterate-stream-trans}).
This makes sense because the pattern writes to an array output by reading from a stream input.
A selection of stream translations is shown in \cref{fig:stream-trans-sel}.
It excludes the stream translation for \inlineDPIA{circularBuffer} which is similar to the one for \inlineDPIA{rotateValues}.
\clearpage

\begin{DPIA}[label={fig:iterate-stream-trans}, caption={The acceptor translation of \inlineDPIA{iterateStream} calls $\strT$}]
$\accT$(iterateStream(n, dt1, dt2, f, array), out) =
  $\strT$($\lambda$(next : (i: nat) $\to$ (exp[dt1, read] $\to$ comm) $\to$ comm).
  for(n, $\Lambda$i. next i ($\lambda$(x : exp[dt1, read]).
    $\accT$(f x, out @ i))))
\end{DPIA}



\begin{DPIA}[numbers=left, label={fig:stream-trans-sel}, caption={Selection of stream translations}]
$\strT$(mapStream(n, dt1, dt2, f, input), sk) =
  $\strT$(input, $\lambda$(nextInput : (i: nat) $\to$ (exp[dt1, read] $\to$ comm) $\to$ comm).$\label{line:map-stream-s-input}$
    sk($\Lambda$i. $\lambda$(k : exp[dt2, read] $\to$ comm). $\label{line:map-stream-sk}$
      nextInput i ($\lambda$(x : exp[dt1, read]). $\conT$(f x, k))$\label{line:map-stream-k}$
    ))
$\strT$(rotateValues(a, n, m, dt, wrdt, input), sk) =
  $\strT$(input, $\lambda$(nextInput : (i: nat) $\to$ (exp[dt1, read] $\to$ comm) $\to$ comm).
    new(a, m.dt, $\lambda$rs.$\label{line:rotate-values-alloc}$
      for(m - 1, $\Lambda$i.$\label{line:rotate-values-init-beg}$
        nextInput i ($\lambda$(x : exp[dt, read]). $\accT$(wrdt x, rs.wr @ i)));$\label{line:rotate-values-init-end}$
      sk($\Lambda$i. $\lambda$(k : exp[m.dt, read] $\to$ comm).
         nextInput (i + m - 1) ($\lambda$(x : exp[dt, read]).$\label{line:rotate-values-load-beg}$
           $\accT$(wrdt x, rs.wr @ (m - 1)));$\label{line:rotate-values-load-end}$
         k(rs.rd);$\label{line:rotate-values-return}$
         for(m - 1, $\Lambda$i. $\accT$(wrdt (rs.rd @ (i + 1)), rs.wr @ i))$\label{line:rotate-values-rotate}$
      )))
\end{DPIA}

For \inlineDPIA{mapStream}, the stream translation of the \inlineDPIA{input} is first called in line \ref{line:map-stream-s-input}, giving access to \inlineDPIA{nextInput}.
In line \ref{line:map-stream-sk}, the stream continuation \inlineDPIA{sk} is called to define how to stream the $i$-the output value.
The continuation \inlineDPIA{k} should be called to return this output value.
Finally in line \ref{line:map-stream-k}, \inlineDPIA{nextInput} is called to consume the next \inlineDPIA{input} value \inlineDPIA{x}, which is used to produce the next output value \inlineDPIA{f x}.
\inlineDPIA{$\conT$(f x, k)} is used to translate \inlineDPIA{f x} to imperative.

For \inlineDPIA{rotateValues}, the stream translation is a bit more involved.
A temporary buffer is allocated to hold window values in line \ref{line:rotate-values-alloc}.
The first $m-1$ window values are stored in the buffer in lines \ref{line:rotate-values-init-beg}-\ref{line:rotate-values-init-end}, before \inlineDPIA{sk} allows streaming values.
To stream an output value, which is a window of $m$ values:
\begin{itemize}
  \item The $m$-th window value, or $i+m$-th input value, is loaded in lines \ref{line:rotate-values-load-beg}-\ref{line:rotate-values-load-end}.
  \item The window of $m$ values is returned in line \ref{line:rotate-values-return}, by calling the continuation \inlineDPIA{k}.
  \item The values are rotated in the buffer to prepare for the next iteration in line \ref{line:rotate-values-rotate}.
\end{itemize}
To write values to the buffer, acceptor translations such as \inlineDPIA{$\accT$(wrdt x, rs.wr @ i)} are called, since \inlineDPIA{wrdt x} is a functional program that needs to be translated to imperative.

It is also possible to transform any array into a stream.
If there is no specialised stream translation available, we fallback on a generic one, shown in \cref{fig:stream-trans-generic}.
\begin{DPIA}[label={fig:stream-trans-generic}, caption={Generic stream translation used as a fallback}]
$\strT$(e : exp[n.dt, read], sk) =
  sk($\Lambda$i. $\lambda$(k : exp[dt, read] $\to$ comm). $\conT$(e @ i, k))
\end{DPIA}

Using this newly defined stream translation, the code generated for \cref{fig:storage-folding-rise} corresponds to the desired code from \cref{fig:storage-folding-c}, modulo small syntactic details.
It is interesting to note that no new imperative patterns are required to support storage folding.

\subsection{Limitations of Storage Folding in \Shine{}}
\label{storage-folding-limitations}

The proposed storage folding implementation has two notable limitations.

\paragraph{Stream Type}
The concept of finite streams is introduced, but is not reflected into the type system.
As a result, it is possible to construct invalid programs that would try to use sequential streams as data-parallel arrays, for example:
\begin{rise}[numbers=none]
circularBuffer private 3 3 ($\lambda$x. x) $\then$ mapGlobal ($\lambda$x. x)
\end{rise}
Code generation will fail for such programs, because stream-returning patterns have no $\accT$/$\conT$ translations.
This is good because buggy code will not be generated, but this is bad because well-typed programs may be invalid.
Future work may look into adding a dedicated stream type to make invalid programs ill-typed.

\paragraph{Stencil Stride}
Only stencil strides of 1 are supported (i.e. the \inlineRise{1} in \inlineRise{slide m 1}).
Future work may look into adding support for arbitrary strides, and whether it would be useful.

\subsection{Related Work}

We mentioned that Halide schedules support circular buffering, but not register rotation.
There are many compilers that perform storage folding, but are domain-agnostic or domain-specific.

\paragraph{Storage Folding in Compilers for Stream Processing}
Programs that need to process potentially infinite streams of values with finite buffers make extensive use of optimisations comparable to storage folding.
Optimising compilers for stream programs are an active field of research \cite{thies2002-streamit, 2006-StreamIt, 2014-StreamJIT, 2014-stream-EDSL, tangwongsan2015general, leben2019programming, 2020-bounded-stream-scheduling}.
Circular buffering and related optimisations were studied for Synchronous Data Flow, one way to model stream programs, decades ago \cite{1987-sync-dataflow}.

\paragraph{Storage Folding in Compilers for Stencil Computations}
There exists many domain-specific compilers for stencil computations \cite{basu2015compiler, pouchet2013polyhedral, stock2014framework}.
Storage folding optimisations are sometimes viewed as \emph{streaming} optimisations in this domain \cite{rawat2018domain}.

\paragraph{Storage Folding in Compilers for High-Level Synthesis}
Compilers that synthesise hardware often achieve storage folding through \emph{shift-registers}, also known as \emph{line buffers} \cite{darkroom-2014, chugh2016dsl}.

~\\
Some domain-extensible compilers were extended with storage folding optimisations independently during this thesis.

\paragraph{Storage Folding in Domain-Extensible Compilers}
AnyHLS \cite{ozkan2020anyhls} synthesises FPGA designs, but differs from \Shine{} by following the partial evaluation approach of AnyDSL \cite{anydsl-2018}.
\textsc{Shir} also targets FPGAs, introducing shift-registers through a \code{SlideStm} pattern \cite{2022-schlaak-shir} that is closely related to our \inlineRise{rotateValues} pattern.
\textsc{Shir} differs from \Shine{} as it focuses on greedy optimisation and emits code in VHDL, a hardware description language.
A \code{MapSeqSlide} pattern was added to \Lift{} for 2.5D tiling \cite{stoltzfus2019-stencils}.
\code{MapSeqSlide} is less flexible than our patterns, as it is a monolithic equivalent to combining \inlineRise{rotateValues} with \inlineRise{iterateStream}.

\subsection{Summary}

This section adds \Rise{} patterns to enable explicit storage folding, notably \inlineRise{circularBuffer} and \inlineRise{rotateValues}.
To generate the desired imperative code, a stream translation $\strT$ is added to DPIA on top of the existing $\accT$ and $\conT$ translations.
\Cref{ch:imgproc} will demonstrate how introducing storage folding patterns via rewriting leads to generating faster code than Halide on a case study.


\section{Conclusion}
\label{sec:codegen-conclusion}

This chapter first introduced the \Rise{} language and its \Shine{} compiler \cite{steuwer2022-rise-shine, dpia-2017}, both resulting from collaboration and heavily inspired by the domain-extensible \Lift{} compiler.
\Shine{} rewrites functional programs before generating imperative code, and differs from \Lift{} as it aims to address the controllability challenge by exploring trade-offs between automation and control of rewrite rule applications.


The novel contribution of this chapter is the design and implementation of three important code generation features.
These features are crucial when using \Shine{} to generate faster code than Halide on an image processing case study via controlled rewriting (\cref{ch:imgproc}), and similarly fast code as TVM on a linear algebra case study via semi-automated rewriting (\cref{ch:guided-rewriting}).
\clearpage
\begin{itemize}
  \item We contribute a synchronisation \emph{barrier insertion} algorithm (\barinsalg{}) that does not need to be modified when extending \Rise{} patterns, contrasting with the barrier elimination algorithm of \Lift{} \cite{lift-ir-2017}.
  \barinsalg{} transforms the intermediate imperative DPIA programs.
  The correctness and efficiency of barrier insertion is evaluated on 38 unit tests and 10 benchmarks, mostly taken from prior \Lift{} work.
  We identify 6 differences in the code generated by \Shine{} and \Lift{}, and observe that our algorithm fixes bugs in 13 unit tests and 1 benchmark, where \Lift{} generates incorrect barriers (\cref{fig:barrier-insertion-eval}).
  There is only 1 benchmark where \Shine{} inserts a barrier that \Lift{} eliminates, and we provide a clear pathway to improve our algorithm to generate more efficient barriers than \Lift{} on all 48 unit tests and benchmarks.
\end{itemize}
While barriers are implicit in \Rise{}, and thus not controllable by rewriting, the next two features add new low-level \Rise{} patterns.
Low-level patterns make implementation choices explicit in \Rise{}, and thus controllable during rewriting.
\begin{itemize}
    \item We add the \inlineRise{oclRun} \Rise{} pattern to enable explicit \emph{kernel execution}, where the value of an expression is computed by launching an OpenCL kernel.
    This requires modifying \Shine{} to generate imperative code for multiple OpenCL kernels, as well as the necessary host code to launch them.
    With this feature, 1K lines of handwritten host code are replaced with 1.2K lines of automatically generated code on a relatively simple design space exploration case study (\cref{fig:harris-kernels,fig:harris-host-code}).

    \item We add the \inlineRise{circularBuffer} and \inlineRise{rotateValues} \Rise{} patterns to enable explicit \emph{storage folding} for temporary arrays.
    \Shine{} is modified to generate the desired imperative code when using these patterns, by adding a new stream translation to DPIA.
    \Cref{ch:imgproc} relies on storage folding to generate high performance code.
\end{itemize}

This chapter focused on code generation, at the bottom of the compilation stack.
The following two chapters will focus on rewriting, moving up in the compilation stack (\cref{fig:rise-shine}).


\chapquote{
I love rewriting because that is where and how you discover the story. It's like you have this skeleton, and you get to put flesh on it and hair and clothes and really wonderful jewelry.
}{Caroline Leavitt}

\chapter{Beyond Halide Scheduling with Controlled Rewriting}
\label{ch:imgproc}
In \cref{ch:background}, we saw that \Lift{}
addresses the \emph{extensibility challenge} using an extensible rewrite system \cite{lift-stencil-2018}, but does not address the \emph{controllability challenge} as it fully automates optimisation. 
Halide 
addresses the \emph{controllability challenge} through \emph{schedules} \cite{halide-2012}, but does not address the \emph{extensibility challenge} as it is domain-specific.

\begin{figure}
  \centering
  \includegraphics[height=8em]{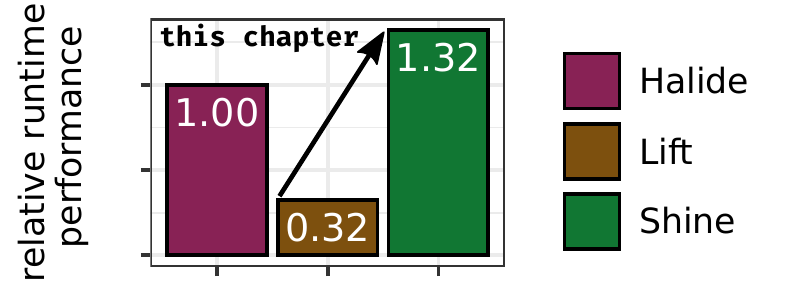}
  \caption{\Lift{} performs poorly compared to an expert Halide schedule when optimising the Harris corner detection image processing pipeline for the Cortex A53 ARM CPU.
  By combining domain-extensibility with controlled rewriting, \Shine{} outperforms Halide by $1.3\times$.
  Extracted from the results of \cref{sec:imgproc-eval} for an image resolution of $1536 \times 2560$ pixels.}
  \label{fig:resultsIntro}
\end{figure}

This chapter demonstrates how both extensibility and controllability are combined in \Shine{} to generate high-performance code.
We optimise the Harris corner detection \cite{harris-1988} -- a standard image processing pipeline already encountered in \cref{sec:host-codegen-eval} -- on 4 ARM CPUs.
Our results on Cortex A53 show that \Shine{} generates 1.32$\times$ faster code than Halide (\cref{fig:resultsIntro}).
By contrast, \Lift{} is missing important optimisations and generates 3.12$\times$ slower code than Halide.

\Cref{sec:imgproc-approach} explains how optimisation is controlled by defining \Elevate{} \emph{rewriting strategies} \cite{hagedorn2020-elevate, hagedorn2020-thesis}, and clarifies the scope of extensibility in \Shine{}.
\Cref{sec:harris-repr} introduces the Harris corner detection case study and how the corresponding algorithm is represented in \Rise{}.
\Cref{sec:harris-opt} describes how 6 well-known optimisations are applied by composing rewrite rules with \Elevate{}.
\Cref{sec:imgproc-eval} presents an experimental evaluation and \cref{sec:imgproc-conclusion} concludes.


\begin{figure}[t]
  \centering
  \includegraphics[width=\linewidth]{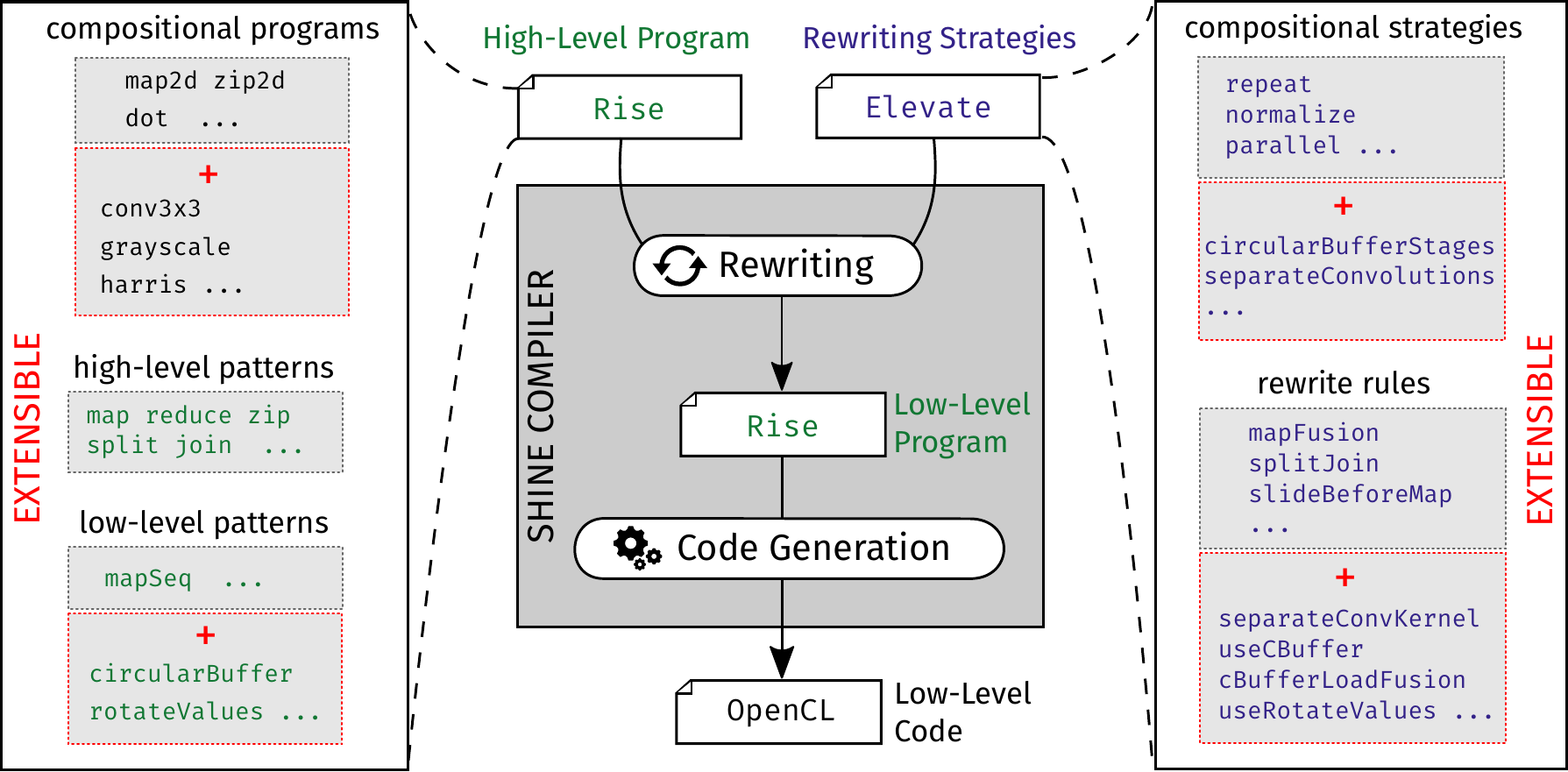}
  \caption{\Shine{} with control of optimisations: computations are composed using extensible patterns in \Rise{}; optimisations are composed using extensible rewrite rules in \Elevate{}.}
  \label{fig:compiler-design}
\end{figure}
\section{Domain-Extensibility with Control of Optimisations}
\label{sec:imgproc-approach}

This chapter focuses on combining domain-extensibility with control of optimisations.
As discussed in \cref{ch:background}, our aim is to enable performance engineers to take control over the optimisation process to achieve their performance goals instead of forcing them to bypass the compiler and resort to manual optimisation.

\paragraph{Control of Optimisations}
Control is enabled in \Shine{} by the use of \Elevate{} rewriting strategies, as in \cref{fig:compiler-design}.
High-level \Rise{} programs describe computations by composing extensible high-level patterns.
Optimisations are encoded by applying rewrite rules, leading to a low-level program from which imperative code such as OpenCL is generated.
The novelty is that \emph{rewriting strategies} are defined in a companion \Elevate{} language, enabling performance engineers to describe optimisations as compositions of rewrite rules (top right of \cref{fig:compiler-design}).

\Elevate{} is the work of Bastian Hagedorn \cite{hagedorn2020-elevate, hagedorn2020-thesis}.
I contributed to its early design and demonstrated that the ideas generalise to image processing \cite{koehler2021-elevate-imgproc}.
The separation between \Rise{} programs and \Elevate{} strategies resembles the separation between algorithms and schedules in domain-specific compilers such as Halide.
There, a \emph{schedule} describes the optimisations to apply to an \emph{algorithm} that defines the functional behavior of the computation.
This fine grained control of optimisations allows performance engineers to steer the compiler to generate highly efficient code that would have been hard to reach automatically.
While Halide schedules are implemented as a set of ad-hoc, predefined APIs exposed by compiler writers, \Elevate{} strategies enable performance engineers to define their own program transformations in an extensible and composable way.

\paragraph{Domain-Extensibility}
The domain-extensibility of \Shine{} is pictured in red on the sides of \cref{fig:compiler-design}.
This domain-extensibility has restricted scope as \Rise{} remains an array language: the extension mechanisms are patterns, rewrite rules, as well as rewriting strategies.

High-level patterns are added, such as \inlineElevateRise{slide} that creates a sliding window and enables defining stencil computations, previously added in \cite{lift-stencil-2018}.
High-level rewrite rules are added, such as \inlineElevateRise{separateConvKernel} that encodes the separability of a convolution kernel.
Lower-level extensions are also possible, and may be target-specific.
Low-level implementation patterns are added, such as \inlineRise{circularBuffer} or \inlineRise{rotateValues}, which also requires implementing code generation for them as discussed in \cref{ch:imperative-code}.
Low-level patterns typically come with rewrite rules such as \inlineElevateRise{useCBuffer} or \inlineElevateRise{useRotateValues} enabling their introduction during rewriting, as we will see later.

Abstractions are built on top of patterns and rewrite rules by composition.
Compositional programs may be defined to abstract over patterns, such as \inlineRise{stencil2d} that is defined in \cref{sec:harris-repr} by composing \inlineRise{map}, \inlineRise{slide} and \inlineRise{transpose}.
Compositional \Elevate{} rewriting strategies may be defined to abstract over rewrite rules, such as \inlineElevateRise{separateConvolutions} that composes \inlineElevateRise{separateConvKernel} with other generic rewrite rules.

Although domain-specific extensions are critical to achieve high performance, we strive to minimise extensions and maximise reuse across algorithms and targets.
It is possible to create a rich space of optimisations by composing simple, specialised rewrite rules with generic ones.
For example, of the 74 rewrite rules applied to the Harris corner detection in this chapter, 2 rewrite rules are inherited from the lambda calculus, 49 rewrite rules express generic properties of high-level patterns, 19 rewrite rules manipulate low-level patterns (4 are for storage folding), and 4 rewrite rules require domain-specific knowledge (for separability and vectorisation).

\paragraph{Reproducing Well-Known Optimisations}
This chapter defines \Elevate{} strategies that reproduce well-known domain-specific optimisations when applied to a given \Rise{} program.
The primary goal is to demonstrate the existence of a rewrite sequence leading to the desired optimised program, which could be found automatically.
\Cref{ch:guided-rewriting} will explore trade-offs between control and automation of optimisations.

\smallskip\noindent
Next, we showcase \Elevate{} strategies by example before giving a language overview.

\subsection{\Elevate{} Strategies by Example}
\label{sec:elevate-rewriting}

To introduce \Elevate{} strategies, we pick up the dot product example from \cref{sec:rewriting}. 
Recall that the high-level \inlineRise{dot} program (\cref{dot-rise}) may be rewritten into the low-level \inlineRise{dotSeq} program (\cref{dot-seq-rise}) by applying the \inlineElevateRise{reduceMapFusion} rewrite rule (\cref{reduceMapFusion}, repeated in \cref{fig:ch4-reduceMapFusion}).
However, many other rewrites are possible, and \cref{sec:rewriting} did not discuss how to decide which rewrite rule to apply where.
Defining an \Elevate{} strategy is one way to make this decision explicit.

\begin{elevate-rise}[numbers=none, xleftmargin=0pt, label={fig:ch4-reduceMapFusion}, caption={\inlineElevateRise{reduceMapFusion} rewrite rule.}]
rule reduceMapFusion = map f $\then$ reduce g init
                    $\rewritesTo$ reduceSeq ($\lambda$acc x. g acc (f x)) init
\end{elevate-rise}

For example, we define the \inlineElevateRise{lowerDot} strategy to rewrite the \inlineRise{dot} program:
\begin{elevate-rise}[numbers=none, xleftmargin=0pt]
strategy lowerDot = topDown(reduceMapFusion)
\end{elevate-rise}

\inlineElevateRise{lowerDot} uses a \inlineElevateRise{topDown} \emph{traversal} combinator to apply the \inlineElevateRise{reduceMapFusion} rewrite rule once, following a depth-first top-down traversal of the \Rise{} Abstract Syntax Tree (AST).
We can visualise the traversal steps on the \inlineRise{dot} program:\footnote{for simplicity, we pretend that $\then$ is an AST node, even though it is syntactic sugar for function application}

\begin{elevate-rise}[xleftmargin=0em, escapechar=|]
|\tikzmark{fa1}|zip a b $\then$ map ($\lambda$x. (fst x) $\times$ (snd x)) $\then$ reduce $+$ 0|\tikzmark{fb1}|
|\tikzmark{fa2}|zip a b|\tikzmark{fb2}| $\then$ map ($\lambda$x. (fst x) $\times$ (snd x)) $\then$ reduce $+$ 0
zip a |\tikzmark{fa3}|b|\tikzmark{fb3}| $\then$ map ($\lambda$x. (fst x) $\times$ (snd x)) $\then$ reduce $+$ 0
|\tikzmark{fa4}|zip a|\tikzmark{fb4}| b $\then$ map ($\lambda$x. (fst x) $\times$ (snd x)) $\then$ reduce $+$ 0
zip |\tikzmark{fa5}|a|\tikzmark{fb5}| b $\then$ map ($\lambda$x. (fst x) $\times$ (snd x)) $\then$ reduce $+$ 0
|\tikzmark{fa6}|zip|\tikzmark{fb6}| a b $\then$ map ($\lambda$x. (fst x) $\times$ (snd x)) $\then$ reduce $+$ 0
zip a b $\then$ |\tikzmark{sa}|map ($\lambda$x. (fst x) $\times$ (snd x)) $\then$ reduce $+$ 0|\tikzmark{sb}|
zip a b $\then$ reduceSeq ($\lambda$acc x. acc + (fst x) $\times$ (snd x)) 0
\end{elevate-rise}%
\begin{tikzpicture}[remember picture, overlay, transform canvas={yshift=0.125em}]
\begin{scope}[line width=2pt, color=red]
    \draw ([yshift=8pt, xshift=-1pt] pic cs:fa1) rectangle ([yshift=-4pt, xshift=1pt]pic cs:fb1);
    \draw ([yshift=8pt, xshift=-1pt] pic cs:fa2) rectangle ([yshift=-4pt, xshift=1pt]pic cs:fb2);
    \draw ([yshift=8pt, xshift=-1pt] pic cs:fa3) rectangle ([yshift=-4pt, xshift=1pt]pic cs:fb3);
    \draw ([yshift=8pt, xshift=-1pt] pic cs:fa4) rectangle ([yshift=-4pt, xshift=1pt]pic cs:fb4);
    \draw ([yshift=8pt, xshift=-1pt] pic cs:fa5) rectangle ([yshift=-4pt, xshift=1pt]pic cs:fb5);
    \draw ([yshift=8pt, xshift=-1pt] pic cs:fa6) rectangle ([yshift=-4pt, xshift=1pt]pic cs:fb6);
\end{scope}
  \begin{scope}[line width=2pt, color=green]
    \draw ([yshift=8pt, xshift=-1pt] pic cs:sa) rectangle ([yshift=-4pt, xshift=1pt]pic cs:sb);
\end{scope}
\end{tikzpicture}

After failing to apply \inlineElevateRise{reduceMapFusion} six times (red rectangles), the traversal succeeds when visiting \inlineElevateRise{map .. $\then$ reduce + 0} (green rectangle).
The left hand side of the rewrite rule is then replaced with its right hand side, producing the low-level \inlineRise{dotSeq} program.

\subsection{\Elevate{} Language Overview}
\label{sec:elevate}

This subsection gives \cref{sec:elevate} an overview of the \Elevate{} strategy language.
A more detailed description is available in \cite{hagedorn2020-elevate}.
\Elevate{} is heavily inspired by earlier works on strategy languages for term rewriting systems \cite{visser1998-strategies, kirchner2015-rewriting}.
Like \Rise{}, \Elevate{} is embedded in Scala.\footnote{the implementation of \Elevate{} is open-source: \url{https://github.com/elevate-lang/elevate}}

\Elevate{} strategies are modeled as functions transforming programs from \Rise{}, or other languages \cite{hagedorn2020-elevate-arxiv}.
The type of a strategy is parameterised by \inlineElevateRise{P}, the type of the rewritten program:
\begin{elevate-rise}[numbers=none]
type Strategy[P] = P $\to$ RewriteResult[P]

enum RewriteResult[P] = Success[P](p: P) | Failure[P](s: Strategy[P])
\end{elevate-rise}

A \inlineElevateRise{RewriteResult} is an applicative error monad \cite{jones1993-monads, liang1995-monad}.
Applied to a program, a strategy may succeed and return the transformed program in \inlineElevateRise{Success}, or alternatively fail and return the unsuccessful strategy in \inlineElevateRise{Failure}.

The simplest examples of strategies are \inlineElevateRise{id} that always succeeds by returning the input program unchanged, and \inlineElevateRise{fail} that always fails by returning itself:

\begin{elevate-rise}[numbers=none]
strategy id = $\lambda$p. Success(p)
strategy fail = $\lambda$p. Failure(fail)
\end{elevate-rise}

In this thesis we use a convenient \inlineElevateRise{strategy} syntax, different from what we would write in the Scala embedding:

\begin{elevate-rise}[numbers=none]
def id[P]: Strategy[P] = (p: P) => Success(p)
def fail[P]: Strategy[P] = (p: P) => Failure(fail)
\end{elevate-rise}

\paragraph{Rewrite Rules as Strategies}
In \Elevate{}, rewrite rules are modelled as strategies.
For example, \inlineElevateRise{reduceMapFusion} succeeds when the input program matches its left-hand side, returning its instantiated right-hand side.
The convenient \inlineElevateRise{rule} syntax used in \cref{fig:ch4-reduceMapFusion} allows defining rules independently from the concept of \Elevate{} strategy.
Separating rules from strategies is also useful because rules need to be trusted or verified, while strategies are correct by composition.
Rules defined using the \inlineElevateRise{rule} syntax have a corresponding definition as \Elevate{} strategies in Scala.
For example, \inlineElevateRise{reduceMapFusion} may be defined by pattern matching over the AST of \Rise{} programs:

\begin{elevate-rise}[numbers=none]
def reduceMapFusion: Strategy[Rise] = {
  case app(app(app(reduce, g), init), app(app(map, f), in)) =>
    Success(app(app(app(reduceSeq, fun(acc => fun(x =>
      app(app(g, acc), app(f, x)))), init), in)))
  case _ => Failure(reduceMapFusion)
}
\end{elevate-rise}


\paragraph{Strategy Combinators}
\emph{Strategy combinators} enable defining strategies as compositions of other strategies.
The sequential combinator (\inlineElevateRise{;}) composes two strategies by performing the second one on the transformed program from the first strategy.
It may be defined in terms of the standard monadic bind combinator.
The \inlineElevateRise{<+} combinator (called \emph{left choice}) composes two strategies by performing the second only if the first strategy fails.
It may be defined in terms of the standard monadic mplus combinator.
More strategy combinators can be defined in terms of these two basic ones, as shown in \cref{elevate-combinators}.
The \inlineElevateRise{try} combinator tries to perform a given strategy, and does nothing if the strategy fails, returning the input program unmodified.
\inlineElevateRise{repeat} performs a strategy repeatedly until it fails.

\begin{elevate-rise}[numbers=none, xleftmargin=0pt, label=elevate-combinators, caption={Selection of \Elevate{} combinator definitions}]
strategy try(s) = s <+ id
strategy repeat(s) = $\lambda$p. try(s; repeat(s))(p)
\end{elevate-rise}


\paragraph{Traversals}
\emph{Traversals} are strategy combinators controlling the AST location at which other strategies are applied.
We have already seen \inlineElevateRise{topDown} that traverses the program using a depth-first top-down AST traversal, performing the given strategy at the first possible location.
Traversals like \inlineElevateRise{topDown} or \inlineElevateRise{bottomUp} are defined recursively in terms of more basic traversal steps called \inlineElevateRise{one}, \inlineElevateRise{some} and \inlineElevateRise{all} that need to be provided for a given program type \inlineElevateRise{P}.
More traversals are defined in \cite{hagedorn2020-elevate}, including \Rise{}-specific traversals.

\paragraph{Normal Forms}
\emph{Normal forms} are often desired when rewriting programs, such as the standard lambda calculus $\beta\eta$ normal form \cite{bezem1993-typed-lambda-calculi} or fusion normal forms \cite{emoto2007-skeleton-fusion}.
Later in this chapter, a custom \inlineElevateRise{reduceFusedForm} is used, combining reduction rules such as $\beta$ and $\eta$ reductions with fusion rules such as map fusion.
To enforce normal forms, the \inlineElevateRise{normalize} combinator applies a given strategy somewhere in the program repeatedly, until it cannot be applied anymore:
\begin{elevate-rise}[numbers=none, xleftmargin=0pt]
strategy normalize(s) = repeat(topDown(s))
\end{elevate-rise}

After performing \inlineElevateRise{normalize(s)}, we know that \inlineElevateRise{s} can no longer be applied to any location in the transformed program.
Termination of normal forms, and \Elevate{} strategies in general, is the performance engineer's responsibility.
\section{The Harris Corner Detection Case Study}
\label{sec:harris-repr}

The image processing domain is a good candidate for a case study.
First, the industrial-strength domain-specific compiler Halide offers a challenging comparison point.
Halide also supports controlling optimisation through schedules, enabling comparison to the control of \Elevate{} strategies.
Second, image processing pipelines present unique optimisation challenges that were not explored in prior \Lift{} work.
To reach high-performance, \Shine{} needs to be extended.

The Harris corner (and edge) detector \cite{harris-1988} is a well established image processing pipeline that we use as a case study in this chapter.
The following subsections first describe how the Harris corner detection is defined in Halide before describing how to represent the Harris corner detection in \Rise{}.

\subsection{The Harris Corner Detection Algorithm in Halide}

While many algorithmic variations of the Harris corner detector exist, we use the algorithm found in the Halide repository as our reference (\cref{halide-harris-algorithm}).
The Harris image processing pipeline is defined by inter-dependent functions.
Each function is defined using an indexed notation and goes from coordinates -- belonging to infinite integer domains -- to value.
This variant does not include padding for the stencil borders, and instead the output image is slightly smaller than the input image.
Image bounds are implicit in the Halide algorithm, they are inferred using interval arithmetic, proceeding from output to input.

\begin{halide}[float, label=halide-harris-algorithm, caption={Harris operator algorithm in Halide, from the official Github repository ({\tiny \url{https://github.com/halide/Halide/blob/c2b6da28843a36e53b1c2cf9fd6fa390afeb5896/apps/harris/harris\_generator.cpp\#L19-L61}}). }, numbers=left,basicstyle=\ttfamily\scriptsize]
Var x, y, c;
Func gray, Iy, Ix, Ixx, Iyy, Ixy, Sxx, Syy, Sxy, det, trace;

gray(x, y) = (0.299f * input(x, y, 0) +
              0.587f * input(x, y, 1) +
              0.114f * input(x, y, 2));

Iy(x, y) = gray(x - 1, y - 1) * (-1.0f / 12) + gray(x - 1, y + 1) * (1.0f / 12) +
           gray(x, y - 1) * (-2.0f / 12) + gray(x, y + 1) * (2.0f / 12) +
           gray(x + 1, y - 1) * (-1.0f / 12) + gray(x + 1, y + 1) * (1.0f / 12);

Ix(x, y) = gray(x - 1, y - 1) * (-1.0f / 12) + gray(x + 1, y - 1) * (1.0f / 12) +
           gray(x - 1, y) * (-2.0f / 12) + gray(x + 1, y) * (2.0f / 12) +
           gray(x - 1, y + 1) * (-1.0f / 12) + gray(x + 1, y + 1) * (1.0f / 12);

Ixx(x, y) = Ix(x, y) * Ix(x, y);
Iyy(x, y) = Iy(x, y) * Iy(x, y);
Ixy(x, y) = Ix(x, y) * Iy(x, y);

Expr sum3x3(Func f, Var x, Var y) {
  return f(x - 1, y - 1) + f(x - 1, y) + f(x - 1, y + 1) +
         f(x, y - 1) + f(x, y) + f(x, y + 1) +
         f(x + 1, y - 1) + f(x + 1, y) + f(x + 1, y + 1);
}

Sxx(x, y) = sum3x3(Ixx, x, y);
Syy(x, y) = sum3x3(Iyy, x, y);
Sxy(x, y) = sum3x3(Ixy, x, y);

det(x, y) = Sxx(x, y) * Syy(x, y) - Sxy(x, y) * Sxy(x, y);
trace(x, y) = Sxx(x, y) + Syy(x, y);
output(x, y) = det(x, y) - 0.04f * trace(x, y) * trace(x, y);
\end{halide}


\begin{figure}
  \centering
  \includegraphics[width=\linewidth]{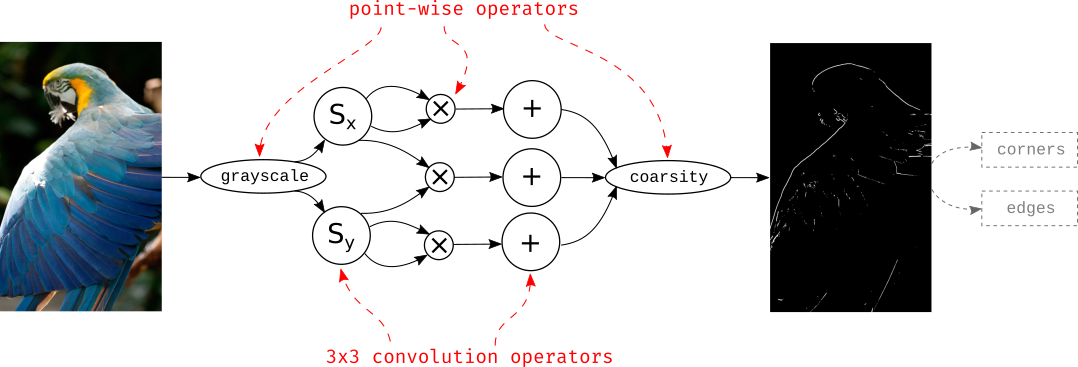}
  \caption{Harris corner detection computation flow, and example image from the Halide repository {(\tiny \url{https://github.com/halide/Halide/blob/c2b6da28843a36e53b1c2cf9fd6fa390afeb5896/apps/images/rgb.png})}. }
  \label{fig:harris}
\end{figure}

\begin{figure}
  \centering
  \includegraphics[width=0.9\linewidth]{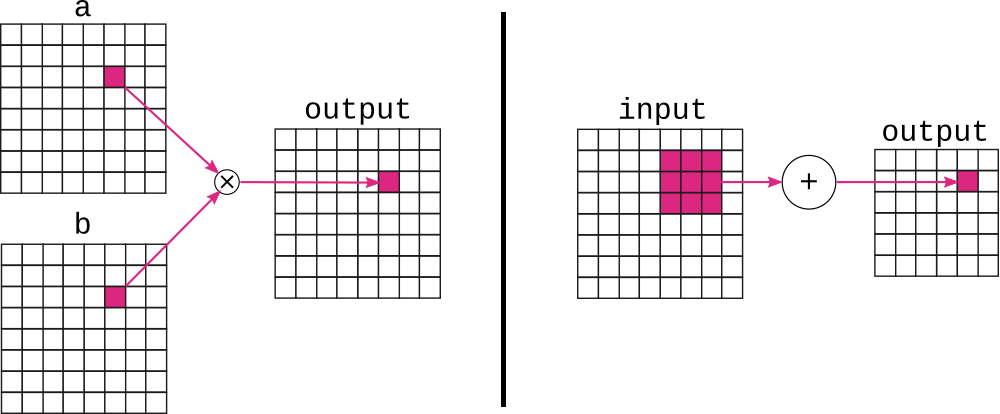}
  \caption{Example of pointwise operator (left) and convolution operator (right).}
  \label{fig:pointwise-and-convolution}
\end{figure}

\paragraph{Computation Flow}
\Cref{fig:harris} visualises the Harris operator computation flow, showing computation nodes and data dependency edges.
Given an image on the left, point-to-point operators (grayscale, multiplications $\times$, coarsity) and $3 \times 3$ convolutions (sobel operators $S_x$ and $S_y$, sums $+$) are combined to detect corners and edges highlighted in the output on the right.
The output of the $S_x$ computation is the \inlineHalide{Ix} image and the output of the coarsity computation is the \inlineHalide{output} image.
The det and trace images from the Halide definition are not explicitly visible in this graph, they are collapsed into the coarsity computation.

\paragraph{Example Operators}
\Cref{fig:pointwise-and-convolution} gives visual examples of pointwise and convolution operators.
On the left, the pointwise operator $\times$ multiplies each pixel of $a$ with a pixel of $b$ to produce a pixel of output.
On the right, the convolution operators $+$ sums each $3 \times 3$ neighbourhood of input pixels to produce a pixel of output.
To avoid reading outside of the image, the output of the convolution ($6 \times 6$) is smaller than its input ($8 \times 8$).

\subsection{Representing the Harris Corner Detection in \Rise{}}

To represent the Harris corner detection as a high-level \Rise{} program, we proceed by composition.
We first look at how to define pointwise operators before looking at convolution operators.
Once we have these building blocks, we can easily assemble them into the complete Harris operator according to the computation flow from \cref{fig:harris}.

\paragraph{Representing Pointwise Operators}
To represent pointwise operators in the high-level \Rise{} language, we do not require any image-specific patterns (\cref{rise-pointwise}).

Where indices are used to access values in Halide, \inlineRise{map} is used to access array values in \Rise{}.
Where values from multiple inputs are freely combined in Halide, \inlineRise{zip} is used to combine arrays in \Rise{}.
Finally, using \inlineRise{transpose} in \Rise{} is similar to swapping indices in Halide.
As done in \Lift{}\cite{lift-rewrite-2015, lift-stencil-2018}, 2D mapping and zipping is defined by composition (\cref{line:map2d,line:zip2d}).

In \cref{line:grayscale}, we define \inlineRise{grayscale} by moving the outer array dimension corresponding to the RGB color channels to the inside where it is consumed by a dot product.
In \cref{line:x2d,line:coarsity}, we define $\times_{2D}$ and \inlineRise{coarsity} by combining pixels of input arrays with \inlineRise{zip2d} before processing each combination with \inlineRise{map2d}.

Note that we use \inlineElevateRise{let x = v ...} as syntactic sugar for \inlineElevateRise{($\lambda$x. ...) v}.

\begin{elevate-rise}[label=rise-pointwise, caption={High-level pointwise operators in \Rise{}.}]
def map2d (f: $s \to t$): $\arr{n}{\arr{m}{s}} \to \arr{n}{\arr{m}{t}}$ =$\label{line:map2d}$
  map (map f)

def zip2d (a: $\arr{n}{\arr{m}{s}}$) (b: $\arr{n}{\arr{m}{t}}$): $\arr{n}{\arr{m}{\tup{s}{t}}}$ =$\label{line:zip2d}$
  zip a b $\then$ map ($\lambda$p. zip (fst p) (snd p))

def grayscale (RGB: $\arr{3}{\arr{n}{\arr{m}{\float{}}}}$): $\arr{n}{\arr{m}{\float{}}}$ =$\label{line:grayscale}$
  RGB $\then$ transpose $\then$ map transpose $\then$
  map2d (dot $\begin{bmatrix}0.299 & 0.587 & 0.114\end{bmatrix}$)

def $\times_{2D}$ (a: $\arr{n}{\arr{m}{\float{}}}$) (b: $\arr{n}{\arr{m}{\float{}}}$): $\arr{n}{\arr{m}{\float{}}}$ =$\label{line:x2d}$
  zip2d a b $\then$ map2d $\times$

def coarsity ($S_{xx}$: $\arr{n}{\arr{m}{\float{}}}$) ($S_{xy}$: $\arr{n}{\arr{m}{\float{}}}$) ($S_{yy}$: $\arr{n}{\arr{m}{\float{}}}$) ($\kappa$: $\float{}$): $\arr{n}{\arr{m}{\float{}}}$ =$\label{line:coarsity}$
  zip2d $S_{xx}$ (zip2d $S_{xy}$ $S_{yy}$) $\then$
  map2d ($\lambda$p.
    let ($s_{xx}$, ($s_{xy}$, $s_{yy}$)) = p
    let det = $s_{xx} \times s_{yy}$ - $s_{xy} \times s_{xy}$
    let trace = $s_{xx}$ + $s_{yy}$
    det - $\kappa~\times$ trace $\times$ trace)
\end{elevate-rise}

\medskip

\paragraph{Representing Convolution Operators}
Where neighbor values are accessed by indexing in Halide, they are collected into arrays using \inlineRise{slide} in \Rise{} (\cref{rise-stencil}).
The \inlineRise{slide} pattern is one-dimensional but can be composed to create multi-dimensional sliding windows \cite{lift-stencil-2018}, \inlineElevateRise{slide2d} is an example shown in \cref{line:slide2d}.
Two-dimensional stencil operators are then constructed by first creating neighborhoods with \inlineElevateRise{slide2d} before processing them using \inlineElevateRise{map2d} as shown in \cref{line:stencil2d}.

In \cref{line:conv3x3}, we define the more specific \inlineElevateRise{conv3x3} in terms of \inlineRise{stencil2d}, since convolutions are stencils.
For each neighbourhood, convolution weights are combined with neighbourhood values using \inlineRise{dot (join weights) (join w)}.
From there, defining $S_x$ and $S_y$ only requires providing the convolution weights (\cref{line:sx,line:sy}).
Finally in \cref{line:p3x3}, $+_{3 \times 3}$ is defined by summing neighbourhood values using \inlineRise{reduce + 0 (join w)}.

\clearpage

\begin{elevate-rise}[label=rise-stencil, caption={High-level stencil operators in \Rise{}.}]
def slide2d ($n_{sz}$: $\nat$) ($n_{sp}$: $\nat$) ($m_{sz}$: $\nat$) ($m_{sp}$): $\nat$ =$\label{line:slide2d}$
  map (slide $n_{sz}$ $n_{sp}$) $\then$ slide $m_{sz}$ $m_{sp}$ $\then$ map transpose

def stencil2d (f: $\arr{N}{\arr{M}{s}} \to t$): $\arr{(n+N-1)}{\arr{(m+M-1)}{s}} \to \arr{n}{\arr{m}{t}}$ =$\label{line:stencil2d}$
  slide2d N 1 M 1 $\then$ map2d f
  
def conv3x3 (weights: $\arr{3}{\arr{3}{\float}}$): $\arr{(n+2)}{\arr{(m+2)}{\float}} \to \arr{n}{\arr{m}{\float}}$ =$\label{line:conv3x3}$
  stencil2d ($\lambda$(w : 3.3.f32). dot (join weights) (join w))

def $S_x$ = conv3x3 (${\scriptscriptstyle\begin{bmatrix}-1&0&1\\ -2&0&2\\ -1&0&1\end{bmatrix}} \times \frac{1}{12}$)$\label{line:sx}$

def $S_y$ = conv3x3 (${\scriptscriptstyle\begin{bmatrix}-1&-2&-1\\ 0&0&0\\ 1&2&1\end{bmatrix}} \times \frac{1}{12}$)$\label{line:sy}$

def $+_{3 \times 3}$ = stencil2d ($\lambda$(w : 3.3.f32). reduce + 0 (join w))$\label{line:p3x3}$
\end{elevate-rise}

\paragraph{Representing the Entire Pipeline}
\Cref{rise-harris} defines the entire Harris image detection pipeline by composing all of the previously defined building blocks.
Note that the final \Rise{} program only contains generic high-level patterns and basic language constructs - no image-specific internal representation is required.
All program abstractions such as \inlineElevateRise{map2d} or \inlineElevateRise{slide2d} can be eliminated by inlining their definition.
In the \Rise{} program, array bounds are inferred as part of type inference and may be explicitly annotated as seen in previous listings.
Here type inference will check that the \inlineRise{RGB} input is of type \inlineRise{3.(n+4).(m+4).f32} and the smaller output of type \inlineRise{n.m.f32}.

\begin{elevate-rise}[label=rise-harris, caption={High-level Harris operator in \Rise{}.}]
def harris (RGB: $\arr{3}{\arr{(n+4)}{\arr{(m+4)}{\float{}}}}$): $\arr{n}{\arr{m}{\float{}}}$ =
  let $I$ = grayscale RGB
  let $I_x$ = $S_x$ $I$
  let $I_y$ = $S_y$ $I$
  let $I_{xx}$ = $\times_{2D}$ $I_x$ $I_x$
  let $I_{xy}$ = $\times_{2D}$ $I_x$ $I_y$
  let $I_{yy}$ = $\times_{2D}$ $I_y$ $I_y$
  let $S_{xx}$ = $+_{3 \times 3}$ $I_{xx}$
  let $S_{xy}$ = $+_{3 \times 3}$ $I_{xy}$
  let $S_{yy}$ = $+_{3 \times 3}$ $I_{yy}$
  coarsity $S_{xx}$ $S_{xy}$ $S_{yy}$ 0.04
\end{elevate-rise}

\clearpage
\section{Optimising the Harris Corner Detection with \Elevate{}}
\label{sec:harris-opt}

We now study how to optimise the \Rise{} Harris corner detection.
As a composition of pointwise and stencil operators, the Harris operator is more complex than its individual parts, which exposes more optimisation opportunities.
Although our case study focuses on the Harris operator, the optimisations that we study are generalisable and applicable to other image processing pipelines that compose pointwise and stencil operators.

The \Lift{} project has been extended to express stencils and overlapped tiling \cite{lift-stencil-2018}, but lacks crucial optimisations for image processing \cite{hlt-lacassagne-2014}, such as operator fusion or circular buffering, that are supported by Halide.
This leads to poor performance, \cref{fig:resultsIntro} showed an example where \Lift{} generates 3.12$\times$ slower code than Halide.
This section uses an optimised Halide schedule of the Harris operator as reference to demonstrate how \Elevate{} is used to perform equivalent and additional optimisations of \Rise{} programs.

\Cref{halide-schedule} shows the Halide schedule describing the optimisations applied to the Harris operator.
The schedule applies multi-threading, vectorisation and describes how the stages interact by storing images in intermediate buffers.
Halide makes some implicit optimisation decisions appropriate for image processing pipelines, such as using circular buffers.

\begin{halide}[label=halide-schedule, caption={Optimised Harris operator schedule from the Halide repository ({\tiny \url{https://github.com/halide/Halide/blob/c2b6da28843a36e53b1c2cf9fd6fa390afeb5896/apps/harris/harris\_generator.cpp\#L86-L90}}). }]
const int vec = natural_vector_size<float>();
output.split(y, y, yi, 32).parallel(y)
  .vectorize(x, vec);
gray.store_at(output, y).compute_at(output, yi)
  .vectorize(x, vec);
Ix.store_at(output, y).compute_at(output, yi)
  .vectorize(x, vec);
Iy.store_at(output, y).compute_at(output, yi)
  .vectorize(x, vec);
Ix.compute_with(Iy, x);
\end{halide}

\begin{figure}
  \centering
  \includegraphics[width=\linewidth]{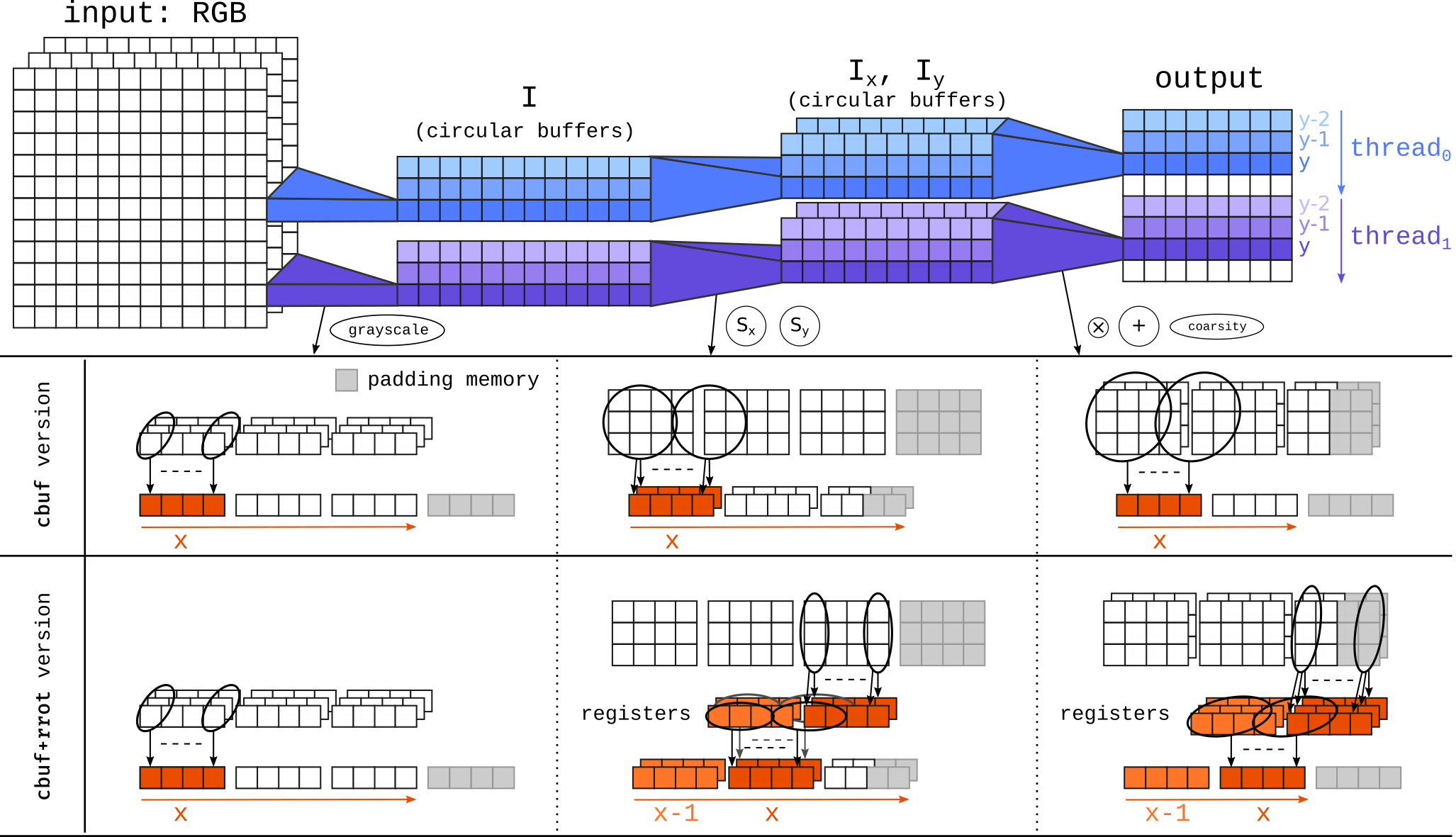}
  \caption{Overview of the optimisations applied on the Harris corner detector.}
  \label{fig:harris-optimisations}
\end{figure}

\Cref{fig:harris-optimisations} visualises the computation with these optimisations applied.
The upper part of \cref{fig:harris-optimisations} shows the input image on the left, where three color channels are combined by \emph{grayscale}.
Grayscale lines are stored in a buffer to be processed by the sobel operators ($S_x$ and $S_y$).
The resulting buffers are then multiplied ($\times$), summed ($+$) and \emph{coarsity} is applied to compute the final output.
\emph{Operator fusion} is applied to the computational flow (\cref{fig:harris}) so that only two intermediate buffers are used.
\emph{Multi-threading} is exploited by parallelising the $y$ dimension and computing chunks of output lines in parallel (thread$_0$ and thread$_1$ annotations on the right).
\emph{Circular buffers} are used for the intermediate results.
Each thread stores three lines in the buffer $I$.
These lines are used to compute two lines and store them in buffers $I_x$ and $I_y$.
Similarly, three lines of both $I_x$ and $I_y$ are used to compute one line of the output.

The lower part of \cref{fig:harris-optimisations} shows two different ways to optimise the computations of individual image lines.
The \inlineElevateRise{cbuf} version is what Halide does: it uses \emph{vectorisation} to process lines one vector at a time.
The \inlineElevateRise{cbuf+rrot} version below is not supported by Halide at the time of writing: it also uses \emph{vectorisation} but further incorporates \emph{convolution separation}, enabling \emph{register rotation} as described in \cite{hlt-lacassagne-2014}.
This is shown in the center and right of the bottom row, where the two-dimensional reductions are decomposed in a vertical reduction followed by a horizontal reduction.
Temporary vector registers are rotated to hold the last vertical reductions that are used for a horizontal reduction.

~\\
The following subsections first show how to replicate the optimisations described by the Halide schedule with extensible \Elevate{} strategies.
This already goes beyond the capabilities of the \Lift{} compiler.
Then, they show how to go beyond the optimisations that Halide performs by incorporating the additional convolution separation and register rotation optimisations.

\subsection{Reproducing the Halide Optimisations with \Elevate{}}
\label{opt-beyond-lift}

\Cref{cbuf-strategy} shows the \Elevate{} rewriting strategy that reproduces the optimisations from the reference Halide schedule through a sequential composition of smaller strategies.
In the following, we discuss the purpose of each smaller strategy and give intuitions on how to define them in \Elevate{} by composing many individual rewrite rule applications.

\clearpage

\begin{elevate-rise}[label=cbuf-strategy, caption={\Elevate{} strategy using circular buffering for the Harris operator.}]
strategy cbufVersion =
  fuseOperators;
  splitPipeline(32); parallel;
  vectorizeReductions(vec);
  harrisIxWithIy;
  circularBufferStages;
  sequentialLines;
  usePrivateMemory; unrollReductions
\end{elevate-rise}

\paragraph{Operator Fusion}

The reference Halide schedule specifies which temporary values should be stored in memory using \inlineHalide{store_at} directives.
Otherwise operators are fused by default, storing temporary results in registers instead of memory.
This transformation is more complex than loop fusion, which is why \Lift{} fails to apply it using its simple \inlineElevateRise{map}-fusion rules.
In, \Elevate{} we define the \inlineElevateRise{fuseOperators} strategy (all strategy definitions can be found in our artefact) transforming the Harris program (\cref{rise-harris}) into a pipeline over image lines:

\begin{elevate-rise}[numbers=none, xleftmargin=0pt, label=multi-threading-goal]
map grayLine  $\then$ slide 3 1 $\then$
map sobelLine $\then$ slide 3 1 $\then$ map coarsityLine
\end{elevate-rise}

\noindent
Where \inlineElevateRise{grayLine} is a function computing a grayscale line, \inlineElevateRise{sobelLine} a function computing a line of sobel convolutions, and \inlineElevateRise{coarsityLine} a function computing a line of output (with multiplications, sums and coarsity fused) as shown in \cref{fig:harris-optimisations}.

\paragraph{Multi-threading}

To take advantage of thread-level parallelism, the Halide schedule splits the output into parallel chunks of 32 lines: \inlineHalide{output.split(y, y, yi, 32).parallel(y)}.
The \Elevate{} strategy \inlineElevateRise{splitPipeline(32); parallel} has the same effect, producing a program that slides over $32+4$ lines of input with step $32$ to compute chunks of size $32$ in parallel:

\begin{elevate-rise}[numbers=none, xleftmargin=0pt]
slide (32+4) 32 $\then$ mapGlobal (
  map grayLine  $\then$ slide 3 1 $\then$
  map sobelLine $\then$ slide 3 1 $\then$
  map coarsityLine
) $\then$ join
\end{elevate-rise}

\noindent
Parallelism is achieved with the low-level \inlineElevateRise{mapGlobal} pattern that applies the nested function in parallel across global threads.
The strategy itself starts by splitting the last map in the pipeline with the \inlineElevateRise{splitJoin} rewrite rule.
Then, it propagates this split to the rest of the pipeline by normalising it with various movement rules.
Finally, all possible map fusions are applied in the pipeline.
All the involved rules are in \cref{multi-threading-strategy}.



\begin{elevate-rise}[label=multi-threading-strategy, caption={Rules involved in the multi-threading optimisation.},float=t]
rule splitJoin(p: $\nat$) =
  map f $\rewritesTo$ split p $\then$ map (map f) $\then$ join

rule slideAfterSplit =
     slide n m $\then$ split p
  $\rewritesTo$ slide (p+n-m) p $\then$ map (slide n m)

rule slideBeforeMap =
  map f $\then$ slide n m $\rewritesTo$ slide n m $\then$ map (map f)

rule slideBeforeSlide =
     slide n 1 $\then$ slide m k
  $\rewritesTo$ slide (m+n-1) k $\then$ map (slide n 1)

rule mapFusion = map f $\then$ map h $\rewritesTo$ map (f $\then$ h)

rule useMapGlobal = map f $\rewritesTo$ mapGlobal f
\end{elevate-rise}

\begin{figure}[t]
  \centering
  \includegraphics[width=0.8\linewidth]{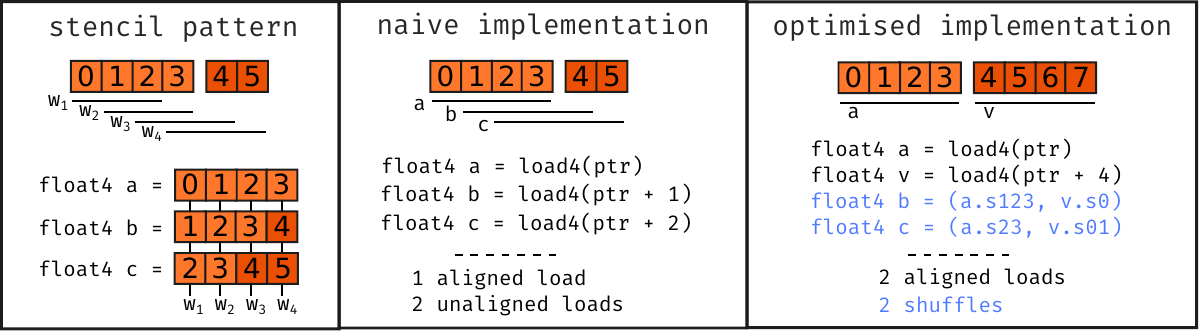}
  \caption{Example of memory loads for a vectorised 1D stencil of size 3. The code is written in pseudo OpenCL syntax.}
  \label{fig:vector-loads}
\end{figure}

\paragraph{Vectorisation}

SIMD parallelism (Single Instruction, Multiple Data) is achieved with vector instructions such as NEON instructions on ARM processors.
In the Halide schedule, multiple \inlineHalide{.vectorize(x, vec)} directives enable this optimisation at multiple spots.

The \inlineElevateRise{vectorizeReductions(vec)} strategy has a similar effect, vectorising all reductions of a program.
To illustrate how the strategy works, we consider a sub-expression found in the Harris operator:

\begin{elevate-rise}[numbers=none, xleftmargin=0pt]
map (reduce + 0) $\then$ map f
\end{elevate-rise}

\noindent
It is vectorised by interpreting the input as a two dimensional array of vectors using \inlineElevateRise{asVector} and computing on vectorised data before going back to scalars using \inlineElevateRise{asScalar}:

\begin{elevate-rise}[numbers=none, xleftmargin=0pt]
transpose $\then$ map (asVector v) $\then$ transpose
  $\then$ map (reduce (mapVec +) (vectorFromScalar 0))
  $\then$ map (mapVec f) $\then$ asScalar
\end{elevate-rise}

\noindent
Where \inlineElevateRise{mapVec} vectorises a scalar function.
Currently, \inlineElevateRise{mapVec} can only be used on functions that consists of easy to vectorise operators such as addition, multiplication, or constants.

\begin{elevate-rise}[label=vectorisation-strategy, caption={Strategy and rules involved in vectorisation},float=t]
strategy vectorize(v: $\nat$) =
  startVectorization(v);
  normalize(vectorizeBeforeMap <+ vectorizeBeforeMapReduce)

rule startVectorization(v: $\nat$) =
  a: $\arr{n \times v}{s}$ $\rewritesTo$ a $\then$ asVector v $\then$ asScalar

rule vectorizeBeforeMap =
  map f $\then$ asVector v $\rewritesTo$ asVector v $\then$ map (mapVec f)

rule vectorizeBeforeMapReduce =
     map (reduce f init) $\then$ asVector v
  $\rewritesTo$ transpose $\then$ map (asVector v) $\then$ transpose $\then$
     map (reduce (mapVec f) (vectorFromScalar init))
\end{elevate-rise}

The desired vectorisation can be defined with an \Elevate{} strategy composing simpler rewrite rules as shown in \cref{vectorisation-strategy}.
In practice, arrays are often not multiples of the vector width.
There are different ways to handle this, but in this case study we round inputs, outputs and temporaries up to a multiple of the vector width -- an option that Halide also provides.





When vectorising stencils the computations are performed on the \inlineElevateRise{w}$_i$ components of three vector values, as shown in the left of \cref{fig:vector-loads}.
The inputs of vectorised stencils are not aligned in memory and can be loaded in different ways.
The naive implementation performs three loads, two of which are not aligned at a vector boundary, as shown in the middle of \cref{fig:vector-loads}.
The optimised implementation, used by \Rise{}, only performs two vector loads followed by vector shuffle instructions, as shown in the right of \cref{fig:vector-loads}.



\paragraph{Circular Buffering}

Circular buffers leverage both the spatial locality of stencils and the temporal locality of sequential execution: only the last $m$ intermediate results need to be stored in memory, and modulo indexing is used: $T[i]$ can be stored in $M[i \mod m]$.
With transparently managed caches, this reduces memory usage and delays cache overflow.

With Halide, \inlineHalide{.store_at(output, y).compute_at(output, yi)} implicitly triggers the use of circular buffers for the introduced temporary.
When combined with the previous multi-threading optimisation, a separate set of circular buffers is used inside each parallel chunk -- where execution is still sequential -- as shown in \cref{fig:harris-optimisations}.

The \Elevate{} strategy \inlineElevateRise{circularBufferStages} has the same effect, producing a program with the shape:

\begin{elevate-rise}[numbers=none, xleftmargin=0pt]
slide (32+4) 32 $\then$ mapGlobal (
  circularBuffer global 3 3 grayLine  $\then$
  circularBuffer global 3 3 sobelLine $\then$
  iterateStream coarsityLine
) $\then$ join
\end{elevate-rise}

\noindent
The \inlineElevateRise{circularBuffer} pattern is a new low-level pattern that we added to \Rise{} in \cref{sec:storage-folding}.
Given an input array, the \inlineElevateRise{circularBuffer} pattern returns an array of sliding windows similar to the \inlineElevateRise{slide} pattern, but the last $m$ values have been loaded into the circular buffer.
The \inlineElevateRise{iterateStream} pattern is used to read sequentially from the circular buffer.
\inlineElevateRise{circularBufferStages} works by rewriting \inlineElevateRise{slide} into the \inlineElevateRise{circularBuffer} pattern, fusing \inlineElevateRise{circularBuffer} and \inlineElevateRise{map}, and introducing the \inlineElevateRise{iterateStream} pattern using the rewrites rules of \cref{circular-buffering-strategy}.


\begin{elevate-rise}[label=circular-buffering-strategy, caption={Rewrite rules involved in circular buffering},float=t]
rule useCBuffer(a: $\addr$) =
  slide m 1 $\rewritesTo$ circularBuffer a m m ($\lambda$x. x)

rule cBufferLoadFusion =
     circularBuffer a alloc m load (map f in)
  $\rewritesTo$ circularBuffer a alloc m ($\lambda$x. load (f x)) in

rule useIterateStream = map f $\rewritesTo$ iterateStream f
\end{elevate-rise}

\paragraph{Other Optimisations}

A couple of other optimisations are encoded as \Elevate{} strategies.
The \inlineElevateRise{harrisIxWithIy} strategy emulates the \inlineHalide{Ix.compute_with(Iy, x)} directive from Halide, fusing the loops computing these two intermediate results. 
The \inlineElevateRise{sequentialLines} strategy makes individual line computations sequential, \inlineElevateRise{usePrivateMemory} stores various temporaries in private memory, and \inlineElevateRise{unrollReductions} unrolls reduction loops.

These transformations are not mentioned in the Halide schedule, the two first ones happen implicitly, while reductions have already been unrolled in the algorithm definition.
All these optimisations are already well supported as rewrites by \Lift{} but have not been encoded as rewriting strategies before.

\subsection{Expressing Optimisations beyond Halide with \Elevate{}}
\label{opt-beyond-halide}

The previous section demonstrates how to express all of the optimisations applied by the Halide schedule from \cref{halide-schedule}, using \Elevate{} strategies.
These optimisations are already beyond the reach of the \Lift{} compiler and its automatic search.
This section further demonstrates how the extensibility of \Elevate{} is leveraged to apply optimisations that are not implemented by Halide, further reducing runtime as shown in \cref{sec:imgproc-eval}.

\Cref{rrot-strategy} shows an \Elevate{} strategy that additionally incorporates convolution separation and register rotation optimisations, on top of the previous optimisations from \cref{opt-beyond-lift}.
These two optimisations are orthogonal to multi-threading and circular buffering in this case study, as they operate on a different dimension.
Separating the convolution is necessary to enable register rotation, and is not expressible in Halide without manually changing the algorithm.
Register rotation is recognised as a worthwhile optimisation by the Halide developers, but is not yet supported by Halide~({\footnotesize\url{https://github.com/halide/Halide/issues/2905}}).

Implementing register rotation in Halide would require non trivial extensions including changes to the scheduling API, resulting in significant work.
We discuss here how both optimisations are applied by defining an \Elevate{} strategy outside of the \Shine{} compiler, leveraging the simple \inlineElevateRise{rotateValues} pattern added in \cref{sec:storage-folding}.

\begin{elevate-rise}[label=rrot-strategy, caption={\Elevate{} strategy applying convolution separation and register rotation to the Harris operator.
These optimisations are not available in Halide.
Changes compared to \cref{cbuf-strategy} are highlighted in pink.}]
strategy cbuf+rrotVersion =
  fuseOperators;
  splitPipeline(32); parallel;
  *separateConvolutions*;
  vectorizeReductions(vec);
  harrisIxWithIy;
  circularBufferStages;
  *rotateValuesAndConsumeLines*;
  usePrivateMemory; unrollReductions
\end{elevate-rise}

\paragraph{Convolution Separation}

The two-dimensional sobel and sum convolutions in the Harris detector are separable into two one-dimensional convolutions following the observation that the convolution kernel matrix is separable into a column and row vector:\footnote{the same principles apply to both column-row and row-column decompositions}
{\small \begin{equation*}\label{convolution-separation-examples}
\begin{bmatrix}-1&0&1\\ -2&0&2\\ -1&0&1\end{bmatrix}
= \begin{bmatrix}1\\2\\1\end{bmatrix} \begin{bmatrix}-1&0&1\end{bmatrix}
\end{equation*}}
This decomposition, sometimes called convolution separation, is used to reduce both memory accesses and arithmetic complexity, but is not possible for arbitrary convolutions as it depends on the weights involved.
Convolution separation is often manually applied \cite{median-filter-separation-1981, anisotropic-gauss-filter-separation-2003, gabor-filter-separation-2005, bilateral-filter-separation-2005}.

With \Elevate{}, such a domain-specific, or even convolution-specific optimisation can be defined outside of the \Shine{} compiler.
This is valuable because incorporating this capability does not require re-engineering the entire compilation stack, something that is challenging in domain-specific compilers as discussed in \cref{domain-specific-compilers}.
In \Rise{}, 2D convolutions typically operate on a vertical neighborhood of lines (\inlineElevateRise{nbhV}).
Multiple 2D neighborhoods are created (\inlineElevateRise{nbh2d}) with \inlineElevateRise{slide} and \inlineElevateRise{transpose} before performing a dot product between the 2D weights and each neighborhood:

\begin{elevate-rise}[numbers=none, xleftmargin=0pt]
nbhV $\then$ map (slide 3 1) $\then$ transpose $\then$ map ($\lambda$nbh2d.
  dot (join weights2d) (join nbh2d))
\end{elevate-rise}

\noindent
The 2D weights are separated into vertical ones (\inlineElevateRise{wV}) and horizontal ones (\inlineElevateRise{wH}), each being used to perform a 1D convolution:

\begin{elevate-rise}[numbers=none, xleftmargin=0pt]
nbhV $\then$ transpose $\then$ map (dot wV)
     $\then$ slide 3 1 $\then$ map (dot wH)
\end{elevate-rise}



The strategy \inlineElevateRise{pushSeparation(separateConvKernel(weights2d, wV, wH))} realises this transformation (\cref{convolution-separation-strategy}).
Given explicit weights, the rewrite rule \inlineElevateRise{separateConvKernel} encodes the decomposition on the dot product.
Then, the \inlineElevateRise{pushSeparation} strategy uses generic rewrite rules to ``push'' the dot product decomposition across the surrounding dimensions.
\inlineElevateRise{separateConvolutions} in \cref{rrot-strategy} uses these components to separate the sobel and sum convolutions of the Harris operator.

\begin{elevate-rise}[label=convolution-separation-strategy, caption={Strategy and rules to separate a convolution through one dimension.},float=t]
rule separateConvKernel(weights2d, wV, wH) =
     dot (join weights2d) (join nbh)
  $\rewritesTo$ nbh $\then$ transpose $\then$ map (dot wV) $\then$ dot wH

strategy pushSeparation(separate) =
  topDown(separate); reducedFissionedForm;
  topDown(mapSlideAfterTranspose); 
  reducedFusedForm; reducedFissionedForm;
  normalize(slideAfterMapMapF)

rule mapSlideAfterTranspose =
     map (slide n m) $\then$ transpose
  $\rewritesTo$ transpose $\then$ slide n m $\then$ map transpose

rule slideAfterMapMapF =
  slide n m $\then$ map (map f) $\rewritesTo$ map f $\then$ slide n m
\end{elevate-rise}



\paragraph{Register Rotation}
As seen in \cref{sec:storage-folding}, register rotation is a storage folding optimisation, just like circular buffering.
In the bottom of \cref{fig:harris-optimisations}, convolution separation is combined with register rotation and vectorisation: vectors of vertical reductions are rotated while computing vectors of horizontal reductions.

If we start from a convolution that was separated as described above:

\begin{elevate-rise}[numbers=none, xleftmargin=0pt]
map (dot wV) $\then$ slide 3 1 $\then$ map (dot wH)
\end{elevate-rise}

\noindent
The \inlineElevateRise{rotateValues} pattern from \cref{sec:storage-folding}  may be introduced in stead of the second \inlineElevateRise{slide}.
As the rotation produces a sequential stream, we must use \inlineElevateRise{iterateStream} afterwards:

\begin{elevate-rise}[numbers=none, xleftmargin=0pt]
map (dot wV) $\then$ rotateValues private 3 ($\lambda$x. x) $\then$ iterateStream (dot wH)
\end{elevate-rise}

\noindent
Given an input array, \inlineElevateRise{rotateValues} returns an array of sliding windows: the last $m$ values that have been stored in registers.
Values are rotated while the array is read sequentially.

The \Elevate{} strategy \inlineElevateRise{rotateValuesAndConsume} shown in \cref{register-rotation-strategy} performs this program transformation.
A similar strategy is used in \cref{rrot-strategy} for the Harris operator.

\begin{elevate-rise}[label=register-rotation-strategy, caption={Strategy and rules involved in register rotation.}]
strategy rotateValuesAndConsume = topDown(useRotateValues(private));
                                  topDown(useIterateStream)

rule useRotateValues(a: $\addr$) = slide m 1 $\rewritesTo$ rotateValues a m ($\lambda$x. x)
\end{elevate-rise}
\section{Evaluation of Runtime Performance}
\label{sec:imgproc-eval}

The previous section discussed how well-known image processing pipeline optimisations are expressed in a composable and extensible way as \Elevate{} rewriting strategies.
The expressed optimisations go beyond what both \Lift{} and Halide support.

This section investigates whether we truly achieve competitive performance, and the performance impact of the additional optimisations.
We report a systematic runtime performance comparison between the \Shine{}, Halide and \Lift{} compilers, the Halide compiler, the \Lift{} compiler.
Additionally, we compare to the popular OpenCV image processing library, providing an additional reference point to validate the benefits of whole-program optimising compilers.

\subsection{Experimental Setup}

Experiments are conducted on two computers with ARM big.LITTLE configuration as these mobile CPUs are often used in image processing applications.
We use an Odroid XU4 board with a 4-core Cortex A7 and a 4-core Cortex A15 as well as an Odroid N2 board with a 2-core Cortex A53 and a 4-core Cortex A73.

The clock frequencies are set to 1.5Ghz for the XU4, and 1.8Ghz for the N2.
Our compiler implementation generates OpenCL kernels that are executed using the POCL~\cite{pocl-2015} open source implementation of OpenCL that is built on top of LLVM.
We use POCL 1.3 with LLVM 8 on the XU4 and POCL 1.5 with LLVM 10 on the N2.

The OpenCL kernels generated with \Shine{} (\cref{app:harris-opt-progs}) are compared against OpenCV, the OpenCL kernels generated with the \Lift{} implementation from \cite{lift-stencil-2018}, and the binaries generated by Halide~(commit c2b6da2 {\footnotesize\url{https://tinyurl.com/rr7awsr}}).
We use OpenCV 4.3 with NEON vector support enabled.
For \Shine{} we use the two \Elevate{} strategies discussed in \cref{sec:harris-opt}, resuling in two different versions called cbuf and cbuf+rrot.
For Halide we use the reference optimised schedule from \cref{halide-schedule}.
Neither the \Shine{}-generated OpenCL code nor the Halide schedule is specialised for each individual processor, but the final assembly will be respectively specialised by the OpenCL implementation and the Halide compiler.

We report the median runtime of 30 executions, which gives reasonably stable results.
To measure the runtime of the OpenCL kernels, we use the OpenCL profiling API.
For Halide we use C++'s \inlineHalide{std::chrono} clocks as it is done with other benchmarks in the Halide repository.

Two input images are used, one with a resolution of $1536 \times 2560$ pixels, and one of $4256 \times 2832$ pixels.
The first one is taken from the Halide repository, and was shown in \cref{fig:harris}.
The second image is taken from the PolyMage repository.\footnote{\tiny \url{https://bitbucket.org/udayb/polymage/src/446f75628c2272651ea002160f50d95f7fbf4b93/images/venice_wikimedia.jpg}}
We verify that the outputs of the different Harris operator implementations are consistent by computing the Mean-Squared Error and PSNR (Peak Signal-to-Noise Ratio) with the reference output from Halide.
The recorded PSNR is always above 170 decibels, indicating a very strong similarity.

\begin{figure}[t]
  \centering
  \includegraphics[width=\linewidth]{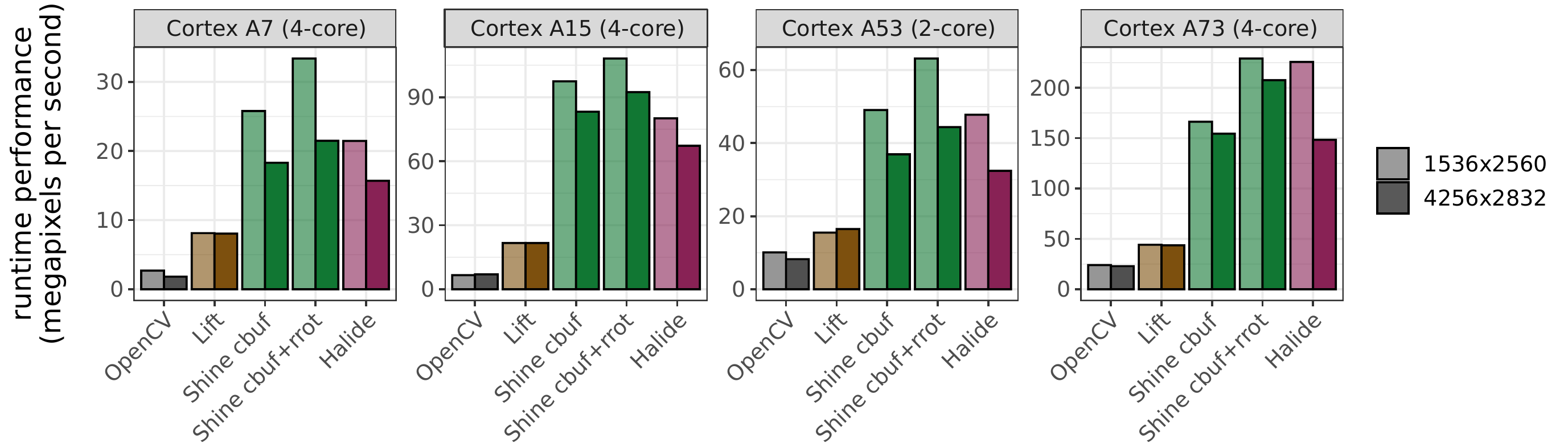}
  \caption{Runtime performance of the Harris corner detection for 4 processors (top legend), 5 implementations (bottom legend) and 2 image resolutions (right legend).}
  \label{fig:imgproc-results}
\end{figure}

\subsection{Performance Results}
\label{perf-results}

\Cref{fig:imgproc-results} shows the runtime performance results.
All 3 compilers -- \Lift{}, \Shine{} and Halide -- outperform OpenCV on all 4 processors and 2 images although OpenCV describes itself as a highly optimised library.
\Shine{} cbuf+rrot outperforms OpenCV by up to 16$\times$ with a geomean of 9.48$\times$.
This highlights the performance benefits brought by whole-program optimisations.

\Shine{} cbuf+rrot outperforms \Lift{} by up to 4.5$\times$ with a geomean speedup of 3.87$\times$.
This is expected because prior \Lift{} work focuses on individual stencil computations and lacks optimisations for image processing pipelines: notably operator fusion and circular buffering.

\Shine{} cbuf is on par with the Halide reference with a geomean speedup of 1.02$\times$.
This demonstrate that \Shine{}'s domain-extensible design is capable of achieving the same performance as Halide, a highly optimised domain-specifc compiler.
While the applied coarse-grain optimisations are the same, small differences in the generated code remain and the resulting performance depends on fine-grain code generation details down to assembly which are out of the scope of this chapter.
At worst, \Shine{} cbuf achieves 0.7$\times$ Halide's performance on the A73 for the $1536\times2560$ image.
At best, \Shine{} cbuf achieves 1.2$\times$ Halide's performance on the A15 for the $4256\times2832$ image.

With convolution separation and register rotation, \Shine{} cbuf+rrot performs much better than without with a geomean speedup of 1.24$\times$ over \Shine{} cbuf and of 1.27$\times$ over Halide.
At best, \Shine{} cbuf+rrot achieves 1.4$\times$ Halide's performance on A7 for the $1536\times2560$ image.
This shows that register rotation is an optimisation worth considering even though it was not implemented in Halide at the time of writing.
Moreover, the \Shine{} cbuf+rrot results demonstrate that a domain-extensible compiler can outperform even a state-of-the-art domain-specific compiler like Halide.
This is achieved by using extensibility to add optimisations that are not built into existing domain-specific compilers.

\subsection{Artifact}

The source code used to produce the performance results from this chapter is available as a peer-reviewed artifact, publicly available on GitHub.\footnote{\url{https://github.com/rise-lang/2021-CGO-artifact}}

\paragraph{Artifact Evaluation}
The artifact supplements the paper \emph{Towards a Domain-Extensible Compiler: Optimizing an Image Processing Pipeline on Mobile CPUs} \cite{koehler2021-elevate-imgproc}, presented at the \emph{International Symposium on Code Generation and Optimization} (CGO) in 2021.
The artifact was evaluated and received the following ACM badges: Artifacts Available, Artifacts Evaluated, and Results Reproduced.
The main goals for artifact evaluation was to use the provided \Shine{} compiler to regenerate the OpenCL kernels used in the experimental evaluation and to reproduce the performance results seen in \cref{fig:resultsIntro} and \cref{fig:imgproc-results}.
For the same processors (or similar enough), we expect the results to show similar performance trends as observed in \cref{perf-results}.

\paragraph{Contents}
The artifact contains the \Rise{}, \Shine{} and \Elevate{} Scala implementations, the Halide compiler, the \Lift{}-generated OpenCL kernels, as well as the benchmarking and plotting programs.
The two input images are provided, and the expected workflow output is CSV and PDF files corresponding to \cref{fig:resultsIntro} and \cref{fig:imgproc-results}.

\paragraph{Requirements}
We recommend using an X86 Linux for the host with at least 2GB of disk space available; and Linux targets with OpenCL support and at least 20MB of disk space available (dependencies excluded).
The software dependencies are listed in the README and we provide an Ubuntu Focal Fossa (20.04 LTS) Dockerfile for convenience.

Reproducing the results reported in \cref{fig:resultsIntro} and \cref{fig:imgproc-results} requires access to ARM Cortex A7, A15, A53 and A73 processors (we used Odroid XU4 and Odroid N2 boards).
Other OpenCL-enabled processors can be used, but may have different performance behavior.
The benchmarks can be run on different processors by writing a small YAML configuration file as long as the benchmark dependencies are available.

\paragraph{Workflow}
Preparing and completing the artifact workflow should take between 2 hours and 1 day, approximately:
\begin{enumerate}[nosep]
\item Install host dependencies
\item Clone repository on the host
\item Generate binaries and OpenCL kernels for each target
\item Configure each target
\item Run benchmarks over ssh for each target
\item Plot figures
\end{enumerate}
More detailed instructions are provided in the README.
\section{Conclusion}
\label{sec:imgproc-conclusion}

This chapter demonstrates how extensibility and controllability are exploited in \Shine{} to generate high-performance code.
Using the Harris corner detection as a case study, our runtime results on four mobile ARM multi-core CPUs and two different image resolutions show that \Shine{} outperforms OpenCV library code by up to 16$\times$ (geomean of 9.48$\times$), outperforms the similarly designed \Lift{} compiler by up to 4.5$\times$ (geomean of 3.87$\times$) and performs up to 1.4$\times$ (geomean of 1.27$\times$) better than the domain-specific compiler Halide (\cref{fig:imgproc-results}).

Optimisations are controlled using \Elevate{} rewriting strategies.
\Elevate{} allows us to reproduce the effect of an optimised Halide schedule applying operator fusion, multi-threading, vectorisation and circular buffering optimisations.
Further, \Elevate{} allows us to go beyond what is possible with Halide schedules, incorporating additional convolution separation and register rotation optimisations.
\Shine{} is extended with new specialised optimisations by adding \Rise{} patterns (\cref{ch:imperative-code}), rewrite rules and \Elevate{} rewriting strategies.

All of the presented optimisations are well-known, and have been studied on the Harris operator before in \cite{hlt-lacassagne-2014}, where they are performed manually.
In general, stencil optimisations such as overlapped tiling \cite{stencil-parallelization-2007, stencil-optimizations-2008, stencil-code-generation-2012, hierarchical-overlapped-tiling-2012} are well-studied and useful beyond image processing.
Similarly, we believe that the presented patterns and rewrite rules are re-usable beyond image processing and across hardware targets, although this remains to be demonstrated.

Automated heuristics and explorations are not always desirable or even feasible as they lack user control, may result in poor performance, and may be too time consuming.
This justifies controlling optimisations with rewriting strategies.
However, \Elevate{} strategies are not easy to write, can be over-detailed and program-specific.
The strategies defined to apply the 6 optimisations consist of more than 600 lines of code defining 57 helper strategies.
To perform all 6 optimisations, thousands of rewrite steps are applied.
The strategies are specialised to our Harris case study and would be challenging to generalise and reuse across image processing pipelines.
The next chapter discusses a novel practical tradeoff between precise control of optimisations (as in this chapter) and full automation of optimisations (as in \Lift{}).

\chapquote{
Every sketch goes through a rewrite stage where a group of writers sits around a table and pitches more jokes and ideas for the piece.
}{Vanessa Bayer}

\chapter{Sketch-Guided Equality Saturation}
\label{ch:guided-rewriting}
The previous chapter combined the control of rewriting strategies with the domain-extensibility of \Shine{} to achieve 6 case study optimisations, and hence generate faster code than Halide schedules that do not support 2 of the optimisations.
However, rewriting strategies are not easy to write.
Deciding when to apply which rewrite is hard: the so-called \emph{phase ordering problem} (\cref{sec:controlling-compilers}).

Equality saturation \cite{tate2009-equality-saturation, willsey2021-egg} is an automated optimisation technique mitigating the phase ordering problem by automatically exploring many possible ways to apply rewrites.
It relies on an efficient representation of many equivalent programs in an e-graph data structure.
Unfortunately, automatically discovering the \Rise{} optimisations applied using rewriting strategies as in \cite{hagedorn2020-elevate} or \cref{ch:imgproc} using equality saturation is prohibitively expensive (\cref{mm-eval}).

This chapter proposes a practical tradeoff between the control of \Elevate{} strategies, and the automation of techniques such as equality saturation.
\Cref{sec:eqsat-background} provides background on equality saturation.
\Cref{sec:sges-motivation} explains our motivations for semi-automatic optimisation.
\Cref{sketching} introduces \emph{sketch-guided equality saturation}, a semi-automatic technique that allows performance engineers to guide rewriting by providing \emph{sketches}: program patterns that leave details unspecified.
\Cref{sec:sges-bindings} explains inefficiencies in the naive encoding of the lambda calculus for equality saturation \cite{willsey2021-egg}, and explores new techniques to efficiently encode a polymorphically typed lambda calculus such as \Rise{}.
Two systematic evaluations of \Rise{} applications are then conducted.
\Cref{lambda-eval} demonstrates that our lambda calculus encoding  reduces the runtime and memory consumption of equality saturation by orders of magnitude when optimising a binomial filter.
\Cref{mm-eval} evaluates sketch-guided equality saturation by reproducing seven realistic optimisations of matrix multiplication from \cite{hagedorn2020-elevate}.
Even with efficient lambda calculus encoding, unguided equality saturation can locate only the two simplest optimisations, the remaining five are undiscovered even with an hour of compilation time and 60GB of RAM.
By specifying three or fewer sketch guides, all seven optimisations are found in seconds of compilation time, using under 1GB of RAM, and generating high performance code.
\Cref{sec:sges-conclusion} concludes.


\section{Background on Equality Saturation}
\label{sec:eqsat-background}

Equality saturation \cite{tate2009-equality-saturation, willsey2021-egg} is a technique for efficiently implementing rewrite-driven compiler optimisations without committing to a single rewrite choice.
Many successful applications of equality saturation sparked from the recent egg library \cite{willsey2021-egg}: optimising linear algebra \cite{wang2020-spores}, shrinking 3D CAD (Computer-Aided Design) models \cite{nandi2020-synthesizing-CAD}, optimising deep learning programs \cite{yang2021-eqsat-tensor, smith2021-access-patterns}, vectorising digital signal processing code \cite{2021-eqsat-vec}, inferring rewrite rules \cite{2021-eqsat-ruler}.

We now demonstrate how equality saturation mitigates the phase ordering problem with a rewriting example where greedily reducing a cost function is not sufficient to find the optimal program.

\subsection{Getting Stuck in a Local Optimum with Greedy Rewriting}
\label{sec:greedy-rewriting}

Rewriting is often used to fuse operators and avoid writing intermediate results to memory, for example:
\begin{align*}
\tag{A} \label{overview-start}
(map~(map~f)) &\circ (transpose \circ (map~(map~g))) \\
&\rewritesTo{}^* \\
\tag{B} \label{overview-goal}
(map~(map~(f &\circ g))) \circ transpose \\
\end{align*}
The initial term (\ref{overview-start}) applies function $g$ to each element of a two-dimensional matrix (using two nested $map$s), transposes the result, and then applies function $f$ to each element.
The optimised term (\ref{overview-goal}) avoids storing an intermediate matrix in memory and transposes the input before applying $g$ and $f$ to each element.
Applying the following rewrite rules in the correct order is sufficient to perform this optimisation:
\begin{align}
\label{overview-move-transpose}
transpose \circ map~(map~a) \longleftrightarrow{}& map~(map~a) \circ transpose\\
\label{overview-comp-assoc}
a \circ (b \circ c) \longleftrightarrow{}& (a \circ b) \circ c\\
\label{overview-map-fusion}
map~a \circ map~b \longleftrightarrow{}& map~(a \circ b)
\end{align}

Rule (\ref{overview-move-transpose}) states that transposing a two-dimensional array before or after applying a function to the elements is equivalent.
Rule (\ref{overview-comp-assoc}) states that function composition is associative.
Finally, rule (\ref{overview-map-fusion}) is the rewrite rule for map fusion.
In this example, minimising the term size results in maximising fused maps and is, therefore, a good cost model. 

If we greedily apply rewrite rules that lower term size, we will only apply rule (\ref{overview-map-fusion}) as this is the only rule that reduces term size.
However, rule (\ref{overview-map-fusion}) cannot be directly applied to term (\ref{overview-start}): it is a local optimum.
The only way to reduce term size further is to first apply the other rewrite rules, which may or may not pay off depending on future rewrites.

\begin{figure}
  \centering
  \begin{subfigure}[b]{0.23\linewidth}
  \includegraphics[width=\linewidth]{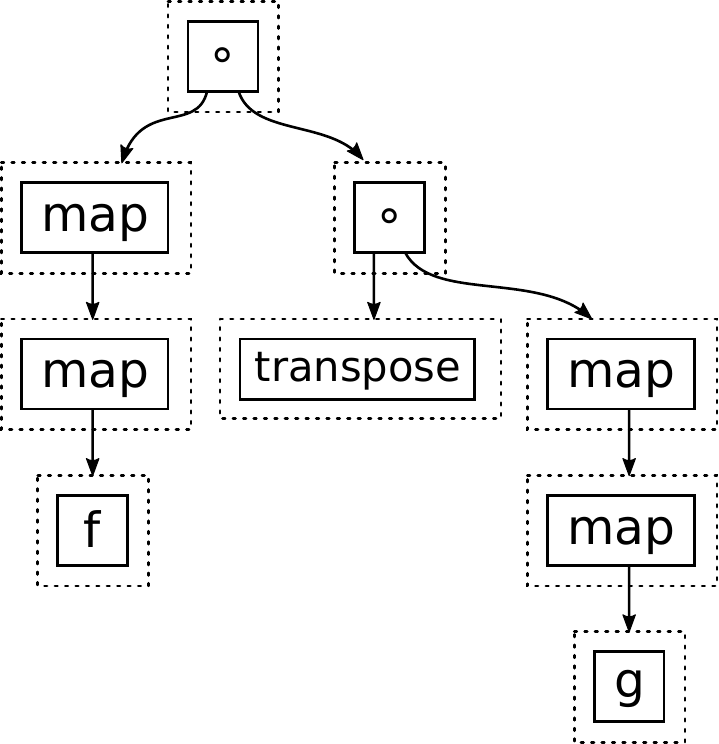}
  \caption{initialisation}
  \label{fig:eqsat-ex-0}
  \end{subfigure}\hfill%
  \begin{subfigure}[b]{0.23\linewidth}
  \includegraphics[width=\linewidth]{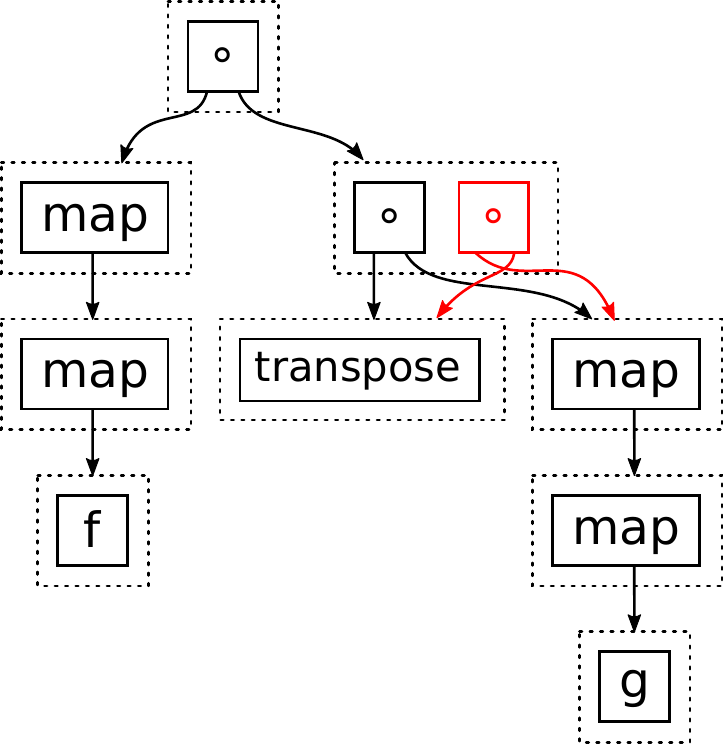}
  \caption{after rewrite (\ref{overview-move-transpose})}
  \label{fig:eqsat-ex-1}
  \end{subfigure}\hfill%
  \begin{subfigure}[b]{0.23\linewidth}
  \includegraphics[width=\linewidth]{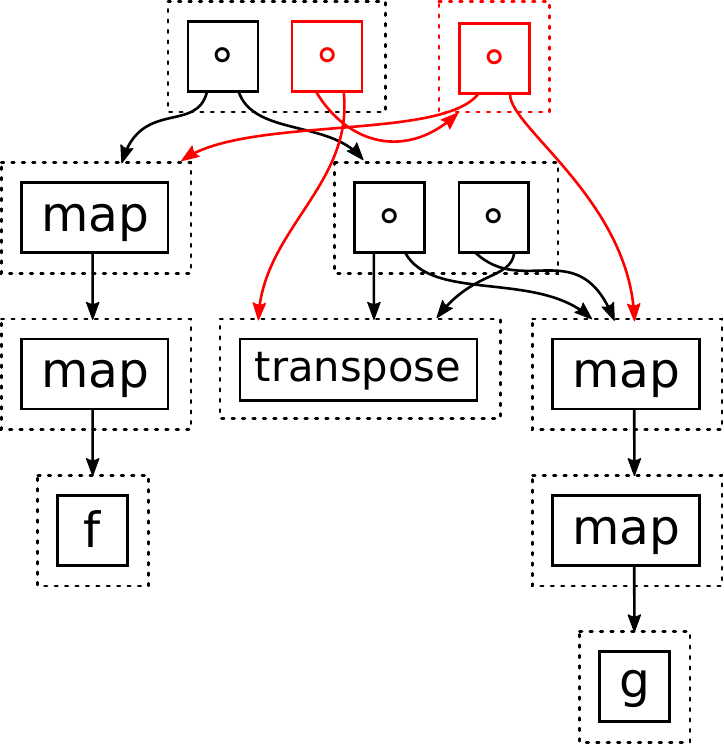}
  \caption{after rewrite (\ref{overview-comp-assoc})}
  \label{fig:eqsat-ex-2}
  \end{subfigure}\hfill%
  \begin{subfigure}[b]{0.29\linewidth}
  \includegraphics[width=\linewidth]{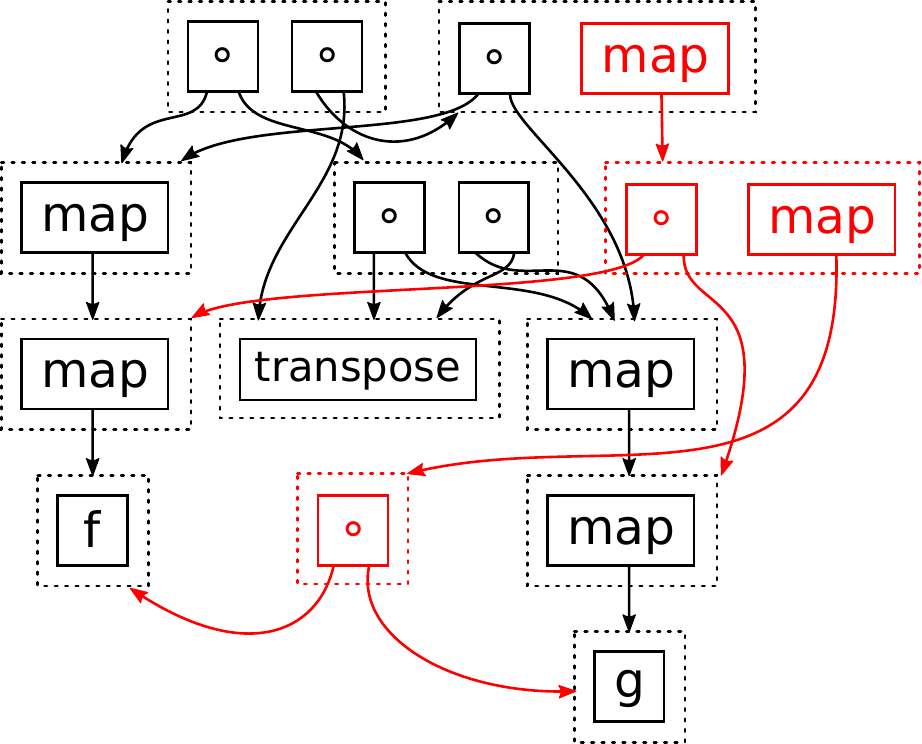}
  \caption{after two rewrites (\ref{overview-map-fusion})}
  \label{fig:eqsat-ex-3}
  \end{subfigure}
  \caption{Growing an e-graph for the term $(map~(map~f)) \circ (transpose \circ (map~(map~g)))$.
  An e-graph is a set of e-classes themselves containing equivalent e-nodes.
  The dashed boxes are e-classes, and the solid boxes are e-nodes.
  New e-nodes and e-classes are shown in red.}
  \label{fig:eqsat-ex}
\end{figure}
\smallskip

\subsection{Exploring Past the Local Optimum with Equality Saturation}

The first phase of equality saturation consists of exploring many possible rewrites.
An e-graph (equivalence graph) is used to efficiently represent and rewrite a set of equivalent programs, intuitively:
\begin{itemize}
\item An \emph{e-graph} is a set of equivalence classes (e-classes) representing all terms represented by its e-classes.
\item An \emph{e-class} is a set of equivalent nodes (e-nodes) representing all terms represented by its e-nodes.
\item An \emph{e-node} $F(e_1, .., e_n)$ is an $n$-ary function symbol ($F$) from the term language, associated with child e-classes ($e_i$).
It represents all terms $F(t_1, .., t_n)$ where each term $t_i$ is represented by the e-class $e_i$.
In this example, symbols are $map$, $transpose$, $\circ$, $f$, $g$.
\end{itemize}

To start the exploration phase, an e-graph representing the initial term (\ref{overview-start}) is constructed (\cref{fig:eqsat-ex-0}).
The e-graph is then iteratively grown by applying rewrite rules non-destructively (\cref{fig:eqsat-ex-1,fig:eqsat-ex-2,fig:eqsat-ex-3}).
On each equality saturation iteration, all possible rewrites are explored in a breadth-first manner.
This contrasts from standard term rewriting where a single possible rewrite is selected in a depth-first manner, requiring careful ordering of rewrite rule applications.
For the sake of simplicity, we only apply a handful of rewrite rules per iteration in \cref{fig:eqsat-ex}.
Rewrite rule applications are considered even if they do not lower cost, which avoids getting stuck in local optima.
When applying a rewrite rule, the equality between its matched left-hand side and its instantiated right-hand side is recorded in the e-graph.
This also contrasts from standard term rewriting that destructively replaces the matched left-hand side with the instantiated right-hand side, producing a new term from the initial one.

The exploration phase terminates, and rewrite rules stop being applied, when a fixed point is reached (saturation), or when another stopping criteria is reached (e.g. timeout or achieved goal).
If saturation is reached, it means that all possible rewrites have been explored.

\begin{figure}
  \begin{minipage}[b]{0.45\linewidth}
    \centering
    \includegraphics[width=0.6\linewidth]{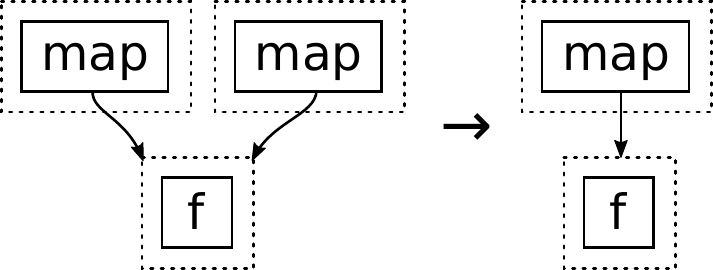}
    \caption{The congruence invariant simplifies the e-graph on the left by merging two identical e-nodes for $map~f$ into a single e-node as shown on the right.}
    \label{fig:congruence-invariant}
  \end{minipage}%
  \hfill
  \begin{minipage}[b]{0.5\linewidth}
    \centering
    \includegraphics[width=0.8\linewidth]{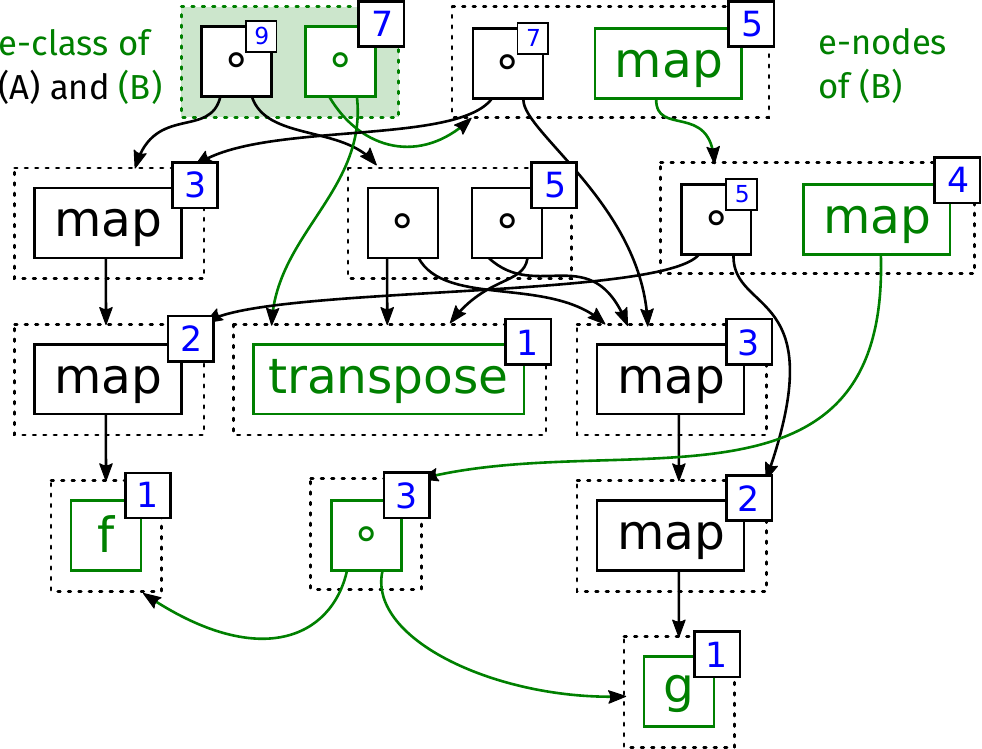}
    \caption{Smallest term size computed during extraction, shown for each e-class in the top-right corner in blue. Where it differs from its e-class value, the smallest term size of e-nodes is also shown. The e-class and e-nodes of the smallest term (\ref{overview-goal}) are shown in green.}
    \label{fig:eqsat-ex-extract}
  \end{minipage}
\end{figure}

Crucially for efficiency, an e-graph is far more compact than a naive set of terms, as equivalent sub-terms are shared.
E-graphs can represent exponentially many terms in polynomial space, and even infinitely many terms in the presence of cycles \cite{willsey2021-egg}.
To maximise sharing, a \emph{congruence invariant} is maintained: intuitively identical e-nodes should not be in different e-classes (\cref{fig:congruence-invariant}).
Later we will see that even extensive sharing does not necessarily prevent e-graph sizes from exploding.

\subsection{Extracting a Global Optimum with Equality Saturation}
\label{sec:eqsat-extraction}

The second and final phase of equality saturation consists of extracting the best term from the e-graph according to a cost function, e.g. the one with smallest term size.
The extracted term is a global optimum if saturation was reached in the exploration phase.

The \emph{extraction} procedure can be a relatively simple bottom-up e-graph traversal if the cost function is local~\cite{panchekha2015-herbie}.
Non-local cost functions require more complex extraction procedures \cite{wang2020-spores, wu2019-carpentry}.

A \emph{local cost function} $c$ can be defined as a function of a term language symbol $F$ and the costs of its children, i.e. $c$ has signature $c(F(k_1: K, .., k_n: K)) : K$ with costs of type $K$.
Term size is a local cost function:
$$ \mathrm{termSize}(F(k_1, .., k_n)) = 1 + \sum_{i} k_i $$
The term sizes computed during a bottom-up extraction procedure are shown in \cref{fig:eqsat-ex-extract}.
The figure reveals that there is a smaller term (\ref{overview-goal}) of size 7 in the same e-class as the original term (\ref{overview-start}) of size 9 (top left in \cref{fig:eqsat-ex-extract}).
Therefore, (\ref{overview-goal}) is extracted as the optimised term.
\section{Motivation for Semi-Automatic Optimisation}
\label{sec:sges-motivation}

This section motivates the need for a semi-automatic optimisation technique, expanding on \cref{sec:controlling-compilers}.
Controlling thousands of rewrite rule applications with \Elevate{} rewriting strategies as in \cite{hagedorn2020-elevate} and \cref{ch:imgproc} is tedious.
Instead, we would like to optimise \Rise{} programs \emph{without having to control individual rewrite rule applications} (\cref{sec:elevate-limitations}).

To automate rewrite rule application, globally exploring all possible rewrites with equality saturation is unfeasible, as the search space is too big.
Novel ways to scale equality saturation are required, and semi-automation is an appealing trade-off (\cref{sec:eqsat-limitations}).

\begin{figure}[h]
  \centering
  \begin{minipage}{0.75\linewidth}\vspace{0pt}%
  \begin{rise}
map ($\lambda$aRow.                    | for m:
  map ($\lambda$bCol.                  |   for n:
    dot aRow bCol)            |     for k:
    (transpose b)) a          |       ...
  \end{rise}
  \end{minipage}
  $$\rewritesTo{}^*$$
  \begin{minipage}{0.75\linewidth}\vspace{0pt}%
  \begin{rise}[escapechar={\#}]
#\tikzmark{mmb-1a}#join (map (map join) (map transpose#\tikzmark{mmb-1b}#
  map                         | for m / 32:
    (map $\lambda$x2.                  |   for n / 32:
      reduceSeq ($\lambda$x3. $\lambda$x4.      |     for k / 4:
        reduceSeq $\lambda$x5. $\lambda$x6.     |       for 4:
          map                 |         for 32:
            (map ($\lambda$x7.         |           for 32:
             #\tikzmark{mmb-2a}#(fst x7) + (fst (snd x7)) $\times$#\tikzmark{mmb-2b}#
              #\tikzmark{mmb-3a}#(snd (snd x7))#\tikzmark{mmb-3b}#)
             #\tikzmark{mmb-4a}#(map ($\lambda$x7. zip (fst x7) (snd x7))#\tikzmark{mmb-4b}#
              #\tikzmark{mmb-5a}#(zip x5 x6))#\tikzmark{mmb-5b}#)
          #\tikzmark{mmb-6a}#(transpose (map transpose#\tikzmark{mmb-6b}#
           #\tikzmark{mmb-7a}#(snd (unzip (map unzip map ($\lambda$x5.#\tikzmark{mmb-7b}#
            #\tikzmark{mmb-8a}#zip (fst x5) (snd x5))#\tikzmark{mmb-8b}#
            #\tikzmark{mmb-9a}#(zip x3 x4))))))#\tikzmark{mmb-9b}#)
        #\tikzmark{mmb-10a}#(generate ($\lambda$x3. generate ($\lambda$x4. 0)))#\tikzmark{mmb-10b}#
        #\tikzmark{mmb-11a}#transpose (map transpose x2)#\tikzmark{mmb-11b}#)
    #\tikzmark{mmb-12a}#(map (map (map (map (split 4))))#\tikzmark{mmb-12b}#
      #\tikzmark{mmb-13a}#(map transpose#\tikzmark{mmb-13b}#
        #\tikzmark{mmb-14a}#(map (map ($\lambda$x2. map (map (zip x2)#\tikzmark{mmb-14b}#
          #\tikzmark{mmb-15a}#(split 32 (transpose b)))))#\tikzmark{mmb-15b}#
            #\tikzmark{mmb-16a}#split 32 a)))#\tikzmark{mmb-16b}#))
  \end{rise}
  \begin{tikzpicture}[remember picture, overlay, transform canvas={yshift=4pt}] 
  \begin{scope}[line width=12pt, color=WildStrawberry, opacity=0.4]
      \draw (pic cs:mmb-1a) -- (pic cs:mmb-1b);
  \end{scope}
  \begin{scope}[line width=12pt, color=OliveGreen, opacity=0.4]
    \draw (pic cs:mmb-2a) -- (pic cs:mmb-2b);
    \draw (pic cs:mmb-3a) -- (pic cs:mmb-3b);
  \end{scope}
  \begin{scope}[line width=12pt, color=brown, opacity=0.4]
    \draw (pic cs:mmb-10a) -- (pic cs:mmb-10b);
  \end{scope}
  \begin{scope}[line width=12pt, color=Cerulean, opacity=0.4]
      \draw (pic cs:mmb-4a) -- (pic cs:mmb-4b);
      \draw (pic cs:mmb-5a) -- (pic cs:mmb-5b);
      \draw (pic cs:mmb-6a) -- (pic cs:mmb-6b);
      \draw (pic cs:mmb-7a) -- (pic cs:mmb-7b);
      \draw (pic cs:mmb-8a) -- (pic cs:mmb-8b);
      \draw (pic cs:mmb-9a) -- (pic cs:mmb-9b);
      \draw (pic cs:mmb-11a) -- (pic cs:mmb-11b);
      \draw (pic cs:mmb-12a) -- (pic cs:mmb-12b);
      \draw (pic cs:mmb-13a) -- (pic cs:mmb-13b);
      \draw (pic cs:mmb-14a) -- (pic cs:mmb-14b);
      \draw (pic cs:mmb-15a) -- (pic cs:mmb-15b);
      \draw (pic cs:mmb-16a) -- (pic cs:mmb-16b);
  \end{scope}
  \end{tikzpicture}
\end{minipage}
\caption{Applying a blocking optimisation to matrix multiplication via rewriting in \Rise{}.
In the initial program (top), a \inlineRise{dot} product is computed between each row of \inlineRise{a} (\inlineRise{aRow}) and column of \inlineRise{b} (\inlineRise{bCol}).
In the final program (bottom), a blocking optimisation has been applied.
Loops characteristic of the optimisation are shown on the right of the \inlineRise{|} symbols, and are not part of the \Rise{} program.
The remaining program regions {\color{Cerulean}reshape input arrays}, {\color{brown}initialise arrays}, {\color{OliveGreen}compute with scalars}, and {\color{WildStrawberry}reshape output arrays}.}
\label{fig:mm-rewrite}
\end{figure}

\begin{figure}
\centering
\begin{minipage}{0.75\linewidth}
\begin{elevate-rise}[escapechar={\#}, caption={\Elevate{} strategy for matrix multiplication blocking \cite{hagedorn2020-elevate}.},label=lst:elevate-blocking]
strategy blocking =
  baseline;
  tileMaps(32,32)   @ outermost(mapNest(2));; #\label{line:elevate-tile}#
  fissionReduceMap  @ outermost(appliedReduce);; #\label{line:elevate-fission}#
  tileReduce(4)     @ innermost(appliedReduce);; #\label{line:elevate-split}#
  reorder(List(1,2,5,6,3,4)) #\label{line:elevate-reorder}#
\end{elevate-rise}
\end{minipage}
\end{figure}

\subsection{Controlling Many Rewrite Steps with Strategies is Tedious}
\label{sec:elevate-limitations}

\Cref{sec:elevate-rewriting} introduced \Elevate{} strategies with a trivial dot product example.
A simple strategy was defined to apply a single rewrite rule using a top-down AST traversal:
\begin{elevate-rise}[numbers=none, xleftmargin=0pt]
strategy lowerDot = topDown(reduceMapFusion)
\end{elevate-rise}

To achieve more complex \Rise{} optimisations leading to high performance, significantly more complex \Elevate{} strategies are defined.
The strategies defined in \Cref{ch:imgproc} to apply the 6 Harris corner detection optimisations consist of more than 600 lines of code defining 57 helper strategies.
Similarly, 36 rewriting strategies were defined in 200 lines of code to apply 7 matrix multiplication optimisations in \cite{hagedorn2020-elevate}.
In both case studies, thousands of rewrite rules are applied by the strategies to achieve the optimisations (including backtracked rule applications).
The \Elevate{} strategy authors estimate having spent between two and five person-weeks developing the strategies for each case study.

We now focus on the limitations of a particular \Elevate{} strategy defined in \cite{hagedorn2020-elevate} to apply a blocking optimisation to the \Rise{} matrix multiplication (\cref{lst:elevate-blocking}).
Blocking (or tiling) is a common loop optimisation that improves data locality, and therefore memory usage \cite{gallivan1988-impact, wolfe1989-tiling}.
The \Elevate{} strategy has for effect to rewrite the \Rise{} program at the top of \cref{fig:mm-rewrite} to the one at the bottom.
Reading the strategy from top to bottom, a \inlineElevateRise{baseline} strategy is applied, rewriting the program into a particular normal form as well as applying a single \inlineElevateRise{reduceMapFusion} rule.
The \inlineElevateRise{tileMaps} strategy is applied to the outermost nest of 2 \inlineRise{map}s, creating 32$\times$32 blocks (or tiles).
Note that while \inlineElevateRise{;} is the straightforward sequential combinator, \inlineElevateRise{;;} is a sequential combinator that also enforces a particular normal form before continuing.
The \inlineElevateRise{fissionReduceMap} rewrite rule is applied at the outermost reduction pattern.
The \inlineElevateRise{tileReduce} rewrite rule is applied to the innermost reduction pattern, creating another block of 4 \cref{line:elevate-split}.
Finally, the \inlineElevateRise{reorder} strategy is applied, reordering nested \inlineRise{map} and \inlineRise{reduce} patterns to create 4$\times$32$\times$32 blocks and hence produce the loop nest at the bottom of \cref{fig:mm-rewrite}.

\Elevate{} enables the development of abstractions that help write concise strategies (\cref{sec:elevate}), such as \inlineElevateRise{tileMaps}, \inlineElevateRise{reorder}, \inlineElevateRise{@}, \inlineElevateRise{outermost} and \inlineElevateRise{innermost} in this example.
Unfortunately, these abstractions are often program specific and complex to implement.
\Cref{fig:elevate-blocking-limitations} shows how many lines of code and internal strategies are defined for \inlineElevateRise{tileMaps} and \inlineElevateRise{reorder}, as well as their limited applicability.
Both strategies are defined recursively.

\begin{table}
  \begin{tabular}{|c|c|c|p{0.7\linewidth}|}
  \hline
  \textbf{Strategy} & \textbf{LOC} & \textbf{IS} & \textbf{Limited Applicability} \\
  \hline
  \inlineElevateRise|tileMaps| & 32 & 6 & only works for perfect \inlineRise|map| nests, e.g.\newline \inlineRise|map f|, \inlineRise|map (map f)|, \inlineRise|map (map (map f))|, etc. \\
  \hline
  \inlineElevateRise|reorder| & 43 & 8 & not capable of reordering arbitrary loop nests;\newline depends on recursively lifting a \inlineRise|reduce| pattern outside of a \inlineRise|map| pattern which is hard-coded for specific cases. \\
  \hline
  \end{tabular}
  \caption{Main strategies backing \cref{lst:elevate-blocking}, their lines of code (LOC), internal strategies (IS), and limited applicability.}
  \label{fig:elevate-blocking-limitations}
\end{table}


Overall, developing generic \Elevate{} strategies is difficult because small syntactic differences in \Rise{} programs require adjustments to the rewrite sequence.
While \Elevate{} empowers performance engineers to manually control the rewrite process, it also delegates them the problem of ordering thousands of rewrites in order to achieve their goals (9K rewrite rules are applied when executing \cref{lst:elevate-blocking}).
To mitigate the phase ordering problem, automation techniques such as equality saturation are compelling.

\subsection{Semi-Automating Many Rewrite Steps with Equality Saturation}
\label{sec:eqsat-limitations}

Equality saturation (\cref{sec:eqsat-background}) has many successful applications, but its applicability remains limited by scaling issues.
As the e-graph grows, iterations become slower and require more memory.
The growth rate is aggravated by some combinations of useful rewrite rules, such as associativity and commutativity, that generate an exponential number of equivalent permutations \cite{wang2020-spores, nandi2020-synthesizing-CAD, willsey2021-egg}.
This makes exploring long rewrite sequences inherently hard, as the breadth-first exploration of all possible rewrites leads to an exponential increase of the e-graph size, despite its compact representation of terms.
In many applications, expecting to reach saturation, and therefore to find optimal solutions according to the cost function, is unrealistic.
We encounter these issues when attempting to optimise matrix multiplication in \Rise{} using equality saturation, as we discuss in \cref{mm-eval}.

One way to reduce e-graph growth is to limit the number of rules applied \cite{wang2020-spores, willsey2021-egg}, but this risks not finding optimisations that require an omitted rule.
An alternative is to use an external solver to speculatively add equivalences \cite{nandi2020-synthesizing-CAD}, but this requires the identification of sub-tasks that can benefit from being delegated.
It is also possible to trade-off between the exploitation of greedy rewriting and the exploration of equality saturation  \cite{kourta2022-caviar}, but this requires a good enough heuristic cost function to make local decisions.

This chapter proposes another approach, sketch-guiding, that factors unfeasible equality saturations into a sequence of smaller, and hence feasible, equality saturations (\cref{sketching}).
The appeal of sketch-guiding is that it still allows performance engineers to control the rewrite process, but saves them from ordering thousands of rewrites.
Additionally, an efficient encoding of the $\lambda$ calculus is introduced to reduce the sizes of the e-graphs produced when optimising \Rise{} programs (\cref{sec:sges-bindings}).

\section{Sketch-Guided Equality Saturation}
\label{sketching}

This section introduces \emph{sketch-guided equality saturation}, a novel semi-automated process offering a trade-off between manually defined rewriting strategies and automated equality saturation.
The performance engineer guides multiple equality saturations by specifying a sequence of sketches that describe how a program should evolve during optimisation.
In this process, the intent is to break down complex optimisations into a sequence of simpler rewrites, each sufficiently simple to be found by equality saturation.
Intuitively, a rewrite is \emph{complex} if achieving it requires applying many inter-dependent rewrite rules.
While rewriting strategies require ordering many rewrite rules, sketch-guided equality saturation enables ordering few sketches instead.
For example, instead of specifying how to apply thousands of rules to achieve matrix multiplication blocking, specifying just two sketches suffices.

\subsection{The Intuition for Sketches}

When designing optimisations, performance engineers often visualise the desired \emph{shape} of the optimised program.
The effect of rewriting strategies or schedules is often explained with program snippets \cite{koehler2021-elevate-imgproc, halide-2013, tvm-2018, halide-autosched-2019, sioutas2020-schedule, ikarashi2021-guided-scheduling, anderson2021-autosched}.
Indeed, this is how we have explained the loop nest blocking optimisation in \cref{fig:mm-rewrite}, as well as many optimisations in \cref{ch:imgproc}.

Our \emph{key new insight} is that explanatory program snippets can be formalised as sketches and used to guide an optimisation search.
\emph{Sketches} are program patterns that capture program shape intuitions.
The guided optimisation search is still based on semantic preserving rules, allowing to provide sketches that leave details unspecified without sacrificing correctness.

\begin{figure}
\centering
\begin{minipage}{0.75\linewidth}
\begin{center}
\begin{rise-sketch}[caption={A sketch for the \goalStyle{baseline} matrix multiplication goal}, label={mm-baseline-sketch}]
containsMap(m,                | for m:
 containsMap(n,               |  for n:
  containsReduceSeq(k,        |   for k:
   containsAddMul)))          |     .. + .. $\times$ ..
\end{rise-sketch}
\begin{rise-sketch}[caption={A sketch for the \goalStyle{blocking} matrix multiplication goal}, label={mm-blocking-sketch}]
containsMap(m / 32,           | for m / 32:
 containsMap(n / 32,          |  for n / 32:
  containsReduceSeq(k / 4,    |   for k / 4:
   containsReduceSeq(4,       |    for 4:
    containsMap(32,           |     for 32:
     containsMap(32,          |      for 32:
      containsAddMul))))))    |        .. + .. $\times$ ..
\end{rise-sketch}
\begin{rise-sketch}[caption={A sketch guide specifying how to split loops for \goalStyle{blocking}}, label={mm-blocking-split-sketch}]
containsMap(m / 32,           | for m / 32:
 containsMap(32,              |  for 32:
  containsMap(n / 32,         |   for n / 32:
   containsMap(32,            |    for 32:
    containsReduceSeq(k / 4,  |     for k / 4:
     containsReduceSeq(4,     |      for 4:
      containsAddMul))))))    |        .. + .. $\times$ ..
\end{rise-sketch}
\end{center}
\end{minipage}
\end{figure}

We illustrate by presenting sketches for matrix multiplication blocking.
\Cref{mm-baseline-sketch} shows a sketch for the unoptimised \goalStyle{baseline} goal, specifying the desired program structure as two nested \inlineRise{map} patterns and a nested \inlineRise{reduce}, with innermost addition and multiplication operations. The formal definitions of \inlineRiseSketch{containsMap}, \inlineRiseSketch{containsReduceSeq} and \inlineRiseSketch{containsAddMul} are in \cref{sketch-def}.
The comments on the right show the equivalent nested \inlineRise{for} loops, using the same intuition as in \cref{fig:mm-rewrite}.
\cref{mm-blocking-sketch} shows a sketch for the \goalStyle{blocking} goal, corresponding to the optimised program where the "loop nests" have been split and reordered to chunk the iteration space into blocks of $4 \times 32 \times 32$, processed by the three innermost \inlineRise{for} loops.

Searching for the \goalStyle{blocking} goal can be made more tractable by specifying intermediate \emph{sketch guides}, and \cref{mm-blocking-split-sketch} is an example.
This sketch guide describes a program shape where the \inlineRise{map} and \inlineRise{reduce} patterns have been split but not yet reordered.

\Elevate{} strategies, as in \cref{lst:elevate-blocking}, are detailed \emph{imperative} specifications of how to rewrite the program. In contrast, a sketch is a \emph{declarative} specification of the optimisation goal, and equality saturation is used to search for a sequence of rewrites to achieve that goal.
A sequence of sketches (e.g., \cref{mm-blocking-split-sketch} followed by \cref{mm-blocking-sketch}) may be used to achieve a desired optimisation when the equality saturation search with a single sketch as a goal does not succeed.



\subsection{Defining Sketches}
\label{sketch-def}


\newcommand{\sLang}{\textsc{SketchBasic}}
\newcommand{\sRepr}[1]{\mathcal{R}\llbracket #1 \rrbracket}

Sketches are specified in a \sLang{} language with just four constructors.
The syntax of \sLang{} and the set of terms that the constructors represent are defined in \cref{fig:sketch-lang-rep}.
A sketch $s$ represents a set of terms $\sRepr{s}$, such that $\sRepr{s} \subset T$ where $T$ denotes all terms in the language we rewrite.
We say that any $t \in \sRepr{s}$ satisfies sketch $s$.

The $\mathord{?}$ sketch is the least precise, representing all terms in the language.
The $F(s_1, .., s_n)$ sketch represents all terms that match a specific $n$-ary function symbol $F$ from the term language, and whose $n$ children satisfy sketches $s_i$.
The $\textit{contains}(s)$ sketch represents all terms containing a term that satisfies sketch $s$: the greatest solution to the recursive $\sRepr{\textit{contains}(s)}$ equation.
Finally, the $s_1 \lor s_2$ sketch represents terms satisfying either $s_1$ or $s_2$.

\begin{figure}
  \centering
  \begin{equation*}
  S \Coloneqq \mathord{?} \mid F(S, .., S) \mid \textit{contains}(S) \mid S \lor S
  \end{equation*}
  \vspace{-2em}
  \begin{align*}
    \sRepr{\mathord{?}} &= T = \{ F(t_1, .., t_n) \} \\
    \sRepr{F(s_1, .., s_n)} &= \{ F(t_1, .., t_n) \mid t_i \in \sRepr{s_i} \} \\
    \sRepr{\textit{contains}(s)} &= \sRepr{s} \cup \{ F(t_1, .., t_n) \mid \exists t_i \in \sRepr{\textit{contains}(s)} \} \\
    \sRepr{s_1 \lor s_2} &= \sRepr{s_1} \cup \sRepr{s_2}\\[-2em]
  \end{align*}
  \caption{Grammar of \sLang{} (top) and terms represented by \sLang{} (bottom).}
  \label{fig:sketch-lang-rep}
\end{figure}

When rewriting terms in a typed language, sketches may be annotated with a type sketch ($s :: s_t$) constraining the type of terms.
If $\sRepr{s_t}$ denotes the set of terms satisfying the type sketch $s_t$, then $\sRepr{s :: s_t} = \sRepr{s} \cap \sRepr{s_t}$.
The grammar of type sketches depends on the language we rewrite.
We elide type sketches from our definition of \sLang{} as they are just a convenience to define sketches over the language $T'$, the explicitly typed version of $T$ where types would be embedded in terms.

\paragraph{Sketch Abstractions}
Sketch abstractions are defined by combining generic constructs from \sLang{} with type annotations from the term language.
To illustrate, \cref{rise-sketch-abs} shows some sketch abstractions used for our \Rise{} matrix multiplication case study.
Recall that \inlineRise{$\to$} is a function type, and \inlineRise{n.dt} an array type of \inlineRise{n} elements of domain type \inlineRise{dt}.
The type annotations restrict the iteration domains of \inlineRise{map} and \inlineRise{reduceSeq}: the input arrays must have type \inlineRiseSketch{n.?dt}, and therefore such patterns will iterate over \inlineRiseSketch{n} elements.


\begin{figure}
\centering
\begin{minipage}{0.75\linewidth}
\begin{center}
\begin{rise-sketch}[label={rise-sketch-abs}, caption={Some sketch abstractions used in this chapter.}]
def containsMap(n: NatSketch, f: Sketch): Sketch =
  contains((map :: ?t $\to$ n.?dt $\to$ ?t) f)
def containsReduceSeq(n: NatSketch, f: Sketch): Sketch =
  contains((reduceSeq :: ?t $\to$ ?t $\to$ n.?dt $\to$ ?t) f)
def containsAddMul: Sketch =
  contains(? $+$ contains($\times$))
\end{rise-sketch}
\end{center}
\end{minipage}
\end{figure}

\paragraph{Sketch Precision}
Writing a useful sketch to guide an optimisation search requires striking a balance between being too precise and too vague.

An overly precise sketch may exclude valid optimised programs with a slightly different structure.
Replacing \inlineRiseSketch{containsAddMul} with \inlineRiseSketch{contains($\lambda$x. x + (fst x) $\times$ (snd x))} in the blocking sketch (\cref{mm-blocking-sketch}) would prevent the search from finding any suitable program in the experiment of \cref{mm-eval}.

An overly vague sketch may lead to finding undesirable programs. Removing the array sizes from the splitting sketch guide (\cref{mm-blocking-split-sketch}) would result in finding a program with the following undesired loop nest in the experiment of \cref{mm-eval}:
\begin{rise-sketch}
for m:
 for n / 32 / 32:
  for 32:
   for 32:
    for k / 4:
     for 4:
      ..
\end{rise-sketch}

This balance also interacts with the set of rewrite rules used, since programs that may be found by the search are $\sRepr{s} \cap \mathcal{E}_{rules}\llbracket t \rrbracket$ where $\mathcal{E}_{rules}\llbracket t \rrbracket$ represents the set of terms that can be discovered to be equivalent to the initial term $t$ according to the given $rules$.
This means that using a more restricted set of rules generally enables specifying less precise sketches.
How to best select rules and write effective sketches are topics for future work (\cref{ch:discussion}).

\subsection{Sketch-Guided Equality Saturation}
\label{sketching-sub}

\begin{figure}
  \centering
  \includegraphics[width=\linewidth]{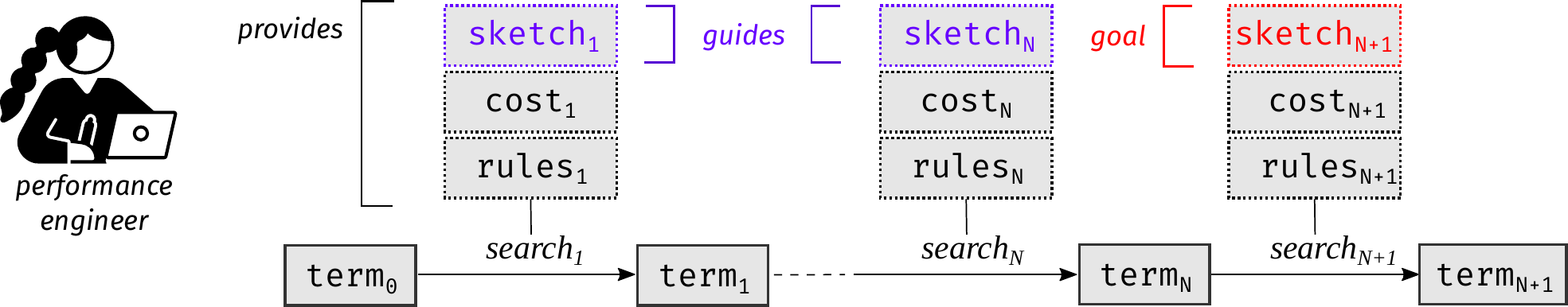}
  \caption{Sketch-Guided Equality Saturation.
  The performance engineer provides $N$ intermediate sketch guides and $1$ final goal sketch.
  Starting from the input term, $N+1$ consecutive equality saturation searches attempt to find a term satisfying each sketch, using the associated cost models and sets of rules (\cref{fig:search-diagram}).}
  \label{fig:sketching}
\end{figure}

\begin{figure}
\centering
\includegraphics[width=0.6\linewidth]{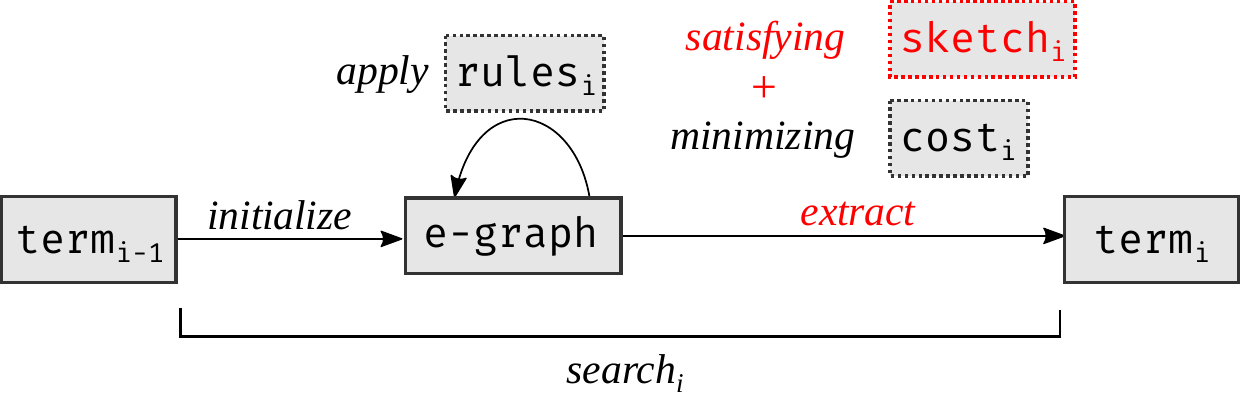}
\caption{Sketch-Satisfying Equality Saturation implementing each $search_i$.
Changes made to standard equality saturation are highlighted in red.}\label{fig:search-diagram}
\end{figure}

While sketches have previously been used as a starting point for program synthesis \cite{solar2008-synthesis-sketching}, our work uses sketches in a novel way, as intermediate goals (guides) for program optimisation.

The process of guiding equality saturation with a sequence of sketches is illustrated in \cref{fig:sketching}.
The performance engineer provides a sequence of $N$ intermediate sketch guides and a final goal sketch: $sketch_1,\,..,\,sketch_{N+1}$.
Successive equality saturation searches are performed to find equivalent terms that satisfy each sketch in the sequence. As each sketch may be satisfied by many terms, the performance engineer must also provide a sequence of cost models $cost_1, .., cost_{N+1}$ to select the term to be used as the starting point for the next search. Sets of rewrite rules ($rules_1, .., rules_{N+1}$) are provided to grow the e-graph in each search.
The cost model and set of rules may be identical for many or all of the searches, but we show in \cref{mm-eval} how restricting the set of rules can reduce search runtime. \Cref{fig:search-diagram} shows how each search is performed, and how these searches differ from standard equality saturation.

\begin{figure}
\centering
\begin{minipage}{0.75\linewidth}
\begin{lstlisting}[numbers=left, xleftmargin=2em, label={fig:sketching-alg}, caption={Sketch-Guided Equality Saturation Algorithm}, escapechar=|, mathescape=true, morekeywords={def, if, then, else},captionpos=b,abovecaptionskip={.5\baselineskip},moredelim={*[is][\rmfamily\itshape\color{black!50}]{~}{~}}]
def SGES(term, params): Option[Term] = |\label{line:sges}|
  if params.isEmpty
  then Some(term)
  else search(term, params.head) |\label{line:sges-search}|
       .and_then($\lambda$t. SGES(t, params.tail)) |\label{line:sges-rec}|
   
def search(term, param): Option[Term] =
  (sketch, cost, rules) = param
  g = ~create empty e-graph~
  normTerm = normalize(term) |\label{line:normalize}|
    ~using a configurable normal form~
  e = g.add(normTerm)
  ~grow~ g ~using~ rules ~until~ found(g, e, sketch) 
  if found(g, e, sketch) then |\label{line:found}|
    Some(extract(g, e, sketch, cost)) |\label{line:beam-extraction}|
  else
    None
\end{lstlisting}
\end{minipage}
\end{figure}

The pseudo-code for the sketch-guided equality saturation algorithm is shown in \cref{fig:sketching-alg}.
The entry point is the \inlineRise{SGES} function (line \ref{line:sges}) that takes  a \inlineRise{term} and a sequence of sketches, cost models and rewrite rules (\inlineRise{params}). It repeatedly \inlineRise{search}es (line \ref{line:sges-search}) for each sketch using the associated cost model and rewrite rules, and outputs a term if found, otherwise nothing.
At the beginning of each \inlineRise{search}, we may normalise the input term (line \ref{line:normalize}) to apply destructive rewrites that are always desired before starting a purely additive equality saturation.
For our matrix multiplication running example we use a $\beta\eta$ normal form.
The \inlineRise{extract} function (line \ref{line:beam-extraction}) is used to extract a term from the e-graph that satisfies the specified sketch while minimising the specified cost model, and we describe it in the next subsection.
In this paper, we terminate equality saturation as soon as a program satisfying the current sketch is found, whether or not the cost could be further improved by a longer search.
This is because we give more value to satisfying the sketch than to minimising the cost.
Other applications of sketch-guided equality saturation could use different stopping criteria.
The \inlineRise{found} function (line \ref{line:found}) is used to stop growing the e-graph by checking whether \inlineRise{extract} would succeed.

Note that sketch-guided equality saturation does not provide any guarantee of optimality or completeness of the search.
Without sketch-guidance, reaching saturation is unfeasible in many use cases (\cref{sec:eqsat-limitations}), and terminating equality saturation before saturation also provides no such guarantees.

Not finding a term satisfying a given sketch in the given resource budget may be for one of two reasons.
The search may be too difficult, and the performance engineer may fix the issue by providing additional or better sketch guides.
It may be impossible to construct an equivalent term satisfying the sketch with the given set of rules, and the performance engineer may fix the issue by fixing an error in the sketch or  providing missing rewrite rules.

\subsection{Sketch-Satisfying Extraction}
\label{beam-extraction}


To \inlineRise{extract} the best program that satisfies a \sLang{} sketch $s$ from an e-graph $g$ we define a helper function $E(c, s, g)$, where $c$ is a cost function that must be monotonic and local.
With costs of type $K$, $c$ is local if it can be defined as a function of a term language symbol $F$ and the cost of its children, i.e. it has signature $c(F(k_1: K, .., k_n: K)) : K$
The helper $E$ returns a map from e-classes to optional tuple values of type $\textit{Option}[(K, \textit{Term})]$.
\inlineRise{extract} uses $E$ to return a term if possible, failing otherwise:
$$ \text{extract}(g, e, s, c) = t \quad \text{if} \quad E(c, s, g)[e] = \textit{Some}\;(\_, t) $$
We write $E(c,s, g)[e]$ for indexing into the map returned by $E$.
$E$ is memoized for efficiency, and recursively defined over the 4 \sLang{} cases as follows.


\paragraph{Case 1:} $E(c, \bm{\mathord{?}}, g)$.
This case is equivalent to extracting the programs minimising $c$ from the e-graph (\cref{sec:eqsat-extraction}).
Such an extraction procedure can be implemented using an e-class analysis \cite{willsey2021-egg}.
An \emph{e-class analysis} propagates analysis data of type $D$ in a bottom-up fashion, and can be used for extraction when the cost function is local.
An e-class analysis is defined by providing two functions: one to $\textit{make}$ the analysis data from an $n$-ary symbol $F$ combined with the data $d_i$ of its child e-classes; and one to $\textit{merge}$ the analysis data of e-nodes in the same e-class.
The domain of the analysis data together with the $\textit{merge}$ operation should form a semilattice.

$$ \textit{make}(F(d_1 : D, .., d_n : D)) : D $$
$$ \textit{merge}(d_1 : D, d_2 : D) : D $$

We implement $E(c, \mathord{?}, g)$ as an e-class analysis with data type $D = \text{Option}[(K, \textit{Term})]$ and the following $\textit{make}$ and $\textit{merge}$ functions.
$\textit{make}$ computes the best cost of an e-node and the corresponding best term, based on children analysis data, if available.
$\textit{merge}$ returns the analysis data with best cost, if available.

\begin{align*}
\textit{make}(F(d_1, .., d_n)) = &\begin{cases}
  \textit{Some}\;\left(\begin{array}{l}
    c(F(k_1, .., k_n)),\\
    F(t_1, .., t_n)
  \end{array}\right) & \forall i.~d_i = \textit{Some}\;(k_i, t_i) \\
  \textit{None} & \text{otherwise}
\end{cases}\\
\textit{merge}(d_1, d_2) = &\begin{cases}
  \text{if } k_1 \leq k_2 \text{ then } d_1 \text{ else } d_2 & \forall i.~d_i = \textit{Some}\;(k_i, \_) \\
  d_i & \exists i.~d_i = \textit{Some}\;(k_i, \_) \\
  \textit{None} & \text{otherwise}
\end{cases}
\end{align*}

\paragraph{Case 2:} $E(c, \bm{F(s_1, .., s_n)}, g)$.
We consider each e-class $e$ containing $F(e_1, .., e_n)$ e-nodes and the terms that should be extracted for each child e-class $e_i$:
\begin{align*}
E(c, F(s_1, .., s_n), g)[e] = \begin{cases}
  \textit{Some}\;\left(\begin{array}{l}
    c(F(k_1, .., k_n)),\\
    F(t_1, .., t_n)
  \end{array}\right) & \forall i.~ E(c, s_i, g)[e_i] = \textit{Some}\;(k_i, t_i) \\
  \textit{None} & \text{otherwise}
\end{cases}
\end{align*}

\paragraph{Case 3:} $E(c, \bm{\textit{\textbf{contains}}(s_2)}, g)$.
We use another e-class analysis and initialise the analysis data to $E(c, s_2, g)$ corresponding to the base case where $\sRepr{s_2} \subset \sRepr{\textit{contains}(s_2)}$.
To $\textit{merge}$ the analysis data, we do the same as for $s = \mathord{?}$.
To $make$ the analysis data we consider all terms that would contain terms from $s_2$ and keep the best by folding them using $\textit{merge}$:
\begin{align*}
\textit{make}(F(d_1, .., d_n)) =~&\textit{foldl}~\textit{merge}~\textit{None} \\
&\left\{
  \textit{Some}\;\left(\begin{array}{l}
    c(F(k_1, .., k_j, .., k_n)),\\
    F(t_1, .., t_j, .., t_n)
  \end{array}\right) \middle\vert
  \begin{array}{c}
    i \neq j,\\
    E(c,\mathord{?},g)[e_i] = \textit{Some}\;(k_i, t_i),\\
    d_j = \textit{Some}\;(k_j, t_j)
  \end{array}
\right\}
\end{align*}

\paragraph{Case 4:} $E(c, \bm{s_1 \lor s_2}, g)$. We $\textit{merge}$ the results from $s_1$ and $s_2$:
$$ E(c, s_1 \lor s_2, g)[e] = \textit{merge}(E(c, s_1, g)[e], E(c, s_2, g)[e]) $$


\section{Efficient Equality Saturation for the Lambda Calculus}
\label{sec:sges-bindings}

The egg library \cite{willsey2021-egg} implements a partial evaluator for the lambda calculus using a naive lambda calculus encoding based on explicit substitution.
Although conceptually simple, this encoding is not efficient enough to optimise \Rise{} programs (\cref{lambda-eval}).
Glenside \cite{smith2021-access-patterns} proposes \emph{access patterns} to increase efficiency by avoiding the need for binding structures when representing tensor programs.
However, using this technique would require re-designing the \Rise{} language, including its patterns and rewrite rules.

This section instead explores how to efficiently encode a polymorphically typed lambda calculus such as \Rise{}, for the purposes of equality saturation.
A set of design choices are realised for the \Rise{} language in the new \kles{} implementation that is heavily inspired by the egg library \cite{willsey2021-egg}.
Our optimised encoding reduces the runtime and memory consumption of equality saturation over lambda terms by orders of magnitude (\cref{lambda-eval}).

First, applying equality saturation to lambda calculus terms requires encoding them as terms of shape $F(t_1, .., t_n)$.
A naive encoding is shown in \cref{fig:rise-node-repr}.
Lambda abstraction is encoded as a unary symbol, lambda application as a binary symbol, and variables as constant symbols.
Variable names are not modeled directly as terms, but as symbol metadata: 'lam x', 'lam y', 'var x' and 'var y' are all treated as distinct symbols.

\begin{table}
\centering
\begin{tabular}{|c|c|c|}
\hline
\textbf{$\bm{\lambda}$ calculus} & $\bm{F}$ & $\bm{t_1, .., t_n}$ \\
\hline
\begin{rise}[numbers=none, frame=none]
$\lambda$x. e
\end{rise}%
& lam x & e \\
\hline
\begin{rise}[numbers=none, frame=none]
e$_1$ e$_2$
\end{rise}%
& app & e$_1$, e$_2$ \\
\hline
\begin{rise}[numbers=none, frame=none]
x
\end{rise}%
& var x & \\
\hline
\end{tabular}
\caption{Naive encoding of $\lambda$ calculus terms as $F(t_1, .., t_n)$ terms for equality saturation.}
\label{fig:rise-node-repr}
\end{table}

\begin{figure}
\centering
\begin{align*}
    \tag{$\beta$-reduction}\label{beta-reduction}
    ({\color{cyan}\lambda x.}~b)~e \rewritesToM{}& {\color{purple}b[e/x]} \\
    \tag{$\eta$-reduction}\label{eta-reduction}
    {\color{cyan}\lambda x.}~f~x \rewritesToM{}& f &{\color{orange}\text{if}~x~\text{not free in}~f} \\
    \tag{map-fusion}\label{map-fusion}
    map~f~{\color{gray}(}map~g~{\color{gray}arg)} \rewritesToM{}& map~{\color{gray}({\color{cyan}\lambda x.}}~f~{\color{gray}(}g~{\color{gray}x))~arg} \\
    \tag{map-fission}\label{map-fission}
    map~{\color{gray}({\color{cyan}\lambda x.}}~f~gx) \rewritesToM{}& {\color{gray}{\color{cyan}\lambda y.}}~map~f~{\color{gray}(}map~{\color{gray}({\color{cyan}\lambda x.}}~gx{\color{gray})~y)}
    &{\color{orange}\text{if}~x~\text{not free in}~f}
\end{align*}
\caption{\Rise{} rewrite rules using {\color{purple}substitution}, {\color{cyan}name bindings} (lambda abstractions), and {\color{orange} freshness predicates}.}
\label{fig:example-rules}
\end{figure}

Second, applying equality saturation to lambda calculus terms requires the efficient support of standard operations and rewrites.
\Cref{fig:example-rules} shows the standard rules of \ref{beta-reduction} and \ref{eta-reduction} that use substitution, name bindings and freshness predicates.
The other two rules encode \ref{map-fusion} and \ref{map-fission}, and are interesting because they introduce new name bindings on their right-hand-side.

The following subsections first discuss how to efficiently implement substitution, name bindings and freshness predicates.
Then, they discuss how polymorphic types are added and how external, user-friendly rewrite rules are compiled into internal rewrite rules.

\subsection{Substitution}
\label{substitution}

The \ref{beta-reduction} rewrite rule requires substituting $b[e/x]$.
Standard term substitution cannot be used directly during equality saturation, as the $b$ and $e$ pattern variables are not matched by terms, but by e-classes.
A simple way to address this is to use \emph{explicit substitution} as in egg's lambda calculus example \cite{willsey2021-egg}.
A syntactic constructor is added to represent substitution, with rewrite rules to encode its small-step behavior:
\begin{align*}
(a~b)[e/v] &\rewritesToM{} (a[e/v]~b[e/v]) \\
v[e/v] &\rewritesToM{} e
\end{align*}

Unfortunately, explicit substitution adds all intermediate substitution steps to the e-graph, quickly exploding its size.
\Cref{lambda-eval} shows that this is a major problem in practice, making relatively simple rewrites involving \ref{map-fusion} and \ref{map-fission} unfeasible.
To avoid adding intermediate substitution steps to the e-graph, we propose \emph{extraction-based substitution} that works as follows.
\begin{enumerate}
    \item extract a term for each e-class involved in the substitution (i.e $b$ and $e$);
    \item perform standard term substitution;
    \item add the resulting term to the e-graph.
\end{enumerate}

\begin{figure}
    \centering
    \includegraphics[width=0.9\linewidth]{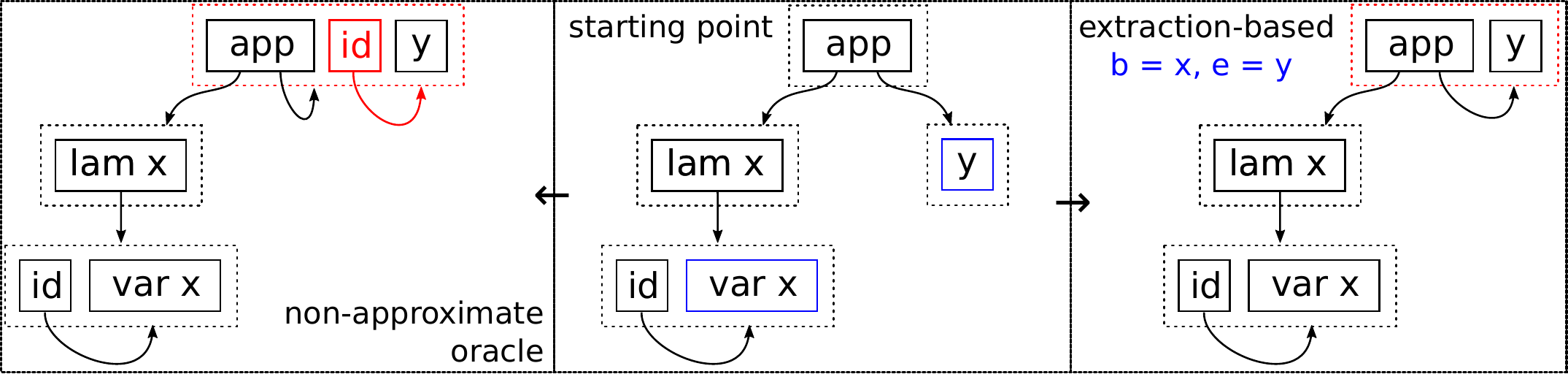}
    \caption{Example of $\beta$-reduction with extraction-based substitution (right).
    The initial e-graph (middle) represents $(\lambda x.~id~x)~y$.
    After extraction-based $\beta$-reduction, the e-graph does not represent $id~y$ because $x$ has been extracted for $b$ in the rewrite rule; ignoring $id~x$.}
    \label{fig:extract-subs}
\end{figure}

\Cref{lambda-eval} demonstrates that extraction-based substitution is far more efficient than explicit substitution.
Extraction-based substitution is, however, an approximation as it computes the substitution for a subset of the terms represented by $b$ and $e$, and ignores the rest.
\Cref{fig:extract-subs} shows an example.
The initial e-graph is in the middle, the e-graph after a non-approximate oracle substitution is on the left, and the e-graph after extraction-based substitution with $b=x$ and $e=y$ is on the right.
This particular choice results in an e-graph lacking the $id~y$ program.

In practice, we have not observed the approximation to be an issue when optimising \Rise{} programs (\cref{lambda-eval,mm-eval}), and believe that two main reasons account for this.
First, the substitution is computed on each equality saturation iteration, where different terms may be extracted, increasing coverage of the set of terms represented by $b$ and $e$.
Second, many of the ignored equivalences are recovered either by e-graph congruence, or by applying further rewrite rules.
Future work may investigate alternative substitution implementations to balance efficiency with non-approximation.

\subsection{Name Bindings}
\label{name-bindings}

During equality saturation, inappropriate handling of name bindings easily leads to serious efficiency issues.
Consider rewrite rules like \ref{map-fusion} that create a new lambda abstraction on their right-hand side.
What name should be introduced when they are applied?
In standard term rewriting, generating a fresh name using a global counter (aka. gensym) is a common solution \cite{1994-gensym}.
But if a new name is generated each time the rewrite rule is applied, the e-graph is quickly burdened with many $\alpha$-equivalent terms\footnote{Two $\lambda$ terms are $\alpha$-equivalent if one can be made equivalent to the other simply by renaming variables.}.

Fewer $\alpha$-equivalent terms are introduced if fresh names are generated as a function of the matched e-class identifiers.
However as the e-graph grows and e-classes are merged, e-class identifiers change, and $\alpha$-equivalent terms are still generated and duplicated in the e-graph.

De Bruijn indices~\cite{DEBRUIJN1972381} are a standard technique for representing lambda calculus terms without naming the bound variables, and avoid the need for $\alpha$ conversions.
If De Bruijn indices enable two $\alpha$-equivalent terms to be structurally equivalent, the standard e-graph congruence invariant prevents their duplication, by ensuring that equivalent e-nodes are not allocated to different e-classes.
Hence we translate our terms and rewrite rules to use De Bruijn indices instead of names, and achieve significant efficiency gains (\cref{lambda-eval}).


The following paragraphs detail various aspects of using De Bruijn indices, and discuss the alternative choice of avoiding name bindings entirely using combinators.

\paragraph{True Equality Modulo $\alpha$-renaming}
While De Bruijn indices give a significant performance improvement, they do not provide equality modulo $\alpha$-renaming for sub-terms.
Consider  $f~(\lambda x.~f) = \%0~(\lambda.~\%1)$, where $\%i$ are De Bruijn indices.
Although $\%0$ and $\%1$ are structurally different, they both correspond to the same variable $f$.
In practice, we have not observed this to be a significant issue when optimising \Rise{} programs, but it does require care when comparing sub-terms that have a different number of surrounding lambdas.
Future work may investigate alternatives to De Bruijn indices, for example through hashing modulo $\alpha$-renaming \cite{maziarz2021-hash-mod-alpha-eq}, nominal rewriting techniques \cite{fernandez2007-nominal-rewriting}, or hierarchical abstract syntax graphs \cite{ghica2021-asg}.

\paragraph{Translating Name-based Rules into Index-based Rules}
Using De Bruijn indices means that rewrite rules must manipulate terms with De Bruijn indices.
Thankfully, more user-friendly name-based rewrite rules can be automatically translated to the index-based rules used internally \cite{bonelli2000-bruijn-rewriting}.
An example demonstrating this is given in \cref{rewrite-example}.

\paragraph{Explicit or Extraction-based Substitution}
Both explicit substitution and extraction-based substitution are compatible with De Bruijn indices, and for explicit substitution we use the $\lambda s$ calculus \cite{kamareddine1995-lambda-v}.

\paragraph{Shifting De Bruijn Indices}
De Bruijn indices must be shifted when a term is used with different surrounding lambdas (example in \cref{rewrite-example}).
As for substitution, shifting can be implemented with explicit rewrite rules, or with \emph{extraction-based index shifting}:
\begin{enumerate}
    \item extract a term from the e-class whose indices need shifting;
    \item perform index shifting on the term;
    \item add the resulting term to the e-graph.
\end{enumerate}
In \kles{} we use extraction-based index shifting when extraction-based substitution is used.

\paragraph{Avoiding Name Bindings using Combinators}
It is also possible to avoid name bindings entirely \cite{smith2021-access-patterns}.
For example, it is possible to introduce a function composition combinator `$\circ$' as in \cref{sec:eqsat-background}, greatly simplifying the \ref{map-fusion} and \ref{map-fission} rules:
\begin{align}
    \tag{$\circ$-intro}
    f~(g~x) &\rewritesToM{} (f \circ g)~x \\
    \tag{map-fusion$_2$}
    map~f \circ map~g &\rewritesToM{} map~(f \circ g) \\
    \tag{map-fission$_2$}
    map~(f \circ g) &\rewritesToM{} map~f \circ map~g
\end{align}

However, this approach has its own downsides.
Associativity rules are required, which increases the growth rate of the e-graph \cite{willsey2021-egg}.
Only using a left-/right-most associativity rule reduces growth, but requires other rewrite rules to take this convention into account, making their definition more difficult and their matching more expensive.
In general, matching modulo associativity or commutativity are algorithmically hard problems \cite{benanav1987-complexity-matching}.

The function composition $\circ$ combinator on its own is also not sufficient to remove the need for name bindings.
At one extreme, combinatory logic could be used as any lambda calculus term can be represented, replacing function abstraction by a limited set of combinators.
However, translating a lambda calculus term into combinatory logic results in a term of size $O(n^3)$ in the worst case, where $n$ is the size of the initial term \cite{lachowski2018-complexity}.
Translating existing rewrite systems to combinatory logic would be challenging in itself.

\subsection{Freshness Predicates}
\label{predicates}

Handling predicates is not trivial in equality saturation.
The \ref{eta-reduction} rewrite rule has the side condition "$\text{if } x \text{ not free in } f$", but in an e-graph $f$ is an e-class and not a term.

The predicate could be precisely handled by filtering $f$ into $f' = \{ t \mid t \in f\text{, }x~\text{not free in}~t \}$, and using $f'$ on the right-hand-side of the rule.
However, current e-graph and equality saturation frameworks do not allow discriminating between different e-class subsets in such a way due to their goal of maximal sharing.

The design of \kles{} makes the engineering trade-off to only apply the \ref{eta-reduction} rewrite rule if $\forall t \in f.~x~\text{not free in}~t$, following egg's lambda calculus example \cite{willsey2021-egg}.
Advantages are that this predicate is efficient to compute using an e-class analysis, and that there is no need to discriminate between different e-class subsets.
The disadvantage is that it is an approximation that ignores some valid terms.

\begin{figure}
    \centering
    \includegraphics[width=0.65\linewidth]{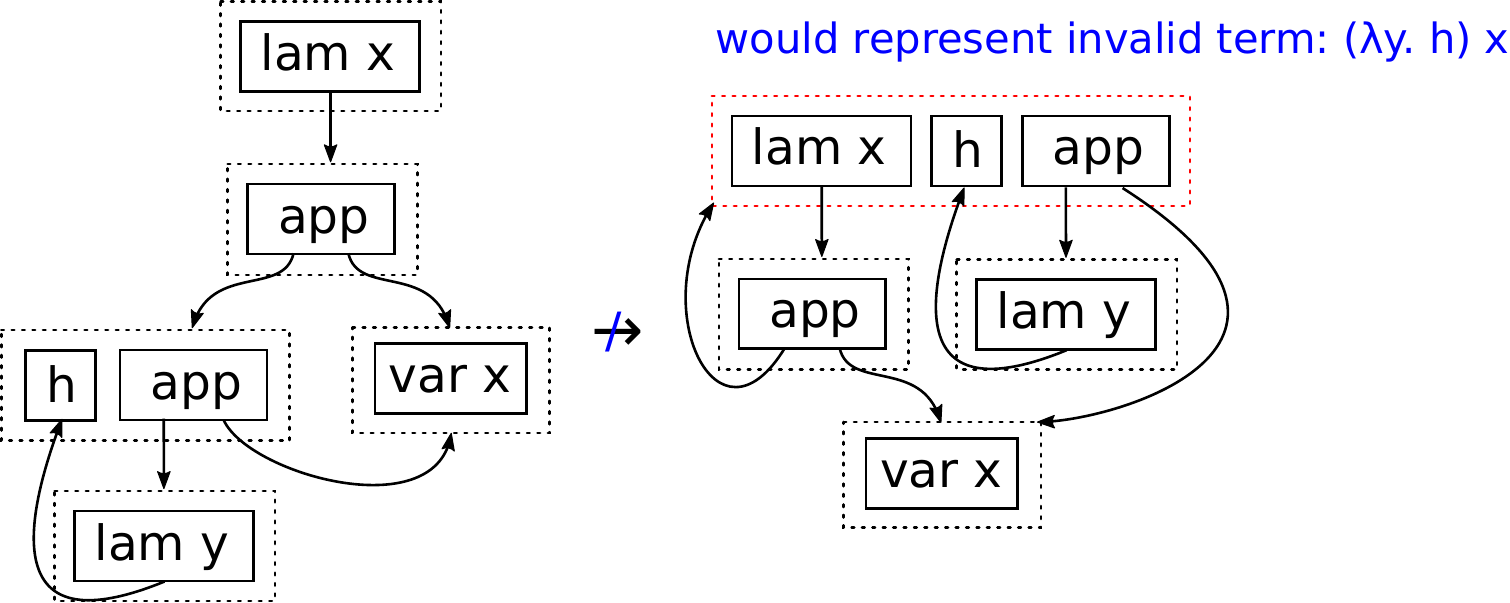}
    \caption{Example where $\eta$-reduction is not applied.
    The initial e-graph represents $\lambda x.~h~x$, but $\eta$-reduction is not triggered because $h = (\lambda y.~h)~x$ where $x$ is free.
    Using $\exists$ instead of $\forall$ in our predicate, we would obtain the e-graph on the right which represents $h$, but also invalid terms such as $(\lambda y.~h)~x$ where $x$ is no longer bound.}
    \label{fig:missed-eta-example}
\end{figure}

\Cref{fig:missed-eta-example} shows an example where \ref{eta-reduction} is not applied.
In practice, we have not observed the approximation to be an issue, e.g.\ for the results presented in \cref{lambda-eval,mm-eval}.

\subsection{Adding Polymorphic Types}
\label{types}

A key consideration is how to add types to the e-graph, as typed lambda calculi are pervasive and lay the foundations of almost all functional languages.

More specifically, we look at how polymorphic types interact with e-classes.
If types can be computed in a bottom-up fashion, an e-class analysis can be used, similar to how the size and shape of tensors is computed in \cite{wang2020-spores}.
However, if polymorphic types are monomorphised, their instantiation is context-dependent and cannot be computed bottom-up.
For example, consider the terms $(\lambda x.~x)~(0: \text{i32})$ and $(\lambda x.~x)~(0.0: \text{f32})$.
In \Rise{}, the identify function is monomorphised and has two different type instantiations that should live in different e-classes: $\lambda x.~x : \text{i32} \rightarrow \text{i32}$ and $\lambda x.~x : \text{f32} \rightarrow \text{f32}$.
Hence, in \kles{} instantiated types are embedded in the e-graph instead of computed by an analysis.
Each e-class is associated with a type that all of its e-nodes must satisfy.

In an e-graph there is a tension between sharing and the availability of contextual information in a given e-class.
Type instantiation prevents sharing, as the same polymorphic expression produces a different e-class at each type. However, instantiation provides additional contextual information, as each e-class is associated with a precise monotype.

\paragraph{Hash-consing Types}
Since types are duplicated many times in the e-graph, and since structural type equality is often required, we hash-cons types for efficiency \cite{filliatre2006-hashconsing}, representing each unique type with a unique identifier and leveraging structural sharing.
More specifically for \Rise{}, we use separate hash-conses for the different type kinds such as \inlineRise{nat} and \inlineRise{data}, to clearly separate their identifiers in the implementation.
For \inlineRise{nat}, hash-consing is combined with arithmetic simplification, e.g. $x \times 0$ will be given the identifier of $0$.
For \inlineRise{addr}, hash-consing is not necessary since there are only constant addresses (e.g. \inlineRise{global}) and no address constructors to hash-cons.

Alternatively, types could be stored as e-graph terms to provide equational reasoning at the type level.

\subsection{Compiling user-defined rewrite rules}
\label{rewrite-example}

To avoid explicit typing or explicit use of De Bruijn indices in user-defined rewrite rules, user-friendly name-based and partially typed rewrite rules are compiled internally as required.

Types are inferred with \Rise{} type inference: both sides of rewrite rules can be seen as terms where free variables correspond to pattern variables.
After inferring the types on the left-hand-side, we check that the right-hand-side is well-typed for any well-typed left-hand-side match.
When applied, typed rewrite rules match (deconstruct) types with their left-hand-side, and construct types on their right-hand-side.
Type annotations can be used to constrain the inferred types.

Bound variables are replaced with their corresponding De Bruijn index, and indices are shifted as required for terms with differing numbers of surrounding lambdas.
We illustrate with examples.

\paragraph{Example 1:  $\eta$-reduction}

\begin{align*}
\lambda x.~f~x \quad\rewritesToM{}\quad& f &\text{if}~x~\text{not free in}~f
\end{align*}

\kles{} translates this rule into:
\begin{align*}
(\lambda. f~(\%0 :\ t_0)) :\ t_0 \to\ t_1 \quad\rewritesToM{}\quad& (\varphi^{-1}_1 f) :\ t_0 \to\ t_1 &\text{if}~\%0~\text{not free in}~f
\end{align*}

The transformed rule uses a De Bruijn index $\%0$ for the bound variable, and pattern variables otherwise: $f$, $t_0$ and $t_1$. It provides index shifting through $\varphi^{-1}_1 f$ that shifts all indices $\ge 1$ by $-1$, because a surrounding $\lambda$ has been removed.
Some types are matched on the left-hand-side ($t_0$ and $t_1$), and used to construct types on the right-hand-side ($t_0 \to t_1$).
Some types, like the type of $f$, are not matched on the left-hand-side as \kles{} avoids matching on redundant type information to some extent, assuming that the terms matched are well-typed.

\paragraph{Example 2: $\eta$-abstraction}

\begin{equation*}
f :\ t_0 \to t_1 \quad\rewritesToM{}\quad \lambda x.~f~x
\end{equation*}
$\eta$-abstraction illustrates how type annotations may be used, and are sometimes required.
\kles{} translates this rule into:
\begin{equation*}
f :\ t_0 \to t_1 \quad\rewritesToM{}\quad (\lambda.~(((\varphi^1_0 f) : t_0 \to t_1)~(\%0 : t_0)) :\ t_1) :\ t_0 \to t_1
\end{equation*}
A type error occurs if the type of $f$ is not annotated as in $f :\ t_0 \to t_1$ on the left-hand-side, since the right-hand-side requires it to be a function type. 
\section{Evaluation of Lambda Calculus Encoding}
\label{lambda-eval}

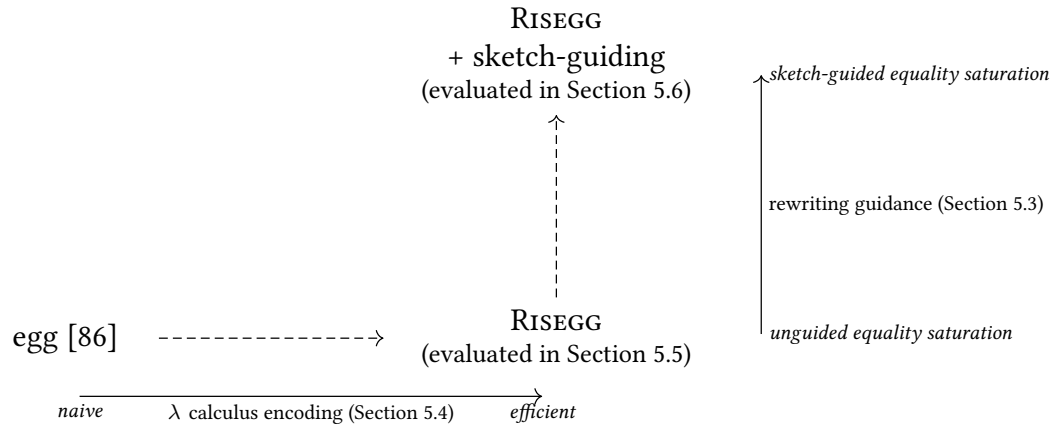
\begin{figure}
    \centering
\[\begin{tikzcd}[sep=tiny]
	&&&&&&&&&&& \text{\parbox{10em}{\centering{}\kles{}\\ + sketch-guiding\\\footnotesize(evaluated in \cref{mm-eval})}} & {} \\
	\\
	\\
	{} \\
	& {} &&&& {} & {} \\
	\\
	& {} \\
	\\
	{} & {} & \text{\parbox{5em}{\centering{}egg~\cite{willsey2021-egg}}} &&&&&&&&& \text{\parbox{10em}{\centering{}\kles{}\\\footnotesize(evaluated in \cref{lambda-eval})}} & {} \\
	{} & {} & {} &&&& {} &&&&& {} & {}
	\arrow["\textit{unguided equality saturation}"' at start, "\text{rewriting guidance (\cref{sketching})}"', "\textit{sketch-guided equality saturation}"' at end, from=9-13, to=1-13]
	\arrow["\textit{naive}"' at start, "\lambda~\text{calculus encoding (\cref{sec:sges-bindings})}"', "\textit{efficient}"' at end, from=10-3, to=10-12]
	\arrow[dashed, from=9-12, to=1-12]
	\arrow[dashed, from=9-3, to=9-12]
\end{tikzcd}\]
\caption{We first evaluate our efficient $\lambda$ calculus encoding before evaluating sketch-guided equality saturation using this encoding.}
    \label{fig:eval-overview}
\end{figure}

\Cref{fig:eval-overview} illustrates our process to evaluate our two proposed techniques for scaling equality saturation to the \Rise{} optimisations previously applied using \Elevate{} rewriting strategies.
Starting from the naive lambda calculus encoding used in egg's example \cite{willsey2021-egg}, we first investigate the effectiveness of the new encoding from \cref{sec:sges-bindings}, implemented in \kles{}, for unguided equality saturation (this section).
The naive lambda calculus encoding uses explicit substitution and variable names.
The efficient encoding uses extraction-based substitution to avoid intermediate substitution steps, and De Bruijn indices to avoid duplicating $\alpha$-equivalent terms.
Thereafter, we adopt the new lambda calculus encoding in \kles{}, and compare unguided and sketch-guided equality saturation to reproduce seven realistic matrix multiplication optimisations previously applied with \Elevate{} rewriting strategies in \cite{hagedorn2020-elevate} (\cref{mm-eval}).

This section evaluates the efficiency of equality saturation for the lambda calculus by attempting to discover three rewrite goals using four combinations of the substitution and name binding techniques outlined in \cref{sec:sges-bindings}.
\emph{Discovering a rewrite goal} means that it is feasible to grow an e-graph starting from the initial program until the goal program is represented in the e-class of the initial program.
In other words, it is feasible to discover that the initial program is equal to the goal program.

\subsection{Experimental Setup}

The alternate encodings are realised in an early prototype of \kles{}.\footnote{https://github.com/Bastacyclop/egg-rise}
This prototype is implemented in Rust using the egg library, and re-implements an untyped subset of \Rise{} (originally implemented in Scala) which is sufficient for quick prototyping.

To measure search runtime, we use egg's built-in mechanisms, falling back to the GNU time utility in case of an out-of-memory exception.
Maximum memory residency is measured with  the GNU time utility.
The experiments are run on a laptop with an AMD Ryzen 5 PRO 2500U processor, and we limit the available RAM to 2 GB.  
The results are reported from a single run since we dot not care about small variations but rather about orders of magnitude.

\paragraph{Rewrite Rule Scheduling}
By default, the egg library uses a \code{BackoffScheduler} preventing specific rules from being applied too often, and reducing e-graph growth in the presence of ``explosive'' rules such as associativity and commutativity.
Our experience with \Rise{} optimisation is that using the \code{BackoffScheduler} is counterproductive as the desired optimisation depends on some explosive rules.
For this reason, and to make result analysis easier, \kles{} does
not use a rewrite rule scheduler.

\paragraph{Rewrite Goals}
We compare the lambda calculus encodings using three rewrite goals with increasing complexity.

\def\reductionGoal{\goalStyle{reduction}}
\def\fission{\goalStyle{fission}}
\def\binomial{\goalStyle{binomial}}

The \reductionGoal{} rewrite goal in \cref{fig:reduction-rewrite} is based on a unit test from egg's lambda calculus example, and simply uses \ref{eta-reduction} and \ref{beta-reduction} rules to normalise a term.
The lambda calculus examples from egg are relatively simple, as the rewrite rules involved do not introduce new names on their right-hand side and in most cases do not increase term size: the e-graph size does not grow explosively.

The \fission{} rewrite goal in \cref{fig:fission-rewrite} adds the use of \ref{map-fusion} and \ref{map-fission} rewrite rules to perform a \inlineRise{map} fission that is more complex than a single \ref{map-fission} rule, as it goes through 3 chained functions.
The \ref{map-fusion} and \ref{map-fission} rewrite rules introduce new name bindings on their right-hand-side, and interact with each other as well as \ref{beta-reduction} to create many possibilities: the e-graph size starts to explode.

The \binomial{} rewrite goal from \cref{fig:binom-rewrite} corresponds to a convolution separation optimisation as seen in \cref{ch:imgproc}, where it was used to reduce both memory accesses and arithmetic complexity.
A binomial filter is an essential component of many image processing pipelines, where it reduces noise or detail.
For example, it is sometimes used as part of the Harris corner detection instead of the 3$\times$3 '$+$' convolution used in \cref{ch:imgproc}.
The purpose of the rewrite is to separate the 2D convolution into two 1D convolutions according to the well-known convolution kernel equation:

{\small \begin{equation*}\label{binom-weights}
    \left[\begin{matrix}
        1 & 2 & 1 \\
        2 & 4 & 2 \\
        1 & 2 & 1 \\
    \end{matrix}\right]
    =
    \left[\begin{matrix}
        1 \\
        2 \\
        1 \\
    \end{matrix}\right]
    \times
    \left[\begin{matrix}
        1 & 2 & 1 \\
    \end{matrix}\right]
\end{equation*}}

The \binomial{} goal adds the use of 6 more rewrite rules: 3 rules for \inlineRise{slide} interactions, 1 rule for the \inlineRise{transpose (transpose x)} identity, and 2 rules for dot product decompositions.
The total 10 rules have many possible interactions, aggravating e-graph growth rate.

The author personally contributed to \cite{koehler2021-elevate-imgproc} by demonstrating how to achieve this optimisation via an \Elevate{} strategy applying a sequence of 30 rewrite rules, including 17 $\eta$/$\beta$-reductions.
Although more complex than the \reductionGoal{} and \fission{} rewrite goals, the \binomial{} rewrite goal is still relatively simple compared to complete Harris corner detection and matrix multiplication optimisations.
We suggest that unguided equality saturation should at least scale to the \binomial{} goal and its relatively low complexity, to be useful in practice for \Rise{}.

\begin{figure}[ht]
\centering
\begin{rise}[belowskip=1ex]
($\lambda$comp.
 ($\lambda$add1.
  comp add1 (comp add1 (comp add1 (comp add1 (comp add1 (comp add1 add1)))))
 ) ($\lambda$y. y + 1)
) ($\lambda$f. $\lambda$g. $\lambda$x. f (g x))
\end{rise}
$$\rewritesTo{}^*$$%
\begin{rise}[aboveskip=0.5ex, belowskip=1ex]
$\lambda$x. ((((((x + 1)+ 1) + 1) + 1) + 1) + 1) + 1
\end{rise}
\caption{\textbf{\reductionGoal{} rewrite goal.}
The initial program creates a \inlineRise{comp} combinator for function composition and uses it to compose \inlineRise{add1} with itself 7 times.
All uses of \inlineRise{comp} and \inlineRise{add1} are $\beta$-reduced in the final program, which simply applies \inlineRise{+ 1} to its input value 7 times.}
\label{fig:reduction-rewrite}
\end{figure}

\begin{figure}[ht]
\centering
\begin{rise}[belowskip=1ex]
map ($\lambda$x. f$_5$ (f$_4$ (f$_3$ (f$_2$ (f$_1$ x)))))
\end{rise}
$$ \rewritesTo{}^* $$
\begin{rise}[aboveskip=0.5ex, belowskip=1ex]
$\lambda$y. map ($\lambda$x. f$_5$ (f$_4$ (f$_3$ x))) (map ($\lambda$x. (f$_2$ (f$_1$ x))) y)
\end{rise}
\caption{\textbf{\fission{} rewrite goal.}
The initial program successively applies \inlineRise!f$_1$! to \inlineRise!f$_5$! inside a \inlineRise{map} pattern.
The final program first applies \inlineRise!f$_1$! and \inlineRise!f$_2$! inside one \inlineRise{map} pattern, before applying \inlineRise!f$_3$! to \inlineRise!f$_5$! inside another \inlineRise{map} pattern.}
\label{fig:fission-rewrite}
\end{figure}

\begin{figure}[ht]
  \centering
  \begin{rise}[belowskip=1ex]
map (map $\lambda$nbh. dot (join weights2d) (join nbh))
 (map transpose (slide 3 1 (map (slide 3 1) input)))
  \end{rise}
  $$ \rewritesTo{}^* $$
  \begin{rise}[aboveskip=0.5ex, belowskip=1ex]
map ($\lambda$nbhL. map ($\lambda$nbhH. dot weightsH nbhH)
  (slide 3 1 (map ($\lambda$nbhV. dot weightsV nbhV) transpose nbhL)))
 (slide 3 1 input)
  \end{rise}
  \caption{\textbf{\binomial{} rewrite goal.}
  The initial \Rise{} program iterates over 2D neighborhoods (\inlineRise{nbh}).
  A \inlineRise{dot} product is computed between the 2D weights and each 2D neighborhood.
  The final program iterates over two 1D neighborhoods combined with 1D weights instead.}
  \label{fig:binom-rewrite}
\end{figure}

\subsection{Performance Results}

\begin{table}
\centering
\small
\begin{tabular}{|c|c|c|c|r|r|r|r|r|} \hline
\multicolumn{2}{|c|}{$\lambda$ calculus encoding} & \multirow{2}{*}{\textbf{goal}} & \multirow{2}{*}{\textbf{found?}} & \multirow{2}{*}{\textbf{runtime}} & \multirow{2}{*}{\textbf{RAM}} & \multirow{2}{*}{\textbf{rules}} & \multicolumn{2}{c|}{e-graph size} \\ \cline{1-2}\cline{8-9}
\textbf{\footnotesize extraction?} & \textbf{\footnotesize De Bruijn?} &&&&&& \textbf{\footnotesize e-nodes} & \textbf{\footnotesize e-classes} \\ \hline \hline
\no & \no & \reductionGoal{} & \yes & 0.02s & 3 MB & 0.5K & 0.5K & 0.1K \\ \hline
\no & \no & \fission{} & \color{red}\no & 16s & $>$2000 MB &  &  &  \\ \hline
\no & \no & \binomial{} & \color{red}\no & 15s & $>$2000 MB &  &  &  \\ \hline \hline

\no & \yes & \reductionGoal{} & \yes & 0.2s & 35 MB & 28K & 53K & 25K \\ \hline
\no & \yes & \fission{} & \yes & 0.3s & 36 MB & 39K & 21K & 10K \\ \hline
\no & \yes & \binomial{} & \color{red}\no & 30s & $>$2000 MB &  &  &  \\ \hline \hline

\yes & \no & \reductionGoal{} & \yes & 0.004s & 3 MB & 0.1K & 0.3K & 0.2K \\ \hline
\yes & \no & \fission{} & \yes & 0.006s & 3 MB & 0.2K & 1K & 0.7K \\ \hline
\yes & \no & \binomial{} & \color{red}\no & 20s & $>$2000 MB &  &  &  \\ \hline \hline

\rowcolor{green!45}
\yes & \yes & \reductionGoal{} & \yes & 0.002s & 3 MB & 0.1K & 0.2K & 0.1K \\ \hline
\rowcolor{green!45}
\yes & \yes & \fission{} & \yes & 0.006s & 3 MB & 0.6K & 0.6K & 0.3K \\ \hline
\rowcolor{green!45}
\yes & \yes & \binomial{} & \yes & 0.1s & 8 MB & 5K & 3K & 1K \\ \hline
\end{tabular}
\caption{Evaluating the efficiency of lambda calculus encoding techniques on three rewrite goals.
Combining extraction-based substitution and De Bruijn indices minimises runtime and memory consumption (green background), and is the only encoding that finds the \binomial{} rewrite goal.}
\label{fig:lambda-res}
\end{table}

\Cref{fig:lambda-res} compares the performance of equality saturation using different combinations of substitution (explicit/extraction-based) and name binding (named/De Bruijn) techniques.
It reports whether the goal is found, the search runtime, memory consumption, number of applied rewrite rules, and e-graph size.
The simple \reductionGoal{} goal is found by all combinations, although explicit substitution with De Bruijn indices is less efficient with 25K rewrite rules, 25K e-classes, and occupying 35 MB.
The \fission{} goal is not found if explicit substitution is used with named variables, exhausting the 2 GB memory and showing that this encoding is particularly inefficient.

The \binomial{} goal is only found by combining extraction-based substitution and De Bruijn indices.
With this encoding, all three rewrite goals are found by applying less than 5K rewrite rules, producing e-graphs with fewer than 3K e-nodes and 1K e-classes, and occupying less than 8 MB of memory.
For all three rewrite goals, this combination provides the fastest searches and most compact e-graphs, often by orders of magnitude.

We conclude that \emph{combining extraction-based substitution and De Bruijn indices gives an efficient encoding of lambda calculus for equality saturation}, and adopt this encoding in \kles{}.
This experiment also demonstrates that we can discover relatively simple \Rise{} rewrite goals using unguided equality saturation.
In the following experiment, we seek to build upon this success to discover significantly more complex \Rise{} rewrite goals using sketch-guided equality saturation.


\section{Evaluation of Sketch Guidance}
\label{mm-eval}

This section compares unguided equality saturation to the new sketch-guided equality saturation to achieve complex optimisation goals.
In the evaluation, \emph{both} equality saturation techniques use the efficient lambda calculus encoding from \cref{lambda-eval}.
We evaluate the 7 matrix multiplication optimisations described in the TVM \cite{tvm-2018} manual\footnote{\url{https://tvm.apache.org/docs/how_to/optimize_operators/opt\_gemm.html}} and reproduced in~\cite{hagedorn2020-elevate} using \Elevate{} strategies.
Matrix multiplication is selected as the case study as it allows us to compare against published \Elevate{} strategies that specify optimisations equivalent to TVM schedules~\cite{hagedorn2020-elevate}.
The TVM schedules were not automatically discovered, but written by performance engineers, similar to the Halide schedule discussed in \cref{ch:imgproc} for Harris corner detection.
The \Elevate{} strategies \cite{hagedorn2020-elevate} express the optimisations performed by TVM as compositions of rewrites and achieve the same high performance as TVM.
The code generated by \Shine{} is at worst 0.29$\times$ slower and on average 1.14$\times$ faster than the code produced by TVM on Intel core i5-4670K~\cite{hagedorn2020-elevate}.
The optimisations are typical compiler optimisations, including loop blocking, loop permutation, vectorisation, and multi-threading.

In this evaluation, we compare how much runtime and memory are required for unguided equality saturation and for sketch-guided equality saturation.
For both guided and unguided equality saturation the optimisation goal is specified as a sketch, that acts as the stopping criteria.
This is less restrictive than the searches for a goal program in the previous subsection as the sketch goal may be satisfied by many programs.

We validate that the result of each optimisation goal is high performance code as follows.
When a program satisfying the optimisation goal sketch is found, we check that the generated C code is equivalent, modulo variable names, to the code obtained using \Elevate{} strategies (\cref{app:matmul-opt-progs}).


\subsection{Experimental Setup}

The full version of \kles{}\footnote{\url{https://github.com/rise-lang/shine/tree/sges/src/main/scala/rise/eqsat}} is implemented in Scala, allowing it to leverage the existing \Rise{} codebase.
We decided to reimplement the features of the egg library in Scala instead of dealing with Rust-Scala interoperability.
The standard Java utilities are used for measurements: \code{System.nanoTime()} to measure search runtime, and the \code{Runtime} API to approximate maximum heap memory residency with regular sampling.

The experiments are performed on two platforms.
For \Elevate{} strategies and our sketch-guided equality saturation, we use a less powerful AMD Ryzen 5 PRO 2500U with 4 GB of RAM available to the JVM.
For unguided equality saturation, we use a more powerful Intel Xeon E5-2640 v2 with 60 GB of RAM available to the JVM.
The results are reported from a single run since we dot not care about small variations but rather about orders of magnitude.

\def\baseline {\goalStyle{baseline}}
\def\blocking {\goalStyle{blocking}}
\def\vectorisation {\goalStyle{vectorisation}}
\def\loopperm {\goalStyle{loop-perm}}
\def\arraypacking {\goalStyle{array-packing}}
\def\cacheblocks {\goalStyle{cache-blocks}}
\def\parallel {\goalStyle{parallel}}

\paragraph{Optimisation Goals}
Each optimisation goal incrementally adds more optimisations.
\Cref{app:matmul-opt-progs} shows the corresponding C code generated by \Shine{}.

The \baseline{} optimisation goal uses 3 straightforward nested loops to perform the matrix multiplication.
The \blocking{} optimisation goal adds a blocking (or tiling) optimisation for improved data locality, resulting in 6 nested loops where the 3 innermost ones process $4 \times 32 \times 32$ blocks.
The \vectorisation{} optimisation goal adds parallelism by vectorising the innermost loop over 32 elements.
The \loopperm{} optimisation goal changes the order of the 6 nested loops, for improved data locality.
The \arraypacking{} optimisation goal adds intermediate storage for the transposed \inlineRise{b} matrix, improving memory access patterns.
The \cacheblocks{} optimisation goal unrolls the inner reduction loop.
The \parallel{} optimisation goal adds parallelism by multi-threading the outermost loop.

\subsection{Runtime and Memory Consumption of (Un)Guided Search}

\paragraph{Unguided Equality Saturation}
\Cref{fig:s2} shows the runtime and memory consumption required to find the optimisation goals with unguided equality saturation.
The search terminates when the sketch describing the optimisation goal is found in the e-graph.

The 5 most complex optimisation goals are not found before exhausting the 60 GB of available memory.
Only the \baseline{} and \blocking{} goals are found, and the search for \blocking{} requires more than 1h and about 35 GB of RAM.
Millions of rewrite rules are applied, and the e-graph contains millions of e-nodes and e-classes.
More complex optimisations involve more rewrite rules, creating a richer space of equivalent programs but exhausting memory faster.
As examples, \vectorisation{} and \loopperm{} use vectorisation rules, while \arraypacking{}, \cacheblocks{}, and \parallel{} use rules for optimising memory storage.

\paragraph{Sketch-Guided Equality Saturation}
\Cref{fig:s3} shows the runtime and memory consumption for sketch-guided equality saturation, where sketch guides are used to break a single equality saturation search into multiple.

All optimisations are found in less than 10s, using less than 0.5 GB of RAM.
Interestingly, the number of rewrite rules applied by sketch-guided equality saturation is in the same order of magnitude as for the manual \Elevate{} strategies reported in~\cite{hagedorn2020-elevate}.
On one hand, equality saturation applies more rules than necessary because of its explorative nature.
On the other hand, \Elevate{} strategies apply more rules than necessary because they re-apply the same rule to the same sub-expression and do not necessarily orchestrate the shortest possible rewrite path.
The e-graphs contain no more than $10^{4}$ e-nodes and e-classes, two orders of magnitude less than the $10^{6}$ required for \blocking{} without sketch-guidance.

\begin{table}[th]
\centering
\small
\begin{tabular}{|l|c|r|r|r|r|r|}
\hline
\textbf{goal} & \textbf{found?} & \textbf{runtime} & \textbf{RAM} & \textbf{rules} & \textbf{e-nodes} & \textbf{e-classes} \\
\hline
\baseline{} & \yes & 0.5s & 0.02 GB & 2 & 51 & 49 \\
\hline
\blocking{} & \yes & >1h & 35 GB & 5M & 4M & 2M \\
\hline
\vectorisation{} & \color{red} \no & >1h & >60 GB &  &  &  \\
\hline
\loopperm{} & \color{red} \no & >1h & >60 GB &  &  &  \\
\hline
\arraypacking{} & \color{red} \no & 35mn & >60 GB &  &  &  \\
\hline
\cacheblocks{} & \color{red} \no & 35mn & >60 GB &  &  &  \\
\hline
\parallel{} & \color{red} \no & 35mn & >60 GB &  &  &  \\
\hline
\end{tabular}
\caption{Runtime and memory consumption for \textbf{unguided equality saturation} with efficient lambda calculus encoding.
Only the \baseline{} and \blocking{} optimisation goals are found, with other optimisations exceeding 60 GB.}
\label{fig:s2}
\end{table}

\begin{table}[th]
\centering
\small
\begin{tabular}{|l|c|c|r|r|r|r|r|}
\hline
\textbf{goal} & \textbf{sketch guides} & \textbf{found?} & \textbf{runtime} & \textbf{RAM} & \textbf{rules} & \textbf{e-nodes} & \textbf{e-classes} \\
\hline
\baseline{} & 0 & \yes & 0.5s & 0.02 GB & 2 & 51 & 49 \\
\hline
\blocking{} & 1 & \yes & 7s & 0.3 GB & 11K & 11K & 7K \\
\hline
\vectorisation{} & 2 & \yes & 7s & 0.4 GB & 11K & 11K & 7K \\
\hline
\loopperm{} & 2 & \yes & 4s & 0.3 GB & 6K & 10K & 7K \\
\hline
\arraypacking{} & 3 & \yes & 5s & 0.4 GB & 9K & 10K & 7K \\
\hline
\cacheblocks{} & 3 & \yes & 5s & 0.5 GB & 9K & 10K & 7K \\
\hline
\parallel{} & 3 & \yes & 5s & 0.4 GB & 9K & 10K & 7K \\
\hline
\end{tabular}
\caption{Runtime and memory consumption for \textbf{sketch-guided equality saturation} with efficient lambda calculus encoding.
All optimisations are found in seconds using less than 0.5 GB of memory, and requiring at most 3 sketch guides.}
\label{fig:s3}
\end{table}

\begin{figure}
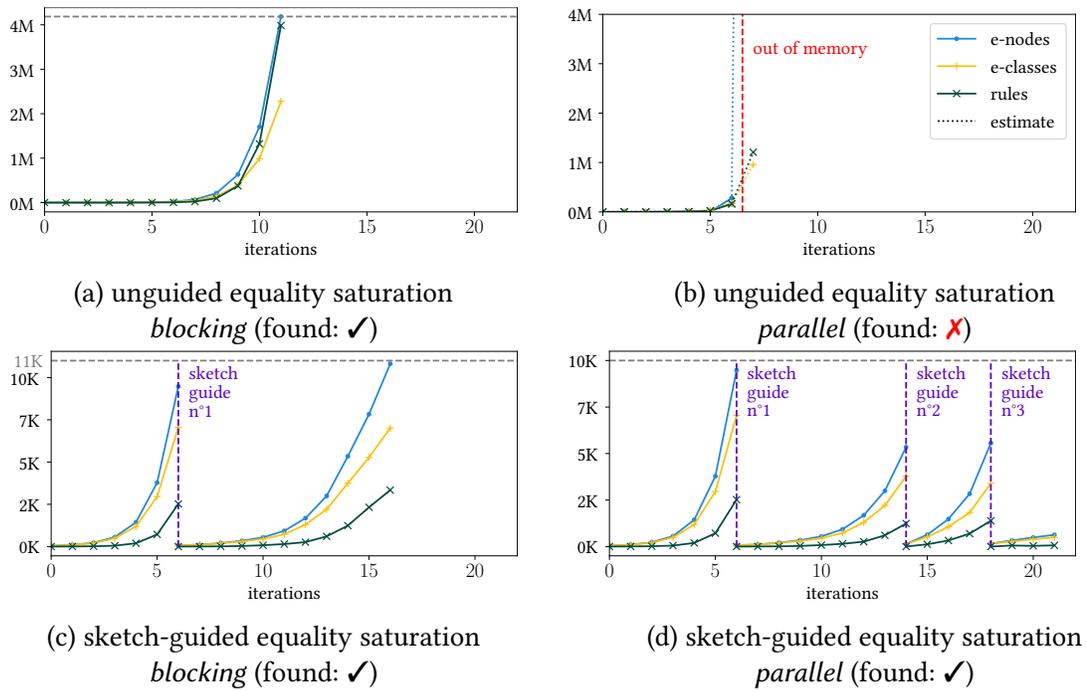

\captionsetup{justification=centering}
    \centering
    \begin{subfigure}[b]{0.49\linewidth}
    \centering
    \resizebox{!}{8em}{\input{media/unguided-blocking.pgf}}
    \caption{unguided equality saturation\\ \blocking{} (found: \yes)}
    \end{subfigure}
    \begin{subfigure}[b]{0.49\linewidth}
    \resizebox{!}{8em}{\input{media/unguided-parallel.pgf}}
    \caption{unguided equality saturation\\ \parallel{} (found: {\color{red}\no})}
    \end{subfigure}\\
    \begin{subfigure}[b]{0.49\linewidth}
    \centering
    \resizebox{!}{8em}{\input{media/guided-blocking.pgf}}
    \caption{sketch-guided equality saturation\\ \blocking{} (found: \yes)}
    \end{subfigure}
    \begin{subfigure}[b]{0.49\linewidth}
    \resizebox{!}{8em}{\input{media/guided-parallel.pgf}}
    \caption{sketch-guided equality saturation\\ \parallel{} (found: \yes)}
    \end{subfigure}
\captionsetup{justification=justified, singlelinecheck=off}
    \caption{The evolution of the e-graph, and the number of rewrite rules applied, during searches for two optimisation goals.
    Sketch guides are depicted with purple vertical lines.
    Note that the scale of the y-axes for unguided graphs (a) and (b) is millions, while for guided graphs (c) and (d) it is thousands.}
    \label{fig:evolution-over-iterations}
\end{figure}

\subsection{E-Graph Evolution in (Un)Guided Search}

\Cref{fig:evolution-over-iterations} plots the growth of the e-graphs during unguided and sketch-guided searches for the \blocking{} and \parallel{} optimisation goals from \cref{fig:s2,fig:s3}.
The e-graphs produced by unguided equality saturation grow exponentially with each search iteration.
The e-graph contains millions of e-nodes and e-classes after applying millions of rules within a small number of iterations (less than 10).
Such rapid growth limits the scalability of unguided search, for example in the 7th iteration of the \parallel{} search the e-graph exhausts 60GB memory.

While the e-graphs produced with sketch-guidance typically also grow exponentially with each iteration, sketches are satisfied within a small number of iterations thanks to an appropriate selection of sketch guides.
The number of rewrites and the maximum e-graph size is \emph{three orders of magnitudes smaller} than for unguided search: no more than 11K in our example searches.
Once a program satisfying a sketch guide is found, a new search is started for the next sketch using that program, growing a fresh e-graph.
Hence sketch-guidance enables scaling to more complex \Rise{} optimisations, such as \parallel{}.
Conceptually there is no limit on the complexity of the optimisations that may be searched for, as optimisations may be factored into as many sketch-guided searches as necessary.

The search for the final \parallel{} sketch goal shows linear rather than exponential growth, as the rewrite rules selected for the search have little interaction.

\subsection{Provided Guidance}

\paragraph{Sketches Guiding the Search}
\Cref{fig:sketch-logical} shows how each optimisation goal is achieved by logical steps, each corresponding to a sketch describing the program after the step is applied.
It transpires that the \goalStyle{split} sketch in \cref{mm-blocking-split-sketch} is a useful first guide for all goals.
While the sketch sizes range from 7 to 12, programs are of size 90 to 124, showing that a sketches elides around 90\% of the program.
Even when 4 sketches must be written, the total sketch size is still small: the largest total being 38.
\Cref{app:sketches} contains all handwritten sketches as well as examples of discovered \Rise{} programs.
Intricate program aspects never need to be specified in the sketches, for example array reshaping patterns such as \inlineRise{split}, \inlineRise{join} and \inlineRise{transpose}.

\begin{table}[ht]
  \centering
  \small
\begin{tabular}{|l|l|l|r|r|} \hline
\textbf{goal} & \textbf{sketch guides} & \textbf{sketch goal} & \textbf{sketch sizes} & \textbf{program size}\\ \hline
\blocking{} & \goalStyle{split} & \goalStyle{reorder$_1$} & 7 & 90 \\ \hline
\vectorisation{} & \goalStyle{split} + \goalStyle{reorder$_1$} & \goalStyle{lower$_1$} & 7 & 124 \\ \hline
\loopperm{} & \goalStyle{split} + \goalStyle{reorder$_2$} & \goalStyle{lower$_2$} & 7 & 104 \\ \hline
\arraypacking{} & \goalStyle{split} + \goalStyle{reorder$_2$} + \goalStyle{store} & \goalStyle{lower$_3$} & 7-12 & 121 \\ \hline
\cacheblocks{} & \goalStyle{split} + \goalStyle{reorder$_2$} + \goalStyle{store} & \goalStyle{lower$_4$} & 7-12 & 121 \\ \hline
\parallel{} & \goalStyle{split} + \goalStyle{reorder$_2$} + \goalStyle{store} & \goalStyle{lower$_5$} & 7-12 & 121 \\ \hline
\end{tabular}
\caption{Decomposition of each optimisation goal into logical steps. A sketch is defined for each logical step. In this table, sketch size counts operators such as \inlineRiseSketch{containsMap}, program size counts operators such as \inlineRise{map}, lambdas, variables and constants: not $\lambda$ applications.}
\label{fig:sketch-logical}
\end{table}

\paragraph{Choice of Rules and Cost Model}
Besides the sketches, performance engineers also specify the rules used in each search and a cost model.
For the \goalStyle{split} sketch, 8 rules explain how to split \inlineRise{map} and \inlineRise{reduce}.
The \goalStyle{reorder} sketches require 9 rules that swap various nestings of \inlineRise{map} and \inlineRise{reduce}.
The \goalStyle{store} sketch requires 4 rules and the \goalStyle{lower} sketches 10 rules including \ref{map-fusion}, 6 rules for vectorisation, 1 rule for loop unrolling and 1 rule for loop parallelisation.
If we naively use all rules for the blocking search, the search runtime increases by about 25$\times$, still finding the goal in minutes but showing the importance of selecting a small set of rules.

We use a simple cost model that minimises weighted term size.
For example, we may give a penalty to using \inlineRise{mapPar}, to avoid implicit multi-threading if it is not explicit in the sketch.
Rules and cost models may be reused and packaged into libraries for recurring logical steps.


\section{Conclusion}
\label{sec:sges-conclusion}

This chapter contributes \emph{sketch-guided equality saturation}, a semi-automated optimisation technique that offers a practical trade-off between the painstaking control of rewriting strategies, and the automated, but often unsuccessful, equality saturation.
Performance engineers guide rewriting by describing how a program should evolve using a sequence of sketches, factoring an infeasible equality saturation search into a sequence of feasible equality saturation searches.

Sketch-guiding leverages the observation that program optimisations are often explained with potentially incomplete program snippets, as in \cref{fig:mm-rewrite} and \cref{ch:imgproc}.
Sketch-guiding enables performance engineers to focus on what the optimised program should look like, rather than on individual program transformation steps.

We demonstrate that sketch-guiding enables seven complex optimisations of matrix multiplication to be applied within seconds in the \Rise{} functional language, using less than 1 GB of RAM (\cref{fig:s3}), using no more than three sketch guides, each 10 times smaller than the complete program  (\cref{fig:sketch-logical}).
By contrast, traditional unguided equality saturation cannot discover the five most complex optimisations even with an hour of runtime and 60 GB of RAM (\cref{fig:s2}).
For each optimisation, the generated code is identical to the high-performance code generated by ordering
thousands of rewrites via the definition of 36 \Elevate{} rewriting strategies in 200 lines of code.
The generated code is at worst 0.29$\times$ slower and on average 1.14$\times$ faster than the code produced by the state-of-the-art TVM compiler on Intel core i5-4670K~\cite{hagedorn2020-elevate}.

This chapter also explores engineering design choices to effectively encode a polymorphically typed lambda calculi like \Rise{} for equality saturation.
The key innovations are extraction-based substitution and representing identifiers as De Bruijn indices.
Combining the techniques reduces the runtime and memory consumption of equality saturation over lambda terms by orders of magnitude, and is necessary to enable unguided equality saturation to discover even relatively simple \Rise{} optimisation goals, such as convolution separation (\cref{fig:lambda-res}).

\chapquote{There is no real ending. It's just the place where you stop the story.}{Frank Herbert}

\chapter{Discussion}
\label{ch:discussion}

\section{Summary}

Optimising programs is challenging, even for skilled performance engineers.
Modern compilers targeting heterogeneous architectures face two major challenges.
First, domain-specific compilers such as Halide for image processing and TVM for machine learning are difficult to extend with the new optimisations required by new algorithms and hardware.
Second, automatic optimisation is often unable to achieve the required performance, and performance engineers often fall back to painstaking manual optimisation.

To mitigate these challenges, this thesis shows the potential of the novel \Shine{} compiler to achieve domain-extensibility, controllable automation, and generate high performance code.
Domain-extensibility facilitates adaptation to new algorithms and hardware.
Controllable automation enables performance engineers to gradually take control of the optimisation process, with the assistance of the compiler.
In \Shine{}, optimisations are applied by rewriting functional programs in the \Rise{} array language, before generating imperative code.

This thesis makes the following research contributions:
\begin{enumerate}
    \item \textbf{Enhancing Code Generation in a Domain-Extensible Compiler} (\cref{ch:imperative-code}).\\
    Three important code generation features are added to \Shine{}, enabling the generation of high-performance code in \Cref{ch:imgproc,ch:guided-rewriting}.
    \begin{itemize}
      \item We contribute a synchronisation \emph{barrier insertion} algorithm that does not need to be modified when extending \Rise{} patterns, as opposed to the barrier elimination algorithm of \Lift{} \cite{lift-ir-2017}.
      The correctness and efficiency of barrier insertion is evaluated on 38 unit tests and 10 benchmarks, mostly taken from prior \Lift{} work.
      We identify 6 differences in the code generated by \Shine{} and \Lift{}, and observe that our algorithm fixes bugs in 13 unit tests and 1 benchmark, where \Lift{} generates incorrect barriers (\cref{fig:barrier-insertion-eval}).
      There is only 1 benchmark where \Shine{} inserts a barrier that \Lift{} eliminates, and we provide a clear pathway to improve our algorithm to generate more efficient barriers than \Lift{} on all unit tests and benchmarks.
    \end{itemize}
    While barrier insertion is implicit and not controllable by rewriting, the next two features add new \Rise{} patterns in order to expose implementation choices to be controlled during rewriting, allowing design space exploration.
    \begin{itemize}
        \item We add the \inlineRise{oclRun} \Rise{} pattern to represent \emph{kernel execution} explicitly, computing the value of an expression by launching an OpenCL kernel.
        This requires modifying the \Shine{} compiler to generate imperative code for multiple OpenCL kernels, as well as the necessary host code to launch them.
        With this feature, 1K lines of handwritten host code are replaced with 1.2K lines of automatically generated code on a relatively simple design space exploration case study (\cref{fig:harris-kernels,fig:harris-host-code}).

        \item We add the \inlineRise{circularBuffer} and \inlineRise{rotateValues} \Rise{} patterns to enable explicit \emph{storage folding} for temporary arrays.
        This requires modifying \Shine{} to generate the desired imperative code when using these patterns.
        \Cref{ch:imgproc} relies on storage folding to generate high performance code.
    \end{itemize}

    \item \textbf{Going Beyond Halide Scheduling with Controlled Rewriting} \cite{koehler2021-elevate-imgproc} (\cref{ch:imgproc}).\\
    Domain-extensibility is combined with controlled rewriting to optimise a standard image processing pipeline: the Harris corner detection \cite{harris-1988}.
    Optimisations are controlled in \Shine{} using \Elevate{} rewriting strategies \cite{hagedorn2020-elevate} that compose rewrite rules.
    First, \Elevate{} allows us to reproduce the effect of an optimised Halide schedule applying 4 standard optimisations: operator fusion, multi-threading, vectorisation and circular buffering.
    Second, \Elevate{} allows us to apply 2 additional optimisations that are not supported by Halide schedules: convolution separation and register rotation.
    Circular buffering and convolution separation in particular leverage the low-level \Rise{} patterns exposed in \cref{ch:imperative-code}, that are introduced via controlled applications of rewrite rules.
    
    Our results on four mobile ARM multi-core CPUs and two different image resolutions show that, with these 6 optimisations, \Shine{} generates code (\cref{app:harris-opt-progs}) up to 16$\times$ (geomean of 9.48$\times$) faster than OpenCV library code, up to 4.5$\times$ (geomean of 3.87$\times$) faster than the similarly designed \Lift{} compiler, and up to 1.4$\times$ (geomean of 1.27$\times$) faster than Halide (\cref{fig:imgproc-results}).

    However, we also observe that controlling rewriting with \Elevate{} strategies is tedious.
    The strategies defined to apply the 6 optimisations are specialised to our Harris case study, and consist of more than 600 lines of code defining 57 helper strategies.
    To perform all 6 optimisations, thousands of rewrite steps are applied.
    It is unclear how to generalise the strategies for reuse across diverse image processing pipelines.
    This motivates the following chapter that aims to lower performance engineer effort through semi-automatic optimisation.
    
    \item \textbf{Proposing a Novel Semi-Automatic Optimisation Technique} \cite{koehler2021-sketch-guided} (\cref{ch:guided-rewriting}).\\
    A new semi-automatic optimisation technique called \emph{sketch-guided equality saturation} is developed, offering a practical trade-off between the painstaking control of rewriting strategies, and the automated, but often unsuccessful, equality saturation.
    Sketch-guiding allows performance engineers to guide program rewriting by specifying rewrite goals as \emph{sketches}: program patterns that leave details unspecified.
    Sketch-guiding leverages the observation that program optimisations are often explained with incomplete program snippets, as in \cref{fig:mm-rewrite} and \cref{sec:harris-opt}.
    Sketch-guiding enables performance engineers to focus on what the optimised program should look like, rather than on individual program transformation steps.

    \Cref{ch:guided-rewriting} evaluates sketch-guided equality saturation by applying 7 realistic optimisations of matrix multiplication in the \Rise{} language.
    Unguided equality saturation alone does not scale to the 5 most complex optimisations, even given an hour and 60GB of RAM (\cref{fig:s2}).
    With the guidance of at most 3 sketch guides, each 10 times smaller than the complete program (\cref{fig:sketch-logical}, \cref{app:sketches}), the compiler applies the optimisations in seconds using less than 1GB of RAM (\cref{fig:s3}).
    For each optimisation, the generated code (\cref{app:matmul-opt-progs}) is identical to the high-performance code generated by manually ordering thousands of rewrites via the definition of 36 rewriting strategies in 200 lines of code.
    The runtime performance of this code on Intel core i5-4670K is at worst 0.71$\times$ and on average 1.14$\times$ that of the code produced by the state-of-the-art TVM compiler~\cite{hagedorn2020-elevate}.

    In addition, \Cref{ch:guided-rewriting} demonstrates how to efficiently encode a polymorphically typed lambda calculus such as \Rise{} for equality saturation.
    The key innovations are extraction-based substitution and representing identifiers as De Bruijn indices.
    Combining the techniques reduces the runtime and memory consumption of equality saturation over lambda terms by orders of magnitude, and is necessary for unguided equality saturation to discover even relatively simple \Rise{} optimisation goals, such as convolution separation (\cref{fig:lambda-res}).
\end{enumerate}

Overall, this thesis demonstrates how extensible rewriting systems are a powerful approach to build domain-extensible compilers with controllable automation of optimisations, that generate high-performance code.
We envision a future were compilers adapt to the rapid pace of change in algorithms and hardware in collaboration with performance engineers.
As new algorithms and hardware are developed, performance engineers will be able to extend compilers with simple, specialised transformations expressed as rewrite rules, to create and explore complex optimisation spaces.
Depending on performance requirements and engineering budgets, performance engineers and compilers will cooperate to explore complex optimisation spaces, for example via rewriting strategies or sketch-guidance.


\section{Limitations}

\paragraph{Code Generation}
The intention is that functional program rewriting should be extensible and controllable, and imperative code generation should be reusable and predictable.
The benefit is that main optimisation choices are encoded in \Rise{} via rewriting, without worrying about imperative details or side effects.
The drawback is that imperative code generation is not easily extensible and controllable by performance engineers.
First, adding new imperative patterns to the intermediate DPIA language requires modifying internal compiler code, e.g. for barrier insertion (\cref{sec:barrier-insertion}).
Second, exposing an implementation choice to rewriting requires encoding that choice into the functional \Rise{} language, e.g. as done with \inlineRise{rotateValues} (\cref{sec:storage-folding}).

\paragraph{Memory Re-use}
In \Shine{}, memory is allocated in a simplistic way when translating functional DPIA to imperative DPIA.
Therefore, there is potential for memory re-use or in-place computation which is not exploited.
For example, there is currently no way to compute \inlineRise{mapSeq f x} in-place by overriding the memory for the array \inlineRise{x}.

\paragraph{Arithmetic Expressions}
In \Shine{}, unrestricted symbolic arithmetic expressions are used to represent array sizes and indices.
Type inference relies on symbolic unification, and code generation on symbolic simplification of arithmetic expressions.
Therefore, type inference success and code generation quality are impacted by the best-effort heuristics used to solve non-linear integer arithmetic formulas in \Shine{}, an undecidable problem in general \cite{kremer2016-generalised}.

\paragraph{Custom Types}
The design of \Shine{} does not facilitate adding new types to \Rise{}.
For example, adding a stream type for storage folding patterns (\cref{sec:storage-folding}) would require modifying internal compiler code, such as the code for type inference.

\paragraph{Property Reasoning} 
When rewriting reductions, it is necessary to reason about associativity and commutativity of the reduction operator.
Currently, \Shine{} does not abstract well over such program properties.
For example, the operator of a \inlineRise{reduce} pattern and the \inlineRise{add} function are considered associative and commutative, but there is no mechanism to assess whether an arbitrary function is associative or commutative.
Similarly, the layout and alignment of arrays in memory matters for the Harris corner detection case study (\cref{ch:imgproc}).
Currently, the desired layout is achieved via careful control using \Elevate{} strategies.
Achieving the desired layout using sketch-guiding would require the ability to constrain memory layouts in sketches.


\paragraph{\Elevate{} Strategies}
The \Elevate{} strategies developed to optimise the Harris corner detection (\cref{ch:imgproc}) are specialised to that case study.
By contrast, the Halide scheduling primitives are more generic, and are reused across diverse image processing pipelines.

\paragraph{Sketch-Guided Equality Saturation}
Sketch-guided equality saturation is a promising technique, but it remains limited.
Currently, there is no clear methodology to guide performance engineers in using this technique.
The strength of sketch-guiding is to allow a declarative specification of the optimisation goal, however in some cases imperative specifications of how to rewrite the program like rewriting strategies might be more suitable.
The expressivity of sketches could be improved, for example we would not know how to support a $\land$ \sLang{} constructor.
Some limitations are inherited from equality saturation, as it is used as the search method.
For example, lambda calculus support remains unsatisfactory as our proposed encoding relies on approximations.
Follow-up work is discussed in the next section.

\paragraph{User Study}
This thesis is motivated by the need to reduce performance engineering efforts, but does not claim to quantify this effort.
While the need for compiler extensibility and controllability is motivated by common knowledge and prior work (\cref{ch:background}), no user study was conducted.
Conducting user studies could give insights into the needs of performance engineers, and inform the design of future semi-automatic compilers for best impact.

\section{Ongoing \& Future Work}


\paragraph{Case Studies}

This thesis focuses on two non-trivial case studies, rather than many trivial case studies.
The Harris corner detection is a standard benchmark for image processing pipeline optimisation (\cref{ch:imgproc}).
Matrix multiplication is a standard benchmark for linear algebra optimisation (\cref{ch:guided-rewriting}).
Achieving competitive performance compared to state-of-the-art domain-specific compilers like Halide and TVM on these case studies is a promising first step.
To demonstrate broader applicability of the techniques explored in this thesis, future work should consider more algorithms and more heterogenous hardware architectures.

The author already developed high-level \Rise{} programs for more diverse image processing pipelines from the Halide benchmark suite (camera pipeline, local laplacian, multiscale interpolation, unsharp mask), as well as deep neural networks subgraphs (e.g. fusing matrix multiplication with activation functions).
The challenge is now to optimise them via rewriting.
We expect such case studies to stress test the \Shine{} compiler design, and to inspire novel ideas, just like they inspired the author to develop sketch-guided equality saturation.

The author also reproduced some of the Harris corner detection optimisations from \cref{ch:imgproc} using sketch-guided equality saturation.
Sketches closely resembling the program snippets from \cref{opt-beyond-lift} were written for the outcome of operator fusion, \inlineElevateRise{harrisIxWithIy}, multi-threading, circular buffering, \inlineElevateRise{sequentialLines} and \inlineRise{unrollReductions}.
Except for operator fusion, where the search is currently failing, sketch-guiding successfully finds programs as desired.
To fully reproduce the \Shine{} cbuf version of Harris corner detection using sketch-guiding, only the operator fusion and vectorisation transformation steps are missing.



\paragraph{Optimisation Assistant}

\Elevate{} rewriting strategies and sketch-guided equality saturation are tools for performance engineers to optimise programs via rewriting.
For increased productivity, future work may integrate such rewriting tools into an interactive, rewrite-based optimisation assistant.
Rewriting strategies and sketch-guiding can be combined into hybrid solutions: it is possible to expose sketch-guided equality saturation as a strategy, i.e. a \inlineElevateRise{P $\to$ RewriteResult[P]} function.
Such an assistant could serve as a testbed for many other rewriting tools: \Lift{}-style stochastic search, SPIRAL-style hardware-aware search, cost estimation feedback, rewrite rule or even sketch synthesis, etc.
Multiple interactive optimisation assistants were recently developed independently, showing that this research direction is valued by the community.
Roly-poly supports performance engineers when developing Halide schedules \cite{ikarashi2021-guided-scheduling}.
DLOOPT is an optimisation assistant for TVM \cite{hsieh2022-dloopt}.
OptiTrust allows developing high-performance C code via series of source-to-source transformations \cite{chargueraud2022-optitrust}.

\paragraph{Sketch-Guided Rewriting}

This thesis only scratches the surface of what is possible with the novel idea of sketch-guided rewriting.
Future work could:
\begin{enumerate}
  \item Develop a methodology for performance engineers to come up with appropriate sketches and sets of rewrite rules.
  We imagine that performance engineers would start by writing their sketch goal, and include all potentially useful rewrite rules for the search.
  Then, if the search is unsuccessful, performance engineers can incrementally write sketch guides to make the search easier, but also to debug the search.
  For example, if tiling fails but intermediate splitting succeeds, then the problem lies in the reordering search.
  \item Investigate how to compute the intersection between the set of programs represented by a sketch, and the set of programs represented by an e-graph.
  This could be used to recover optimality and search completeness guarantees for use cases where reaching saturation is feasible throughout the entire sketch-guided equality saturation process.
  \item Attempt to remove the need for intermediate sketch goals by synthesizing or inferring them.
  For example, given a tiling sketch (splitting + reordering), could an intermediate splitting sketch be automatically inferred?
  Recent related work constructs a search space by generating program sketches \cite{2022-tensorIR}.
  \item Develop reusable transformation libraries.
  For example, would it be possible to provide a complete set of rewrite rules for splitting and reordering the loops of any \Rise{} program?
  How would such libraries interact with domain-extensibility and the addition of new \Rise{} patterns?
  \item Investigate how to design useful generic sketches, that may be used for multiple programs.
  An example use-case would be to eliminate a certain pattern from any given program, by introducing a \inlineRiseSketch{notContains} sketch construct.
  \item Combine sketch-guiding with other search methods (e.g. polyhedral modeling, reinforcement learning, stochastic search, greedy search, beam search).
  This could be used to trade-off between exploration and exploitation of the search space \cite{kourta2022-caviar}, and to provide optimised search strategies for specific transformations.
  \item Compare sketch-guiding to program synthesis tools like Sketch \cite{solar2008-synthesis-sketching} or Rosette \cite{torlak2013-rosette}.
  Is incremental sketch-guiding beneficial on standard program synthesis benchmarks compared to traditional counterexample guided inductive synthesis?
  Is writing a program to start rewriting from easier than writing a program specification to check against?
\end{enumerate}

We hope that the community will be inspired to apply sketch-guided equality saturation, or sketch-guiding, to more diverse applications.
To this end, the author has already released a Rust library implementing sketch-guided equality saturation on top of the egg library, it can be used for any term language compatible with egg ({\footnotesize\url{https://github.com/Bastacyclop/egg-sketches}}).


\paragraph{Domain-Extensible Compilers}

We advocate that domain-extensible compilers should be developed to cope with the rapid pace of change in algorithms and hardware, and contribute to this relatively recent research direction.
More work is required to establish domain-extensible compilers as a viable, production-ready alternative to domain-specific compilers.
A sensible next step would be to investigate compiling frontend languages through \Shine{} (\cref{fig:rise-shine}), for example by translating Halide algorithms into \Rise{} programs, or by translating Fortran code as done in TyTraCL \cite{2015-TyTra, nabi2019automatic}.
With the aim to build the next generation of domain-extensible compilers, we are currently exploring multiple potential collaborations.

We are engaging with the authors of the recent Exo language \cite{2022-exo}, independently developed to help performance engineers write, optimise, and target high-performance computing kernels onto new hardware accelerators.
\emph{Exocompilation} is about externalising target-specific code generation support and optimisation policies to user-level code, and is rooted in similar motivations as domain-extensibility and controllable automation.
In \cite{2022-exo}, performance engineers can add support for custom imperative instructions, and control when they are introduced using a scheduling language.
Exo inherits ideas from both Halide schedules and \Elevate{} strategies, its scheduling language is implemented by composing rewrite rules over imperative programs.

We are engaging with the authors of AnyDSL \cite{anydsl-2018}, that are currently working on the next generation of their IR: Thorin 2.
To provide additional flexibility, Thorin 2 will be based on the calculus of constructions \cite{coquand1986-coc}, which is also the basis of proof assistants like Coq \cite{huet1997-coq}.
Of particular interest is the possibility to embed the \Rise{} language into Thorin 2, and to bring some of our program rewriting techniques into this ecosystem, where they could be combined with powerful partial evaluation techniques.

We are engaging with users and developers of MLIR~\cite{mlir-2020} (the Multi-Layer Intermediate Representation).
MLIR is a novel approach for building reusable and extensible compiler infrastructures that was independently developed during this thesis.
MLIR allows compiler writers to define their own specialised \emph{dialects} that can interact with other MLIR dialects.
Martin Lücke has already embedded a subset of \Rise{} as an MLIR dialect \cite{lucke2021-mlir}, allowing to interact with a higher-level TensorFlow machine learning dialect, and a lower-level polyhedral affine dialect.
There is also interest in bringing our program rewriting techniques into the MLIR ecosystem, so that they can be applied to any dialect.

\paragraph{Formal Verification}

This thesis uses rewrite rules that must be semantics preserving, but makes no attempt at formally verifying rewrite rule correctness.
Steuwer's thesis \cite{steuwer2015-thesis} and Qin's work\footnote{\url{https://github.com/XYUnknown/individual-project}}, mentioned in \cite{hagedorn2020-elevate}, provide proofs for multiple rewrite rules, however we use many additional rules.
In \Shine{}, rewrite rules are treated as axioms, and trusted to be semantics preserving without being verified.
Code generation is also not formally verified in \Shine{}.

Future work could attempt to formally verify all or parts of the \Shine{} compiler.
During this thesis, a paper was independently written on formally verified rewriting for tensor program optimisation \cite{2022-POPL-verified-tensor-rewrite}.
We expect combining domain-extensibility, controllable automation and formal verification to be highly rewarding, but also highly challenging. 
For example, formally verifying the patterns and rewrite rules added to \Rise{} \& \Shine{} by performance engineers would provide extension safety.
In Exo \cite{2022-exo}, performance engineers can introduce their own instructions by pairing them with a semantic model, allowing to automatically check for semantic preservation of code replacements using effect analyses and SMT solving.

The connection between theorem proving and program optimisation is intriguing, and may lead to promising research.
While rewriting strategies specify how to transform a program, proof tactics \cite{gordon1979-edinburgh} specify how to transform a proof state. 
Just as program sketches specify partial programs and can guide optimisation, 
proof sketches specify partial proofs and can guide theorem proving \cite{wiedijk2003-formal-proof-sketches, corbineau2007-declarative-coq}.
Although equality saturation exploits e-graphs for program optimisation~\cite{tate2009-equality-saturation}, e-graphs were originally designed for efficient congruence closure in theorem provers \cite{nelson1980-techniques, de2008-z3}.


\paragraph{Numerical Analysis}
This thesis treats floating points as real numbers, ignoring accuracy problems coming from rounding errors \cite{goldberg1991-floats}.
Similarly, underflow and overflow problems are ignored for integers.
Future work may incorporate numerical analyses \cite{darulova2017-realc, panchekha2015-herbie, becker2022-cakeml} in \Shine{} to reason about the accuracy of number representations.

\appendix
\chapter{Optimised Harris Corner Detection}
\label{app:harris-opt-progs}

This appendix contains the OpenCL programs generated by \Shine{} after rewriting the \Rise{} program for Harris corner detection using the two \Elevate{} strategies from \cref{cbuf-strategy,rrot-strategy}.

To keep the code readable and compact, we may apply cosmetic changes to the code such as renaming variables and removing unnecessary parentheses, brackets, or space.
For example, we may rewrite the code from \cref{fig:before-cosmetic-harris} into the code from \cref{fig:after-cosmetic-harris}.


\begin{opencl}[label={fig:before-cosmetic-harris}, caption={Code sample before cosmetic changes.}, basicstyle=\ttfamily\scriptsize, breaklines=true, postbreak=\mbox{{$\hookrightarrow$}\space}]
{
  float x145;
  x145 = 0.0f;
  /* comment */
  x145 = (x145 + 1.0f);
  vstore(x145, (&(x234[((((((2 * n0) * n1) + (4 * i1)) + (8 * n1)) + ((32 * gid) * n1)) + (i0 * n1))])));
}
\end{opencl}

\begin{opencl}[label={fig:after-cosmetic-harris}, caption={Code sample after cosmetic changes.}, basicstyle=\ttfamily\scriptsize, breaklines=true, postbreak=\mbox{{$\hookrightarrow$}\space}]
float x2 = 0.0f;
x2 += 1.0f;
vstore(x2, &x1[2*n0*n1 + 4*i1 + 8*n1 + 32*gid*n1 + i0*n1]);
\end{opencl}

\section{Comparable to Halide Reference}

The OpenCL code generated by \Shine{} after rewriting the \Rise{} Harris corner detection from \cref{rise-harris} (\cref{sec:harris-opt}) using the cbuf \Elevate{} strategy from \cref{cbuf-strategy} (\cref{sec:harris-opt}).

\begin{opencl}[caption={OpenCL code generated by \Shine{} for the cbuf Harris corner detection (\cref{cbuf-strategy}).}, basicstyle=\ttfamily\tiny, breaklines=true, postbreak=\mbox{{$\hookrightarrow$}\space}]
struct Record_float4_float4 {
  float4 a;
  float4 b;
};

__kernel
void harris(global float* restrict output, int n0, int n1, const global float* restrict x0,
            global float* restrict t1, global float* restrict t2, global float* restrict t3) {
  for (int gid = get_global_id(0);(gid < (n0 / 32));gid = (gid + get_global_size(0))) {
    for (int i0 = 0;(i0 < 2);i0 = (1 + i0)) {
      for (int i1 = 0;(i1 < (n1 / 4));i1 = (1 + i1)) {
        float4 t4 = (float4)(0.0f);
        t4 = t4 + (float4)(0.299f) * vload4(0, &x0[4*i1 + 32*gid*n1 + i0*n1]);
        t4 = t4 + (float4)(0.587f) * vload4(0, &x0[4*i1 + 4*n1 + 32*gid*n1 + i0*n1 + n0*n1]);
        t4 = t4 + (float4)(0.114f) * vload4(0, &x0[2*n0*n1 + 4*i1 + 8*n1 + 32*gid*n1 + i0*n1]);
        vstore4(t4, 0, &t3[3*n1*get_global_id(0) + 4*i0 + 4*i1 + 12*get_global_id(0) + i0*n1]);
      }
    }
    
    for (int i2 = 0;(i2 < 2);i2 = (1 + i2)) {
      for (int i3 = 0;(i3 < (n1 / 4));i3 = (1 + i3)) {
        float4 t4 = (float4)(0.0f);
        t4 = t4 + ((float4)(0.299f) * vload4(0, &x0[((((2 * n1) + (4 * i3)) + ((32 * gid) * n1)) + (i2 * n1))]));
        t4 = t4 + ((float4)(0.587f) * vload4(0, &x0[(((((4 * i3) + (6 * n1)) + ((32 * gid) * n1)) + (i2 * n1)) + (n0 * n1))]));
        t4 = t4 + ((float4)(0.114f) * vload4(0, &x0[((2 * n0) * n1) + (4 * i3) + (10 * n1) + ((32 * gid) * n1) + (i2 * n1)]));
        vstore4(t4, 0, &t3[3*n1*get_global_id(0) + 4*i3 + 4*((2 + i2) % 3) + 12*get_global_id(0) + n1*((2 + i2) % 3)]);
      }
      
      for (int i4 = 0;(i4 < (n1 / 4));i4 = (1 + i4)) {
        float4 t5[6];
        t5[0] = vload4(0, &t3[((3 * n1) * get_global_id(0)) + (4 * i2) + (4 * i4) + (12 * get_global_id(0)) + (i2 * n1)]);
        t5[1] = vload4(0, &t3[(4 + (3 * n1 * get_global_id(0))) + (4 * i2) + (4 * i4) + (12 * get_global_id(0)) + (i2 * n1)]);
        t5[2] = vload4(0, &t3[4 + n1 + 3*n1*get_global_id(0) + 4*i2 + 4*i4 + 12*get_global_id(0) + i2*n1]);
        t5[3] = vload4(0, &t3[8 + n1 + 3*n1*get_global_id(0) + 4*i2 + 4*i4 + 12*get_global_id(0) + i2*n1]);
        t5[4] = vload4(0, &t3[3*n1*get_global_id(0) + 4*i4 + 4*((2 + i2) % 3) + 12*get_global_id(0) + n1*((2 + i2) % 3)]);
        t5[5] = vload4(0, &t3[4 + 3*n1*get_global_id(0) + 4*i4 + 4*((2 + i2) % 3) + 12*get_global_id(0) + n1*((2 + i2) % 3)]);
        
        float4 t6 = (float4)(0.0f);
        t6 = (t6 + ((float4)-0.083333336f * (float4)(t5[0].s0, t5[0].s1, t5[0].s2, t5[0].s3)));
        t6 = (t6 + ((float4)0.0f * (float4)(t5[0].s1, t5[0].s2, t5[0].s3, t5[1].s0)));
        t6 = (t6 + ((float4)0.083333336f * (float4)(t5[0].s2, t5[0].s3, t5[1].s0, t5[1].s1)));
        t6 = (t6 + ((float4)-0.16666667f * (float4)(t5[2].s0, t5[2].s1, t5[2].s2, t5[2].s3)));
        t6 = (t6 + ((float4)0.0f * (float4)(t5[2].s1, t5[2].s2, t5[2].s3, t5[3].s0)));
        t6 = (t6 + ((float4)0.16666667f * (float4)(t5[2].s2, t5[2].s3, t5[3].s0, t5[3].s1)));
        t6 = (t6 + ((float4)-0.083333336f * (float4)(t5[4].s0, t5[4].s1, t5[4].s2, t5[4].s3)));
        t6 = (t6 + ((float4)0.0f * (float4)(t5[4].s1, t5[4].s2, t5[4].s3, t5[5].s0)));
        t6 = (t6 + ((float4)0.083333336f * (float4)(t5[4].s2, t5[4].s3, t5[5].s0, t5[5].s1)));
        vstore4(t6, 0, &t2[((((((3 * n1) * get_global_id(0)) + (4 * i2)) + (4 * i4)) + (12 * get_global_id(0))) + (i2 * n1))]);
          
        float4 t7 = (float4)(0.0f);
        t7 = (t7 + ((float4)-0.083333336f * (float4)(t5[0].s0, t5[0].s1, t5[0].s2, t5[0].s3)));
        t7 = (t7 + ((float4)-0.16666667f * (float4)(t5[0].s1, t5[0].s2, t5[0].s3, t5[1].s0)));
        t7 = (t7 + ((float4)-0.083333336f * (float4)(t5[0].s2, t5[0].s3, t5[1].s0, t5[1].s1)));
        t7 = (t7 + ((float4)0.0f * (float4)(t5[2].s0, t5[2].s1, t5[2].s2, t5[2].s3)));
        t7 = (t7 + ((float4)0.0f * (float4)(t5[2].s1, t5[2].s2, t5[2].s3, t5[3].s0)));
        t7 = (t7 + ((float4)0.0f * (float4)(t5[2].s2, t5[2].s3, t5[3].s0, t5[3].s1)));
        t7 = (t7 + ((float4)0.083333336f * (float4)(t5[4].s0, t5[4].s1, t5[4].s2, t5[4].s3)));
        t7 = (t7 + ((float4)0.16666667f * (float4)(t5[4].s1, t5[4].s2, t5[4].s3, t5[5].s0)));
        t7 = (t7 + ((float4)0.083333336f * (float4)(t5[4].s2, t5[4].s3, t5[5].s0, t5[5].s1)));
        vstore4(t7, 0, &t1[((((((3 * n1) * get_global_id(0)) + (4 * i2)) + (4 * i4)) + (12 * get_global_id(0))) + (i2 * n1))]);
      }
    }
    
    for (int i5 = 0;(i5 < 32);i5 = (1 + i5)) {
      for (int i6 = 0;(i6 < (n1 / 4));i6 = (1 + i6)) {
        float4 t8 = (float4)(0.0f);
        t8 = (t8 + ((float4)(0.299f) * vload4(0, &x0[4*i6 + 4*n1 + 32*gid*n1 + i5*n1])));
        t8 = (t8 + ((float4)(0.587f) * vload4(0, &x0[4*i6 + 8*n1 + 32*gid*n1 + i5*n1 + n0*n1])));
        t8 = (t8 + ((float4)(0.114f) * vload4(0, &x0[2*n0*n1 + 4*i6 + 12*n1 + 32*gid*n1 + i5*n1])));
        vstore4(t8, 0, &t3[3*n1*get_global_id(0) + 4*i6 + 4*((1 + i5) % 3) + 12*get_global_id(0) + n1*((1 + i5) % 3)]);
      }
      
      for (int i7 = 0;(i7 < (n1 / 4));i7 = (1 + i7)) {
        float4 t5[6];
        t5[0] = vload4(0, &t3[3*n1*get_global_id(0) + 4*i7 + 4*((2 + i5) % 3) + 12*get_global_id(0) + n1*((2 + i5) % 3)]);
        t5[1] = vload4(0, &t3[4 + 3*n1*get_global_id(0) + 4*i7 + 4*((2 + i5) % 3) + 12*get_global_id(0) + n1*((2 + i5) % 3)]);
        t5[2] = vload4(0, &t3[3*n1*get_global_id(0) + 4*i7 + 4*(i5 % 3) + 12*get_global_id(0) + n1*(i5 % 3)]);
        t5[3] = vload4(0, &t3[4 + 3*n1*get_global_id(0) + 4*i7 + 4*(i5 % 3) + 12*get_global_id(0) + n1*(i5 % 3)]);
        t5[4] = vload4(0, &t3[3*n1*get_global_id(0) + 4*i7 + 4*((1 + i5) % 3) + 12*get_global_id(0) + n1*((1 + i5) % 3)]);
        t5[5] = vload4(0, &t3[4 + 3*n1*get_global_id(0) + 4*i7 + 4*((1 + i5) % 3) + 12*get_global_id(0) + n1*((1 + i5) % 3)]);
        
        float4 t6 = (float4)(0.0f);
        t6 = (t6 + ((float4)-0.083333336f * (float4)(t5[0].s0, t5[0].s1, t5[0].s2, t5[0].s3)));
        t6 = (t6 + ((float4)0.0f * (float4)(t5[0].s1, t5[0].s2, t5[0].s3, t5[1].s0)));
        t6 = (t6 + ((float4)0.083333336f * (float4)(t5[0].s2, t5[0].s3, t5[1].s0, t5[1].s1)));
        t6 = (t6 + ((float4)-0.16666667f * (float4)(t5[2].s0, t5[2].s1, t5[2].s2, t5[2].s3)));
        t6 = (t6 + ((float4)0.0f * (float4)(t5[2].s1, t5[2].s2, t5[2].s3, t5[3].s0)));
        t6 = (t6 + ((float4)0.16666667f * (float4)(t5[2].s2, t5[2].s3, t5[3].s0, t5[3].s1)));
        t6 = (t6 + ((float4)-0.083333336f * (float4)(t5[4].s0, t5[4].s1, t5[4].s2, t5[4].s3)));
        t6 = (t6 + ((float4)0.0f * (float4)(t5[4].s1, t5[4].s2, t5[4].s3, t5[5].s0)));
        t6 = (t6 + ((float4)0.083333336f * (float4)(t5[4].s2, t5[4].s3, t5[5].s0, t5[5].s1)));
        vstore4(t6, 0, &t2[3*n1*get_global_id(0) + 4*i7 + 4*((2 + i5) % 3) + 12*get_global_id(0) + n1*((2 + i5) % 3)]);
          
        float4 t7 = (float4)(0.0f);
        t7 = (t7 + ((float4)-0.083333336f * (float4)(t5[0].s0, t5[0].s1, t5[0].s2, t5[0].s3)));
        t7 = (t7 + ((float4)-0.16666667f * (float4)(t5[0].s1, t5[0].s2, t5[0].s3, t5[1].s0)));
        t7 = (t7 + ((float4)-0.083333336f * (float4)(t5[0].s2, t5[0].s3, t5[1].s0, t5[1].s1)));
        t7 = (t7 + ((float4)0.0f * (float4)(t5[2].s0, t5[2].s1, t5[2].s2, t5[2].s3)));
        t7 = (t7 + ((float4)0.0f * (float4)(t5[2].s1, t5[2].s2, t5[2].s3, t5[3].s0)));
        t7 = (t7 + ((float4)0.0f * (float4)(t5[2].s2, t5[2].s3, t5[3].s0, t5[3].s1)));
        t7 = (t7 + ((float4)0.083333336f * (float4)(t5[4].s0, t5[4].s1, t5[4].s2, t5[4].s3)));
        t7 = (t7 + ((float4)0.16666667f * (float4)(t5[4].s1, t5[4].s2, t5[4].s3, t5[5].s0)));
        t7 = (t7 + ((float4)0.083333336f * (float4)(t5[4].s2, t5[4].s3, t5[5].s0, t5[5].s1)));
        vstore4(t7, 0, &t1[3*n1*get_global_id(0) + 4*i7 + 4*((2 + i5) % 3) + 12*get_global_id(0) + n1*((2 + i5) % 3)]);
      }
      
      for (int i8 = 0;(i8 < (n1 / 4));i8 = (1 + i8)) {
        struct Record_float4_float4 t9[6];
        t9[0].a = vload4(0, &t2[3*n1*get_global_id(0) + 4*i8 + 4*(i5 % 3) + 12*get_global_id(0) + n1*(i5 % 3)]);
        t9[0].b = vload4(0, &t1[3*n1*get_global_id(0) + 4*i8 + 4*(i5 % 3) + 12*get_global_id(0) + n1*(i5 % 3)]);
        t9[1].a = vload4(0, &t2[4 + 3*n1*get_global_id(0) + 4*i8 + 4*(i5 % 3) + 12*get_global_id(0) + n1*(i5 % 3)]);
        t9[1].b = vload4(0, &t1[4 + 3*n1*get_global_id(0) + 4*i8 + 4*(i5 % 3) + 12*get_global_id(0) + n1*(i5 % 3)]);
        t9[2].a = vload4(0, &t2[3*n1*get_global_id(0) + 4*i8 + 4*((1+i5) % 3) + 12*get_global_id(0) + n1*((1+i5) % 3)]);
        t9[2].b = vload4(0, &t1[3*n1*get_global_id(0) + 4*i8 + 4*((1+i5) % 3) + 12*get_global_id(0) + n1*((1+i5) % 3)]);
        t9[3].a = vload4(0, &t2[4 + 3*n1*get_global_id(0) + 4*i8 + 4*((1+i5) % 3) + 12*get_global_id(0) + n1*((1+i5) % 3)]);
        t9[3].b = vload4(0, &t1[4 + 3*n1*get_global_id(0) + 4*i8 + 4*((1+i5) % 3) + 12*get_global_id(0) + n1*((1+i5) % 3)]);
        t9[4].a = vload4(0, &t2[3*n1*get_global_id(0) + 4*i8 + 4*((2+i5) % 3) + 12*get_global_id(0) + n1*((2+i5) % 3)]);
        t9[4].b = vload4(0, &t1[3*n1*get_global_id(0) + 4*i8 + 4*((2+i5) % 3) + 12*get_global_id(0) + n1*((2+i5) % 3)]);
        t9[5].a = vload4(0, &t2[4 + 3*n1*get_global_id(0) + 4*i8 + 4*((2+i5) % 3) + 12*get_global_id(0) + n1*((2+i5) % 3)]);
        t9[5].b = vload4(0, &t1[4 + 3*n1*get_global_id(0) + 4*i8 + 4*((2+i5) % 3) + 12*get_global_id(0) + n1*((2+i5) % 3)]);

        float4 t10 = (float4)(0.0f);
        t10 += (float4)(t9[0].a.s0,t9[0].a.s1,t9[0].a.s2,t9[0].a.s3) * (float4)(t9[0].b.s0,t9[0].b.s1,t9[0].b.s2,t9[0].b.s3);
        t10 += (float4)(t9[0].a.s1,t9[0].a.s2,t9[0].a.s3,t9[1].a.s0) * (float4)(t9[0].b.s1,t9[0].b.s2,t9[0].b.s3,t9[1].b.s0);
        t10 += (float4)(t9[0].a.s2,t9[0].a.s3,t9[1].a.s0,t9[1].a.s1) * (float4)(t9[0].b.s2,t9[0].b.s3,t9[1].b.s0,t9[1].b.s1);
        t10 += (float4)(t9[2].a.s0,t9[2].a.s1,t9[2].a.s2,t9[2].a.s3) * (float4)(t9[2].b.s0,t9[2].b.s1,t9[2].b.s2,t9[2].b.s3);
        t10 += (float4)(t9[2].a.s1,t9[2].a.s2,t9[2].a.s3,t9[3].a.s0) * (float4)(t9[2].b.s1,t9[2].b.s2,t9[2].b.s3,t9[3].b.s0);
        t10 += (float4)(t9[2].a.s2,t9[2].a.s3,t9[3].a.s0,t9[3].a.s1) * (float4)(t9[2].b.s2,t9[2].b.s3,t9[3].b.s0,t9[3].b.s1);
        t10 += (float4)(t9[4].a.s0,t9[4].a.s1,t9[4].a.s2,t9[4].a.s3) * (float4)(t9[4].b.s0,t9[4].b.s1,t9[4].b.s2,t9[4].b.s3);
        t10 += (float4)(t9[4].a.s1,t9[4].a.s2,t9[4].a.s3,t9[5].a.s0) * (float4)(t9[4].b.s1,t9[4].b.s2,t9[4].b.s3,t9[5].b.s0);
        t10 += (float4)(t9[4].a.s2,t9[4].a.s3,t9[5].a.s0,t9[5].a.s1) * (float4)(t9[4].b.s2,t9[4].b.s3,t9[5].b.s0,t9[5].b.s1);
        
        float4 t12 = (float4)(0.0f);
        t12 += (float4)(t9[0].b.s0,t9[0].b.s1,t9[0].b.s2,t9[0].b.s3) * (float4)(t9[0].b.s0,t9[0].b.s1,t9[0].b.s2,t9[0].b.s3);
        t12 += (float4)(t9[0].b.s1,t9[0].b.s2,t9[0].b.s3,t9[1].b.s0) * (float4)(t9[0].b.s1,t9[0].b.s2,t9[0].b.s3,t9[1].b.s0);
        t12 += (float4)(t9[0].b.s2,t9[0].b.s3,t9[1].b.s0,t9[1].b.s1) * (float4)(t9[0].b.s2,t9[0].b.s3,t9[1].b.s0,t9[1].b.s1);
        t12 += (float4)(t9[2].b.s0,t9[2].b.s1,t9[2].b.s2,t9[2].b.s3) * (float4)(t9[2].b.s0,t9[2].b.s1,t9[2].b.s2,t9[2].b.s3);
        t12 += (float4)(t9[2].b.s1,t9[2].b.s2,t9[2].b.s3,t9[3].b.s0) * (float4)(t9[2].b.s1,t9[2].b.s2,t9[2].b.s3,t9[3].b.s0);
        t12 += (float4)(t9[2].b.s2,t9[2].b.s3,t9[3].b.s0,t9[3].b.s1) * (float4)(t9[2].b.s2,t9[2].b.s3,t9[3].b.s0,t9[3].b.s1);
        t12 += (float4)(t9[4].b.s0,t9[4].b.s1,t9[4].b.s2,t9[4].b.s3) * (float4)(t9[4].b.s0,t9[4].b.s1,t9[4].b.s2,t9[4].b.s3);
        t12 += (float4)(t9[4].b.s1,t9[4].b.s2,t9[4].b.s3,t9[5].b.s0) * (float4)(t9[4].b.s1,t9[4].b.s2,t9[4].b.s3,t9[5].b.s0);
        t12 += (float4)(t9[4].b.s2,t9[4].b.s3,t9[5].b.s0,t9[5].b.s1) * (float4)(t9[4].b.s2,t9[4].b.s3,t9[5].b.s0,t9[5].b.s1);
        
        float4 t14 = (float4)(0.0f);
        t14 += (float4)(t9[0].a.s0,t9[0].a.s1,t9[0].a.s2,t9[0].a.s3) * (float4)(t9[0].a.s0,t9[0].a.s1,t9[0].a.s2,t9[0].a.s3);
        t14 += (float4)(t9[0].a.s1,t9[0].a.s2,t9[0].a.s3,t9[1].a.s0) * (float4)(t9[0].a.s1,t9[0].a.s2,t9[0].a.s3,t9[1].a.s0);
        t14 += (float4)(t9[0].a.s2,t9[0].a.s3,t9[1].a.s0,t9[1].a.s1) * (float4)(t9[0].a.s2,t9[0].a.s3,t9[1].a.s0,t9[1].a.s1);
        t14 += (float4)(t9[2].a.s0,t9[2].a.s1,t9[2].a.s2,t9[2].a.s3) * (float4)(t9[2].a.s0,t9[2].a.s1,t9[2].a.s2,t9[2].a.s3);
        t14 += (float4)(t9[2].a.s1,t9[2].a.s2,t9[2].a.s3,t9[3].a.s0) * (float4)(t9[2].a.s1,t9[2].a.s2,t9[2].a.s3,t9[3].a.s0);
        t14 += (float4)(t9[2].a.s2,t9[2].a.s3,t9[3].a.s0,t9[3].a.s1) * (float4)(t9[2].a.s2,t9[2].a.s3,t9[3].a.s0,t9[3].a.s1);
        t14 += (float4)(t9[4].a.s0,t9[4].a.s1,t9[4].a.s2,t9[4].a.s3) * (float4)(t9[4].a.s0,t9[4].a.s1,t9[4].a.s2,t9[4].a.s3);
        t14 += (float4)(t9[4].a.s1,t9[4].a.s2,t9[4].a.s3,t9[5].a.s0) * (float4)(t9[4].a.s1,t9[4].a.s2,t9[4].a.s3,t9[5].a.s0);
        t14 += (float4)(t9[4].a.s2,t9[4].a.s3,t9[5].a.s0,t9[5].a.s1) * (float4)(t9[4].a.s2,t9[4].a.s3,t9[5].a.s0,t9[5].a.s1);
        
        vstore4(t14*t12 - t10*t10 - (float4)(0.04f) * (t14 + t12) * (t14 + t12), 0, &output[4*i8 + 32*gid*n1 + i5*n1]);
      }
    }
  }
}
\end{opencl}

\section{Beyond Halide Reference}

The OpenCL code generated by \Shine{} after rewriting the \Rise{} Harris corner detection from \cref{rise-harris} (\cref{sec:harris-opt}) using the cbuf+rrot \Elevate{} strategy from \cref{rrot-strategy} (\cref{sec:harris-opt}).

\begin{opencl}[caption={OpenCL code generated by \Shine{} for the cbuf+rrot Harris corner detection (\cref{rrot-strategy}).}, basicstyle=\ttfamily\tiny, breaklines=true, postbreak=\mbox{{$\hookrightarrow$}\space}]
struct Record_float4_float4 {
  float4 a;
  float4 b;
};

struct Record_float4__float4_float4_ {
  float4 a;
  struct Record_float4_float4 b;
};

__kernel
void harris(global float* restrict output, int n0, int n1, const global float* restrict x0,
            global float* restrict t1, global float* restrict t2, global float* restrict t3){
  for (int gid = get_global_id(0);(gid < (n0 / 32));gid = (gid + get_global_size(0))) {
    for (int i0 = 0;(i0 < 2);i0 = (1 + i0)) {
      for (int i1 = 0;(i1 < (n1 / 4));i1 = (1 + i1)) {
        float4 t4 = (float4)(0.0f);
        t4 = (t4 + ((float4)(0.299f) * vload4(0, &x0[4*i1 + 32*gid*n1 + i0*n1))]);
        t4 = (t4 + ((float4)(0.587f) * vload4(0, &x0[4*i1 + 4*n1 + 32*gid*n1 + i0*n1 + n0*n1])));
        t4 = (t4 + ((float4)(0.114f) * vload4(0, &x0[2*n0*n1 + 4*i1 + 8*n1 + 32*gid*n1 + i0*n1])));
        vstore4(t4, 0, &t3[3*n1*get_global_id(0) + 4*i0 + 4*i1 + 12*get_global_id(0) + i0*n1]);
      }
    }
    
    for (int i2 = 0;(i2 < 2);i2 = (1 + i2)) {
      for (int i3 = 0;(i3 < (n1 / 4));i3 = (1 + i3)) {
        float4 t4 = (float4)(0.0f);
        t4 = (t4 + ((float4)(0.299f) * vload4(0, &x0[2*n1 + 4*i3 + 32*gid*n1 + i2*n1])));
        t4 = (t4 + ((float4)(0.587f) * vload4(0, &x0[4*i3 + 6*n1 + 32*gid*n1 + i2*n1 + n0*n1])));
        t4 = (t4 + ((float4)(0.114f) * vload4(0, &x0[2*n0*n1 + 4*i3 + 10*n1 + 32*gid*n1 + i2*n1])));
        vstore4(t4, 0, &t3[3*n1*get_global_id(0) + 4*i3 + 4*((2 + i2) % 3) + 12*get_global_id(0) + n1*((2 + i2) % 3)]);
      }
      
      struct Record_float4_float4 t16[2];

      float4 t5 = (float4)(0.0f);
      t5 += (float4)(1.0f) * vload4(0, &t3[3*n1*get_global_id(0) + 4*i2 + 12*get_global_id(0) + i2*n1]);
      t5 += (float4)(2.0f) * vload4(0, &t3[4 + n1 + 3*n1*get_global_id(0) + 4*i2 + 12*get_global_id(0) + i2*n1]);
      t5 += (float4)(1.0f) * vload4(0, &t3[3*n1*get_global_id(0) + 4*((2+i2) % 3) + 12*get_global_id(0) + n1*((2+i2) % 3)]);
      t16[0].a = t5;
    
      float4 t8 = (float4)(0.0f);
      t8 += (float4)(-1.0f) * vload4(0, &t3[3*n1*get_global_id(0) + 4*i2 + 12*get_global_id(0) + i2*n1]);
      t8 += (float4)(0.0f) * vload4(0, &t3[4 + n1 + 3*n1*get_global_id(0) + 4*i2 + 12*get_global_id(0) + i2*n1]);
      t8 += (float4)(1.0f) * vload4(0, &t3[3*n1*get_global_id(0) + 4*((2+i2) % 3) + 12*get_global_id(0) + n1*((2+i2) % 3)]);
      t16[0].b = t8;
        
      for (int i4 = 0;(i4 < (n1 / 4));i4 = (1 + i4)) {
        float4 t9 = (float4)(0.0f);
        t9 += (float4)(1.0f) * vload4(0, &t3[4 + 3*n1*get_global_id(0) + 4*i2 + 4*i4 + 12*get_global_id(0) + i2*n1]);
        t9 += (float4)(2.0f) * vload4(0, &t3[8 + n1 + 3*n1 * get_global_id(0) + 4*i2 + 4*i4 + 12*get_global_id(0) + i2*n1]);
        t9 += (float4)(1.0f) * vload4(0, &t3[4 + 3*n1*get_global_id(0) + 4*i4 + 4*((2+i2) % 3) + 12*get_global_id(0) + n1*((2+i2) % 3)]);
        t16[1].a = t9;
    
        float4 t12 = (float4)(0.0f);
        t12 += (float4)(-1.0f) * vload4(0, &t3[4 + 3*n1*get_global_id(0) + 4*i2 + 4*i4 + 12*get_global_id(0) + i2*n1]);
        t12 += (float4)(0.0f) * vload4(0, &t3[8 + n1 + 3*n1*get_global_id(0) + 4*i2 + 4*i4 + 12*get_global_id(0) + i2*n1]);
        t12 += (float4)(1.0f) * vload4(0, &t3[4 + 3*n1*get_global_id(0) + 4*i4 + 4*((2+i2) % 3) + 12*get_global_id(0) + n1*((2+i2) % 3)]);
        t16[1].b = t12;
          
        float4 t13 = (float4)(0.0f);
        t13 = (t13 + ((float4)(-0.083333336f) * (float4)(t16[0].a.s0, t16[0].a.s1, t16[0].a.s2, t16[0].a.s3)));
        t13 = (t13 + ((float4)(0.0f) * (float4)(t16[0].a.s1, t16[0].a.s2, t16[0].a.s3, t16[1].a.s0)));
        t13 = (t13 + ((float4)(0.083333336f) * (float4)(t16[0].a.s2, t16[0].a.s3, t16[1].a.s0, t16[1].a.s1)));
        vstore4(t13, 0, &t2[3*n1*get_global_id(0) + 4*i2 + 4*i4 + 12*get_global_id(0) + i2*n1]);

        float4 t14 = (float4)(0.0f);
        t14 = (t14 + ((float4)(0.083333336f) * (float4)(t16[0].b.s0, t16[0].b.s1, t16[0].b.s2, t16[0].b.s3)));
        t14 = (t14 + ((float4)(0.16666667f) * (float4)(t16[0].b.s1, t16[0].b.s2, t16[0].b.s3, t16[1].b.s0)));
        t14 = (t14 + ((float4)(0.083333336f) * (float4)(t16[0].b.s2, t16[0].b.s3, t16[1].b.s0, t16[1].b.s1)));
        vstore4(t14, 0, &t1[3*n1*get_global_id(0) + 4*i2 + 4*i4 + 12*get_global_id(0) + i2*n1]);
          
        t16[0].a = t16[1].a;
        t16[0].b = t16[1].b;
      }
    }
    
    for (int i5 = 0;(i5 < 32);i5 = (1 + i5)) {
      for (int i6 = 0;(i6 < (n1 / 4));i6 = (1 + i6)) {
        float4 t15 = (float4)(0.0f);
        t15 = (t15 + ((float4)(0.299f) * vload4(0, &x0[4*i6 + 4*n1 + 32*gid*n1 + i5*n1])));
        t15 = (t15 + ((float4)(0.587f) * vload4(0, &x0[4*i6 + 8*n1 + 32*gid*n1 + i5*n1 + n0*n1])));
        t15 = (t15 + ((float4)(0.114f) * vload4(0, &x0[2*n0*n1 + 4*i6 + 12*n1 + 32*gid*n1 + i5*n1])));
        vstore4(t15, 0, &t3[3*n1*get_global_id(0) + 4*i6 + 4*((1 + i5) % 3) + 12*get_global_id(0) + n1*((1 + i5) % 3)]);
      }
      
      struct Record_float4_float4 t16[2];

      float4 t5 = (float4)(0.0f);
      t5 += (float4)(1.0f) * vload4(0, &t3[3*n1*get_global_id(0) + 4*((2+i5) % 3) + 12*get_global_id(0) + n1*((2+i5) % 3)]);
      t5 += (float4)(2.0f) * vload4(0, &t3[3*n1*get_global_id(0) + 4*(i5 % 3) + 12*get_global_id(0) + n1*(i5 % 3)]);
      t5 += (float4)(1.0f) * vload4(0, &t3[3*n1*get_global_id(0) + 4*((1+i5) % 3) + 12*get_global_id(0) + n1*((1+i5) % 3)]);
      t16[0].a = t5;
    
      float4 t18 = (float4)(0.0f);
      t18 += (float4)(-1.0f) * vload4(0, &t3[3*n1*get_global_id(0) + 4*((2+i5) % 3) + 12*get_global_id(0) + n1*((2+i5) % 3)]);
      t18 += (float4)(0.0f) * vload4(0, &t3[3*n1*get_global_id(0) + 4*(i5 % 3) + 12*get_global_id(0) + n1*(i5 % 3)]);
      t18 += (float4)(1.0f) * vload4(0, &t3[3*n1*get_global_id(0) + 4*((1+i5) % 3) + 12*get_global_id(0) + n1*((1+i5) % 3)]);
      t16[0].b = t18;
        
      for (int i7 = 0;(i7 < (n1 / 4));i7 = (1 + i7)) {
        float4 t9 = (float4)(0.0f);
        t9 += (float4)(1.0f) * vload4(0, &t3[4 + 3*n1*get_global_id(0) + 4*i7 + 4*((2 + i5) % 3) + 12*get_global_id(0) + n1*((2 + i5) % 3)]);
        t9 += (float4)(2.0f) * vload4(0, &t3[4 + 3*n1*get_global_id(0) + 4*i7 + 4*(i5 % 3) + 12*get_global_id(0) + n1*(i5 % 3)]);
        t9 += (float4)(1.0f) * vload4(0, &t3[4 + 3*n1*get_global_id(0) + 4*i7 + 4*((1 + i5) % 3) + 12*get_global_id(0) + n1*((1 + i5) % 3)]);
        t16[1].a = t9;
        
        float4 t20 = (float4)(0.0f);
        t20 = (t20 + ((float4)(-1.0f) * vload4(0, &t3[4 + 3*n1*get_global_id(0) + 4*i7 + 4*((2 + i5) % 3) + 12*get_global_id(0) + n1*((2 + i5) % 3)])));
        t20 = (t20 + ((float4)(0.0f) * vload4(0, &t3[4 + 3*n1*get_global_id(0) + 4*i7 + 4*(i5 % 3) + 12*get_global_id(0) + n1*(i5 % 3)])));
        t20 = (t20 + ((float4)(1.0f) * vload4(0, &t3[4 + 3*n1*get_global_id(0) + 4*i7 + 4*((1 + i5) % 3) + 12*get_global_id(0) + n1*((1 + i5) % 3)])));
        t16[1].b = t20;
        
        float4 t13 = (float4)(0.0f);
        t13 = (t13 + ((float4)(-0.083333336f) * (float4)(t16[0].a.s0, t16[0].a.s1, t16[0].a.s2, t16[0].a.s3)));
        t13 = (t13 + ((float4)(0.0f) * (float4)(t16[0].a.s1, t16[0].a.s2, t16[0].a.s3, t16[1].a.s0)));
        t13 = (t13 + ((float4)(0.083333336f) * (float4)(t16[0].a.s2, t16[0].a.s3, t16[1].a.s0, t16[1].a.s1)));
        vstore4(t13, 0, &t2[3*n1*get_global_id(0) + 4*i7 + 4*((2 + i5) % 3) + 12*get_global_id(0) + n1*((2 + i5) % 3)]);
          
        float4 t14 = (float4)(0.0f);
        t14 = (t14 + ((float4)(0.083333336f) * (float4)(t16[0].b.s0, t16[0].b.s1, t16[0].b.s2, t16[0].b.s3)));
        t14 = (t14 + ((float4)(0.16666667f) * (float4)(t16[0].b.s1, t16[0].b.s2, t16[0].b.s3, t16[1].b.s0)));
        t14 = (t14 + ((float4)(0.083333336f) * (float4)(t16[0].b.s2, t16[0].b.s3, t16[1].b.s0, t16[1].b.s1)));
        vstore4(t14, 0, &t1[3*n1*get_global_id(0) + 4*i7 + 4*((2 + i5) % 3) + 12*get_global_id(0) + n1*((2 + i5) % 3)]);
          
        t16[0].a = t16[1].a;
        t16[0].b = t16[1].b;
      }
      
      struct Record_float4__float4_float4_ t21[2];
        
      float4 t33 = (float4)(0.0f);
      t33 = (t33 + (vload4(0, &t2[3*n1*get_global_id(0) + 4*(i5 % 3) + 12*get_global_id(0) + n1*(i5 % 3)]) * vload4(0, &t2[3*n1*get_global_id(0) + 4*(i5 % 3) + 12*get_global_id(0) + n1*(i5 % 3)]);
      t33 = (t33 + (vload4(0, &t2[3*n1*get_global_id(0) + 4*((1 + i5) % 3) + 12*get_global_id(0) + n1*((1 + i5) % 3)]) * vload4(0, &t2[3*n1*get_global_id(0) + 4*((1 + i5) % 3) + 12*get_global_id(0) + n1*((1 + i5) % 3)])));
      t33 = (t33 + (vload4(0, &t2[3*n1*get_global_id(0) + 4*((2 + i5) % 3) + 12*get_global_id(0) + n1*((2 + i5) % 3)]) * vload4(0, &t2[3*n1*get_global_id(0) + 4*((2 + i5) % 3) + 12*get_global_id(0) + n1*((2 + i5) % 3)])));
      t21[0].a = t33;
        
      float4 t22 = (float4)(0.0f);
      t22 = (t22 + (vload4(0, &t2[3*n1*get_global_id(0) + 4*(i5 % 3) + 12*get_global_id(0) + n1*(i5 % 3)]) * vload4(0, &t1[3*n1*get_global_id(0) + 4*(i5 % 3) + 12*get_global_id(0) + n1*(i5 % 3)])));
      t22 = (t22 + (vload4(0, &t2[3*n1*get_global_id(0) + 4*((1 + i5) % 3) + 12*get_global_id(0) + n1*((1 + i5) % 3)]) * vload4(0, &t1[3*n1*get_global_id(0) + 4*((1 + i5) % 3) + 12*get_global_id(0) + n1*((1 + i5) % 3)])));
      t22 = (t22 + (vload4(0, &t2[3*n1*get_global_id(0) + 4*((2 + i5) % 3) + 12*get_global_id(0) + n1*((2 + i5) % 3)]) * vload4(0, &t1[3*n1*get_global_id(0) + 4*((2 + i5) % 3) + 12*get_global_id(0) + n1*((2 + i5) % 3)])));
      t21[0].b.a = t22;
    
      float4 t23 = (float4)(0.0f);
      t23 = (t23 + (vload4(0, &t1[3*n1*get_global_id(0) + 4*(i5 % 3) + 12*get_global_id(0) + n1*(i5 % 3)]) * vload4(0, &t1[3*n1*get_global_id(0) + 4*(i5 % 3) + 12*get_global_id(0) + n1*(i5 % 3)])));
      t23 = (t23 + (vload4(0, &t1[3*n1*get_global_id(0) + 4*((1 + i5) % 3) + 12*get_global_id(0) + n1*((1 + i5) % 3)]) * vload4(0, &t1[3*n1*get_global_id(0) + 4*((1 + i5) % 3) + 12*get_global_id(0) + n1*((1 + i5) % 3)])));
      t23 = (t23 + (vload4(0, &t1[3*n1*get_global_id(0) + 4*((2 + i5) % 3) + 12*get_global_id(0) + n1*((2 + i5) % 3)]) * vload4(0, &t1[3*n1*get_global_id(0) + 4*((2 + i5) % 3) + 12*get_global_id(0) + n1*((2 + i5) % 3)])));
      t21[0].b.b = t23;
        
      for (int i8 = 0;(i8 < (n1 / 4));i8 = (1 + i8)) {
        float4 t24 = (float4)(0.0f);
        t24 += vload4(0, &t2[4 + 3*n1*get_global_id(0) + 4*i8 + 4*(i5 % 3) + 12*get_global_id(0) + n1*(i5 % 3)]) * vload4(0, &t2[4 + 3*n1*get_global_id(0) + 4*i8 + 4*(i5 % 3) + 12*get_global_id(0) + n1*(i5 % 3)]);
        t24 += vload4(0, &t2[4 + 3*n1*get_global_id(0) + 4*i8 + 4*((1+i5) % 3) + 12*get_global_id(0) + n1*((1+i5) % 3)]) * vload4(0, &t2[4 + 3*n1*get_global_id(0) + 4*i8 + 4*((1+i5) % 3) + 12*get_global_id(0) + n1*((1+i5) % 3)]);
        t24 += vload4(0, &t2[4 + 3*n1*get_global_id(0) + 4*i8 + 4*((2+i5) % 3) + 12*get_global_id(0) + n1*((2+i5) % 3)]) * vload4(0, &t2[4 + 3*n1*get_global_id(0) + 4*i8 + 4*((2+i5) % 3) + 12*get_global_id(0) + n1*((2+i5) % 3)]);
        t21[1].a = t24;
          
        float4 t25 = (float4)(0.0f);
        t25 += vload4(0, &t2[4 + 3*n1*get_global_id(0) + 4*i8 + 4*(i5 % 3) + 12*get_global_id(0) + n1*(i5 % 3)]) * vload4(0, &t1[4 + 3*n1*get_global_id(0) + 4*i8 + 4*(i5 % 3) + 12*get_global_id(0) + n1*(i5 % 3)]);
        t25 += vload4(0, &t2[4 + 3*n1*get_global_id(0) + 4*i8 + 4*((1+i5) % 3) + 12*get_global_id(0) + n1*((1+i5) % 3)]) * vload4(0, &t1[4 + 3*n1*get_global_id(0) + 4*i8 + 4*((1+i5) % 3) + 12*get_global_id(0) + n1*((1+i5) % 3)]);
        t25 += vload4(0, &t2[4 + 3*n1*get_global_id(0) + 4*i8 + 4*((2+i5) % 3) + 12*get_global_id(0) + n1*((2+i5) % 3)]) * vload4(0, &t1[4 + 3*n1*get_global_id(0) + 4*i8 + 4*((2+i5) % 3) + 12*get_global_id(0) + n1*((2+i5) % 3)]);
        t21[1].b.a = t25;
          
        float4 t26 = (float4)(0.0f);
        t26 += vload4(0, &t1[4 + 3*n1*get_global_id(0) + 4*i8 + 4*(i5 % 3) + 12*get_global_id(0) + n1*(i5 % 3)]) * vload4(0, &t1[4 + 3*n1*get_global_id(0) + 4*i8 + 4*(i5 % 3) + 12*get_global_id(0) + n1*(i5 % 3)]);
        t26 += vload4(0, &t1[4 + 3*n1*get_global_id(0) + 4*i8 + 4*((1+i5) % 3) + 12*get_global_id(0) + n1*((1+i5) % 3)]) * vload4(0, &t1[4 + 3*n1*get_global_id(0) + 4*i8 + 4*((1+i5) % 3) + 12*get_global_id(0) + n1*((1+i5) % 3)]);
        t26 += vload4(0, &t1[4 + 3*n1*get_global_id(0) + 4*i8 + 4*((2+i5) % 3) + 12*get_global_id(0) + n1*((2+i5) % 3)]) * vload4(0, &t1[4 + 3*n1*get_global_id(0) + 4*i8 + 4*((2+i5) % 3) + 12*get_global_id(0) + n1*((2+i5) % 3)]);
        t21[1].b.b = t26;
          
        float4 t27 = (float4)(0.0f);
        t27 = (t27 + (float4)(t21[0].b.a.s0, t21[0].b.a.s1, t21[0].b.a.s2, t21[0].b.a.s3));
        t27 = (t27 + (float4)(t21[0].b.a.s1, t21[0].b.a.s2, t21[0].b.a.s3, t21[1].b.a.s0));
        t27 = (t27 + (float4)(t21[0].b.a.s2, t21[0].b.a.s3, t21[1].b.a.s0, t21[1].b.a.s1));
          
        float4 t29 = (float4)(0.0f);
        t29 = (t29 + (float4)(t21[0].b.b.s0, t21[0].b.b.s1, t21[0].b.b.s2, t21[0].b.b.s3));
        t29 = (t29 + (float4)(t21[0].b.b.s1, t21[0].b.b.s2, t21[0].b.b.s3, t21[1].b.b.s0));
        t29 = (t29 + (float4)(t21[0].b.b.s2, t21[0].b.b.s3, t21[1].b.b.s0, t21[1].b.b.s1));
          
        float4 t31 = (float4)(0.0f);
        t31 = (t31 + (float4)(t21[0].a.s0, t21[0].a.s1, t21[0].a.s2, t21[0].a.s3));
        t31 = (t31 + (float4)(t21[0].a.s1, t21[0].a.s2, t21[0].a.s3, t21[1].a.s0));
        t31 = (t31 + (float4)(t21[0].a.s2, t21[0].a.s3, t21[1].a.s0, t21[1].a.s1));
              
        vstore4(t31*t29 - t27*t27 - (float4)(0.04f) * (t31 + t29) * (t31 + t29), 0, &output[4*i8 + 32*gid*n1 + i5*n1]);
        
        t21[0].a = t21[1].a;
        t21[0].b.a = t21[1].b.a;
        t21[0].b.b = t21[1].b.b;
      }
    }
  }
}
\end{opencl}
\chapter{Optimised Matrix Multiplication}
\label{app:matmul-opt-progs}

This appendix contains the C programs generated by \Shine{} after rewriting the \Rise{} program for matrix multiplication using the 7 \Elevate{} strategies from \cite{hagedorn2020-elevate}.
We apply trivial cosmetic changes to keep the code readable and compact, as in \cref{app:harris-opt-progs}.

The same programs are generated using sketch-guided equality saturation in \cref{mm-eval}, modulo variable names.
The handwritten sketches from \cref{fig:sketch-logical} of \cref{mm-eval} are included in \cref{app:sketches}.

\bigskip

\noindent
\begin{minipage}{\linewidth}
\begin{c-code}[caption={C code generated by \Shine{} for the \baseline{} mat-mul in  \cref{mm-eval}.}, basicstyle=\ttfamily\scriptsize, breaklines=true, postbreak=\mbox{{$\hookrightarrow$}\space}]
void baseline(float* output, float* x0, float* x1) {
  for (int i0 = 0;(i0 < 1024);i0 = (1 + i0)) {
    for (int i1 = 0;(i1 < 1024);i1 = (1 + i1)) {
      float t1 = 0.0f;
      for (int i2 = 0;(i2 < 1024);i2 = (1 + i2)) {
        t1 += x0[(i2 + (1024 * i0))] *
              x1[(i1 + (1024 * i2))];
      }
    
      output[(i1 + (1024 * i0))] = t1;
    }
  }
}
\end{c-code}
\end{minipage}

\noindent
\begin{minipage}{\linewidth}
\begin{c-code}[caption={C code generated by \Shine{} for the \blocking{} mat-mul in  \cref{mm-eval}.}, basicstyle=\ttfamily\scriptsize, breaklines=true, postbreak=\mbox{{$\hookrightarrow$}\space}]
void blocking(float* output, float* x0, float* x1) {
  for (int i0 = 0;(i0 < 32);i0 = (1 + i0)) {
    for (int i1 = 0;(i1 < 32);i1 = (1 + i1)) {
      float t1[1024];
      for (int i2 = 0;(i2 < 32);i2 = (1 + i2)) {
        for (int i3 = 0;(i3 < 32);i3 = (1 + i3)) {
          t1[(i3 + (32 * i2))] = 0.0f;
        }
      }
        
      for (int i4 = 0;(i4 < 256);i4 = (1 + i4)) {
        float t2[1024];
        for (int i5 = 0;(i5 < 32);i5 = (1 + i5)) {
          for (int i6 = 0;(i6 < 32);i6 = (1 + i6)) {
            t2[(i6 + (32 * i5))] = t1[(i6 + (32 * i5))];
          }
        }
            
        for (int i7 = 0;(i7 < 4);i7 = (1 + i7)) {
          for (int i8 = 0;(i8 < 32);i8 = (1 + i8)) {
            for (int i9 = 0;(i9 < 32);i9 = (1 + i9)) {
              t2[i9 + 32*i8] += x0[i7 + 4*i4 + 1024*i8 + 32768*i0] *
                                x1[i9 + 32*i1 + 1024*i7 + 4096*i4];
            }
          }
        }
            
        for (int i10 = 0;(i10 < 32);i10 = (1 + i10)) {
          for (int i11 = 0;(i11 < 32);i11 = (1 + i11)) {
            t1[(i11 + (32 * i10))] = t2[(i11 + (32 * i10))];
          }
        }
      }
        
      for (int i12 = 0;(i12 < 32);i12 = (1 + i12)) {
        for (int i13 = 0;(i13 < 32);i13 = (1 + i13)) {
          output[(((i13 + (32 * i1)) + (1024 * i12)) + (32768 * i0))] = t1[(i13 + (32 * i12))];
        }
      }
    }
  }
}
\end{c-code}
\end{minipage}

\noindent
\begin{minipage}{\linewidth}
\begin{c-code}[caption={C code generated by \Shine{} for the \vectorisation{} mat-mul in  \cref{mm-eval}.}, basicstyle=\ttfamily\scriptsize, breaklines=true, postbreak=\mbox{{$\hookrightarrow$}\space}]
void vectorization(float* output, float* x0, float* x1) {
  for (int i0 = 0;(i0 < 32);i0 = (1 + i0)) {
    for (int i1 = 0;(i1 < 32);i1 = (1 + i1)) {
      float t1[1024];
      for (int i2 = 0;(i2 < 32);i2 = (1 + i2)) {
        for (int i3 = 0;(i3 < 32);i3 = (1 + i3)) {
          t1[(i3 + (32 * i2))] = 0.0f;
        }    
      }
        
      for (int i4 = 0;(i4 < 256);i4 = (1 + i4)) {
        float t2[1024];
        for (int i5 = 0;(i5 < 32);i5 = (1 + i5)) {
          for (int i6 = 0;(i6 < 32);i6 = (1 + i6)) {
            t2[(i6 + (32 * i5))] = t1[(i6 + (32 * i5))];
          }      
        }
            
        for (int i7 = 0;(i7 < 4);i7 = (1 + i7)) {
          for (int i8 = 0;(i8 < 32);i8 = (1 + i8)) {
            #pragma omp simd
            for (int i9 = 0;(i9 < 32);i9 = (1 + i9)) {
              t2[i9 + 32*i8] += x0[i7 + 4*i4 + 1024*i8 + 32768*i0] *
                                x1[i9 + 32*i1 + 1024*i7 + 4096*i4];
            }
          }
        }
            
        for (int i10 = 0;(i10 < 32);i10 = (1 + i10)) {
          for (int i11 = 0;(i11 < 32);i11 = (1 + i11)) {
            t1[(i11 + (32 * i10))] = t2[(i11 + (32 * i10))];
          }
        }
      }
        
      for (int i12 = 0;(i12 < 32);i12 = (1 + i12)) {
        for (int i13 = 0;(i13 < 32);i13 = (1 + i13)) {
          output[(((i13 + (32 * i1)) + (1024 * i12)) + (32768 * i0))] = t1[(i13 + (32 * i12))];
        }
      }
    }
  }
}
\end{c-code}
\end{minipage}

\noindent
\begin{minipage}{\linewidth}
\begin{c-code}[caption={C code generated by \Shine{} for the \loopperm{} mat-mul in  \cref{mm-eval}.}, basicstyle=\ttfamily\scriptsize, breaklines=true, postbreak=\mbox{{$\hookrightarrow$}\space}]
void loop_permutation(float* output, float* x0, float* x1) {
  for (int i0 = 0;(i0 < 32);i0 = (1 + i0)) {
    for (int i1 = 0;(i1 < 32);i1 = (1 + i1)) {
      float t1[1024];
      for (int i2 = 0;(i2 < 32);i2 = (1 + i2)) {
        for (int i3 = 0;(i3 < 32);i3 = (1 + i3)) {
          t1[(i3 + (32 * i2))] = 0.0f;
        }    
      }
          
      for (int i4 = 0;(i4 < 256);i4 = (1 + i4)) {
        for (int i5 = 0;(i5 < 32);i5 = (1 + i5)) {
          float t2[32];
          for (int i6 = 0;(i6 < 32);i6 = (1 + i6)) {
            t2[i6] = t1[(i6 + (32 * i5))];
          }
                
          for (int i7 = 0;(i7 < 4);i7 = (1 + i7)) {
            #pragma omp simd
            for (int i8 = 0;(i8 < 32);i8 = (1 + i8)) {
              t2[i8] += x0[i7 + 4*i4 + 1024*i5 + 32768*i0] *
                        x1[i8 + 32*i1 + 1024*i7 + 4096*i4];
            }      
          }
                
          for (int i9 = 0;(i9 < 32);i9 = (1 + i9)) {
            t1[(i9 + (32 * i5))] = t2[i9];
          }        
        }      
      }
          
      for (int i10 = 0;(i10 < 32);i10 = (1 + i10)) {
        for (int i11 = 0;(i11 < 32);i11 = (1 + i11)) {
          output[(((i11 + (32 * i1)) + (1024 * i10)) + (32768 * i0))] = t1[(i11 + (32 * i10))];
        }    
      }
    }
  }   
}  
\end{c-code}
\end{minipage}

\noindent
\begin{minipage}{\linewidth}
\begin{c-code}[caption={C code generated by \Shine{} for the \arraypacking{} mat-mul in  \cref{mm-eval}.}, basicstyle=\ttfamily\scriptsize, breaklines=true, postbreak=\mbox{{$\hookrightarrow$}\space}]
void array_packing(float* output, float* x0, float* x1) {
  float t1[1048576];
  #pragma omp parallel for
  for (int i0 = 0;(i0 < 32);i0 = (1 + i0)) {
    for (int i1 = 0;(i1 < 1024);i1 = (1 + i1)) {
      #pragma omp simd
      for (int i2 = 0;(i2 < 32);i2 = (1 + i2)) {
        t1[((i2 + (32 * i1)) + (32768 * i0))] = x1[((i2 + (32 * i0)) + (1024 * i1))];
      }
    }
  }
      
  for (int i3 = 0;(i3 < 32);i3 = (1 + i3)) {
    for (int i4 = 0;(i4 < 32);i4 = (1 + i4)) {
      float t2[1024];
      for (int i5 = 0;(i5 < 32);i5 = (1 + i5)) {
        for (int i6 = 0;(i6 < 32);i6 = (1 + i6)) {
          t2[(i6 + (32 * i5))] = 0.0f;
        }      
      }
            
      for (int i7 = 0;(i7 < 256);i7 = (1 + i7)) {
        for (int i8 = 0;(i8 < 32);i8 = (1 + i8)) {
          float t3[32];
          for (int i9 = 0;(i9 < 32);i9 = (1 + i9)) {
            t3[i9] = t2[(i9 + (32 * i8))];
          }
                  
          for (int i10 = 0;(i10 < 4);i10 = (1 + i10)) {
            #pragma omp simd
            for (int i11 = 0;(i11 < 32);i11 = (1 + i11)) {
              t3[i11] += x0[i10 + 4*i7 + 1024*i8 + 32768*i3] *
                         t1[i11 + 32*i10 + 128*i7 + 32768*i4];
            }
          }
                  
          for (int i12 = 0;(i12 < 32);i12 = (1 + i12)) {
            t2[(i12 + (32 * i8))] = t3[i12];
          }
        }
      }
            
      for (int i13 = 0;(i13 < 32);i13 = (1 + i13)) {
        for (int i14 = 0;(i14 < 32);i14 = (1 + i14)) {
          output[(((i14 + (32 * i4)) + (1024 * i13)) + (32768 * i3))] = t2[(i14 + (32 * i13))];
        }
      }
    }
  }
}
\end{c-code}
\end{minipage}

\noindent
\begin{minipage}{\linewidth}%
\begin{c-code}[caption={C code generated by \Shine{} for the \cacheblocks{} mat-mul in  \cref{mm-eval}.}, basicstyle=\ttfamily\scriptsize, breaklines=true, postbreak=\mbox{{$\hookrightarrow$}\space}]
void cache_blocks(float* output, float* x0, float* x1) {
  float t1[1048576];
  #pragma omp parallel for
  for (int i0 = 0;(i0 < 32);i0 = (1 + i0)) {
    for (int i1 = 0;(i1 < 1024);i1 = (1 + i1)) {
      #pragma omp simd
      for (int i2 = 0;(i2 < 32);i2 = (1 + i2)) {
        t1[((i2 + (32 * i1)) + (32768 * i0))] = x1[((i2 + (32 * i0)) + (1024 * i1))];
      }
    }
  }
  for (int i3 = 0;(i3 < 32);i3 = (1 + i3)) {
    for (int i4 = 0;(i4 < 32);i4 = (1 + i4)) {
      float t2[1024];
      for (int i5 = 0;(i5 < 32);i5 = (1 + i5)) {
        for (int i6 = 0;(i6 < 32);i6 = (1 + i6)) {
          t2[(i6 + (32 * i5))] = 0.0f;
        }
      }
      for (int i7 = 0;(i7 < 256);i7 = (1 + i7)) {
        for (int i8 = 0;(i8 < 32);i8 = (1 + i8)) {
          float t3[32];
          for (int i9 = 0;(i9 < 32);i9 = (1 + i9)) {
            t3[i9] = t2[(i9 + (32 * i8))];
          }
                  
          #pragma omp simd
          for (int i11 = 0;(i11 < 32);i11 = (1 + i11)) {
            t3[i11] += x0[4*i7 + 1024*i8 + 32768*i3] * t1[i11 + 128*i7 + 32768*i4];
          }
          #pragma omp simd
          for (int i11 = 0;(i11 < 32);i11 = (1 + i11)) {
            t3[i11] += x0[1 + 4*i7 + 1024*i8 + 32768*i3] * t1[32 + i11 + 128*i7 + 32768*i4];
          }
          #pragma omp simd
          for (int i11 = 0;(i11 < 32);i11 = (1 + i11)) {
            t3[i11] += x0[2 + 4*i7 + 1024*i8 + 32768*i3] * t1[64 + i11 + 128*i7 + 32768*i4];
          }
          #pragma omp simd
          for (int i11 = 0;(i11 < 32);i11 = (1 + i11)) {
            t3[i11] += x0[3 + 4*i7 + 1024*i8 + 32768*i3] * t1[96 + i11 + 128*i7 + 32768*i4];
          }
                  
          for (int i12 = 0;(i12 < 32);i12 = (1 + i12)) {
            t2[(i12 + (32 * i8))] = t3[i12];
          }
        }
      }
      for (int i13 = 0;(i13 < 32);i13 = (1 + i13)) {
        for (int i14 = 0;(i14 < 32);i14 = (1 + i14)) {
          output[(((i14 + (32 * i4)) + (1024 * i13)) + (32768 * i3))] = t2[(i14 + (32 * i13))];
        }
      }
} } }
\end{c-code}%
\end{minipage}%

\noindent
\begin{minipage}{\linewidth}%
\begin{c-code}[caption={C code generated by \Shine{} for the \parallel{} mat-mul in  \cref{mm-eval}.}, basicstyle=\ttfamily\scriptsize, breaklines=true, postbreak=\mbox{{$\hookrightarrow$}\space}]
void parallel(float* output, float* x0, float* x1) {
  float t1[1048576];
  #pragma omp parallel for
  for (int i0 = 0;(i0 < 32);i0 = (1 + i0)) {
    for (int i1 = 0;(i1 < 1024);i1 = (1 + i1)) {
      #pragma omp simd
      for (int i3 = 0;(i3 < 32);i3 = (1 + i3)) {
        t1[((i3 + (32 * i1)) + (32768 * i0))] = x1[((i3 + (32 * i0)) + (1024 * i1))];
      }   
    }  
  }
  #pragma omp parallel for
  for (int i4 = 0;(i4 < 32);i4 = (1 + i4)) {
    for (int i5 = 0;(i5 < 32);i5 = (1 + i5)) {
      float t2[1024];
      for (int i6 = 0;(i6 < 32);i6 = (1 + i6)) {
        for (int i_50146 = 0;(i_50146 < 32);i_50146 = (1 + i_50146)) {
          t2[(i_50146 + (32 * i6))] = 0.0f;
        }
      }
      for (int i7 = 0;(i7 < 256);i7 = (1 + i7)) {
        for (int i8 = 0;(i8 < 32);i8 = (1 + i8)) {
          float t3[32];
          for (int i9 = 0;(i9 < 32);i9 = (1 + i9)) {
            t3[i9] = t2[(i9 + (32 * i8))];
          }
                
          #pragma omp simd
          for (int i11 = 0;(i11 < 32);i11 = (1 + i11)) {
            t3[i11] += x0[4*i7 + 1024*i8 + 32768*i4] * t1[i11 + 128*i7 + 32768*i5];
          }
          #pragma omp simd
          for (int i11 = 0;(i11 < 32);i11 = (1 + i11)) {
            t3[i11] += x0[1 + 4*i7 + 1024*i8 + 32768*i4] * t1[32 + i11 + 128*i7 + 32768*i5];
          }  
          #pragma omp simd
          for (int i11 = 0;(i11 < 32);i11 = (1 + i11)) {
            t3[i11] += x0[2 + 4*i7 + 1024*i8 + 32768*i4] * t1[64 + i11 + 128*i7 + 32768*i5];
          }
          #pragma omp simd
          for (int i11 = 0;(i11 < 32);i11 = (1 + i11)) {
            t3[i11] += x0[3 + 4*i7 + 1024*i8 + 32768*i4] * t1[96 + i11 + 128*i7 + 32768*i5];
          }
                
          for (int i12 = 0;(i12 < 32);i12 = (1 + i12)) {
            t2[(i12 + (32 * i8))] = t3[i12];
          }
        }
    , }
      for (int i13 = 0;(i13 < 32);i13 = (1 + i13)) {
        for (int i14 = 0;(i14 < 32);i14 = (1 + i14)) {
          output[(((i14 + (32 * i5)) + (1024 * i13)) + (32768 * i4))] = t2[(i14 + (32 * i13))];
        }
      }
} } }
\end{c-code}%
\end{minipage}%
\chapter{Handwritten Sketches and Selection of Discovered Programs}
\label{app:sketches}

This appendix contains the handwritten sketches from \cref{fig:sketch-logical} of \cref{mm-eval}, used to guide matrix multiplication optimisation (\cref{sec:matmul-opt-sketches}).
It also contains the \Rise{} programs found by sketch-guided equality saturation for the \parallel{} optimisation goal in  \cref{mm-eval} (\cref{sec:matmul-opt-parallel-guided}).

\section{Matrix Multiplication Sketches}
\label{sec:matmul-opt-sketches}

The handwritten sketches from \cref{fig:sketch-logical} of \cref{mm-eval}.

~\\
\noindent
\begin{minipage}{\linewidth}%
\begin{rise-sketch}[caption={A sketch for the \goalStyle{baseline} goal (\cref{mm-baseline-sketch}).}, xleftmargin=0.15\linewidth, xrightmargin=0.15\linewidth]
containsMap(m,
 containsMap(n,
  containsReduceSeq(k,
   containsAddMul)))
\end{rise-sketch}%
\end{minipage}%

\noindent
\begin{minipage}{\linewidth}%
\begin{rise-sketch}[caption={\goalStyle{split} sketch specifying how to split loops for all 7 goals (\cref{mm-blocking-split-sketch}).}, xleftmargin=0.15\linewidth, xrightmargin=0.15\linewidth]
containsMap(m / 32,
 containsMap(32,
  containsMap(n / 32,
   containsMap(32,
    containsReduceSeq(k / 4,
     containsReduceSeq(4,
      containsAddMul))))))
\end{rise-sketch}
\end{minipage}

\noindent
\begin{minipage}{\linewidth}%
\begin{rise-sketch}[caption={\goalStyle{reorder$_1$} sketch for the \goalStyle{blocking} and \vectorisation{} goals (\cref{mm-blocking-sketch}).}, xleftmargin=0.15\linewidth, xrightmargin=0.15\linewidth]
containsMap(m / 32,
 containsMap(n / 32,
  containsReduceSeq(k / 4,
   containsReduceSeq(4,
    containsMap(32,
     containsMap(32,
      containsAddMul))))))
\end{rise-sketch}
\end{minipage}

\noindent
\begin{minipage}{\linewidth}%
\begin{rise-sketch}[caption={\goalStyle{reorder$_2$} sketch for the \loopperm{}, \arraypacking{}, \cacheblocks{}, and \parallel{} goals.}, xleftmargin=0.15\linewidth, xrightmargin=0.15\linewidth]
containsMap(m / 32,
 containsMap(n / 32,
  containsReduceSeq(k / 4,
   containsMap(32,
    containsReduceSeq(4,
     containsMap(32,
      containsAddMul))))))
\end{rise-sketch}
\end{minipage}

\noindent
\begin{minipage}{\linewidth}%
\begin{rise-sketch}[caption={\goalStyle{lower$_1$} sketch goal for the \vectorisation{} goal.}, xleftmargin=0.15\linewidth, xrightmargin=0.15\linewidth]
containsMap(m / 32,
 containsMap(n / 32,
  containsReduceSeq(k / 4,
   containsReduceSeq(4,
    containsMap(32,
     containsMap(1,
      containsAddMulVec))))))
\end{rise-sketch}
\end{minipage}

\noindent
\begin{minipage}{\linewidth}%
\begin{rise-sketch}[caption={\goalStyle{lower$_2$} sketch goal for the \loopperm{} goal.}, xleftmargin=0.15\linewidth, xrightmargin=0.15\linewidth]
containsMap(m / 32,
 containsMap(n / 32,
  containsReduceSeq(k / 4,
   containsMap(32,
    containsReduceSeq(4,
     containsMap(1,
      containsAddMulVec))))))
\end{rise-sketch}
\end{minipage}

\noindent
\begin{minipage}{\linewidth}%
\begin{rise-sketch}[caption={\goalStyle{store} sketch for the \arraypacking{}, \cacheblocks{}, and \parallel{} goals.}, xleftmargin=0.15\linewidth, xrightmargin=0.15\linewidth]
containsMap(m / 32,
 containsMap(n / 32,
  containsReduceSeq(k / 4,
   containsMap(32,
    containsReduceSeq(4,
     containsMap(32,
      containsAddMul))))),
 containsToMem(n.k.f32,
  containsMap(n / 32,
   containsMap(k,
    containsMap(32.f32, ?)))))
\end{rise-sketch}
\end{minipage}

\noindent
\begin{minipage}{\linewidth}%
\begin{rise-sketch}[caption={\goalStyle{lower$_3$} sketch goal for the \arraypacking{} goal.}, xleftmargin=0.15\linewidth, xrightmargin=0.15\linewidth]
containsMap(m / 32,
 containsMap(n / 32,
  containsReduceSeq(k / 4,
   containsMap(32,
    containsReduceSeq(4,
     containsMap(1,
      containsAddMulVec))))),
 containsToMem(n.k.f32,
  containsMap(n / 32,
   containsMap(k,
    containsMap(1.<32>f32, ?)))))
\end{rise-sketch}
\end{minipage}

\noindent
\begin{minipage}{\linewidth}%
\begin{rise-sketch}[caption={\goalStyle{lower$_4$} sketch goal for the \cacheblocks{} goal.}, xleftmargin=0.15\linewidth, xrightmargin=0.15\linewidth]
containsMap(m / 32,
 containsMap(n / 32,
  containsReduceSeq(k / 4,
   containsMap(32,
    containsReduceSeqUnroll(4,
     containsMap(1,
      containsAddMulVec))))),
 containsToMem(n.k.f32,
  containsMapPar(n / 32,
   containsMap(k,
    containsMap(1.<32>f32, ?)))))
\end{rise-sketch}
\end{minipage}

\noindent
\begin{minipage}{\linewidth}%
\begin{rise-sketch}[caption={\goalStyle{lower$_5$} sketch goal for the \parallel{} goal.}, xleftmargin=0.15\linewidth, xrightmargin=0.15\linewidth]
containsMapPar(m / 32,
 containsMap(n / 32,
  containsReduceSeq(k / 4,
   containsMap(32,
    containsReduceSeqUnroll(4,
     containsMap(1,
      containsAddMulVec))))),
 containsToMem(n.k.f32,
  containsMapPar(n / 32,
   containsMap(k,
    containsMap(1.<32>f32, ?)))))
\end{rise-sketch}
\end{minipage}

\section{\Rise{} Programs for the \parallel{} Matrix Multiplication}
\label{sec:matmul-opt-parallel-guided}

This section shows the \Rise{} programs that are found using sketch-guided equality saturation for the \parallel{} matrix multiplication optimisation goal in \cref{mm-eval}.

The initial program is shown in \cref{lst:mm-init}.
The intermediate programs are shown in \cref{lst:mm-after-split,lst:mm-after-reorder2,lst:mm-after-store}, and each satisfy a corresponding sketch guide.
The final program satisfying the sketch goal is shown in \cref{lst:mm-after-lower4}.

Additionally, a final automatic transformation is applied to obtain a valid low-level program that can be translated through DPIA, i.e. implying valid read-write annotations (\cref{sec:codegen}).
Sequential loops and memory copies are inserted where required, and let expressions are hoisted as much as possible, resulting in \cref{lst:mm-parallel}.

~\\
\noindent
\begin{minipage}{\linewidth}%
\begin{rise}[caption={Initial \Rise{} program for matrix multiplication.}, label={lst:mm-init}]
$\Lambda$n0:nat. $\Lambda$n1:nat. $\Lambda$n2:nat. $\lambda$x0. $\lambda$x1.
map ($\lambda$x2.
 map ($\lambda$x3.
  reduce add 0 (map ($\lambda$x4. (mul (fst x4) (snd x4))) (zip x2 x3)))
  (transpose x1))
 x0
\end{rise}
\end{minipage}

\noindent
\begin{minipage}{\linewidth}%
\begin{rise}[caption={\Rise{} program satisfying the \goalStyle{split} sketch guide.}, label={lst:mm-after-split}]
$\Lambda$n0:nat. $\Lambda$n1:nat. $\Lambda$n2:nat. $\lambda$x0. $\lambda$x1.
join (
 map (
  map ($\lambda$x2.
   join (
    map (
     map ($\lambda$x3.
      reduceSeq ($\lambda$x4. $\lambda$x5.
       add x4 (reduceSeq ($\lambda$x6. $\lambda$x7. add x6 (mul (fst x7) (snd x7))) 0 x5))
       0
       (split 4 (zip x2 x3))))
     (split 32 (transpose x1)))))
 (split 32 x0))
\end{rise}
\end{minipage}

\noindent
\begin{minipage}{\linewidth}%
\begin{rise}[caption={\Rise{} program satisfying the \goalStyle{reorder$_2$} sketch guide.}, label={lst:mm-after-reorder2}]
$\Lambda$n0:nat. $\Lambda$n1:nat. $\Lambda$n2:nat. $\lambda$x0. $\lambda$x1.
join (map (map join) (map transpose (
 map (
  map ($\lambda$x2.
   reduceSeq ($\lambda$x3. $\lambda$x4.
    map ($\lambda$x5.
     ($\lambda$x6. reduceSeq ($\lambda$x7. $\lambda$x8.
      map ($\lambda$x9. add (fst x9) (mul (fst (snd x9)) (snd (snd x9))))
       (zip x7 x8))
      (fst x6)
      (transpose (snd x6)))
     (unzip (zip (fst x5) (snd x5))))
     (zip x3 x4))
    (generate ($\lambda$x3. generate ($\lambda$x4. 0)))
    (transpose x2)))
 (map transpose (map (map ($\lambda$x2.
  map transpose (map (map ($\lambda$x3.
   split 4 (zip x2 x3))) (split 32 (transpose x1)))))
  (split 32 x0))))))
\end{rise}
\end{minipage}

\noindent
\begin{minipage}{\linewidth}%
\begin{rise}[caption={\Rise{} program satisfying the \goalStyle{store} sketch guide.}, label={lst:mm-after-store}]
$\Lambda$n0:nat. $\Lambda$n1:nat. $\Lambda$n2:nat. $\lambda$x0. $\lambda$x1.
join (map (map join) (map transpose (
 map (
  map ($\lambda$x2.
   reduceSeq ($\lambda$x3. $\lambda$x4.
    map ($\lambda$x5.
     reduceSeq ($\lambda$x6. $\lambda$x7.
      map ($\lambda$x8.
       add (fst x8) (mul (fst (snd x8)) (snd (snd x8))))
       (zip x6 x7))
     (fst (unzip (zip (fst x5) (snd x5))))
     (transpose (snd (unzip (zip (fst x5) (snd x5))))))
    (zip x3 x4))
   (generate ($\lambda$x3. generate ($\lambda$x4. 0)))
   (transpose x2)))
 (map transpose (map (map ($\lambda$x2.
   map transpose (map (map ($\lambda$x3. split 4 (zip x2 x3)))
    (split 32 (let (toMem (
      join (map transpose (map (map (map ($\lambda$x3. x3)))
       (map transpose (split 32 (transpose x1)))))))
      ($\lambda$x3. x3))))))
  (split 32 x0))))))
\end{rise}
\end{minipage}

\noindent
\begin{minipage}{\linewidth}%
\begin{rise}[caption={\Rise{} program satisfying the \goalStyle{lower$_4$} sketch goal.}, label={lst:mm-after-lower4}]
$\Lambda$n0:nat. $\Lambda$n1:nat. $\Lambda$n2:nat. $\lambda$x0. $\lambda$x1.
join (mapPar ($\lambda$x2.
  map join (transpose (map ($\lambda$x3.
   reduceSeq ($\lambda$x4. $\lambda$x5.
    map ($\lambda$x6. 
     reduceSeqUnroll ($\lambda$x7. $\lambda$x8.
      ($\lambda$x9. asScalar (map ($\lambda$x10.
       add (fst x10) (mul (fst (snd x10)) (snd (snd x10))))
       (zip (asVector 32 (fst (unzip x9)))
        (zip (asVector 32 (fst (unzip (snd (unzip x9)))))
         (asVector 32 (snd (unzip (snd (unzip x9)))))))))
      (zip x7 x8))
      (fst (unzip (zip (fst x6) (snd x6))))
      (transpose (snd (unzip (zip (fst x6) (snd x6))))))
     (zip x4 x5))
    (generate ($\lambda$x4. generate ($\lambda$x5. 0)))
    (transpose x3))
   (transpose x2))))
  (map
   (map ($\lambda$x2. map transpose
      (map (map ($\lambda$x3. split 4 (zip x2 x3)))
        (split 32 (let (toMem (join (
          mapPar ($\lambda$x3.
           transpose (map ($\lambda$x4.
            asScalar (map ($\lambda$x5. x5) (asVector 32 x4)))
            (transpose x3)))
           (split 32 (transpose x1)))))
          ($\lambda$x3. x3))))))
   (split 32 x0)))
\end{rise}
\end{minipage}

\noindent
\begin{minipage}{\linewidth}%
\begin{rise}[caption={\Rise{} program after final lowering.}, label={lst:mm-parallel}]
$\Lambda$n0:nat. n1:nat. $\Lambda$n2:nat. $\lambda$x0. $\lambda$x1.
let (toMem (
 join (mapPar ($\lambda$x18.
  transpose (mapSeq ($\lambda$x19.
   asScalar (mapSeq ($\lambda$x20. x20) (asVector 32 x19)))
   (transpose x18)))
  (split 32 (transpose x1)))))
 ($\lambda$x21. join (
  mapPar ($\lambda$x2.
   map join (transpose (mapSeq ($\lambda$x3.
    mapSeq (mapSeq ($\lambda$x4. x4)) (reduceSeq ($\lambda$x5. $\lambda$x6.
     mapSeq ($\lambda$x7. mapSeq ($\lambda$x8. x8) (
      reduceSeqUnroll ($\lambda$x9. $\lambda$x10.
       asScalar (mapSeq ($\lambda$x11.
        add (fst x11) (mul (fst (snd x11)) (snd (snd x11))))
        (zip (asVector 32 (fst (unzip (zip x9 x10))))
         (zip (asVector 32 (fst (unzip (snd (unzip (zip x9 x10))))))
          (asVector 32 (snd (unzip (snd (unzip (zip x9 x10))))))))))
      (mapSeq ($\lambda$x12. x12) (fst (unzip (zip (fst x7) (snd x7)))))
      (transpose (snd (unzip (zip (fst x7) (snd x7)))))))
     (zip x5 x6))
    (mapSeq (mapSeq ($\lambda$x13. x13)) (generate ($\lambda$x14. generate ($\lambda$x15. 0))))
    (transpose x3)))
   (transpose x2))))
  (map ($\lambda$e3839.
   map ($\lambda$x16.
    map transpose (map (map ($\lambda$x17. split 4 (zip x16 x17))) (split 32 x21)))
    e3839)
   (split 32 x0))))
\end{rise}
\end{minipage}

\backmatter  

\printbibliography[heading=bibintoc]

\end{document}